\documentclass[12,oneside,a4paper]{scrreprt}
\usepackage{amsmath}
\usepackage{bbm}
\usepackage{wasysym}
\usepackage{slashed}
\usepackage{graphicx}
\usepackage{feynarts}
\usepackage{feynmp}
\usepackage{mcite}
\newcommand{\whizard}{\texttt{WHIZARD}}
\newcommand{\comphep}{\texttt{CompHEP}}
\newcommand{\madgraph}{\texttt{MadGraph}}
\newcommand{\looptools}{\texttt{LoopTools}}
\newcommand{\circe}{\texttt{CIRCE}}
\newcommand{\eqn}{equation}
\newcommand{\ol}{\overline}
\newcommand{\wt}{\widetilde}

\newcommand{\chapm}{\tilde{\chi}_1^\pm}
\newcommand{\chbpm}{\tilde{\chi}_2^\pm}
\newcommand{\lb}{\left(}
\newcommand{\rb}{\right)}
\newcommand{\be}{\beta}
\newcommand{\al}{\alpha}
\newcommand{\Sp}{\slashed{p}}
\newcommand{\Sk}{\slashed{k}}
\newcommand{\Sq}{\slashed{q}}
\newcommand{\prog}{\texttt}

\newcommand{\M}{\mathcal{M}}
\newcommand{\ME}{\mathcal{M}}
\newcommand{\mL}{\mathcal{L}}
\newcommand{\mO}{\mathcal{O}}
\newcommand{\vareps}{\varepsilon}

\newcommand{\mP}{\mathcal{P}}
\newcommand{\TeV}{{\ensuremath\rm TeV}}
\newcommand{\GeV}{{\ensuremath\rm GeV}}
\newcommand{\ab}{{\ensuremath\rm ab}}
\newcommand{\ora}{\overrightarrow}
\newcommand{\cham}{\widetilde{\chi}_1^-}
\newcommand{\chap}{\widetilde{\chi}_1^+}
\newcommand{\chbp}{\widetilde{\chi}_2^+}
\newcommand{\chp}{\widetilde{\chi}^+}
\newcommand{\chm}{\widetilde{\chi}^-}
\newcommand{\feynarts}{\texttt{FeynArts}}
\newcommand{\formcalc}{\texttt{FormCalc}}
\newcommand{\oMega}{\texttt{O'Mega}}
\begin{document}
%{\pagestyle{empty}
%\bibliographystyle{unsrt}
\bibliographystyle{prsty-mod}
 \begin{titlepage}
\rightline{DESY-THESIS-2006-029}
    \vspace*{2cm}
    \begin{center}
      \Huge \bfseries Event Generation for Next to Leading Order Chargino Production at the International Linear Collider
    \end{center}
    \vspace{2cm}
    \begin{center}\Large \bfseries
      Dissertation\\[2mm]
      zur Erlangung des Doktorgrades\\[2mm]
      des Departments Physik\\[2mm]
      der Universit\"{a}t Hamburg
    \end{center}
    \vspace{2.5cm}
    \begin{center}\large
      vorgelegt von\\[2mm]
      Tania Robens\\[2mm]
      aus Antwerpen
    \end{center}
    \vspace{\fill}
    \begin{center}
      Hamburg\\[2mm]
      2006
    \end{center}
  \end{titlepage}
%\cleardoublestandardpage
\newpage
\thispagestyle{empty}
%\mbox{}
\vspace*{\fill}
  \begin{tabular}{ll}
    Gutachter des Dissertation: & Prof.~Dr.~W.~Kilian\\[1mm]
                                & Prof.~Dr.~J.~Bartels\\[1mm]
    Gutachter der Disputation: &  Prof.~Dr.~W.~Kilian\\[1mm]
                               &  Prof.~Dr.~B.~Kniehl\\[1mm]
    Datum der Disputation:     & 24.~10.~2006\\[1mm]
    Vorsitzender des Pr\"{u}fungsausschusses: & Prof.~Dr.~C.~Hagner\\[1mm]
    Vorsitzender des Promotionsausschusses: & Prof.~Dr.~G.~Huber\\[1mm]
    Dekan der Fakult\"{a}t MIN: & Prof.~Dr.~A.~Fr\"{u}hwald
  \end{tabular}
  \pagebreak
\newpage
\thispagestyle{empty}
%\pagebreak
%\thispagestyle{empty}
%\pagebreak
%\newpage
%\thispagestyle{empty}
  \bigskip
  \begin{center}
    {\large \bfseries\sffamily Abstract}
   
    \vspace{1cm}
    \begin{minipage}{\textwidth}
      At the International Linear Collider (ILC), parameters of supersymmetry (SUSY) can be determined with an experimental accuracy matching the precision of next-to-leading order (NLO) and higher-order theoretical predictions. Therefore,  these contributions need to be included in the analysis of the parameters.\\
We present a Monte-Carlo event generator for simulating chargino
  pair production at the ILC at
  next-to-leading order in the electroweak couplings. We 
consider two approaches of including photon radiation. A
strict fixed-order approach allows for comparison and consistency
checks with published semianalytic results in the literature.  A
version with soft- and hard-collinear resummation of photon radiation, which combines photon resummation with the inclusion of the NLO matrix element for the production process,
avoids
  negative event weights, so the program can simulate physical
  (unweighted) event samples.  Photons are explicitly generated
  throughout the range where they can be experimentally resolved. In addition, it includes further higher-order corrections unaccounted for by the fixed-order method.
  Inspecting the dependence on the cutoffs separating the soft and
  collinear regions, we evaluate the systematic errors due to soft and
  collinear approximations for NLO and higher-order contributions.  In the resummation approach, the residual
  uncertainty can be brought down to the per-mil level, coinciding
  with the expected statistical uncertainty at the ILC. We closely investigate the two-photon phase space for the resummation method. We present results for cross sections and event generation for both approaches. 
    \end{minipage}
  \end{center}
\vspace{0.5cm}
\newpage
\thispagestyle{empty}
\mbox{}
\newpage
\thispagestyle{empty}
\bigskip
  \begin{center}
    {\large \bfseries\sffamily Zusammenfassung}
  
    \vspace{1cm}
    \begin{minipage}{\textwidth}
     Am Internationalen Linearbeschleuniger (International Linear Collider, ILC) k\"onnen die Parameter supersymmetrischer Theorien (SUSY) mit einer experimentellen Genauigkeit gemessen werden, die der Pr\"azision von theoretischen Vorhersagen von n\"achstf\"uhrenden (next-to-leading, NLO) und h\"oheren Ordnungen entspricht. Daher m\"ussen diese Beitr\"age in die Analyse der Paramter eingeschlossen werden.\\
Wir pr\"asentieren einen Monte Carlo Ereignis Generator f\"ur die Simulation von Chargino-Paarproduktion am ILC in NLO in den elektroschwachen Kopplungen. Wir betrachten zwei Ans\"atze f\"ur den Einschluss von Photon-Abstrahlung. Ein strikter Ansatz fester Ordnung (fixed-order) erlaubt den Vergleich und Konsistenztests mit publizierten semianalytischen Ergebnissen in der Literatur. Eine Version mit weicher und harter kollinearer Resummation von Photonabstrahlung , welcher die Resummation von Photonen mit dem Einschluss des NLO Matrix Elements f\"ur den Produktionsprozess kombiniert, vermeidet Ereignisse mit negativem Gewicht, sodass das Programm physikalische (ungewichtetete) Ereignissamples simulieren kann. Photonen werden in dem Bereich, in dem sie experimentell aufgel\"ost werden k\"onnen, expliziert generiert. Ausserdem enth\"alt die Methode weitere Korrekturen h\"oherer Ordnung, die in der fixed-order Methode nicht eingeschlossen sind. Durch die \"Uberpr\"ufung der Abh\"angigkeit von den Cutoffs, die den weichen und den kollinearen Bereich abtrennen, evaluieren wir die systematischen Fehler, die infolge der weichen und kollinearen N\"aherung auftreten, f\"ur NLO Betr\"age sowie Beitr\"age h\"oherer Ordnungen. Im Resummationsansatz kann die restliche Ungenauigkeit auf Promille-Niveau reduziert werden, welches der experimentellen statistischen Ungenauigkeit am ILC entspricht. Wir untersuchen den zwei-Photon Phasenraum der Resummationsmethode genau.  Wir pr\"asentieren Ergebnisse f\"ur Wirkungsquerschnitte und Ereignisgeneration f\"ur beide Ans\"atze. 
    \end{minipage}
  \end{center}
  \newpage
\thispagestyle{empty}
\mbox{}
\setcounter{page}{0}
\tableofcontents
\newpage
\chapter{Introduction}
In this section, we will give a short overview of supersymmetry (SUSY) and its minimal realization, expectations, and results from experiments for SUSY particles, and the available computer tools. We also present an outline of the structure of this thesis.
\section{Standard Model (SM) and minimal supersymmetric extension}
\subsubsection*{The Standard Model}
\begin{table}[b]
  \begin{equation*}
    \begin{array}{l|c|c|cl}
&&SU(3)_{c}&SU(2)\,\times\,U(1)\\
\hline 
\text{leptons}&(e,\,\mu,\,\tau)_{L}&&\checked&\\   
 &(e,\,\mu,\,\tau)_{R}&&\checked&(U(1)\,\text{only})\\
 &(\nu_{e},\,\nu_{\mu},\,\nu_{\tau})_{L}&&\checked&\\
\text{quarks}&(u,d,c,s,t,b)_{L}&\checked&\checked&\\
&(u,d,c,s,t,b)_{R}&\checked&\checked&(U(1)\,\text{only})\\
\text{gluons ($SU(3)$ gauge bosons)}&&\checked&\\
\text{W,\,Z ($SU(2)\,\times\,U(1)$ gauge bosons)}&&&\checked&
    \end{array}
  \end{equation*}
  \caption{Standard model particle content}
  \label{tab:smpart}
\end{table}
The Standard Model (SM) of particle physics \cite{Glashow:1961tr,Weinberg:1967tq,Salam:1968rm, Glashow:1970gm, Fritzsch:1973pi} is a $SU(3)_{c}\,\times\,SU(2)\,\times\,U(1)$ gauge theory with the particle content listed in Table \ref{tab:smpart}. The $SU(2)\,\times\,U(1)$ part of the theory is spontaneously broken by the nonzero vacuum expectation value of the Higgs boson \cite{Higgs:1964ia, Englert:1964et, Higgs:1964pj,Higgs:1966ev,Kibble:1967sv} which gives masses to three of the $SU(2)\,\times\,U(1)$ gauge bosons. It describes current experimental data with high accuracy \cite{Yao:2006px}. However, it suffers from a number of theoretical drawbacks. In general, the Standard Model can only be an effective low-energy theory as it does not describe gravity and is therefore valid at most up to the Planck scale. In addition, it suffers from the fine-tuning or naturalness problem: in the Standard Model, corrections to the mass of the Higgs boson are quadratically divergent. They can be regularized by a finite cutoff-parameter which can maximally be set to the Planck scale as the highest scale of the theory. However, the Higgs mass is theoretically bounded \cite{Lee:1977eg,Lee:1977yc,Cabibbo:1979ay,Hambye:1996wb,Isidori:2001bm}. Therefore, extreme finetuning is needed to obtain the physical Higgs mass from the bare mass in the theory \cite{Witten:1981nf}. Similarly, the Standard Model alone does not provide an explanation for dark matter. From a more aesthetic point of view, it also does not allow for gauge coupling unification of all gauge groups \cite{Georgi:1974sy}. Further problems are the origin of particle masses, possible symmetries between the leptonic and the quark sector, and so on.\\
\subsubsection*{Supersymmetry}
Supersymmetric theories \cite{Haag:1974qh,Wess:1992cp} are one of the most promising candidates for the description of physics beyond the SM. They introduce a new symmetry as an extension of the Poincar\'e group which connects fermionic and bosonic degrees of freedom. This symmetry transforms the bosons/ fermions of the Standard Model into their fermionic/ bosonic superpartner with a different spin but otherwise identical quantum numbers. As the superpartners have not been directly observed in experiments, SUSY has to be broken such that all new particles obtain higher masses. The breaking most likely takes place in a ``hidden sector'' invisible to the Standard Model gauge groups and is then transferred to the visible sector. There are numerous suggestions for SUSY breaking mechanisms, including (minimal) supergravity \cite{Chamseddine:1982jx,Nilles:1983ge,Hall:1983iz}, gauge-mediation \cite{Dine:1993yw,Dine:1994vc, Giudice:1998bp}, and anomaly-mediation \cite{Randall:1998uk,Giudice:1998xp}. An additional symmetry (R-parity) only allows for the simultaneous creation/ annihilation of even numbers of SUSY particles. R-parity violation is strongly limited by experimental results from proton decay. \\
Supersymmetric theories address two main problems of the Standard Model: they allow for a natural solution for the fine-tuning problem. In addition, if R-parity is exactly conserved, the lightest supersymmetric particle (LSP) is a good dark matter candidate. Furthermore, they easily allow for the embedding of the $SU(3)_{c}\,\times\,SU(2)\,\times\,U(1)$ group in a grand unified theory and the unification of gauge couplings.\\
 For more technical details on the construction of supersymmetric theories, cf. Appendix \ref{app:susy}.
\subsection*{Minimal Supersymmetric Standard Model}
The Minimal Supersymmetric Standard Model (MSSM) is the minimal supersymmetric extension of the Standard Model \cite{Dimopoulos:1981zb, Nilles:1983ge,Haber:1985rc, Barbieri:1987xf}. It contains a superpartner for each SM particle and an extended Higgs sector with the two Higgs doublets $H_{u}={H^{+}_{u} \choose H^{0}_{u}}, H_{d}={H^{0}_{d} \choose H^{-}_{d}}$. The Lagrangian at the electroweak scale is given by
\begin{\eqn*}
\mL_{tot}\,=\,\mL_{kin}\,+\,\mL_{gauge}\,+\,\mL_{V}\,+\,\mL_{soft},
\end{\eqn*} 
where
$\mL_{kin}$ contains the kinetic terms of the free theory, 
$\mL_{gauge}$ are terms arising when imposing the gauge symmetry of the SM,
$\mL_{V}$ is the Lagrangian part derived from the superpotential, and
$\mL_{soft}$ are the soft SUSY breaking terms. For more details and the complete MSSM Lagrangian, cf. Appendix \ref{app:susy}. While $\mL_{kin}$ and $\mL_{gauge}$ only depend on SM parameters, new SUSY related parameters appear in the superpotential and the soft breaking terms. The unconstrained model contains in total 105 new parameters. The number of free parameters, however, can be constrained by imposing lepton number conservation, suppression of flavor changing neutral currents, and applying experimental bounds on CP violation. The assumption of a specific breaking mechanism can reduce the number of new parameters even further. In the mSugra scenario of the MSSM, all parameters can be derived from 5 new parameters at the SUSY breaking scale. The parameters at the electroweak scale are then determined using renormalization group equations.\\
For recent reviews, see \cite{Sohnius:1985qm,Drees:1996ca,Martin:1997ns}.
\subsection*{The Chargino and Neutralino Sector of the MSSM}
A solid prediction of the MSSM is the existence of charginos
$\chapm,\chbpm$ and neutralinos $\wt{\chi}^{0}_{1},\,\wt{\chi}^{0}_{2},\,\wt{\chi}^{0}_{3},\,\wt{\chi}^{0}_{4}$. These are the superpartners of the $W^\pm$ and charged/ neutral Higgs bosons
$H^\pm,\,H^{0}$ rotated into their mass eigenstates. In grand unified theories (GUTs) the chargino masses tend to be near the lower
edge of the superpartner spectrum, since the absence of strong
interactions precludes large positive renormalizations of their
effective masses. The precise measurement of chargino parameters (masses, mixing of
$\chapm$ with $\chbpm$, and couplings) is a key for uncovering any of
the fundamental properties of the MSSM.
These values give a handle for verifying supersymmetry in the Higgs and
gauge-boson sector and thus the cancellation of power divergences.
Charginos decay either directly or via short cascades into the LSP, which is the dark-matter candidate of the MSSM in case of R-parity conservation. Thus, a precise knowledge of masses and mixing parameters in the
chargino/neutralino sector is the most important ingredient for
predicting the dark-matter content of the universe. Finally, 
all SUSY parameters in the gaugino and higgsino sector of the MSSM can be determined from the measurements of the chargino production cross sections and masses and the lightest neutralino mass \cite{Allanach:2004ud}.  The
high-scale evolution of the mass parameters should point to a
particular supersymmetry-breaking scenario, if the context of a GUT
model is assumed~(cf.~\cite{Blair:2002pg, Aguilar-Saavedra:2005pw}). In all these cases, a knowledge of
parameters with at least percent-level accuracy is necessary. 
\section{SUSY at colliders: discovery and precision}
In this section, we give a quick overview on SUSY searches at past and future colliders. We refer to \cite{Schmitt:2004gz} and \cite{Weiglein:2004hn,Allanach:2006fy} and references therein for details.
\subsection*{Experimental bounds}
Direct experimental searches for SUSY particles have been conducted at both LEP and the Tevatron. Neither experiment can claim a direct discovery for a SUSY particle, but both have determined lower mass limits \cite{Schmitt:2004gz}. While the combined LEP runs provide the most stringent mass limits for sleptons and gauginos, limits for squarks and gluinos can be obtained from CDF at the Tevatron and ALEPH at LEP. Most lower limits for visible SUSY particles (ie, not the LSP) are $\mO(100\,\GeV)$. However, limits also depend strongly on the underlying SUSY scenarios and breaking mechanism. For a recent review for the mSugra parameter space, cf. \cite{Djouadi:2006be}. Results from HERA give limits on R-parity violating scenarios \cite{Adloff:2001at,Kuze:2002vb,Adloff:2003jm}.\\
Indirect constraints on the SUSY parameter space are in general given by (loop-induced) effects in electroweak precision data. The most stringent restrictions arise from the mass of the lightest Higgs boson. Further observables include the W boson mass, $\sin^{2}\,\theta_{W}$, the flavour changing neutral current $b\,\rightarrow\,s\,\gamma$, and the anomalous magnetic moment of the muon $(g_{\mu}\,-\,2)$ \cite{Ellis:2006ix}.\\
Additional restrictions on the SUSY parameter space can be obtained from cosmological data, as dark matter searches. 
\subsection*{LHC: discovery}
The LHC as a hadron-hadron collider with a cm energy of $\sqrt{s}\,=\,14\,\TeV$ serves as a discovery machine for SUSY particles in most SUSY scenarios. The most dominant channels are pair production or associated production of gluinos and squarks.  As supersymmetric particles usually decay via long decay chains, mass determination involves reconstruction from combined mass distributions. Analyses are parameter-point specific. For the mSugra point SPS1a(')\cite{Allanach:2002nj}(cf. Appendix \ref{app:sps1asl}), analyses for chargino, neutralino, squark, gluino, and slepton masses have been done \cite{Weiglein:2004hn, Nojiri:2004hp}. Relative errors for masses and cross sections are usually expected to lie in the $\%$ regime.
\subsection*{International Linear Collider (ILC): precision}
The International Linear Collider (ILC) is a planned $e^{+}\,e^{-}$ machine with a cm energy of $500\,\GeV\,(1\,\TeV)$. It provides much cleaner production
channels and decay signatures with lower background than the LHC. In addition, direct pair production of sparticles in the slepton and chargino-/ neutralino sector is easily observable. If kinematically accessible, the sparticle content of the MSSM can be studied with high precision \cite{Aguilar-Saavedra:2001rg, Weiglein:2004hn}. Combining sparticle measurements from the ILC and the LHC leads to errors in the percent to per mille regime and significantly reduces the errors based on LHC measurements only \cite{Desch:2003vw,Weiglein:2004hn}. The same accuracy is reached in the respective fitting routines Sfitter \cite{Lafaye:2004cn} and Fittino \cite{Bechtle:2005ns} for the reconstruction of the Lagrangian parameters at the weak scale. \\
Due to its low mass in most SUSY scenarios, the lighter chargino $\chapm$ is likely to be
pair-produced with a sizable cross section at a first-phase ILC with c.m.\ energy of
$500\;\GeV$. In many models, including supergravity-inspired scenarios such as
SPS1a/SPS1a'~\cite{Allanach:2002nj}, the second chargino $\chbpm$ will
also be accessible at the ILC, at least if the c.m.\ energy is
increased to $1\;\TeV$.  Similar arguments hold for the
neutralinos.% Therefore, the masses and cross sections can be obtained with much higher precision \cite{Aguilar-Saavedra:2001rg, Weiglein:2004hn}. Combining results from the ILC and the LHC lead to errors in the $\%$ to $\permil$ regime \cite{Desch:2003vw}. The same accuracy is reached in the respective fitting routines Sfitter \cite{Lafaye:2004cn} and Fittino \cite{Bechtle:2005ns} for the reconstruction of the Lagrangian parameters at the weak scale. 
\section{Next-to-leading order at Monte Carlo Generators}
Several event generators \cite{Kilian:2001qz,Gleisberg:2003xi, Paige:2003mg,Sjostrand:2006za} already include the particle content of the MSSM and allow for event generation including sparticles at tree level.
In Ref.~\cite{Hagiwara:2005wg}, some of these have been presented and
verified against each other, both for the SM and the MSSM. \\
To match the experimental
accuracy for (SUSY) processes at (future) colliders, theoretical predictions with
next-to-leading order (NLO) accuracy have to be implemented in the simulation tools (see e.g. \cite{Allanach:2006fy}).
The inclusion of higher-order corrections in Monte Carlo event generators has already been discussed for electroweak precision physics at LEP (cf. \cite{Kleiss:1989de, Was:1992mm,lepewwg}). For the determination of the W-mass, Monte-Carlo event generators including higher-order corrections \cite{Denner:2000bj,Jadach:2001mp} can reduce the theoretical uncertainties down to approximately $0.5\%$ \cite{Denner:1999dt,Denner:2000tw, Yao:2006px}.\\
NLO calculations for the production of charginos and neutralinos at an $e^{+}e^{-}$ collider are within the percent regime \cite{Fritzsche:2004nf, Oller:2004br, Oller:2005xg}. If SUSY is discovered, the experimental precision at the ILC requires to equally include these processes at next-to-leading order in a Monte Carlo event generator. 
 Furthermore, it is essential for the simulation of physical (i.e., unweighted)
event samples that the effective matrix elements are
positive semidefinite over the whole accessible phase space.  The QED
part of radiative corrections does not meet this requirement in some
phase space regions.  This usually requires higher-order resummation for soft (and collinear) photonic contributions to the process. Methods for dealing with this problem have been
developed in the LEP1 era~\cite{Nicrosini:1986sm, Bonvicini:1988vv, Kleiss:1989de}.  
Since the ILC precision actually exceeds the one achieved in LEP
experiments, these higher-order effects from resummation can have the same order of magnitude as the experimental errors.\\
Similar techniques concern the event simulation of hadronic processes including partonic showers. The higher-order parton showers have to be matched with the exact NLO matrix element. Work along these lines has been done for $e^{+}e^{-}$ \cite{Nagy:2005aa, Kramer:2005hw} and hadron colliders \cite{Catani:2001cc,Frixione:2002ik, Frixione:2006he}. 
\section{Outline}
In this work, which is the extension of a recent publication \cite{Kilian:2006cj}, we present an extension of the Monte Carlo Event generator \whizard~ \cite{Kilian:2001qz} which includes the $\mO(\al)$ electroweak and SUSY corrections to chargino production at an $e^{+}\,e^{-}$ collider. The fixed-order version, which we will discuss first, is limited to first order photon emission and suffers the well-known problem of negative weights for low cuts on the photon energy. We then introduce a method which combines the photon resummation already discussed at LEP with NLO matrix elements for the production process. It avoids negative events and therefore allows for lower energy cuts. In addition, it includes further higher-order corrections unaccounted for by the fixed-order method. We evaluate the systematic errors due to soft and collinear photon approximations and the cut-dependence of higher-order contributions. We present results for the cross section and event simulation for both methods. \\
\\
In Chapter \ref{chap:tree}, we discuss chargino production at an $e^{+}\,e^{-}$ collider at tree level, presenting the total and differential cross sections for any helicity combination. In Chapter \ref{chap:NLO}, we present the analytic form of the infrared and cut-independent cross section for chargino production at next-to-leading order and the inclusion of resummed soft and virtual photonic contributions. We sketch the treatment of $2\,\rightarrow\,n$ processes including chargino decay in the double pole approximation.\\
In Chapter \ref{chap:fixed}, we present a method to include the fixed $\mO(\al)$ result in the Monte Carlo event generator \whizard. We give a brief overview on the technicalities of the inclusion and discuss the energy-cut-dependent problem of negative weights in certain points of phase space. We present first results for cutoff-independence, cross section integration and event generation in the allowed cut parameter regions.\\
In Chapter \ref{chap:resum}, we present our method of combining the completely resummed soft and virtual photonic contributions with the exact next-to-leading order contributions to chargino production. We explicitly discuss the description of collinear photonic contributions up to $\mO(\al^{2})$ and analytic differences between the resummation and the fixed order method. We present the results of our work in Chapter \ref{chap:resumres}. In Chapter \ref{chap:sumup}, we summarize and give an outlook on future work.\\
\\
The appendices basically contain conventions, derivations of SUSY Feynman rules and the approximations used for the photonic contributions. We start with a general discussion of supersymmetric theories and the MSSM in Appendix \ref{app:susy}. The chargino- and neutralino-sector, especially the derivation of Feynman rules, is discussed in Appendix \ref{app:chneutr}. Appendix \ref{app:helmass} gives an overview on the treatment of helicity states for massive particles needed for the discussion of the helicity-dependent Born cross section.\\
Appendix \ref{app:diagramms} shows all generic diagrams contributing to chargino production at NLO. Appendix \ref{app:photons} sketches the derivation of the collinear photon approximation and the initial state radiation (ISR) structure function \cite{Skrzypek:1990qs}. We present the mSugra point SPS1a' in Appendix \ref{app:sps1asl} and list some useful functions in Appendix \ref{app:trafos}. Appendix \ref{app:refs} lists the references for all computer programs mentioned in this work. 
\chapter{Chargino production at tree level}\label{chap:tree}
In the MSSM, charginos $\wt{\chi}^{\pm}$ are superpositions of superpartners of the charged gauge bosons $W^{\pm}$ and the charged Higgs fields $H^{\pm}$. In the following, we use the mass-diagonalization matrices 
\begin{\eqn*}
U_{L,R}\,{\wt{W}^{-} \choose \wt{H}^{-}}\,=\,{\wt{\chi}^{-}_{1} \choose \wt{\chi}^{-}_{2}}_{L,R}
\end{\eqn*}
and the parameterization
\begin{\eqn*}
U_{L,R}\,=\,\lb\begin{array}{cc} \cos\Phi_{L,R}&\sin\Phi_{L,R}\\
-\sin\Phi_{L,R}&\cos\Phi_{L,R}\end{array}\rb
\end{\eqn*}
for the case with no CP violation. For more details, cf. Appendix \ref{app:chneutr}.
\section{Born matrix element}
\begin{figure}[bt]
%  \vspace{5mm}
  \begin{center}
    \includegraphics[width=60mm]{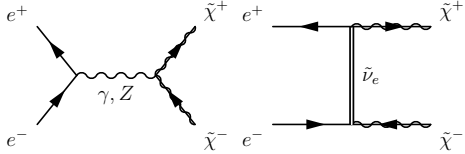}
  \end{center}
  \caption{Feynman graphs for chargino pair production at the ILC.}
  \label{fig:LO-graphs}
\end{figure}

At an $e^{+}\,e^{-}$ collider, charginos are produced by $\gamma$ and $Z$ exchange in the $s$-channel and $\tilde{\nu}_{e}$ exchange in the $t$-channel as shown in Figure \ref{fig:LO-graphs}.
Neglecting the electron mass $m_{e}$, we can perform a Fierz transformation on the t-channel $\tilde{\nu}_{e}$ exchange. The sum of all contributions is then given by \cite{Choi:2000ta}:
\begin{\eqn}\label{eq:mborn}
\M=\frac{e^{2}}{s}\,Q_{\al\be}\left[\bar{v}(e^{+})\gamma_{\mu}P_{\al}u(e^{-})\right]\left[\bar{u}(\wt{\chi}^{-}_{j})\gamma^{\mu}P_{\be}v(\wt{\chi}^{+}_{i})\right],
\end{\eqn}
with $\al, \be\,=\,L,R$ and $P_{L/R}\,=\,\frac{1}{2}\,(1\,\mp\,\gamma_{5})$. The factors $Q_{\al\be}$ have been explicitly calculated in \cite{Choi:2000ta} and are given by
\begin{eqnarray*}
Q_{LL}\,=\,D_{L}\,\mp\,F_{L}\,\cos\,2\,\Phi_{L}&,&Q_{RL}\,=\,D_{R}\,\mp\,F_{R}\,\cos\,2\,\Phi_{L},\\
Q_{LR}\,=\,D'_{L}\,\mp\,F'_{L}\,\cos\,2\,\Phi_{R}&,&Q_{RR}\,=\,D_{R}\,\mp\,F_{R}\,\cos\,2\,\Phi_{R},\\
\end{eqnarray*}
for the production of $\wt{\chi}^{+}_{1}\,\wt{\chi}^{-}_{1}$ and $\wt{\chi}^{+}_{2}\,\wt{\chi}^{-}_{2}$ and
\begin{eqnarray*}
Q_{LL}\,=\,F_{L}\,\sin\,2\,\Phi_{L}&,&Q_{RL}\,=\,F_{R}\,\sin\,2\,\Phi_{L},\\
Q_{LR}\,=\,F'_{L}\,\sin\,2\,\Phi_{R}&,&Q_{RR}\,=\,F_{R}\,\sin\,2\,\Phi_{R}\\
\end{eqnarray*}
for the production of $\wt{\chi}^{+}_{1}\,\wt{\chi}^{-}_{2}$ and $\wt{\chi}^{+}_{2}\,\wt{\chi}^{-}_{1}$ pairs. We here use the notations \cite{Choi:2000ta}
\begin{eqnarray}\label{eq:fds}
&&D_{L}\,=\,1+\frac{D_{Z}}{s^{2}_{W}\,c^{2}_{W}}\,\lb s^{2}_{W}-\frac{1}{2}\rb\,\lb s^{2}_{W}-\frac{3}{4}\rb\;,\;F_{L}\,=\,\frac{D_{Z}}{4\,s^{2}_{W}\,c^{2}_{W}}\,\lb s^{2}_{W}-\frac{1}{2}\rb, \nonumber\\
&&D_{R}\,=\,1+\frac{D_{Z}}{c^{2}_{W}}\,\lb s^{2}_{W}-\frac{3}{4}\rb\;,\;F_{R}\,=\,\frac{D_{Z}}{4\,c^{2}_{W}}\;,\;
D'_{L}\,=\,D_{L}\,+\,\frac{D_{\tilde{\nu}}}{4\,s^{2}_{W}},\nonumber\\
&&\;F'_{L}\,=\,F_{L}-\frac{D_{\tilde{\nu}}}{4\,s^{2}_{W}}\;,\;D_{Z}\,=\,\frac{s}{s-m_{Z}^{2}+\imath\,\Gamma_{Z}}\;,\;D_{\tilde{\nu}}\,=\,\frac{s}{t-m^{2}_{\tilde{\nu}}},
\end{eqnarray}
where $s_{W}^{2}$ and $c_{W}^{2}$ denote the squared sine and cosine of the weak mixing angle.
\section{Helicity amplitudes}
The matrix element $\M$ given by Eq. (\ref{eq:mborn}) can be decomposed into helicity amplitudes. We use the formalism introduced in \cite{Hagiwara:1985yu} (see Appendix \ref{app:helmass} for a short review on helicity eigenstates and an introduction to the used method). We define the $\hat{e}_{z}$ direction as the flight direction of the electron and obtain
\begin{eqnarray}\label{eq:pinandout}
p_{1}(e^{+})=\lb\begin{array}{c} p_{e}\\0\\0\\-p_{e}\end{array}\rb&,&p_{2}(e^{-})=\lb\begin{array}{c}p_{e}\\0\\0\\p_{e}\end{array}\rb,\nonumber\\
p_{3}(\wt{\chi}^{-})=\lb\begin{array}{c} E_{\chi}\\p_{\chi} \sin\theta \\0\\p_{\chi} \cos \theta\end{array}\rb&,&p_{4}(\wt{\chi}^{+})=\lb\begin{array}{c} E_{\chi}\\-p_{\chi} \sin\theta \\0\\-p_{\chi} \cos \theta \end{array}\rb,
\end{eqnarray}
where $p_{e}$ and $p_{\chi}$ are the magnitude of the electron/ positron and chargino three-momentum, respectively. $\theta$ is the angle between the electron and the negatively charged chargino.\\
$\M(\sigma;\lambda_{i}\,\lambda_{j})\,=\,2\,\pi\,\al\, \langle \sigma;\lambda_{i}\,\lambda_{j}\rangle$ defines the matrix element for an electron with helicity $\sigma$ and $\wt{\chi}^{\mp}$ with helicity $\lambda_{i/j}$. The helicity amplitudes are then given by \cite{Choi:1998ei}:
\begin{eqnarray*}
\langle +;++ \rangle &=&-\left[ Q_{RR}\,\sqrt{1-\eta^{2}_{+}}+Q_{RL}\,\sqrt{1-\eta^{2}_{-}}\right]\sin\theta,\\
\langle +;+-\rangle&=&-\left[ Q_{RR}\,\sqrt{(1+\eta_{+})\,(1+\eta_{-})}+Q_{RL}\,\sqrt{(1-\eta_{+})\,(1-\eta_{-})}\,\right ]\,\lb 1+\cos\theta \rb,\\
\langle +;-+\rangle&=&\left [Q_{RR}\,\sqrt{(1-\eta_{+})\,(1-\eta_{-})}+Q_{RL}\,\sqrt{(1+\eta_{+})\,(1+\eta_{-})}\,\right]\,\lb 1-\cos\theta \rb,\\
\langle +;--\rangle&=&\left[ Q_{RR}\,\sqrt{1-\eta^{2}_{-}}+Q_{RL}\,\sqrt{1-\eta^{2}_{+}}\right] \sin\theta,\\
\end{eqnarray*}
\begin{eqnarray*}
\langle -;++ \rangle &=&-\left[ Q_{LR}\,\sqrt{1-\eta^{2}_{+}}+Q_{LL}\,\sqrt{1-\eta^{2}_{-}}\right]\sin\theta,\\
\langle -;+-\rangle&=&\left[ Q_{LR}\,\sqrt{(1+\eta_{+})\,(1+\eta_{-})}+Q_{LL}\,\sqrt{(1-\eta_{+})\,(1-\eta_{-})}\,\right] \,\lb 1-\cos\theta \rb,\\
\langle -;-+\rangle&=&-\left[ Q_{LR}\,\sqrt{(1-\eta_{+})\,(1-\eta_{-})}+Q_{LL}\,\sqrt{(1+\eta_{+})\,(1+\eta_{-})}\,\right]\,\lb 1+\cos\theta \rb,\\
\langle -;--\rangle&=&\left[ Q_{LR}\,\sqrt{1-\eta^{2}_{-}}+Q_{LL}\,\sqrt{1-\eta^{2}_{+}}\right] \sin\theta,\\
\end{eqnarray*}
with
\begin{\eqn*}
\eta_{\pm}\,=\,\sqrt{\lambda}\,\pm(\mu_{i}^{2}-\mu_{j}^{2})\;;\;\lambda\,=\,\left[1-(\mu_{i}+\mu_{j})^{2}\right]\,\left[1-(\mu_{i}-\mu_{j})^{2}\right]\;;\;\mu_{i}\,=\,\frac{m^{2}_{\chi_{i}}}{s}.
\end{\eqn*}
\section{Differential and total cross sections}
\begin{figure}
  \begin{center}
    \includegraphics[width=100mm]{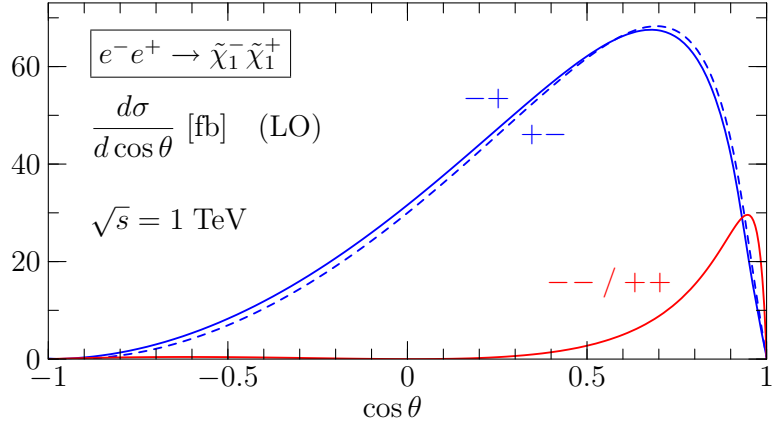}
  \end{center}
  \caption{Chargino pair production at the ILC: Dependence of the
    differential distribution in polar angle~$\cos\theta$ between
    $e^{-}$ and $\cham$ for different helicity
    combinations. The labels indicate $\cham$ and $\chap$ helicity;
    the electron/positron helicity is fixed to $-+$.}
  \label{fig:heldep}
\end{figure}

\begin{figure}
  \begin{center}
    \includegraphics[width=100mm]{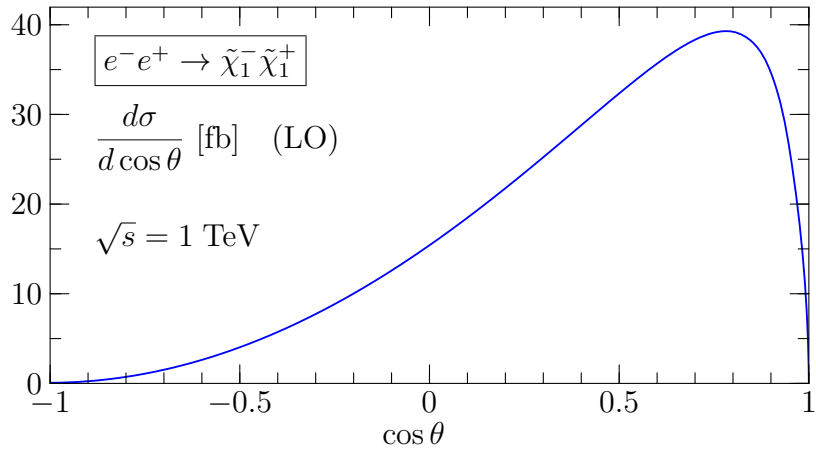}
  \end{center}
\vspace{5mm}
  \caption{Chargino pair production at the ILC: Dependence of the
    spin summed and averaged differential distribution in polar angle~$\cos\theta$ between $e^{-}$ and $\cham$.  }
  \label{fig:allhel}
\end{figure}

\begin{figure}
  \begin{center}
\includegraphics[width=100mm]{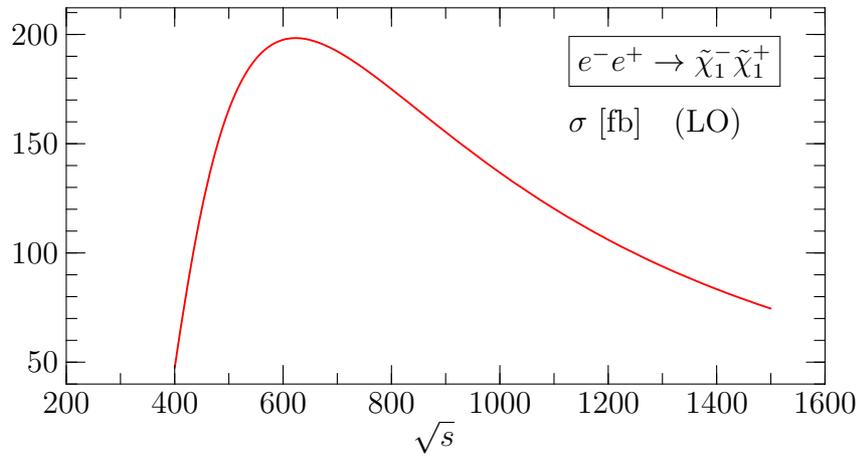}
\vspace{5mm}
  \end{center}
  \caption{Chargino pair production at the ILC: unpolarized total cross section as a function of $\sqrt{s}$ (averaged over initial and summed over final spins)  }
  \label{fig:tots}
\end{figure}

For the calculation of the differential cross section, we use the general formula for $2\,\rightarrow\,2$ particle processes: 
\begin{equation*}
d\sigma = \frac{1}{2w} \frac{d^{3}k_{1}}{ 2k^{0}_{1}}\, \frac{1}{(2\pi)^{2}}\,\delta(k^{2}_{2}-m'^{2}_{2})\,\Theta(k^{0}_{2}) \left|\cal M\right|^{2},
\end{equation*}
where $k_{i},\,k^{0}_{i},\,m'_{i}$ are the momenta/ energy/ mass of the outgoing particles. The flux factor
\begin{\eqn*}
w\,(s,m^{2}_{1},m^{2}_{2})\,=\,\sqrt{(s-m_{1}^{2}-m_{2}^{2})^{2}\,-4\,m_{1}^{2}\,m_{2}^{2}}
\end{\eqn*}
depends on the center of mass energy $\sqrt{s}$ and the masses of the incoming particles. For particles with very small masses with respect to the cm energy, $w\,=\,s$.\\
The differential cross section for a specific helicity state is then given by
\begin{\eqn*}
\frac{d\sigma}{d\cos\theta}\,=\,\frac{\pi\,\al^{2}}{8\,s}\,\sqrt{\lambda}\,|\langle \sigma; \lambda_{i}\,\lambda_{j}\rangle|^{2}.
\end{\eqn*}
For chargino pair production, the differential cross section is
\begin{\eqn}\label{eq:diffcs}
\frac{d\sigma}{d\cos\theta}\,=\,\frac{\pi\,\al^{2}}{2\,s}\,\sqrt{\lambda}\,\left\{\left[1-(\mu^{2}_{i}-\mu^{2}_{j})^{2}+\lambda\,\cos^{2}\theta\right]Q_{1}+4\mu_{i}\mu{_j}Q_{2}+2\,\sqrt{\lambda}Q_{3}\,\cos\theta\right\}
\end{\eqn}
with
\begin{eqnarray*}
Q_{1}&=&\frac{1}{4}\,\left[ |Q_{RR}|^{2}\,+\,|Q_{LL}|^{2}\,+\,|Q_{RL}|^{2}\,+\,|Q_{LR}|^{2}\right],\\
Q_{2}&=&\frac{1}{2}\,Re\,\left[ Q_{RR}\,Q^{*}_{RL}\,+\,Q_{LL}\,Q^{*}_{LR}\right],\\
Q_{3}&=&\frac{1}{4}\,\left[ |Q_{RR}|^{2}\,+\,|Q_{LL}|^{2}\,-\,|Q_{RL}|^{2}\,-\,|Q_{LR}|^{2}\right],\\
\end{eqnarray*}
where we averaged over initial and summed over final particle spins.
Figure \ref{fig:heldep} shows the results for the differential cross section for the dominant helicity amplitudes for $\wt{\chi}^{+}_{1}\,\wt{\chi}^{-}_{1}$ production for the point SPS1a' (cf. Appendix \ref{app:sps1asl}). Independently of the SUSY parameters, processes with a left handed electron in the initial state are dominant, as diagrams with a $\gamma$ and $Z$ exchange interfere constructively. The same diagrams give destructive interference between the $\gamma,\,Z$ exchange for righthanded initial electrons. This can easily be seen from Eqs. (\ref{eq:fds}). We obtain
\begin{\eqn*}
D_{L}\,\approx\,1+0.727\,D_{Z},\,D_{R}\,\approx\,1-0.67\,D_{Z},\,F_{L}\,\approx\,-0.01\,D_{Z},\,F_{R}\,\approx\,0.33\,D_{Z}.
\end{\eqn*}
For the cm values where chargino pair production is possible, $D_{Z}\,\geq\,1$. Then, even without a possible enhancement from the $t$-channel sneutrino exchange and independent of the mixing parameters, $|Q_{RL}/Q_{LL}|\,\leq\,0.33$. The ratio of $|Q_{RR}/Q_{LR}|$ strongly depends on the mixing parameters and the sneutrino mass. For the point SPS1a' and a cm energy of 1 $\TeV$ (400 $\GeV$), $|Q_{RL}/Q_{LL}|=3\% (1\%)$ and $|Q_{RL}/Q_{LL}|\,=\,1\,\permil (1\%)$ (neglecting contributions from the sneutrino exchange).
  We therefore only show the results for states with a left-handed electron in the initial state. Amplitudes with two charginos of the same helicity in the final state are suppressed by a factor $\propto\,\frac{m^{2}_{\chi}}{s}$.
Figure \ref{fig:allhel} shows the helicity averaged/ summed differential cross section according to Eq. (\ref{eq:diffcs}), Figure \ref{fig:tots} the $\sqrt{s}$ dependence of the corresponding total cross section.
%Similar results have been reproduced for neutralino production; they agree with the results in \cite{Choi:2001ww}. 
\\
For the same parameter point, we have compared the tree level results for $\wt{\chi}^{-}_{i}\,\wt{\chi}^{+}_{j}$ production using the analytic form \cite{Choi:2000ta} and several computer codes \cite{Kilian:2001qz, Tilman:progr,FormCalc, Hahn:1998yk}. The analytic and numeric results are in complete agreement. The values are given in Table \ref{tab:charginocs}. 
\begin{table}[tb]
  \begin{equation*}
    \begin{array}{c|c}
      \hline
      & \sigma_{tot} [fb]\\
      \hline
      \wt{\chi}^{-}_{1}\wt{\chi}^{+}_{1} & 136.8\\
     \wt{\chi}^{-}_{1}\wt{\chi}^{+}_{2} & 10.97\\
\wt{\chi}^{-}_{2}\wt{\chi}^{+}_{2} & 80.07\\
      \hline
    \end{array}
  \end{equation*}
  \caption{Chargino total production cross sections for the SUSY parameter set SPS1a' and $\sqrt{s}=1$ TeV.}
  \label{tab:charginocs}
\end{table}
\section{Polarized incoming and outgoing particles}
In \cite{Choi:2000ta, Choi:1998ei}, the authors also list the analytic results for the $\wt{\chi}^{-}$ polarization vector as well as the $e^{\pm}$-polarization dependent differential cross sections. Although not pursued further in this work, we reproduced these results and sketch the derivation for completeness.\\
\subsubsection*{Polarized outgoing particles}
The polarization vector of the outgoing $\wt{\chi}^{-}$ in its rest-frame is defined in the following coordinate system:  $z$ is given by the component parallel to the flight-direction of the chargino in the lab-frame, $x$ is defined in the production plane, and $y$ normal to the production plane. The $\wt{\chi}^{-}$ polarization vector is then given by
\begin{\eqn*}
\overrightarrow{\mathcal{P}}=tr(\overrightarrow{\sigma} \rho)=(\mathcal{P_{T},P_{N},P_{L}}),
\end{\eqn*}
where $\overrightarrow{\sigma}$ is the vector made of the Pauli-matrices $\sigma_{i}$, $\rho$ the spin-density matrix, and
\begin{eqnarray*}
\mP_{\mathcal{L}}&=&\frac{1}{4}\,\sum_{\sigma=\pm}\,\left\{|\langle \sigma;\,++\rangle|^{2}\,+\,|\langle \sigma;\,+-\rangle|^{2}\,-\,|\langle \sigma;\,-+\rangle|^{2}\,-\,|\langle \sigma;\,--\rangle|^{2}\right\}/\mathcal{N},\\
\mP_{\mathcal{T}}&=&\frac{1}{2}\,Re\left\{\,\sum_{\sigma=\pm}\,\langle \sigma;\,++\rangle\,\langle \sigma;\,-+\rangle^{*}\,+\,\langle \sigma;\,--\rangle\,\langle \sigma;\,+-\rangle^{*}\right\}/\mathcal{N},\\
\mP_{\mathcal{N}}&=&\frac{1}{2}\,Im\left\{\,\sum_{\sigma=\pm}\,\langle \sigma;\,--\rangle\,\langle \sigma;\,+-\rangle^{*}\,-\,\langle \sigma;\,++\rangle\,\langle \sigma;\,-+\rangle^{*}\right\}/\mathcal{N},\\
\mathcal{N}&=&\frac{1}{4}\,\sum_{\lambda_{i}\,\lambda_{j}}\,\left[|\langle +;\,\lambda_{i}\lambda_{j}\rangle|^{2}\,+\,|\langle -;\,\lambda_{i}\lambda_{j}\rangle|^{2}\right].
\end{eqnarray*}
\subsubsection*{Polarized incoming particles}
For the calculation of the differential and total cross sections depending on the polarization of the incoming particle beams, we start with
\begin{\eqn*}
\mathcal{P}=(\mathcal{P_{T}},0,\mathcal{P_{L}})\;,\mP'=(\mathcal{P'_{T}}\cos\eta,\mathcal{P'_{T}}\sin\eta,\mathcal{-P'_{L}}) 
\end{\eqn*}
for the polarization vectors of the electron and positron, respectively, in a coordinate system where the z-axis is given by the momentum of the electron and the x-axis by the electron's transverse polarization vector. $\eta$ is the azimuthal angle of the positron transverse polarization vector with respect to the x-axis. We transform this into the lab-system, where the x-axis is defined by the scattering plane (cf. Eqs (\ref{eq:pinandout})):
\begin{\eqn*}
\mathcal{P}=(\mathcal{P_{T}}\cos\Phi,-\mathcal{P_{T}}\sin\Phi,\mathcal{P_{L}})\;;\;\mathcal{P}'=(\mathcal{P'_{T}}\cos(\eta-\Phi),\mathcal{P'_{T}}\sin(\eta-\Phi),-\mathcal{P'_{L}}); 
\end{\eqn*}
$\Phi$ is the angle between the transverse polarization vector of the electron and the new x-axis.\\
The squared matrix element including an incoming particle with
possible helicity eigenstates labeled by $\lambda$ is given by \cite{Ladinsky:1992bd}
\begin{\eqn*}
\M^{2}\,=\,\sum_{\lambda,\lambda'}\M(\lambda)\rho_{\lambda\lambda'}\M^{\dagger}(\lambda')
\end{\eqn*}
for particles and 
\begin{\eqn*}
\M^{2}\,=\,\sum_{\lambda,\lambda'}\M(\lambda)\rho_{\lambda'\lambda}\M^{\dagger}(\lambda')
\end{\eqn*}
for antiparticles.
The spin density matrix is given by $\rho\,=\,\frac{1}{2}(1+\overrightarrow{\sigma}\overrightarrow{\mathcal{P}})$, and $\mathcal{P}$ is defined such that $\overrightarrow{p}\,=\,|p|\hat{e}_{z}$. Taking this into account, the electron spin density matrix is 
\begin{\eqn*}
\rho_{\sigma_{1}\sigma_{2}}\,=\,\frac{1}{2}\lb\begin{array}{cc}1+\mathcal{P_{L}} & \mathcal{P_{T}}e^{\imath\Phi}\\\mathcal{P_{T}}e^{-\imath\Phi}&1-\mathcal{P_{L}}\end{array}\rb.
\end{\eqn*}
For the positron, we still have to perform a rotation around the x-axis such that $\overrightarrow{p_{e^{+}}}\,=\,|p_{e^{+}}|\hat{e}_{z,e^{+}}$ leading to
\begin{\eqn*}
P'=(\mathcal{P'_{T}}\cos(\eta-\Phi),-\mathcal{P'_{T}}\sin(\eta-\Phi),\mathcal{P'_{L}})
\end{\eqn*}
which then gives
\begin{\eqn*}
\rho_{\sigma'_{1}\sigma'_{2}}\,=\,\frac{1}{2}\lb\begin{array}{cc}1+\mathcal{P'_{L}} & \mathcal{P'_{T}}e^{\imath(\eta-\Phi)}\\\mathcal{P'_{T}}e^{-\imath(\eta-\Phi)}&1-\mathcal{P'_{L}}\end{array}\rb.
\end{\eqn*}
If we now calculate the sum over the squared helicity amplitudes using
\begin{\eqn*}
\M^{\dagger}\M\,=\sum_{\sigma_{1},\sigma_{2},\sigma'_{1}\sigma'_{2};\lambda_{i}}\M^{\dagger}(\sigma_{1}\sigma'_{1};\lambda_{i})\rho_{\sigma_{1}\sigma_{2}}\rho_{\sigma'_{1}\sigma'_{2}}\M(\sigma_{2}\sigma'_{2};\lambda_{i})
\end{\eqn*}
and 
\begin{\eqn*}
\M(\sigma_{i}\sigma'_{i};\lambda_{j})\,=\,\M(\sigma_{i};\lambda_{j})\delta_{\sigma_{i}\,-\sigma'_{i}},
\end{\eqn*}
we obtain
\begin{\eqn*}
\frac{d\sigma}{d\Omega}\,=\,\frac{\al^{2}}{16\,s}\,\sqrt{\lambda}\,\left[(1-\mP_{\mathcal{L}}\mP'_{\mathcal{L}})\Sigma_{unpol}\,+\,(\mP_{\mathcal{L}}-\mP'_{\mathcal{L}})\,\Sigma_{LL}\,+\,\mP_{\mathcal{T}}\,\mP'_{\mathcal{T}}\cos\lb 2\,\Phi-\eta\rb\,\Sigma_{TT}\right]
\end{\eqn*}
with
\begin{eqnarray*}
\Sigma_{unpol}&=&4\left\{\left[1-(\mu^{2}_{i}-\mu^{2}_{j})^{2}+\lambda\,\cos^{2}\theta\right]\,Q_{1}\,+\,4\,\mu_{i}\,\mu_{j}\,Q_{2}+2\,\sqrt{\lambda}\,Q_{3}\,\cos\theta\right\},\\
\Sigma_{LL}&=&4\left\{\left[1-(\mu^{2}_{i}-\mu^{2}_{j})^{2}+\lambda\,\cos^{2}\theta\right]\,Q'_{1}\,+\,4\,\mu_{i}\,\mu_{j}\,Q'_{2}+2\,\sqrt{\lambda}\,Q'_{3}\,\cos\theta\right\},\\
\Sigma_{TT}&=&-\,4\,\lambda\sin^{2}\theta\,Q_{5},
\end{eqnarray*}
\begin{eqnarray*}
Q'_{1}&=&\frac{1}{4}\,\left[ |Q_{RR}|^{2}\,-\,|Q_{LL}|^{2}\,+\,|Q_{RL}|^{2}\,-\,|Q_{LR}|^{2}\right],\\
Q'_{2}&=&\frac{1}{2}\,Re\,\left[ Q_{RR}\,Q^{*}_{RL}\,-\,Q_{LL}\,Q^{*}_{LR}\right],\\
Q'_{3}&=&\frac{1}{4}\,\left[ |Q_{RR}|^{2}\,-\,|Q_{LL}|^{2}\,-\,|Q_{RL}|^{2}\,+\,|Q_{LR}|^{2}\right],\\
Q_{5}&=&\frac{1}{2}\,Re\left[Q_{LR}\,Q_{RR}^{*}+Q_{LL}\,Q^{*}_{LR}\right]
\end{eqnarray*}
and $\mu_{i,j}$ and $\lambda$ as introduced in the last section. In \cite{Choi:2000ta, Choi:1998ei}, the authors show that the measurement of the lightest chargino mass, the production cross sections, and spin-spin correlations suffice to completely determine the parameters of the chargino system in the tree level approximation.
\chapter{Chargino production at next-to-leading order (NLO)}\label{chap:NLO}
\section{Divergencies and infrared-safe cross sections in higher-order calculations}
Before discussing the $\mO(\al)$ corrections to chargino production at an $e^{+}\,e^{-}$ collider, we list some general features of the calculation of (virtual) higher-order corrections in perturbation theory.\\
In finite order calculations, divergences in the ultraviolet ($E\,\rightarrow\,\infty$) as well as infrared ($E\,\rightarrow\,0$) regime can appear. The UV divergencies have to be regularized, which introduces a regularization parameter $\Lambda$. This leads to a parameter-dependence of the physical quantities with respect to the bare parameters in the Lagrangian. In bare perturbation theory, the bare parameters are then eliminated in relations between physically measurable quantities. If the underlying theory is meaningful, these relations should be cutoff-independent. Alternatively, the theory can be renormalized by absorbing the UV divergencies in a redefinition of the parameters and fields of the theory, thus giving a physical meaning to the parameters in the Lagrangian. The bare Lagrangian is split into the renormalized Lagrangian and a counterterm part,
\begin{\eqn*}
\mL_{bare}\,=\mL_{ren}\,+\mL_{ct},
\end{\eqn*}
where the Feynman rules resulting from $\mL_{ct}$ have to be included in all calculations of physical observables.\\
IR divergences arise in the case of zero-mass virtual gauge boson exchange. When virtual massive gauge bosons with the mass $m_{g}$ are connected to an on-shell particle with the mass $m$, logarithms of the form
\begin{\eqn*}
\log\lb\frac{m_{g}}{m}\rb
\end{\eqn*}
appear. This term gets infinite as $m_{g}\,\rightarrow\,0$.\\
The IR divergence in QED can be regularized by introducing an infinitesimal gauge boson mass $\lambda$. According to the Bloch-Nordsieck theorem \cite{Bloch:1937pw}, the divergencies cancel if the emission of soft real photons is also taken into account. Infrared-safe observables therefore always include terms describing the emission of soft real photons. In the collinear approximation, where the transverse momentum of the emitted photon is neglected, the dominant contributions originating from the multiple emission of soft and virtual photons from the initial particles can be summed up into structure functions. This is discussed in Section \ref{sec:strfunisr}.\\
\\
In addition, collinear divergencies can appear when the (massless) gauge boson is emitted at a small angle relative to the emitting particle. In the collinear approximation the integration over the photon phase space leads to terms of the form
\begin{\eqn*}
\log\frac{k_{\perp, max}}{m}
\end{\eqn*}
which diverge when $m\,=\,0$.
These singularities are regularized by keeping the physical nonzero mass $m$ in this region of phase space, cf. Section \ref{sec:hardcoll}.\\
\\
An infrared-safe total cross section with a cm energy $\sqrt{s}$ and $n$ particles in the final state then includes the following contributions
\begin{itemize}
\item{}Born cross section:
\begin{\eqn}\label{eq:sborn}
\sigma_{Born}(s)\,=\,\int\,d\Gamma_{n}\,|\M_{Born}|^{2},
\end{\eqn}
\item{}interference term between Born and first order terms describing the purely virtual contribution
\begin{equation}\label{eq:svirt}
  \sigma_\text{virt}(s,\lambda^2) = \int d\Gamma_n\left[
  2\mathrm{Re}\left(\ME_\text{Born}(s)^*\,
                    \ME_\text{1-loop}(s,\lambda^2)\right)\right]
\end{equation}
\item{}soft photon contribution
\begin{equation}\label{eq:ssoft}
  \sigma_\text{s}(s,\Delta E_\gamma,\,\lambda) = \int d\Gamma_n\left[
  f_\text{soft}(\Delta E_\gamma,\lambda)\,|\ME_\text{Born}(s)|^2 \right]
\end{equation}
(The soft photonic approximation is explained in Section \ref{sec:softf}).
\end{itemize}
Here, $\lambda$ denotes the photon mass and $\Delta\,E_{\gamma}$ is the soft photon energy cutoff separating the hard from the soft region. $f_{soft}$  (\ref{eq:f-soft}) denotes the soft photon factor.\\ 
\\
The sum of the three contributions (\ref{eq:sborn}), (\ref{eq:svirt}), (\ref{eq:ssoft})
\begin{\eqn*}
\sigma'(s,\Delta E_\gamma)\,=\,\sigma_{Born}(s)\,+\,\sigma_\text{virt}(s,\lambda^2)+\sigma_\text{s}(s,\Delta E_\gamma,\,\lambda)
\end{\eqn*}
is infrared-safe. However, it still depends on the soft photon cutoff $\Delta E_\gamma$ as the soft approximation only takes a part of the $m\,\rightarrow\,n+\gamma$ phase space into account. For a cutoff-independent result, the hard cross section
\begin{equation*}
  \sigma_\text{$m\to (n+1)$}(s,\Delta E_\gamma) =
  \int_{\Delta E_\gamma} d\Gamma_{(n+1)}\,|\ME_{m\to n+1}(s)|^2
\end{equation*}
has to be added. This is usually split into a hard, collinear and a hard, non-collinear part
\begin{equation}\label{eq:shard}
  \sigma_\text{$m\to n+1$}(s,\Delta E_\gamma) =
  \sigma_\text{hard,non-coll}(s,\Delta E_\gamma,\Delta\theta_\gamma) +
  \sigma_\text{hard,coll}(s,\Delta E_\gamma,\Delta\theta_\gamma),
\end{equation}
where the cutoff $\Delta\,\theta_{\gamma}$ separates the collinear from the non-collinear region.
The hard-collinear part is treated in Section \ref{sec:hardcoll}, the non-collinear in Section \ref{sec:hardnoncoll}.\\
The total $\mO(\al)$ cross section
\begin{\eqn}\label{eq:stot}
\sigma_{tot}(s)\,=\,\sigma_{Born}(s)\,+\,\sigma_{v} (s,\lambda)\,+\,\sigma_{s} (s,\Delta E_{\gamma},\lambda)\,+\,\sigma_{m\,\rightarrow\,(n+1)}(s,\Delta E_{\gamma},\Delta \theta_{\gamma})
\end{\eqn}
is then cutoff-independent.
\section{Virtual corrections}\label{sec:virtcorr}
The one-loop corrections to the process $e^-e^+\to\chm_i\chp_j$ with
$i,j=1,2$ have been computed in the SUSY on-shell scheme in Ref.~\cite{Fritzsche:2004nf, thomdiss}.  An independent calculation in the $\overline{DR}$ scheme has been presented in~\cite{Oller:2005xg}.
These calculations include the complete set of virtual diagrams
contributing to the process with both SM and SUSY particles in
the loop.  The collinear singularity for photon radiation off the
incoming electron/positron is regulated by the finite electron
mass~$m_e$.  As an infrared regulator, the calculation introduces a
fictitious photon mass~$\lambda$. Both calculations use the
\feynarts/\formcalc\ package~\cite{FeynArts,FormCalc,Hahn:1998yk,Hahn:2001rv} for the evaluation of
one-loop Feynman diagrams in the MSSM.\\
A complete list of all generic one-loop diagrams (excluding tadpoles) contributing to the process $e^{+}\,e^{-}\,\rightarrow\,\chap\,\cham$ is given in Appendix \ref{app:diagramms}. The total number of diagrams is $\mO(1500)$, with $\mO(1000)$ self-energy diagrams  for all contributing particles. It also includes $e^{+}e^{-}$Higgs couplings which are absent if $m_{e}$ is set to zero.  We refer to \cite{Fritzsche:2004nf, thomdiss, Oller:2005xg} for details of the calculation.
\section{Soft terms}\label{sec:softf}
The soft-photon factor has been derived in~\cite{'tHooft:1978xw,Bohm:1986fg,Berends:1987jm, Denner:1991kt}. We just sketch the derivation and refer to the literature for further details.\\
In the soft photon approximation, the radiation of a photon off an incoming or outgoing charged particle for an arbitrary process is considered in the limit $E_{\gamma}\,\rightarrow\,0$, where only the terms contributing to the infrared singularity are kept.  Then, the radiation can be described by a factor depending on the particle and photon momenta and the matrix element describing the radiation is proportional to the Born matrix element:
\begin{\eqn*}
\M_{m\,\rightarrow\,n+\gamma}\,=\,f(p,k)\,\M_{m\,\rightarrow\,n},
\end{\eqn*} 
where $p_{e^{\pm}}$ ($k$) denote the electron/positron (photon)
four-vectors.
For an incoming fermion, this factor is given by
\begin{\eqn*}
f(p,k)\,=\,-e\,Q_{F}\,\frac{\vareps\,p}{kp}.
\end{\eqn*}
Here, $Q_{F}$ is the charge of the fermion and $\vareps$ the polarization vector of the photon. In addition,
\begin{equation*}
  \omega_{k} = \sqrt{\bf{k}^{2}+\lambda^{2}}
\end{equation*}
is the energy of a photon regularized by the photon mass $\lambda$.\\
 For cross sections, this leads to
\begin{\eqn*}
\sigma_{m\,\rightarrow\,n+\gamma}\,=\,f_{soft}\sigma_{m\,\rightarrow\,n},
\end{\eqn*}
where 
\begin{equation}\label{eq:f-soft}
  f_\text{soft} =-\frac{\al}{2\,\pi}
  \sum_{i,j} \int_{|\bf{k}|\leq\Delta E_\gamma}
  \frac{d^{3}k}{2\omega_{k}}\,\frac{(\pm)\,p_{i}p_{j}Q_{i}Q_{j}}{(p_{i}k)(p_{j}k)}
\end{equation}
is summed over all charged incoming/ outgoing particles. The $(\pm)$ in the numerator depends on the charge flow ($-$ for an incoming and $+$ for an outgoing charge). The integral appearing in $f_{soft}$ has been calculated in \cite{'tHooft:1978xw} and is given in Appendix \ref{app:softphot}. $\sigma_{soft}$ is then given by
\begin{\eqn}\label{eq:sigsoft}
\sigma_{soft}\,=\,\int d\Gamma_2\,f_{soft}
  |\ME_\text{Born}(s)|^2.
\end{\eqn}
\section{Hard-collinear photons}\label{sec:hardcoll}
If a photon is emitted off a particle of mass $m$ with a small transverse momentum $k_{\perp}$, logarithms of the form
\begin{\eqn}\label{eq:collogs}
\log \frac{k_{\perp}}{m}
\end{\eqn} 
appear. Considering photon emission off electrons, these logarithms give rise to divergences if the electron mass is set to zero. Therefore, in order to regulate these collinear divergencies, the electron mass has to be taken into account in these regions of phase space. Similarly, numerical integration becomes tedious or even unreliable for very small transverse photon momenta. It is therefore customary to use an analytic collinear approximation for small $k_{\perp}$ (or, alternatively, small emission angles $\theta_{\gamma}$) when integrating over these regions of the  photon phase space.\\
The explicit expression for the (hard) collinear photon approximation has been derived in \cite{Dittmaier:1993jj, Bohm:1993qx, Dittmaier:1993da}; cf. also Appendix \ref{app:hardcol}.\\
 The hard-collinear contribution to the cross sections from the radiation of photons off one incoming particle is then given by convoluting the Born cross section with the
structure function $f^{\sigma}(x;\Delta\theta_\gamma,\frac{m_e^2}{s})$, with
$x=1-2E_\gamma/\sqrt{s}$ being the energy fraction of the electron after
radiation,
\begin{align}\label{eq:sighardcoll}
  \sigma_\text{hard,coll}(s,\Delta E_\gamma,\Delta\theta_\gamma,m_e^2)
  &= \int_{\Delta E_\gamma,\Delta\theta_\gamma}d\Gamma_3\,|\ME_{2\to 3}(s,m_e^2)|^2
\nonumber\\
  &= \sum_{\sigma\,=\,\pm} \int^{x_0}_{0} dx\, f^{\sigma}(x;\Delta\theta_\gamma,\tfrac{m_e^2}{s})\int d\Gamma_2\,
  |\ME^{\sigma}_\text{Born}(xs,m_e^2)|^2.
\end{align}
The two structure functions $f^+,f^-$ are given by (\ref{eq:hardcollfs}) 
\begin{eqnarray*}
f^{+}(x)&=&\,\frac{\al}{2\,\pi}\,\frac{1+x^{2}}{(1-x)}\,\lb\ln \lb \frac{s\,(\Delta\theta)^{2}}{4\,m^{2}}\rb-1 \rb,\nonumber \\
f^{-}(x)&=&\frac{\al}{2\,\pi}\,(1-x).
\end{eqnarray*}
They correspond to helicity conservation and helicity flip, respectively;
each one is convoluted with the corresponding matrix element.  The
cutoff $\Delta E_\gamma$ is replaced by $x_0=1-2\,\Delta
E_\gamma/\sqrt s$.  In this approximation, positive powers of
$\Delta\theta_\gamma$ are neglected. For radiation off both incoming particles, we have 
\begin{\eqn*}
\sigma_{\text{hard, coll}}\,=\,\sigma_{\text{hard, coll}} (e^{+})+\sigma_{\text{hard, coll}} (e^{-}).
\end{\eqn*}
\section{Hard non-collinear photons}\label{sec:hardnoncoll}
\begin{figure}[tb]
\begin{center}
\includegraphics[width=0.8\textwidth]{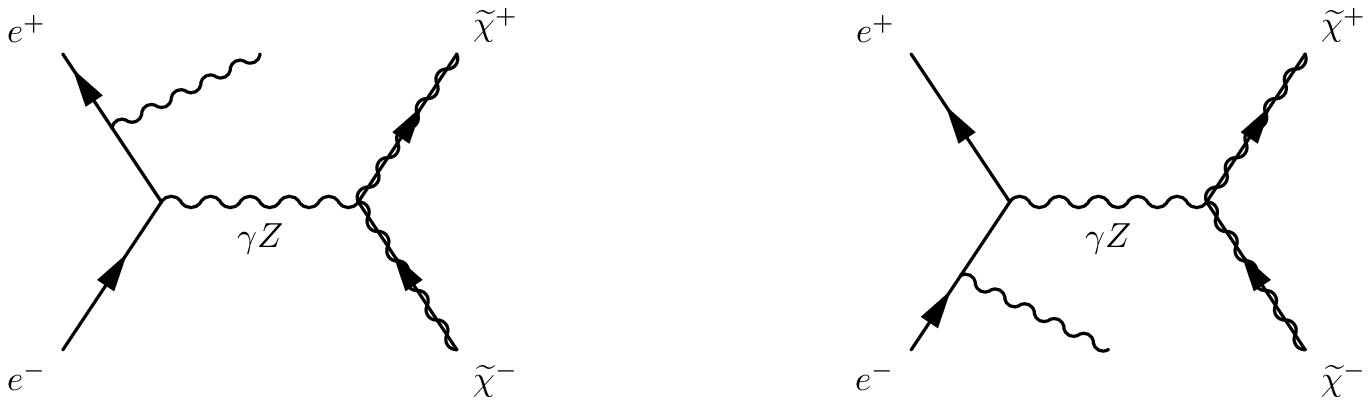}
\vspace{4mm}
\includegraphics[width=0.8\textwidth]{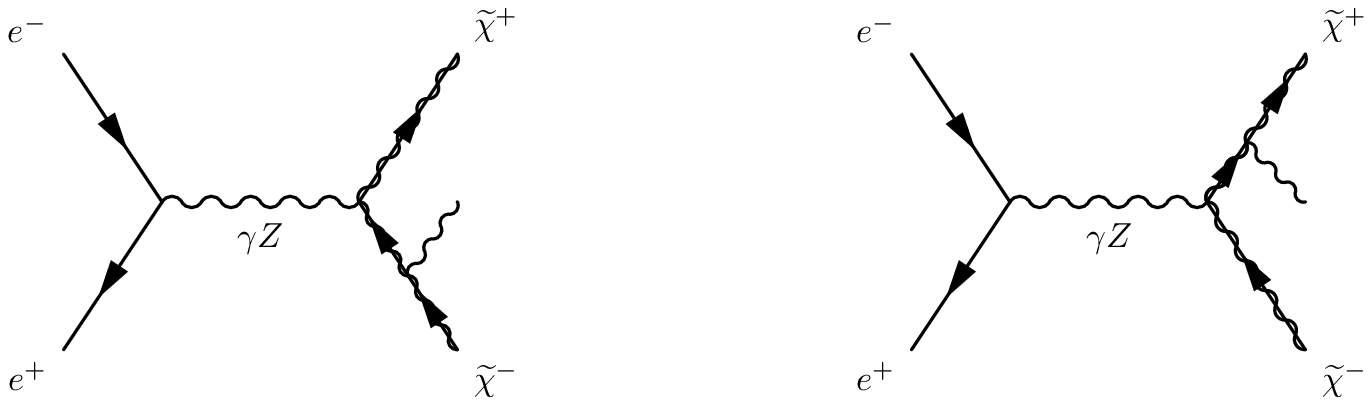}
\vspace{4mm}
\caption{\label{fig:2to3} Matrix elements contributing to the $2\,\rightarrow\,3$ process. We only show the $s$ channel contribution ($t$ channel exchange analogously).}
\end{center}
\end{figure}
The hard non-collinear contributions are added in form of the analytic $e^{+}\,e^{-}\,\rightarrow\,\chp\,\chm\,\gamma$ matrix element. In order to prevent double counting for soft and hard-collinear photons, which are already accounted for  in $\sigma_{soft}$ and $\sigma_{hard,coll}$, lower angular and energy cuts for the explicitly generated photon are set. Then,
\begin{\eqn}\label{eq:sighnc}
\sigma_{hard, non-coll}\,=\,
\int_{\Delta E_\gamma,\Delta\theta_\gamma} d\Gamma_3\,|\ME_{2\to 3}(s)|^2.
\end{\eqn}
In the non-collinear region, logarithms as (\ref{eq:collogs}), which are regulated by a finite photon mass, do no longer appear. For cm energies $\mO(100\,\GeV)$ and larger, we can neglect contributions proportional to the electron mass and set $m_{e}\,=\,0$.\\
 The contributing Feynman diagrams for the $s$ channel exchange are shown in Figure \ref{fig:2to3}.
%\begin{figure}[tb]
%\begin{center}
%\input{2to3}
%\caption{\label{fig:2to3} matrix elements contributing to the $2\,\rightarrow\,3$ process. $s$ channel only is shown ($t$ channel exchange analogously)}
%\end{center}
%\end{figure}
The matrix elements can be easily obtained from Eq. (\ref{eq:mborn}) by substituting
\begin{eqnarray*}
\M\,=\bar{A}u(p)&\longrightarrow&\M'\,=\,-\frac{e}{2\,pk}\,\bar{A}\,(\slashed{p}-\slashed{k}+m)\slashed{\epsilon}u(p),\\
\M\,=\,\bar{A'}v(p')&\longrightarrow&\M'\,=\,-\frac{e}{2\,p'k}\,\bar{A'}\,(\slashed{p'}-\slashed{k}+m)\slashed{\epsilon}v(p'),\\
\end{eqnarray*}
where $u$ or $v$ are the spinors of the particle radiating off the photon with a momentum $k^{\mu}$ and the polarization vector $\epsilon^{\mu}$ and   $A/ A'$ symbolize the part of the matrix element untouched by the radiation.
\section{Fixed $\mO(\al)$ results for total cross section}\label{sec:nloex}
\begin{figure}[tb]
\begin{center}
\includegraphics[width = 100mm]{sigexact.eps}
\vspace{10mm}
\caption{\label{fig:sigexact} Born (LO) and fixed $\mO(\al)$ (Eq. (\ref{eq:sigex})) (NLO) corrected cross sections }
\end{center}
\end{figure}
\begin{figure}
\begin{center}
\includegraphics[width = 100mm]{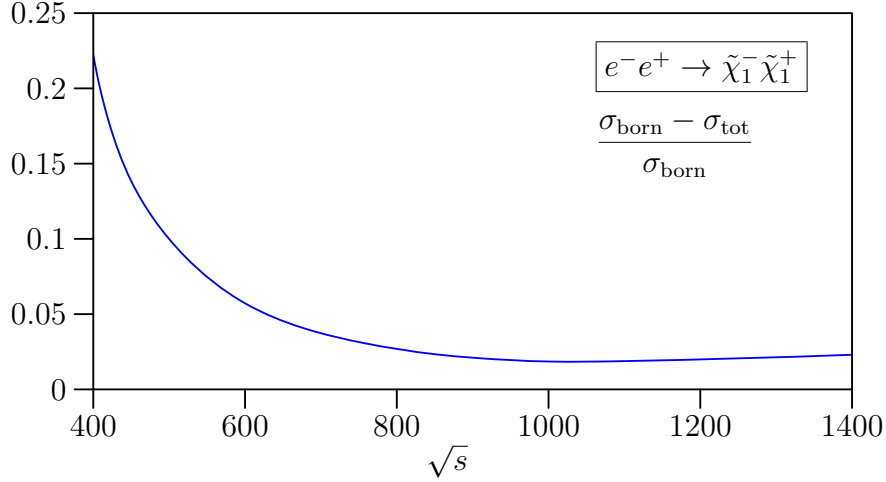}
\vspace{10mm}
\caption{\label{fig:exactreldiff} relative difference between Born and fixed-order result for total cross section: $\frac{\sigma_{Born}-\sigma_{tot}}{\sigma_{Born}}$}
\end{center}
\end{figure}

The total fixed-order $\mO(\al)$ cross section (Eq. (\ref{eq:stot})) for the process $e^{+}\,e^{-}\,\rightarrow\,\chap\,\cham,\,(\gamma)$ is then given by the sum of Eqs. (\ref{eq:sborn}),(\ref{eq:svirt}), (\ref{eq:ssoft}), (\ref{eq:shard}), (\ref{eq:sigsoft})
\begin{eqnarray}\label{eq:sigex}
\sigma_{tot}(s)&=&\sigma_\text{Born}(s)\,+\,\sigma_{soft}(s)\,+\,\sigma_\text{virt}(s,\lambda^2,m_e^2)\,+\,\sigma_\text{hard,coll}(s,\Delta E_\gamma,\Delta\theta_\gamma,m_e^2)\nonumber\\ &+&\sigma_\text{hard,non-coll}(s,\Delta E_\gamma,\Delta\theta_\gamma)\nonumber\\
&=&\int d\Gamma_2\,
  |\ME_\text{Born}(s,\,\cos\theta)|^2\,+\,\int d\Gamma_2\,f_{\text{soft}}
  |\ME_\text{Born}(s,\,\cos\theta)|^2\nonumber\\ &+&\int d\Gamma_2\left[
  2\mathrm{Re}\left(\ME_\text{Born}(s)^*\,
                    \ME_\text{1-loop}(s,\lambda^2,m_e^2)\right)\right]\nonumber\\
&+&\sum_{i=1,2}\,\sum_{\sigma\,=\,\pm} \int^{x_0}_{0} dx_{i}\, f^{\sigma}(x_{i};\Delta\theta_\gamma,\tfrac{m_e^2}{s})\int d\Gamma_2\,
  |\ME^{\sigma}_\text{Born}(x_{i},s,m_e^2)|^2\nonumber\\&+&\int_{\Delta E_\gamma,\Delta\theta_\gamma} d\Gamma_3\,|\ME_{2\to 3}(s)|^2\nonumber\\
&&
\end{eqnarray}
with the contributions introduced in the previous sections. For the mSugra point SPS1a' (cf. Appendix \ref{app:sps1asl}), this leads to corrections $\mO(20\%)$ near the threshold and in the $5\%$ region for $\sqrt{s}\,\geq\,600\,\text{GeV}$; cf. Figures \ref{fig:sigexact} and \ref{fig:exactreldiff}.
\section{Resummation of higher logarithms: Initial state radiation}\label{sec:strfunisr}
\begin{figure}[tb]
\begin{center}
\includegraphics[width = 100mm]{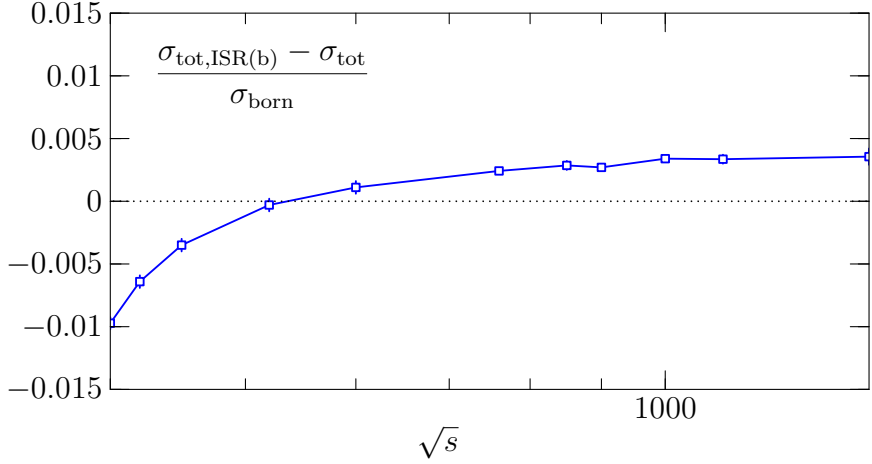}
\vspace{10mm}
\caption{\label{fig:higho} Relative effect of ISR higher-order initial state radiation: $\frac{\sigma_{tot,ISR(b)}-\sigma_{tot}}{\sigma_{Born}}$}
\end{center}
\end{figure}

The logarithms (\ref{eq:collogs}) which arise in the collinear emission of photons can become large for small $m$. In the collinear approximation, where the transverse momentum of a photon with respect to the emitting particle is neglected, the divergencies originating from the emission of real collinear photons as well as their virtual counterparts can be summed up in splitting functions. The electron-electron splitting function in the leading logarithmic approximation is given by
\begin{\eqn}\label{eq:elsplit}
P_{ee}(x)\,=\,\lb\,\frac{1+x^{2}}{1-x}\rb_{+},
\end{\eqn} 
where only terms proportional to the collinear logarithm are kept. The $+$-distribution is given by Eq. (\ref{eq:plusdef}). The differential cross section taking the emission of one real and one virtual photon collinear photon into account then reads
\begin{\eqn*}
d\sigma_{X+\gamma} (p)\,=\,\frac{\al}{2\,\pi}\,\log\lb\frac{Q^{2}}{m^{2}}\rb\,\int^{1}_{0}dx\,P_{ee}(x)\,d\sigma_{X}(xp),
\end{\eqn*}
where $p$ is the electron momentum, $X$ symbolizes the final state of the reaction and $Q$ is the scale of the process. The distribution function
\begin{\eqn*}
f_{e,e}(x,Q^{2})\,=\,\delta(1-x)\,+\,\frac{\al}{2\,\pi}\,\log\lb\frac{Q^{2}}{m^{2}}\rb\,P_{ee}(x)
\end{\eqn*}
gives the probability of finding an electron with the longitudinal momentum fraction $x$ in an incoming electron, when the emittance of photons with the maximum transverse momentum $Q^{2}$ are taken into account. In the collinear limit, $Q$ should be set equal to $k_{\perp}$ and be small compared to $p_{0}$.\\
If the coherent emission of more than one photon is considered, the splitting function is replaced by a distribution function $D^{NS}$. The dominant logarithmic contributions stem from the emission of photons with strong $k_{\perp}$-ordering such that $k_{\perp,1}\,\ll\,k_{\perp,2}\,\ll\,...$. The distribution function $D^{NS}$ then obeys the evolution equation \cite{Gribov:1972rt}
\begin{\eqn}\label{eq:llev}
\frac{\partial}{\partial\eta}D^{NS}(x,\eta)\,=\,\frac{1}{4}\,\int^{1}_{x}\,\frac{dz}{z}\,P(z)D^{NS}\lb\frac{x}{z},\eta \rb,
\end{\eqn}
where
\begin{\eqn*}
\eta\,=\,\int^{Q^{2}}_{m^{2}}\,\frac{ds'}{s'}\frac{2\,\al(s')}{\pi}.
\end{\eqn*}
In a first approximation, we can neglect the running of $\al$ and set it constant. In this case,
\begin{\eqn}\label{eq:eta}
\eta\,=\,\frac{2\,\al}{\pi}\,\log\lb\frac{Q^{2}}{m^{2}}\rb.
\end{\eqn}
The first order solution of Eq. (\ref{eq:llev}) is then
\begin{\eqn*}
D^{NS}(x,\eta)\,=\,\delta(1-x)+\frac{\eta}{4}\,P_{e,e}(x)\,+\,\mO(\eta^{2}),
\end{\eqn*}
where the $\delta(1-x)$ term corresponds to the tree level (= no photon emission) part. higher-order solutions can be found in the literature; cf. \cite{Skrzypek:1990qs,Chen:1990qz,Skrzypek:1992vk}. The exponentiated structure function $f_\text{ISR}(x;\Delta\theta_\gamma,\tfrac{m_e^2}{s})$ (Eq. (\ref{eq:f-ISR})) includes photon radiation to all orders in the soft regime at
leading-logarithmic approximation and, simultaneously, correctly
describes collinear radiation of up to three photons in the hard
regime.  It does not account for the helicity-flip part
$f^-$  of the fixed-order structure function. More details on this can be found in Appendix \ref{app:isr}.\\
Initial state radiation from both incoming particles is then given by
\begin{\eqn}\label{eq:sigbornisr}
\sigma_\text{Born+ISR}(s,\Delta\theta_\gamma,m_e^2) = \int
  dx_{1}\,f_\text{ISR}(x_{1})\,\int
  dx_{2}\,f_\text{ISR}(x_{2})\, \int
  d\Gamma_2\,|\ME_\text{Born}(x_{1},x_{2},s)|^2.
\end{\eqn}
In combining $\sigma_{tot}$ (Eq. \ref{eq:sigex}) and $\sigma_{Born+ISR}$ (Eq. \ref{eq:sigbornisr}), we have to subtract the $\mO(1)$ and $\mO(\al)$ contributions of $\sigma_{Born+ISR}$ as they are already accounted for by $\sigma_{tot}$. This way, the total cross section including all $\mO(\al)$ contributions as well as higher-order initial state radiation, is given by
\begin{eqnarray}\label{eq:sigfin}
\lefteqn{\sigma_{tot,ISR(b)}\,=\,\sigma_{tot}\,+\,\sigma_{Born+ISR}
\,-\,\sigma_{Born}}\nonumber\\&-&\lb\int
  dx_{1}\,f^{\al}_\text{ISR}(x_{1})\,
  \int
  d\Gamma_2\,|\ME_\text{Born}(x_{1},s)|^2 \,+\,\int
  dx_{2}\,f^{\al}_\text{ISR}(x_{2})\,
  \int
  d\Gamma_2\,|\ME_\text{Born}(x_{2},s)|^2 \rb,\nonumber\\
&&
\end{eqnarray}
where $f^{\al}_{ISR}$ is the $\mO(\al)$ contribution of $f_\text{ISR}$. NLO results for total cross sections in the literature are typically presented in the form of Eq. (\ref{eq:sigfin}). The effects of including higher-order ISR are in the per mille regime as can be seen from Figure \ref{fig:higho}. 
\section{Further higher-order contributions}
\begin{figure}
\begin{center}
\includegraphics{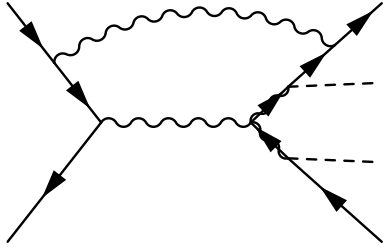}
\end{center}
\caption{\label{fig:nonfac} General example for a non-factorizing NLO contribution to the $2\,\rightarrow\,4$ process with $s$ channel chargino production and unspecified final states}
\end{figure}

Charginos usually decay via decay chains involving (at least) two final state particles. A complete NLO calculation therefore also includes factorizable corrections to the chargino decays as well as non-factorizable corrections as e.g. in Figure \ref{fig:nonfac}.\\
%Here, the propagators of the unstable particles are expanded around their poles, and only leading order contributions are kept \cite{Aeppli:1993rs,Roth:1999kk,Denner:2000bj}.
A first step is the use of the double-pole approximation for the unstable particles. It includes  (i) loop corrections to
the SUSY production and decay processes, (ii) nonfactorizable, but
maximally resonant photon exchange between production and decay, (iii)
real radiation of photons, and (iv) off-shell kinematics for the signal
process.  Recent complete NLO calculations for SM  W pair production at an $e^{+}e^{-}$ collider \cite{Denner:2005es} have explicitly verified the validity
of this approximation in the signal region. A complete analysis also requires the consideration of (v) irreducible background from all other multi-particle SUSY
processes, and (vi) reducible, but experimentally indistinguishable
background from Standard Model processes. So far, no calculation provides all NLO pieces for a process involving
SUSY particles.\\
In this work, we  only consider the extension of the tree-level simulation
of chargino production at the ILC by radiative corrections to the
on-shell process, i.e., we consider (i) in the above list and
consistently include real photon radiation (iii). This is actually a
useful approximation since nonfactorizable NLO contributions are
suppressed by $\mO(\frac{\Gamma}{m})$ \cite{Fadin:1993kt, Fadin:1993dz} and in many MSSM scenarios the widths of charginos, in
particular $\chapm$, are quite narrow (cf.\
Table~\ref{tab:charginos}).\\
Chargino decays (ii), non-factorizing dominant contributions (iii), and finite-width effects (iv) will be covered in future work. NLO corrections to chargino decays for specific decay products are available from \cite{thomdiss}. They are in the $\%$ regime and can easily be combined with the present analysis. The inclusion of background effects (v) and (vi) can easily be done using Monte Carlo Event generators \cite{Hagiwara:2005wg}. Non-factorizing contributions  and finite width effects can be treated in the double-pole approximation \cite{Aeppli:1993rs, Roth:1999kk, Denner:2000bj}. Here the propagators of the unstable particles are expanded around their poles, and only leading order contributions are kept. For W pair production at an $e^{+}e^{-}$ collider, analytic results for non-factorizing contributions \cite{Melnikov:1995fx, Beenakker:1997bp,Beenakker:1997ir, Denner:1997ia} and a full double-pole approximation \cite{Jadach:1996hi,Jadach:1998tz,Beenakker:1998gr, Kurihara:1999ii, Denner:2000bj} are available in the literature.\\
At the production threshold, additional large corrections can arise from the Coulomb singularity \cite{sommi,Sakharov:1948yq}. For on-shell particles, the cross section factorizes according to
\begin{\eqn*}
\sigma_{C}\,=\,\frac{\al\,\pi}{2\,\beta}\,\sigma_{Born},
\end{\eqn*} 
where $\beta$ denotes the relative velocity of the produced particles. This expression diverges for $\beta\,\rightarrow\,0$. At threshold, the Coulomb singularity needs to be resummed using effective field theories. If the produced particles are off-shell, the singularity is cut off at a relative velocity $\beta\,\propto\,\sqrt{\Gamma/m}$. For the masses and widths of charginos given in Table \ref{tab:charginos}, this leads to corrections $\leq\,1\,\%$ at threshold. Similarly, for W pair production \cite{Bardin:1993mc, Fadin:1993kg,Fadin:1994pm,Fadin:1995fp} and slepton production, \cite{Freitas:2001zh},  corrections are in the percent regime.

\begin{table}
  \begin{equation*}
    \begin{array}{c|cc}
      \hline
      & \text{Mass} & \text{Width} \\
      \hline
      \chap & 183.7\;\GeV & 0.077\;\GeV \\
      \chbp & 415.4\;\GeV & 3.1\;\GeV \\
      \hline
    \end{array}
  \end{equation*}
  \caption{Chargino masses and widths for the SUSY parameter set SPS1a'.}
  \label{tab:charginos}
\end{table}

\chapter{Inclusion of NLO corrected matrix elements in \whizard~ (fixed order)}\label{chap:fixed}
\section{Monte Carlo (event) generators}\label{sec:mcs}
In general, Monte Carlo techniques make use of random numbers to numerically determine values of integrals or, given a probability distribution, simulate the outcome of physical events. For a general introduction of Monte Carlo techniques in particle and especially collider physics, see e.g. \cite{James:1980yn, Kleiss:1989de, Dobbs:2004qw}.
\subsection{Monte Carlo integration}\label{sec:mcint}
In Monte Carlo integration, the general idea is to use
\begin{\eqn*}
I\,=\,\int^{x_{2}}_{x_{1}}\,f(x)\,dx\,\approx\,(x_{2}\,-\,x_{1})\,\langle f(x)\rangle,
\end{\eqn*}
where $\langle f(x)\rangle$ is determined by the averaged value of $N$ random calls of $x$:
\begin{\eqn*}
\langle f(x)\rangle\,=\,\frac{1}{N}\,\sum^{N}_{i=1}\,f(x_{i}).
\end{\eqn*}
According to the central limit theorem for large numbers, the error is then $\propto\,1/\sqrt{N}$. The $N$-dependent error can be decreased by importance sampling, where more values of $x$ are chosen in regions where $f(x)$ is largest, or similar techniques.\\
For the numerical calculation of cross sections with $n$ final particles, we need to integrate
\begin{\eqn}\label{eq:sigmc}
\sigma\,=\,\frac{1}{2\,s}\,\int\,|\M|^{2}\,d\Pi_{n},
\end{\eqn}
where
\begin{\eqn*}
\Pi_{n}\,=\,\left[\,\prod^{n}_{i=1}\,\frac{d^{3}p_{i}}{(2\,\pi)^{3}\,(2E_{i})}\right]\,(2\,\pi)^{4}\,\delta^{(4)}\,\lb p_{0}\,-\,\sum^{n}_{i=1}p_{i}\rb
\end{\eqn*}
is the $n$-dimensional final state phase space. In Monte Carlo programs, the matrix element $\M$ is either coded manually or generated by some (internal or external) automatic matrix element generator. Examples for external programs are \comphep~\cite{Pukhov:1999gg}, \madgraph~\cite{Stelzer:1994ta}, or \oMega~\cite{Moretti:2001zz}. The integral depends on $3n-4$ independent variables. Of course, the multi-dimensional integration of phase space is non-trivial and equally requires refined techniques. Differential cross sections can be obtained accordingly.
\subsection {Event generation}
A physical event is defined by the specification of the $n$ four-momenta of the final state particles, which require the generation of $3\,n-4$ random numbers. In a straightforward Monte Carlo integration of $\sigma$ (\ref{eq:sigmc}), all events have the same a-priori probability. To obtain the final result $\sigma$, they are weighted with the corresponding differential cross section $d\sigma/(\prod dp_{i})$.\\
A Monte Carlo Event generator, in contrast, should generate events according to their actual probability. This can be achieved by adapting the a-priori probability to the physical distribution or applying a hit-and-miss technique where each event is assigned the corresponding probability $P_{i}\,=\,d\sigma_{(i)}/d\sigma_{max}$ which is compared with a random number $r$ between $0$ and $1$ and kept if $P_{i}\,\geq\,r$. Notice that this requires that $d\sigma_{i}\,\geq\,0$. Although this condition is always fulfilled in leading order, NLO calculations might cause problems for certain points of phase space \cite{Kleiss:1989de}; cf. Sections \ref{sec:drawback} and \ref{sec:resumlep}.\\
The events generated by the Monte Carlo program provide the same information as experimental data and can be analyzed using the respective detector simulation and analysis tools. Note that this equally allows for plotting of one- or multidimensional partial distributions, correlations, etc. without any further analytic calculation.
\subsection{\whizard}
\whizard~\cite{Kilian:2001qz} is a universal Monte Carlo event generator for multiparticle scattering processes. It interfaces several matrix event generators such as \comphep~, \madgraph~, and \oMega~. In addition, it includes initial state radiation, beamstrahlung using the program \circe~\cite{Ohl:1996fi}, and fragmentation and hadronization routines from \prog{pythia} \cite{Sjostrand:2006za}. It is designed as a $2\,\rightarrow\,n$ particle event generator. In one call, several processes can be combined such that background studies are simplified. Similarly, the results of several matrix element generators can be compared. For SUSY processes, this has recently been used for an extensive comparison \cite{Reuter:2005us}. Currently, it includes the Standard Model, MSSM, little Higgs models, and non-commutative geometry models. Similarly, it allows for user-modified spectra, structure functions, and cuts.
\section{Calculating NLO matrix elements using \feynarts~ and \formcalc}
\feynarts~\cite{Hahn:2000kx,Hahn:2001rv} and \formcalc~\cite{Hahn:1998yk} are Mathematica- and Form-based programs for (higher-order) matrix element generation and the calculation of the respective total and differential cross sections. It includes the SM, the MSSM, and can be extended to any model desired by the user. Furthermore, it can generate Feynman diagrams in a Latex or postscript format. Both programs use \looptools~ \cite{Hahn:1998yk} for the calculation of n-point functions and other loop-related quantities. We will quickly discuss both \feynarts~ and \formcalc~ and refer to the respective manuals \cite{FormCalc,FeynArts} for more details.

\subsubsection{\feynarts}
\feynarts~ is a purely Mathematica-based program. For a given number of in- and outcoming particles and loops, it first generates the corresponding general topologies. After choosing a physical model and specifying the in- and outgoing particles, all amplitudes for the specified process are generated and given analytically (depending on the process, the output might be quite complex). The user can then apply numerous specifications, e.g. diagram selections or renormalization conditions. \feynarts~ equally creates the Latex or postscript output for all created or specifically selected diagrams.

\subsubsection{\formcalc}\label{sec:formcalc}
\formcalc~ is a Form \cite{Vermaseren:1992vn,Vermaseren:2000nd}-based program with a Mathematica interface. It generates a Fortran code corresponding to the \feynarts~ amplitudes. The resulting program numerically integrates the total or (angular) differential cross section for the corresponding process. Currently, it contains the kinematics for $1\,\rightarrow\,2$,  $2\,\rightarrow\,2$, and $2\,\rightarrow\,3$ particle reactions. In addition, it provides an easy input for e.g. mSugra parameters, loops over the cm energy or model parameters, or energy and angular cuts. It equally allows for different choices of multi-dimensional integration routines. The integrations are carried out according to the techniques described in Section \ref{sec:mcint}.\\

Technically, the generated code consists of different process-dependent or independent modules. They contain e.g. routines for general features of the process, the process kinematics,  the code for the Born and one-loop matrix element and, for SUSY processes, code for the SUSY spectrum-generation. The compilation creates libraries for the calculation of the matrix element (\prog{squared$\_$me.a}), the renormalization constants (\prog{renconst.a}), and kinematics-dependent variables (\prog{util.a}) which are linked to the main executable. Furthermore, the \looptools~ library (\prog{libooptools.a}) has to be included. \\
\section{Inclusion of the fixed order NLO contribution using a structure function}\label{sec:fixinclusion}
 In \whizard, there is an interface for
arbitrary structure functions $f(x_{1},x_{2})$ that can be convoluted
with the cross section according to 
\begin{\eqn*}
\sigma(s)\,=\,\int^{1}_{0}\,dx_{1}\,\int^{1}_{0}\,dx_{2}\,f(x_{1},x_{2})\,\sigma(x_{1},x_{2},s),
\end{\eqn*}
where $x_{i}$ is the beam energy fraction. $f(x_{1},x_{2})$ can be the sum of two uncorrelated structure functions (one for each incoming beam) or a correlated structure function for two incoming beams. For more details, cf. \cite{Kilian:2001qz}.\\ We can therefore implement the fixed-order one-loop result $\sigma_{tot}$ (Eq. (\ref{eq:sigex})) using
\begin{align}\label{eq:sigtotimp}
  \sigma_\text{tot}(s,m_e^2) &=
  \int dx\,f_\text{eff}(x_{1},\,x_{2};\Delta E_\gamma,\Delta\theta_\gamma,\tfrac{m_e^2}{s})\,
  \int d\Gamma_2\,|\ME_\text{eff}(s,x_{1},\,x_{2};m_e^2)|^2
\nonumber\\ &\quad
  +
  \int_{\Delta E_\gamma,\Delta\theta_\gamma} d\Gamma_3\,|\ME_{2\to 3}(s)|^2,
\end{align}
with the structure function
\begin{align}\label{eq:effstrfun}
  f_\text{eff}(x_{1},\,x_{2};\Delta E_\gamma,\Delta\theta_\gamma,\tfrac{m_e^2}{s})
  &= \delta(1-x_{1})\,\delta(1-x_{2})
\nonumber\\ &\quad
  + \delta(1-x_{1})\,f(x_{2};\Delta\theta_\gamma,\tfrac{m_e^2}{s})\,\theta(x_0-x_{2}) 
\nonumber\\ &\quad
  + f(x_{1};\Delta\theta_\gamma,\tfrac{m_e^2}{s})\,\delta(1-x_{2})\,\theta(x_0-x_{1})
\end{align}
and the effective squared amplitude
\begin{align}\label{eq:meff}
  |\ME_\text{eff}(s,x_{1},\,x_{2};m_e^2)|^2
  &= \left[1 + f_\text{soft}(\Delta E_\gamma,\lambda^2)\,\theta(x_{1},x_{2}))\right]
  \,|\ME_\text{Born}(s)|^2
\nonumber\\ &\quad
  + 2\mathrm{Re}\left[\ME_\text{Born}(s)\,
                      \ME_\text{1-loop}(s,\lambda^2,m_e^2)\right]
  \theta(x_{1},x_{2})
\end{align}
with $\theta(x_{1},x_{2})\equiv
\theta(x_{1}-x_{0})\,\theta(x_{2}-x_{0})$.
The parameter
\begin{\eqn*}
x_{0}\,=\,1-\frac{2\,\Delta E}{\sqrt{s}}
\end{\eqn*}
separates the hard from the soft photon region.
In addition to the introduction of the structure function $f_\text{eff}$, we  replace the Born matrix element as computed by the
matrix-element generator, \oMega, by the effective matrix
element $\M_{eff}$ (\ref{eq:meff}). The latter is computed by a call to the
\formcalc-generated routine.

 Technically, (\ref{eq:sigtotimp}) is implemented by splitting the structure function into four different regions in the $(x_{1},x_{2})$ space: 
\\
\begin{itemize}
\item{}$x_{1}\,\ge\,x_{0}, x_{2}\,\ge\,x_{0}:$ (soft-soft)\\
\\
This corresponds to the region where $f_{eff}\,=\,\delta(1-x_{1})\,\delta(1-x_{2})$. Both photons are in the soft photon regime; we set $x_{i}\,=\,1$ and calculate the Born+virtual+soft cross sections:
\begin{\eqn*}
\sigma_{Born}+\sigma_{virt}+\sigma_{soft}=\int d\Gamma_2\,\left\{
  (1+f_{soft}) |\ME_\text{Born}(s)|^2\,\,+\,
\left[ 2\mathrm{Re}\left(\ME_\text{Born}(s)^*\,
                    \ME_\text{1-loop}(s)\right)\right]\right\}. 
\end{\eqn*}
All matrix element contributions in this region as well as the soft photon factor are generated by a call to \formcalc. As we are mapping two $\delta$ functions to a finite $(x_{1},x_{2})$ region, we need to divide the result by a normalization factor $N$ such that $\frac{1}{N}\int^{1}_{x_{0}}\,\int^{1}_{x_{0}}dx_{1}\,dx_{2}\,=\,1$.\\
\item{}$x_{i}\ge x_{0}\,,\,x_{j}\,<\,x_{0}$ (soft-hard)\\
\\
This corresponds to the region where $f_{eff}\,=\,\delta(1-x_{i})\,f(x_{j};\Delta\theta_\gamma,\tfrac{m_e^2}{s})$. One of the photons is in the soft regime, while the other one is in the hard regime. For the soft photon, $x$ is again set to 1. The hard-collinear contribution of the hard photon is then generated according to
\begin{eqnarray*}
\sigma_{hard,coll}&=&\int (f^{+}(x)\,\sigma^{+}_{Born}(x,s)\,+\,f^{-}(x)\,\sigma^{-}_{Born}(x,s))\,dx\\
&=&\sum_{\sigma\,=\,\pm} \int^{x_0}_{0} dx\, f^{\sigma}(x;\Delta\theta_\gamma,\tfrac{m_e^2}{s})\int d\Gamma_2\,
  |\ME^{\sigma}_\text{Born}(xs,m_e^2)|^2.
\end{eqnarray*}
Here, the matrix elements can be calculated using either \formcalc~ or \oMega~. The helicity-dependent form factors are implemented by a modification of the density matrix. In general,
\begin{\eqn*}
|\M|^{2}\,=\,\M^{\dagger}(\lambda)\,\rho_{\lambda\lambda'}\,\M(\lambda'),
\end{\eqn*} 
where $\rho_{\lambda\lambda'}$ is the helicity-dependent density matrix for the incoming particles. For the inclusion of $f^{\pm}$, it is modified such that
\begin{\eqn*}
\rho'_{\lambda\lambda}\,=\,f^{+}\delta_{\lambda\,,\lambda'}\,\rho_{\lambda'\lambda'}+f^{-}\delta_{\lambda\,,-\lambda'}\,\rho_{\lambda'\lambda'},
\end{\eqn*}
where we assumed that $\lambda, \lambda'$ can take the values of $\pm\,1$ as in fermionic cases. Taking the mapping of the $\delta$ function into account, we again have to divide the result by a normalization factor.\\
\item{}$x_{1}<x_{0}\,,\,x_{2}<x_{0}$ (hard-hard)\\
\\
Both photons are in the hard regime. In principle, this region would describe the radiation of two hard photons. As this is a second order effect, we set $f_{eff}\,=\,0$ in this region.\\
\end{itemize}
As the soft photon approximation requires $x\,\approx\,1$,
we have to artificially split the interval $[0:1]$ such that 
%\begin{\eqn*}
$x\,\ge\,x_{0}$
%\end{\eqn*}
is reached sufficiently often. This can be done by the introduction of an additional $x_{\text{del}}$ and the projection
\begin{\eqn*}
[0:x_{\text{del}}]\,\longrightarrow\,[0:x_{0}]\;;\;[x_{\text{del}}:1]\,\longrightarrow\,[x_{0}:1].
\end{\eqn*}
We then have to multiply all results by the Jacobian of the transformation.\\
\\
The hard, non-collinear part is added by a separate run of \whizard\vspace{1mm} as given by eq. (\ref{eq:sighnc}) with the explicitly generated matrix element $\M_{2\,\rightarrow\,3}$  applying the respective $\Delta E_{\gamma},\,\Delta\theta_{\gamma}$ cuts. However, in \whizard~it is equally possible to combine different processes in one run which then reproduces $\sigma_{tot}$ as given in Eq. (\ref{eq:sigex}).\\

Implementing this algorithm in \whizard, we construct an unweighted
event generator.  With separate runs for the $2\to 2$ and $2\to 3$
parts, the program first adapts the phase space sampling and
calculates a precise estimate of the cross section.  The built-in
routines apply event rejection based on the effective weight and thus
generate unweighted event samples.\\

On the technical side, for the actual implementation we have
carefully checked that all physical parameters and, in particular, the
definition of helicity states are correctly matched between the
conventions~\cite{Hagiwara:1985yu} used by \oMega~/\whizard~
and those used by \formcalc\, (cf.~e.g.~\cite{Dittmaier:1998nn}). These differ by a complex phase in the definition of the two-component helicity eigenstates.
\section{Technicalities of the implementation}
The inclusion of the \formcalc~ generated NLO contributions described in Section \ref{sec:fixinclusion} implies the modification of the \whizard~ as well as the \formcalc~ code. Here, we just sketch the general approach and refer to \cite{howto} for more details. \\
We used the \formcalc~ code for chargino production at NLO available from \cite{Fritzsche:2004nf, thomdiss}. \whizard~ already includes calls to other external programs, such as CompHep, MadGraph, and \oMega. Here, libraries are created from the respective programs, which are then linked to the \whizard~ executable. The same is done here for the inclusion of the NLO \formcalc~ matrix element.\\
We therefore modify, in the \formcalc~ code, those modules calculating the matrix element as well as the soft photon factor (to create a \whizard~-compatible in- and output) and the kinematics (in \formcalc, the routines setting the kinematics equally set the soft photon factor and give values to internal variables needed for the matrix element evaluation). In addition, we create a new interface setting the SM and MSSM variables to the values used by \whizard~. We can therefore make use of the internal \whizard~ data reading routines and can equally use input in the Les Houches accord format \cite{Skands:2003cj}. A library \prog{libformcalc.a} is then created and linked. The library equally contains the (modified) \formcalc~ libraries for the squared matrix element (\prog{squared$\_$me.a}), some kinematic variables (\prog{util.a}), and the renormalization constants (\prog{renconst.a}).\\
In \whizard, we add a routine \prog{set$\_$pars$\_$formcalc} setting the SM and MSSM values used by \whizard~ in the \formcalc~ routines as well as a call to the (NLO)-generated matrix element subroutines \prog{SquaredME} which depends on the cm energy and the four-vectors of the outgoing particles. \\
\\
To summarize, we
\begin{itemize}
\item{}modify the file \prog{2to2.F} from \formcalc, containing the $2\,\rightarrow\,2$ kinematics, for in- and output of the \whizard~ generated particle momenta,
\item{}modify the file \prog{squared$\_$me.F} such that the helicity-dependent subroutine\\
\prog{SquaredME(result, spins, reset, matrix$\_$Born, 
     $\&$ matrix$\_$loop, old$\_$angles, old$\_$s)} 
can explicitly be called from \whizard,
\item{}generate a new file \prog{transer.f} for SM and MSSM value transfer,
\item{}create the library \prog{libformcalc.a} from the modified code, including also the unmodified libraries \prog{renconst.a} and \prog{util.a},
\item{}link \prog{libformcalc.a} and \prog{libooptools.a} to \whizard.
\end{itemize}  
In \whizard~, we
\begin{itemize}
\item{}add the respective routine \prog{set$\_$pars$\_$formcalc} for variable transfer in the process-file,
\item{}call the subroutines setting the case-to-case kinematic variables \prog{SetEnergy} and \prog{SquaredME}  each time we need the \formcalc~ one loop matrix element and soft photon factor. 
\end{itemize}
In addition, we included checks guaranteeing that the \formcalc~ loop-related routines are only called when necessary. For example, $\sqrt{s}$ dependent quantities are recalculated only if $\sqrt{s}$ actually changes.\\
The difference between the helicity bases for fermions used in \whizard~ and \formcalc~ (cf. Section \ref{sec:fixinclusion}) needs to be accounted for by the introduction of an additional phase of the matrix element. For the Born contribution, this has been checked for all possible helicity combinations. \\
The soft and collinear cuts $\Delta\,E_{\gamma},\,\Delta\theta_{\gamma}$ are added as variables in the file \prog{cutpars.dat} in the \prog{results} subdirectory of \whizard.
\section{Drawback of the fixed-order method}\label{sec:drawback}
\begin{figure}
  \begin{center}
    \includegraphics[width=.95\textwidth]{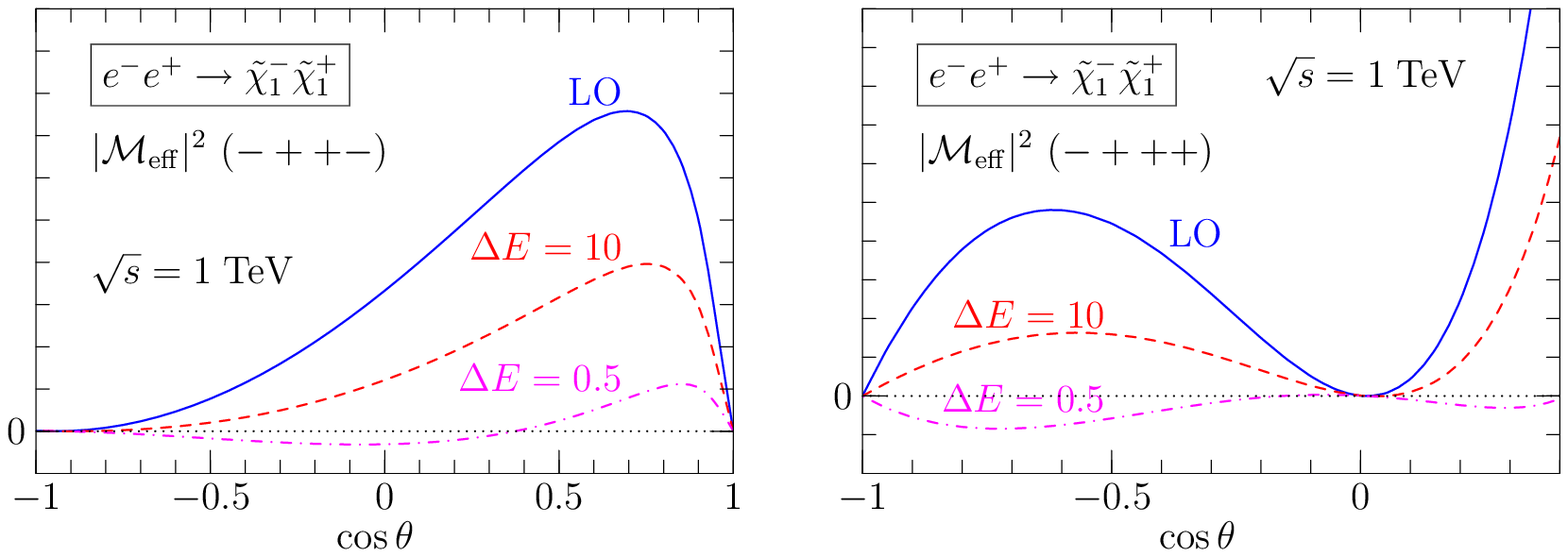}
  \end{center}
  \caption{Effective squared matrix element (arbitrary units) for
    $e^-e^+\to\cham\chap$ as a function of the polar scattering
    angle~$\theta$ at $\sqrt{s}=1\;\TeV$.  Left figure: Helicity
    combination $-++-$; right figure: $-+++$.  Solid line: Born term;
    dashed: including virtual and soft contributions for $\Delta
    E_\gamma=10\;\GeV$; dash-dots: same with $\Delta E_\gamma=0.5\;\GeV$.}
  \label{fig:twohelis}
\end{figure}

For any fixed helicity combination and chargino scattering angle
the differential cross section is positive if we include all contributions defined in Eq. (\ref{eq:sigex}). If the integration and simulation is split into a $2\,\rightarrow\,2$ and $2\,\rightarrow\,3$ part as implied by $\sigma_{tot}$ (\ref{eq:stot}), however, the fixed-order approach runs into the well-known problem of negative event
weights~\cite{Bohm:1986fg,Alexander:1987be, Kleiss:1989de}: The effective $2\,\rightarrow\,2$ matrix element in the soft-soft integration region,
\begin{\eqn}\label{eq:meff1}
|\M_{eff}|^{2}\,=\,(1+f_{soft})\,|\M_{Born}(s)|^{2}+2\,Re\,(\M_{Born}(s)\,\M^{*}_{1-loop}(s)), 
\end{\eqn} 
is no longer positive definite if $\Delta
E_\gamma$ becomes sufficiently small. If we lower the cutoff, this becomes negative within some range
of scattering angle. We will investigate this effect closer and concentrate on the dominant helicity contribution with
\begin{\eqn}\label{eq:domhel}
\lambda_{e^{+}}\,=\,1\;,\;\lambda_{e^{-}}\,=\,-1\;,\;\lambda_{\wt{\chi}^{-}}\,=\,1\;,\;\lambda_{\wt{\chi}^{+}}\,=\,-1,
\end{\eqn} 
and consider the $\lambda,\,\Delta E_{\gamma}$ dependencies of $|\M_{eff}|^{2}$.\\
In the effective squared matrix element (\ref{eq:meff1}), we have
\begin{\eqn*}
f_{soft}\,=\,f_{soft}(\lambda,\Delta E_{\gamma})\;;\;\M_{1-loop}\,=\,\M_{1-loop}(\lambda).
\end{\eqn*} 
The $\Delta E_{\gamma}$ dependence enters only in the soft factor
\begin{\eqn*}
f_{soft}\,\propto\,\ln \lb \frac{\Delta E_{\gamma}}{\lambda}\rb.
\end{\eqn*}
In the combination of virtual and soft photons in $\M_{eff}$, the $\log\lambda$ dependence cancels exactly.\\
As an example, we consider the angular behavior for a fixed helicity state with $\frac{\Delta E_{\gamma}}{\sqrt{s}}\,=\,0.005$ (high cut) and $\frac{\Delta E_{\gamma}}{\sqrt{s}}\,=\,0.0005$ (low cut), respectively. Figures \ref{fig:meffsq, de5} and \ref{fig:meffsq, de0.5} show all contributions to $|\M_{eff}|^{2}$ in these cases. $\M_{Born}$ and $\M_{1-loop}(\lambda)$ are $\Delta E_{\gamma}$ independent and therefore do not change (for a fixed value for $\lambda$). 
With the high cut, we obtain
%\begin{\eqn*}
$|\M_{eff}|^{2}\,>\,0$
%\end{\eqn*}
for all values of $\theta$, while for the lower cut,
%\begin{\eqn*}
$|\M_{eff}|^{2}\,<\,0$
%\end{\eqn*}
for  $\cos\theta\,\le\,0.4$. In these regions of phase space, $\frac{\Delta E_{\gamma}}{\sqrt{s}}$ is small enough such that the virtual photon contributions are not sufficiently canceled by soft real photon contributions. Figure \ref{fig:twohelis} shows the behavior of the $|\M_{eff}|^{2}$ for the helicity combination (\ref{eq:domhel}) as well as the subdominant contribution with $\lambda_{\wt{\chi}^{-}}\,=\,\lambda_{\wt{\chi}^{+}}\,=\,1$. \\
\begin{figure}
\begin{center}
\includegraphics[width=9cm]{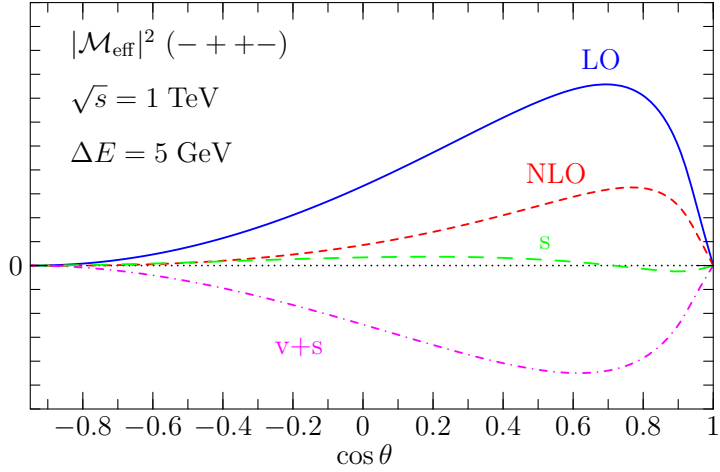}
\vspace{10mm}
\caption{$|\M_{eff}|^{2}$ contributions for $\frac{\Delta E_{\gamma}}{\sqrt{s}}\,=\,5\cdot\,10^{-3}$, LO: Born, NLO: $|\wt{\M}_{eff}|^{2}$, v denotes the virtual and s the  soft contribution}
\label{fig:meffsq, de5}
\end{center}
\end{figure}
\begin{figure}
\begin{center}
\includegraphics[width=9cm]{pmpmp2.eps}
\vspace{10mm}
\caption{$|\M_{eff}|^{2}$ contributions for $\frac{\Delta E_{\gamma}}{\sqrt{s}}\,=\,5\cdot\,10^{-4}$, LO: Born, NLO: $|\wt{\M}_{eff}|^{2}$, v denotes the virtual and s the  soft contribution}
\label{fig:meffsq, de0.5}
\end{center}
\end{figure}
\begin{figure}
\begin{center}
\includegraphics[width=10cm]{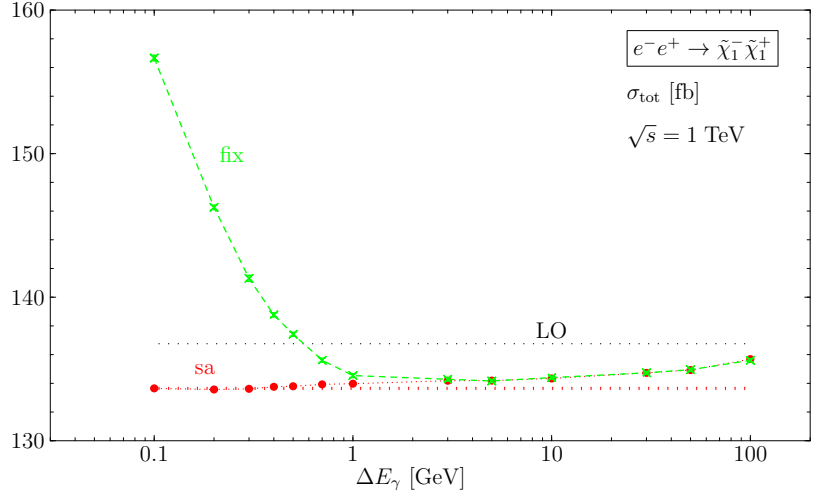}
\vspace{7mm}
\caption{ Total cross section as a function of the energy cutoff $\Delta
    E_\gamma$ using different calculational methods: (green, dashed) = fixed-order semianalytic result using
    \feynarts/\formcalc;  (red, dotted) = fixed-order
    Monte-Carlo result~ using \whizard; semianalytic result for lowest numerically reachable $\Delta E_{\gamma}$. Effects from setting $|\M_{eff}|^{2}\,=\,0$ where it becomes negative are visible for $\Delta E_{\gamma}/\sqrt{s}\,\leq\,10^{-3}$. 
    Statistical Monte-Carlo integration errors are shown.  For the
   Monte-Carlo results, the collinear cutoff has been fixed to
    $\Delta\theta_\gamma=1^\circ$. The rise of the fixed order result reflects the breakdown of the soft photon approximation; see Section \ref{sec:fixedres}.}
\label{fig:edepfixed}
\end{center}
\end{figure}
\begin{figure}
\begin{center}
\includegraphics[width=10cm]{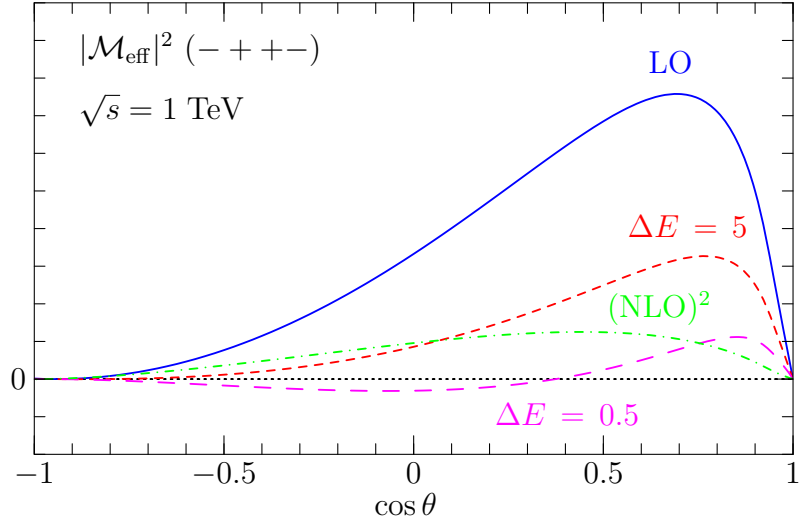}
\vspace{10mm}
\caption{Effective squared matrix element and squared NLO contribution $|\M_{v+s}|^{2}$  as a function of the polar scattering
    angle~$\theta$ at $\sqrt{s}=1\;\TeV$.  Helicity
    combination $-++-$; Solid line: Born term;
    dashed: including virtual and soft contributions for $\Delta
    E_\gamma=5\;\GeV$; long dashed: same with $\Delta E_\gamma=0.5\;\GeV$; dash-dotted: $|\M_{v+s}|^{2}$, $\lambda\,=\,5\,\cdot\,10^{-5}$  \GeV}
\label{fig:meffsq and m1lsq}
\end{center}
\end{figure}
If we insist on a positive weight Monte Carlo generator, an ad-hoc solution for the fixed $\mO(\al)$ contribution would now be to set $\M_{eff}^{2}\,=\,0$ in the respective regions of phase space. However, for too low soft cuts, this leads to wrong results for the total and differential cross section, cf. Figure \ref{fig:edepfixed}. An alternative approach uses
subtractions in the integrand to eliminate the singularities before
integration~\cite{Catani:1996jh,Catani:1996vz,Dittmaier:1999mb, Denner:1999gp,Roth:1999kk,Nagy:2003qn}. The subtracted pieces are integrated
analytically and added back or canceled against each other where
possible.  However, the subtracted integrands do not necessarily satisfy
positivity conditions either.\\
Alternatively, we can include negative event weights. Such event samples are not a
possible outcome of a physical experiment and imply further modification of the detector and analysis tools.  
\\
In general, experimental resolution at the ILC can well reach $\frac{\Delta E_{\gamma}}{\sqrt{s}}\,=\,10^{-3}$ and lower values. From Figure \ref{fig:meffsq and m1lsq}, we see that, for a fixed $\lambda$, the 1-loop contribution $|\M_{1-loop}|^{2}$ is nearly the same order of magnitude as $|\M_{eff}|^{2}$. However, the soft contribution here is treated with the infrared cutoff $\lambda$; setting
%\begin{\eqn*}
$\lambda\,\rightarrow\,0$
%\end{\eqn*}
will always lead to the well-known infrared divergence 
%\begin{\eqn*}
$\sigma_{soft}\,\rightarrow\,\infty$.
%\end{\eqn*}
In principle, this can be canceled by the terms describing the emission of two soft photons equally regulated with the photon mass $\lambda$. Therefore, if we want to construct an event generator reaching the experimental cut requirements, we should take second and higher-order contributions into account.\\
For low enough soft energy cuts, the problem of negative event weights and negative soft cross sections remains as long as only finite order photon emissions are taken into account (\cite{Kleiss:1989de} and references therein). This signals a breakdown of perturbation theory in this region of phase space and requires the inclusion of summed ISR contributions discussed in Section \ref{sec:strfunisr}. We will present a method to combine this with the fixed-order corrections $\sigma_{virt}$ in Chapter \ref{chap:resum}.
\section{Results}\label{sec:fixedres}
\subsection{Cut dependencies}\label{sec:excuts}
In the kinematical ranges below the soft and collinear cutoffs,
several approximations are made.  In particular, the method neglects
contributions proportional to positive powers of $\Delta E_\gamma$ and
$\Delta\theta_\gamma$, so the cutoffs must not be increased into the
region where these effects could become important.  On the other hand, when 
decreasing cutoffs too much we can enter a region where the limited
machine precision induces numerical instabilities.  Therefore, we have
to check the dependence of the total cross section as calculated by
adding all pieces and identify parameter ranges for $\Delta E_\gamma$
and $\Delta\theta_\gamma$ where the result is stable but does not
depend significantly on the cutoff values.\\
In the following, we will compare
\begin{enumerate}
\item{}{\bf fixed order:} The implementation of $\sigma_{tot}$ according to Eq. (\ref{eq:sigtotimp}) in \whizard~ with both soft ($\Delta E_{\gamma}$) and collinear ($\Delta \theta_{\gamma}$) cuts,
\item{}{\bf semianalytic:} the NLO calculation presented in \cite{Fritzsche:2004nf} using \formcalc~ in combination with the $2\,\rightarrow\,3$ part from \whizard~. Here, only a soft cut $\Delta E_{\gamma}$ is applied.
\end{enumerate}
Both programs use the same routine for the calculation of $\M_{virt}$ and $f_{soft}$ as well as the same SM and MSSM input parameters. Therefore, differences are due to the use of the collinear approximation and implementation differences described in Section \ref{sec:mcs}.\\
Throughout this section, we set the process energy to
$\sqrt{s}=1\;\TeV$ and refer to the SUSY parameter point SPS1a'.  All
$2\to 2$ and $2\to 3$ contributions are included, so the results would
be cutoff-independent if there were no approximations involved.
\subsubsection*{Energy cutoff dependence}
\begin{figure}
\begin{center}
\includegraphics[width = 10cm]{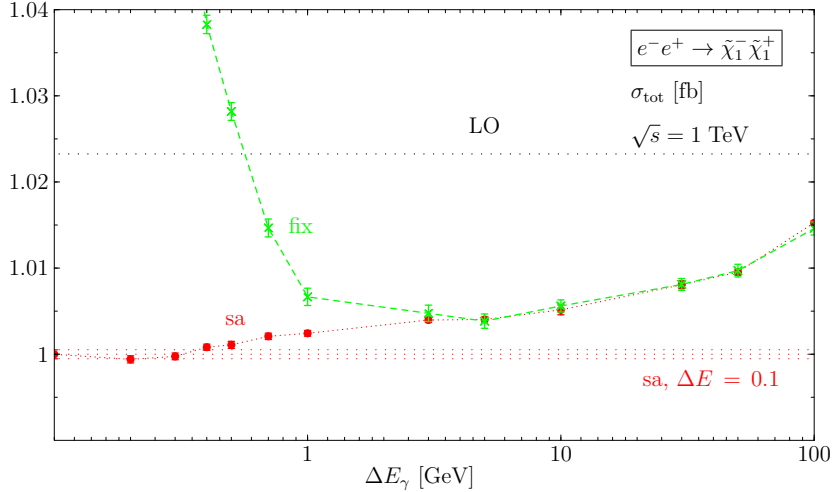}
\vspace{10mm}
\caption{\label{fig:edepfixedrel} 
Total cross section as a function of the energy cutoff $\Delta
    E_\gamma$ using different calculational methods: fixed-order semianalytic result;  fixed-order
    Monte-Carlo result~ using \whizard. Relative change with respect to $\sigma_{tot}$ for $\Delta E_{\gamma}\,=\,0.1\,\GeV$, no angular cut}
\end{center}
\end{figure}
In Figures \ref{fig:edepfixed} and \ref{fig:edepfixedrel}, we compare the numerical results obtained using
the semianalytic calculation with our Monte-Carlo
integration in the fixed-order scheme. The semianalytic result is not exactly
cutoff-independent. Instead, it exhibits
a slight rise of the calculated cross section with increasing cutoff;
for $\Delta E_\gamma=1\;\GeV$ ($10\;\GeV$) the shift is about
$2\,\permil$ ($5\,\permil$) of the total cross section, respectively.  This is an effect of the soft photon approximation, where the energy fraction $x$ of the incoming electron/ positron is set to 1 in the soft regime. Therefore, for $x\,\approx\,1$  the error is $\mO(\Delta E_{\gamma}/\sqrt{s})$ with respect to $\sigma_{Born}$, cf. Figure \ref{fig:edepfixedrel}.  For $\Delta E_{\gamma}/\sqrt{s}\,\le\,10^{-5}$, we run into numerical problems with the exact $2\,\rightarrow\,3$ process. Otherwise, the errors of the semianalytic calculation are in the per mille regime and smaller.\\
The fixed-order Monte-Carlo result agrees with the semianalytic
result, as it should be the case, as long as the cutoff is greater
than a few $\GeV$.  For smaller cutoff values the Monte-Carlo result
drastically departs from the semianalytic one because the virtual
correction exceeds the LO term there, and therefore the $2\to 2$
effective squared matrix element becomes negative in part of phase
space as discussed in Section \ref{sec:drawback}. There, the integrand is set to zero.

\subsubsection*{Collinear cutoff dependence}
\begin{figure}
\begin{center}
\includegraphics[width = 10cm]{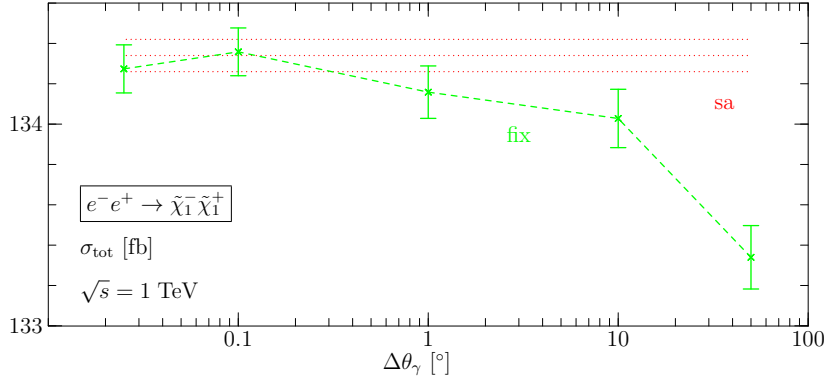}
\vspace{7mm}
\caption{\label{fig:thdepfixed} 
Total cross section dependence on the collinear cutoff
    $\Delta \theta_\gamma$ using different calculational
    methods: {\rm sa} (red, dotted) = fixed-order semianalytic result
     {\rm fix} (green, dashed) = fixed-order
    Monte-Carlo result $\sigma_{tot}$ (\ref{eq:sigex}); The soft cutoff has been fixed to $\Delta
    E_\gamma=10\;\GeV,\,\sqrt{s}\,=\,1\,\TeV$ }
\end{center}
\end{figure}

The collinear cutoff $\Delta\theta_\gamma$ separates the region where,
in the collinear approximation, higher-order radiation is resummed
from the region where only a single photon is included, but treated
with exact kinematics.  We show the dependence of the result on this cutoff in
Fig.~\ref{fig:thdepfixed}. We see that, for $\Delta\,\theta\,\leq\,1^{\circ}$, the collinear approximation holds. For  $\Delta\,\theta\,=\,1^{\circ}$, its errors are less than per mille. For larger $\Delta\theta$, the collinear approximation breaks down. Similar results are found in \cite{thomdiss,Oller:2005xg}. 

\subsubsection*{Photon mass dependence}
The infinitesimal photon mass $\lambda$ is used in the FormCalc matrix element calculation for the regularization of infrared divergencies. The effective matrix element $|\M_{eff}|^{2}$ (Eq. \ref{eq:meff1})
should then be independent of $\lambda$. Numerically, this has been tested for
\begin{\eqn*}
5\,\times\,10^{-5}\,\mbox{GeV}\,\le\,\lambda\,\le\,10^{10}\,\mbox{GeV}
\end{\eqn*}
for the FormCalc integration routines. In these regions, while the photon mass
remains a parameter in the matrix element code, the result does not
numerically depend on it, regardless which method has been used. It is of course more meaningful to choose small photon masses.
\subsection{Total cross section}
Fixing the cutoffs to $\Delta E_{\gamma}/\sqrt{s}\,=\,5\,\cdot\,10^{-3},\,\Delta\theta\,=\,1^{\circ}$, we can use the integration
part of the Monte-Carlo generator to evaluate the total cross section
at NLO for various energies. They exactly reproduce the semianalytic results modulo the $\Delta\,E_{\gamma}$-dependent $\permil$ errors discussed in Section \ref{sec:excuts}. For a discussion on the general behaviour of the fixed NLO cross section, cf. Chapter \ref{chap:NLO}. The numerical results qualitatively agree with \cite{Fritzsche:2004nf,Oller:2005xg}. However, we did not compare all SUSY parameters used here.
\subsection{Event generation}
\begin{figure}
\begin{center}
 \includegraphics[width=10cm]{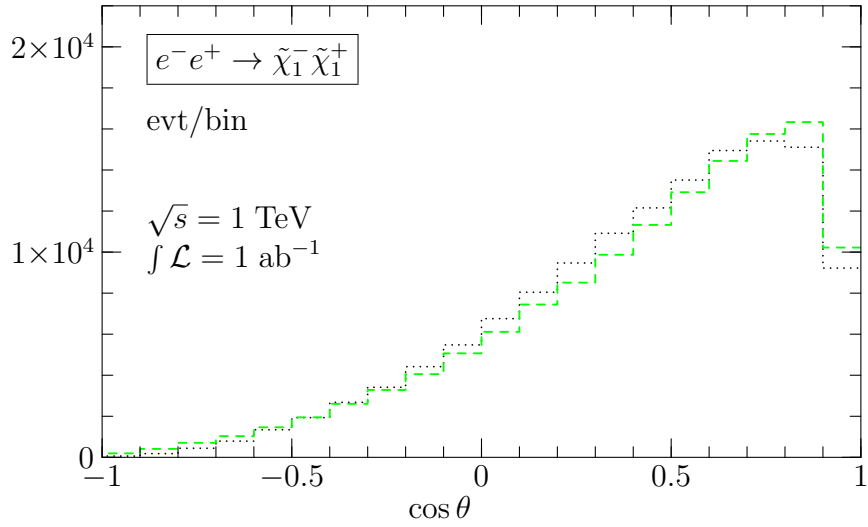}
\end{center}
\vspace{5mm}
  \caption{Polar scattering angle distribution for an integrated
    luminosity of $1\;\ab^{-1}$ at $\sqrt{s}=1\;\TeV$. Total
    number of events per bin, Born (dotted, black) and fixed order (dashed, green) result}
  \label{fig:histthex1}
\end{figure}

\begin{figure}
\begin{center}
\includegraphics[width=10cm]{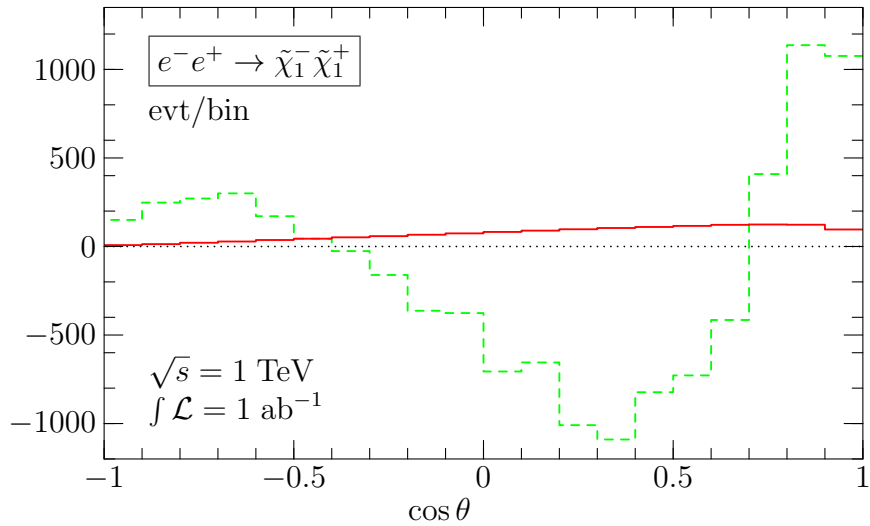}
\end{center}
\vspace{5mm}
  \caption{Polar scattering angle distribution for an integrated
    luminosity of $1\;\ab^{-1}$ at $\sqrt{s}=1\;\TeV$. 
    Number of events per bin, difference between fixed order and Born (dashed, green). For comparison, the standard deviation from the Born result is also shown (solid, red)}
  \label{fig:histthex2}
\end{figure}

The strength of the Monte-Carlo method lies not in the ability to
calculate total cross sections, but to simulate physical event
samples.  We have used the \whizard\ event generator augmented by the
effective matrix element (\ref{eq:meff1}) and structure function (\ref{eq:effstrfun}) to generate unweighted event samples for chargino production.\\
To evaluate the importance of the NLO improvement, in
Figures \ref{fig:histthex1} and \ref{fig:histthex2} we show the binned distribution of the chargino
production angle as obtained from a sample of unweighted events
corresponding to $1\;\ab^{-1}$ of integrated luminosity for $\sqrt{s}\,=\,1\;\TeV$ and the SUSY parameter point
SPS1a'. With cutoffs
$\Delta\theta_\gamma=1^\circ$ and $\Delta E_\gamma=3\;\GeV$ we are not
far from the expected experimental resolution, while for the
fixed-order approach negative event weights do not yet pose a problem.\\
The histograms illustrate the fact that NLO corrections in chargino
production are not just detectable, but rather important for an
accurate prediction, given the high ILC luminosity.  The correction
cannot be approximated by a constant proportionality factor K between the leading and next-to-leading order cross section ($K$-factor) but takes a different
shape than the LO distribution.  The correction is positive
in the forward and backward directions, but negative in the central
region. For comparison, Figure \ref{fig:histthex2} also shows the 1 $\sigma$ error from the Born result.
%\section{Exact method: summary}
\chapter{Resumming photons}\label{chap:resum}
\section{NLO and resummation at LEP and LHC}\label{sec:resumlep}
The principle of analytic photon resummation has already been discussed in Section \ref{sec:strfunisr}. We will here shortly sketch the implementation of photon resummation in Monte Carlo generators at LEP as well as NLO matrix elements for hadronic machines and refer to the literature \cite{Kleiss:1989de,Dobbs:2004qw} for more details.
\subsubsection*{Photon resummation at LEP}
The problem of negative event weights discussed in Section \ref{sec:drawback} was, at LEP times, known as the $k_{0}$ problem. When slicing the Monte Carlo generator phase space in a $2\,\rightarrow\,2$ (Born, virtual, soft) and a  $2\,\rightarrow\,3$ (hard) part as in the fixed order approach, event weights in the $2\,\rightarrow\,2$ part can become negative for $k_{0}\,\rightarrow\,0$. This holds true in any finite-order treatment of soft real and virtual photons. The solution is to sum photons up to an infinite order. In \cite{Kleiss:1989de}, different methods are proposed in order to include exponentiation in Monte Carlo event generators: ad hoc-exponentiation which estimates the energy loss due to photon radiation and modifies the Monte Carlo code accordingly, the use of (positive definite) structure functions convoluted with the Born cross section at a reduced cm energy, and YFS exponentiation \cite{Yennie:1961ad} which correctly simulates the hard part, while using a structure function for soft and virtual photons.  In Section \ref{sec:strfunisr} and in the following, we will use the structure-function method proposed by Jadach and Skrzypek \cite{Skrzypek:1990qs}. Note, however, that none of these approaches includes non-photonic NLO contributions as $\sigma_{virt}$ (Eq. \ref{eq:svirt}).
\subsubsection*{NLO and parton showers for LHC}
Recently, several computer codes attempt to include exact NLO matrix elements in Monte Carlo generators containing partonic showers, which are the QCD equivalent of the infinitely summed up photon contributions described above. These are basically directed to experiments at hadronic machines such as the LHC, cf. \cite{Catani:2001cc,Frixione:2002ik,Dobbs:2004qw,Frixione:2006he}. A similar code treating jet production at $e^{+}\,e^{-}$ colliders was proposed in \cite{Nagy:2005aa, Kramer:2005hw}. Basically, the NLO contributions are added by hand, while the corresponding first order parts from the parton showers are subtracted. In the following, we will pursue a similar approach for the inclusion of electroweak NLO corrections.
\section{Resummation method}\label{sec:resum}
The shortcomings of the fixed-order approach described in Chapter \ref{chap:fixed} are
associated with the soft-collinear region $E_\gamma<\Delta E_\gamma$,
$\theta_\gamma<\Delta\theta_\gamma$, where the appearance of double
logarithms $\frac{\alpha}{\pi}\ln\frac{E_\gamma^2}{s}\ln\theta_\gamma$
invalidates the perturbative series.  However, in that region
higher-order radiation can be resummed~\cite{Gribov:1972rt, Gribov:1972ri, Kuraev:1985hb}.  The
exponentiated structure function
$f_\text{ISR}$ (\ref{eq:f-ISR}) that resums initial-state radiation,
\begin{equation*}
  \sigma_\text{Born+ISR}(s,\Delta\theta_\gamma,m_e^2) = \int
  dx\,f_\text{ISR}(x;\Delta\theta_\gamma,\tfrac{m_e^2}{s}) \int
  d\Gamma_2\,|\ME_\text{Born}(xs)|^2,
\end{equation*}
includes photon radiation to all order in the soft regime at
leading-logarithmic approximation and, simultaneously, correctly
describes collinear radiation of up to three photons in the hard
regime.  It does not account for the helicity-flip part
$f^-$~(\ref{eq:hardcollfs}) of the fixed-order structure function; this may
either be added separately or just be dropped since it is subleading.\\
\\
We now combine the ISR-resummed LO result with the additional NLO
contributions $\sigma_{virt}$ given by Eq. (\ref{eq:svirt}).  To achieve this and avoid double counting, we
first subtract from the effective squared matrix element, for each
incoming particle, the contribution of one soft real and virtual collinear photon that is
contained in the ISR structure function. The soft photon has already been
accounted for in the soft-photon factor, while the virtual part is contained in the interference term.
Then
\begin{align}\label{eq:meffisr}
  |\widetilde\ME_\text{eff}(\hat{s};\Delta E_\gamma,\Delta\theta_\gamma,m_e^2)|^2
  &= \left[1 + f_\text{soft}(\tfrac{\Delta E_\gamma}{\lambda})
           - 2f_\text{soft,ISR}(\Delta E_\gamma,\Delta\theta_\gamma,\tfrac{m_e^2}{s})\right]
  \,|\ME_\text{Born}(\hat{s})|^2
\nonumber\\ &\quad
  + 2\mathrm{Re}\left[\ME_\text{Born}(\hat{s})\,
                      \ME^{*}_\text{1-loop}(\hat{s},\lambda^2,m_e^2)\right]
\end{align}
with $\hat{s}$ being the c.m.\ energy after radiation and $f_\text{soft,ISR}$ the integrated $\mO(\al)$ contribution of $f_{ISR}$.  This
expression contains the Born term, the virtual and soft-collinear
contribution with the leading-logarithmic part of virtual photons and
soft-collinear emission removed, and soft non-collinear radiation of
one photon; it still depends on the cutoff $\Delta E_\gamma$.  Convoluting
this with the resummed ISR structure function,
\begin{align}\label{eq:resummeds}
  \lefteqn{ \sigma_\text{v+s,ISR}(s,\Delta E_\gamma,\Delta\theta_\gamma,m_e^2)}
\nonumber\\
  &=
  \int dx_{1}\,f_\text{ISR}(x_{1};\Delta\theta_\gamma,\tfrac{m_e^2}{s})\,
  \int dx_{2}\,f_\text{ISR}(x_{2};\Delta\theta_\gamma,\tfrac{m_e^2}{s})
  \int d\Gamma_2\,
    |\widetilde\ME_\text{eff}(\hat s;\Delta E_\gamma,\Delta\theta_\gamma,m_e^2)|^2,
\end{align}
we obtain a modified $2\to 2$ part of the total cross section. \\
\\
In this description of the collinear region, there is no explicit
cutoff $\Delta E_\gamma$ involved, and collinear virtual photons
connected to at least one incoming particle are included. The cancellation of infrared singularities between
virtual and real corrections is built-in for collinear photons. The main source of negative event weights is eliminated, and we obtain a better behaviour for the integrand such that smaller energy cuts can be applied; cf. Section \ref{sec:noneg}. Furthermore, the method is exact in $\mO(\al)$ leading log terms. In the following, we will consider the description of one and more photons resulting from $\sigma_\text{v+s,ISR}$ in more detail.\\
\\
In the resummation method, the emission of real and virtual photons is described in various ways:
the soft approximation (cf. Sec. \ref{sec:softf}), initial state radiation (cf. Sec. \ref{sec:strfunisr}), virtual contribution from interference term (cf. Section \ref{sec:virtcorr}), and 
real emission given by exact (hard non-collinear) matrix element 
$\M_{2\,\rightarrow\,3}$ (cf. Sec \ref{sec:hardnoncoll}). We will now consider the respective description of photons resulting from the resummation method in different points of phase space. Here, we go to infinite order for the photon emission from one incoming particle and to second order for the simultaneous photon emission from two incoming particles.\\
\subsubsection*{Radiation off one incoming particle}
\begin{figure}
\begin{center}
\includegraphics{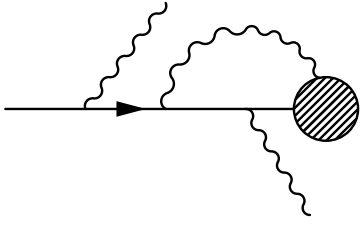}
\end{center}
\caption{\label{fig:mixedemi} Example for mixed real and virtual photon emissions described by $f_\text{ISR}$. All photons are collinear and $k_{\perp}$-ordered }
\end{figure}

We first consider photon radiation from one particle only and use the exponentiated electron structure function $f_{ISR}$ (\ref{eq:f-ISR}). In the following, we ignore the emission of the second photon contained in $f_{soft}$ and only subtract the contribution of $f_{ISR}$ for one photon. The corresponding effective matrix element for the radiation of only one photon is 
\begin{align}\label{eq:mtild1}
  |\widetilde\ME^{(1)}_\text{eff}(\hat{s};\Delta E_\gamma,\Delta\theta_\gamma,m_e^2)|^2
  &= \left[1 + f_\text{soft}(\tfrac{\Delta E_\gamma}{\lambda})
           - f_\text{soft,ISR}(\Delta E_\gamma,\Delta\theta_\gamma,\tfrac{m_e^2}{s})\right]
  \,|\ME_\text{Born}(\hat{s})|^2
\nonumber\\ &\quad
  + 2\mathrm{Re}\left[\ME_\text{Born}(\hat{s})\,
                      \ME^{*}_\text{1-loop}(\hat{s},\lambda^2,m_e^2)\right]
\end{align}
and the total cross section
\begin{\eqn}\label{eq:sigeff1}
  \sigma^{(1)}_\text{v+s,ISR}(s,\Delta E_\gamma,\Delta\theta_\gamma,m_e^2)
  \,=\,
  \int dx\,f_\text{ISR}(x;\Delta\theta_\gamma,\tfrac{m_e^2}{s})\,
  \int d\Gamma_2\,
    |\widetilde\ME^{(1)}_\text{eff}(x,s;\Delta E_\gamma,\Delta\theta_\gamma,m_e^2)|^2.
\end{\eqn}
According to the exponentiation principle discussed in Appendix \ref{app:expo}, we can decompose $(\int)\,f_{ISR}$ in factors $F^{(l)}, F_{l}$ such that 
\begin{eqnarray}\label{eq:fsisr}
\int^{1}_{x_{0}}\,D^{NS}(x)\,dx&=&\sum_{l=0}^{n}\,F^{(l)}(x_{0})+\mO(\eta^{n+1}),\nonumber\\
D^{NS}(x)&=&\sum_{l=0}^{n}F_{l}(x)+\mO(\eta^{n+1})\;\;\;x<x_{0},
\end{eqnarray}
where
%\begin{eqnarray*}
$F^{(n)}(x_{0})$ is the $\mO(\eta^{n})$ contribution to the integrated exact solution of the evolution equation (\ref{eq:llev}) in the soft limit and 
$F_{n}(x)$ is the exact $\mO(\eta^{n})$ perturbative solution to it. In the structure function $f_{ISR}$ (\ref{eq:f-ISR}) used in this work, $F_{n}$ has been calculated up to $n\,=\,3$. 
%\end{eqnarray*}
Using the scale $Q^{2}\,=\,(\Delta\theta\,p_{0})^{2}$, where $p_{0}$ is the energy of the electron, these factors exclusively describe collinear photons. We have
\begin{\eqn}\label{eq:isrFs}
F^{(0)}\,=\,1\;,\;F^{(1)}\,=\,f_{soft,ISR}\,;\,F_{0}\,=\,0.
\end{\eqn}
\\
$|\wt{\M}_{eff}|^{2}$ mixes different orders of $\eta$. For a fixed order $n$, the photon contributions to $\sigma^{(1)}_\text{v+s,ISR}$ are given by
\begin{eqnarray}\label{eq:1galln}
\sigma^{(1)}_{\gamma}&=&\,F^{(n)}(x_{0})\sigma_{Born}(s)+\int^{x_{0}}_{0}\,F_{n}(x)\,\sigma_{Born}(x,s)\,dx+F^{(n-1)}\hat{\sigma}(s)\nonumber\\
&&+\int^{x_{0}}_{0}\,F_{n-1}(x)\,\hat{\sigma}(x,s)dx
-F^{(1)}(x_{0})\,F^{(n-1)}(x_{0})\sigma_{Born}(s)\nonumber\\
&&-F^{(1)}(x_{0})\int^{x_{0}}_{0}\,F_{n-1}(x)\,\sigma_{Born}(x,s)dx,
\end{eqnarray}
where
\begin{\eqn*}
\hat{\sigma}\,=\,\int d\Gamma_2\,f_\text{soft}|\ME_\text{Born}(x,s)|^2\,+\,
   2\mathrm{Re}\left[\ME^{*}_\text{Born}(x,s)\,
                      \ME_\text{1-loop}(x,s)\right]
\end{\eqn*}
contains all contributions from virtual photon emission according to the exact $\M_{\text{1-loop}}$ calculation as well soft photon emission in the soft approximation. 
\\
   We will now consider the results for soft-collinear photons, distinguishing between multiple photon emissions where the last emitted photon does or does not obey the transverse momentum ordering $k_{\perp,n-1}\,\ll\,k_{\perp,n}$:\\
\begin{itemize}
%\\
\item{}$k_{\perp}$ ordering obeyed\\
\\
If the first $n-1$ emitted photons are soft (i.e. their total energy is lower than $\Delta E_{\gamma}$), this part of $\sigma_{\gamma}^{(1)}$ is given by
\begin{\eqn*}
F^{(n)}(x_{0})\sigma_{Born}(s)+F^{(n-1)}\hat{\sigma}(s)
-F^{(1)}(x_{0})\,F^{(n-1)}(x_{0})\sigma_{Born}(s)\,=\,F^{(n-1)}\hat{\sigma}(s),
\end{\eqn*}
and by
\begin{eqnarray*}
&&\int^{x_{0}}_{0}\,\lb F_{n}(x)-F^{(1)}(x_{0})\,F_{n-1}(x)\rb \,\sigma_{Born}(x,s)\,dx+\int^{x_{0}}_{0}\,F_{n-1}(x)\,\hat{\sigma}(x,s)dx\\
&&
\,=\,\int^{x_{0}}_{0}\,F_{n-1}(x)\,\hat{\sigma}(x,s)dx
\end{eqnarray*}
otherwise.
The last soft-collinear photon is described by the soft photon approximation, all others by $f_{ISR}$.\\
\item{}no $k_{\perp}$ ordering\\
\\
The terms $\propto\,F^{(n)},\,F_{n}$, which describe the emission of $n$ $k_{\perp}$-ordered photons, are missing in $\sigma^{(1)}_{\gamma}$. This results in
\begin{\eqn}\label{eq:nokt}
F^{(n-1)}\,\lb\hat{\sigma}(s)
-F^{(1)}(x_{0})\,\sigma_{Born}(s)\rb\,+\int^{x_{0}}_{0}\,F_{n-1}(x)\,\lb \hat{\sigma}(x,s)
-F^{(1)}(x_{0})\,\sigma_{Born}(x,s)\rb dx.
\end{\eqn}
We only obtain differences between the exact and the leading log contribution for the last radiated photon, multiplied with the $f_{ISR}$ contribution for the first $n-1$ photons. \\
\\
\end{itemize}  
If only hard-collinear photons are emitted, they are all described by $f_{ISR}$. For virtual photons, the same relations hold as for soft real photons: in the case of $k_{\perp}$ ordering of the last photon, it is given by the contribution to $M_\text{1-loop}$. Otherwise, the emission of the virtual last photon is again described by difference terms according to Eq. (\ref{eq:nokt}).\\
  Note that the radiation of a non-collinear soft photon as well as a non-collinear virtual photon is not touched by the subtraction mechanism. Here, the respective contribution to $\sigma_{\gamma}^{(1)}$ is simply
\begin{\eqn}\label{eq:1grest}
F^{(n-1)}\hat{\sigma}_{rest}(s)+\int^{x_{0}}_{0}\,F_{n-1}(x)\,\hat{\sigma}_{rest}(x,s)dx,
\end{\eqn}
where $\hat{\sigma}_{rest}$ contains all non-collinear virtual and non-photonic contributions.
\\
In general, $f_{ISR}$ describes a combination of real and virtual photon emissions, e.g. graphs given in Figure \ref{fig:mixedemi}, as the electron splitting function $P_{ee}$ (\ref{eq:elsplit}) combines the radiation of both a real and a virtual photon in order to cancel the IR divergence. Both soft real and virtual photons are contained in the soft contributions to the factors $F^{(n)}$ and $F_{n}$. For example, we can split $F^{(n)}$ according to
\begin{\eqn}\label{eq:splitsoftvirt}
F^{(n)}(x_{0})\,=\,(F^{(1)})^{n}\,=\,(F_{soft}+F_{virt})^{n}.
\end{\eqn}
This describes the emission of n collinear photons in an arbitrary combination of soft and virtual contributions. The only requirement is the $k_{\perp}$ ordering of their transverse momenta and $\sum_{\text{real $\gamma$}}\,E_{\gamma}\,\leq\,\Delta E_{\gamma}$.
\subsubsection*{Radiation off two incoming particles}
\begin{figure}
\begin{center}
\includegraphics[width=0.9\textwidth]{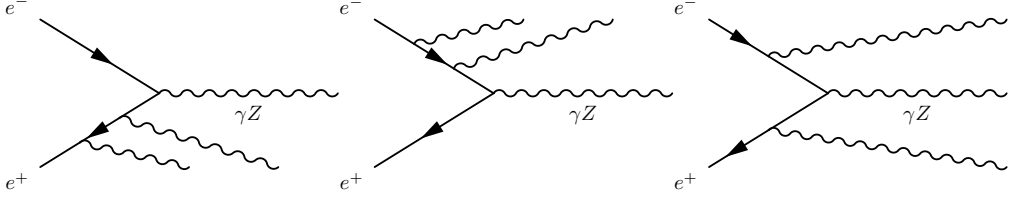}
\end{center}
\caption{\label{fig:emitwophotons} Diagrams contributing to the radiation of two photons, either from one incoming particle, or both incoming particles, for the $s$ channel $\gamma, Z$ exchange. Missing are diagrams from crossing symmetry of radiating photons. The $t$ channel $\tilde{\nu}$-exchange behaves analogously}
\end{figure}

Figure \ref{fig:emitwophotons} shows the diagrams contributing to the emission of two photons off two incoming particles for the $s$ channel $\gamma,\,Z$ exchange (we omitted diagrams resulting from the crossing symmetry for the radiated photons). As discussed in the last section, virtual photons are contained in the soft parts of $f_{ISR}$ (cf. Eq. (\ref{eq:splitsoftvirt})).
We now consider the  contributions to $\sigma_{virt+ISR}$ (\ref{eq:resummeds}) from up to two photon emissions. In the remainder of this section, $\sigma\,=\,\sigma_{Born}$.\\
\\
In general, the photonic $\mO(\al^{n})$ contributions to $\sigma_\text{v+s,ISR}$ are given by
\begin{eqnarray*}
&&\sum^{n}_{l=0}\,\lb \int^{x_{0}}_{0}\,dx_{1}\,\int^{x_{0}}_{0}\,dx_{2}\,F_{l}(x_{1})\,F_{n-l}(x_{2})\,\sigma(x_{1}\,x_{2}\,s)\,+\,\int^{x_{0}}_{0}\,dx_{1} F_{l}(x_{1})\,F^{(n-l)}(x_{2})\,\sigma(x_{1}\,s)\right.\\
&&\left.
+ \int^{x_{0}}_{0}\,dx_{2} F_{n-l}(x_{2})\,F^{(l)}(x_{1})\,\sigma(x_{2}\,s)\,+\,F^{(l)}(x_{1})\,F^{(n-l)}(x_{2})\,\sigma(s) \rb\\
&&+\sum^{n'\,=\,n-1}_{l=0}\,\lb \int^{x_{0}}_{0}\,dx_{1}\,\int^{x_{0}}_{0}\,dx_{2}\,F_{l}(x_{1})\,F_{n'-l}(x_{2})\,\lb \hat{\sigma}(x_{1}\,x_{2}\,s) - 2\,F^{(1)}\,\sigma(x_{1}\,x_{2}\,s)\rb \right.\\&&
\left. \,+\,\int^{x_{0}}_{0}\,dx_{1} F_{l}(x_{1})\,F^{(n'-l)}(x_{2})\,\lb \hat{\sigma}(x_{1}\,s) - 2\,F^{(1)}\,\sigma(x_{1}\,s)\rb \right.\\
&&\left. \,+\,\int^{x_{0}}_{0}\,dx_{2} F_{n'-l}(x_{2})\,F^{(l)}(x_{1})\,\lb \hat{\sigma}(x_{2}\,s) - 2\,F^{(1)}\,\sigma(x_{2}\,s)\rb \right.\\
&& \left. +\,F^{(l)}(x_{1})\,F^{(n'-l)}(x_{2})\,\lb \hat{\sigma}(s) - 2\,F^{(1)}\,\sigma(s)\rb \rb.
\end{eqnarray*}
\\
\\
For $\boldsymbol{n\,=\,0}$, we obtain (recall Eq (\ref{eq:isrFs})) the Born contribution (\ref{eq:sborn}):
\begin{eqnarray*}
&&\int^{x_{0}}_{0}\,dx_{1}\,\int^{x_{0}}_{0}\,dx_{2}\,F_{0}(x_{1})\,F_{0}(x_{2})\,\sigma(x_{1}\,x_{2}\,s)\,+\,\int^{x_{0}}_{0}\,dx_{1} F_{0}(x_{1})\,F^{(0)}(x_{2})\,\sigma(x_{1}\,s)\\
&&
+ \int^{x_{0}}_{0}\,dx_{2} F_{0}(x_{2})\,F^{(0)}(x_{1})\,\sigma(x_{2}\,s)\,+\,F^{(0)}(x_{1})\,F^{(0)}(x_{2})\,\sigma(s)\\
&&\,=\,\sigma(s).
\end{eqnarray*}
\\
 Setting $\boldsymbol{n\,=\,1}$ gives the additional terms
\begin{eqnarray*}
&& \lb \int^{x_{0}}_{0}\,dx_{1}\,\int^{x_{0}}_{0}\,dx_{2}\,F_{1}(x_{1})\,F_{0}(x_{2})\,\sigma(x_{1}\,x_{2}\,s)\,+\,\int^{x_{0}}_{0}\,dx_{1} F_{1}(x_{1})\,F^{(0)}(x_{2})\,\sigma(x_{1}\,s) \right. \\
&&
\left. \,+\,\int^{x_{0}}_{0}\,dx_{1} F_{0}(x_{1})\,F^{(1)}(x_{2})\,\sigma(x_{1}\,s) \,+\,F^{(1)}(x_{1})\,F^{(0)}(x_{2})\,\sigma(s)\rb + \lb x_{1}\,\leftrightarrow\,x_{2} \rb \\
&&+\int^{x_{0}}_{0}\,dx_{1}\,\int^{x_{0}}_{0}\,dx_{2}\,F_{0}(x_{1})\,F_{0}(x_{2})\,\lb \hat{\sigma}(x_{1}\,x_{2}\,s) - 2\,F^{(1)}\,\sigma(x_{1}\,x_{2}\,s)\rb \\&&
\,+\,\int^{x_{0}}_{0}\,dx_{1} F_{0}(x_{1})\,F^{(0)}(x_{2})\,\lb \hat{\sigma}(x_{1}\,s) - 2\,F^{(1)}\,\sigma(x_{1}\,s) \rb\\
&&\,+\,\int^{x_{0}}_{0}\,dx_{2} F_{0}(x_{2})\,F^{(0)}(x_{1})\,\lb \hat{\sigma}(x_{2}\,s) - 2\,F^{(1)}\,\sigma(x_{2}\,s)\rb \\
&& +\,F^{(0)}(x_{1})\,F^{(0)}(x_{2})\,\lb \hat{\sigma}(s) - 2\,F^{(1)}\,\sigma(s)\rb \\
&&\,=\,\int^{x_{0}}_{0}\,dx_{1} F_{1}(x_{1})\,\sigma(x_{1}\,s)\,+\,\int^{x_{0}}_{0}\,dx_{2} F_{1}(x_{2})\,\sigma(x_{2}\,s)\,+\,\hat{\sigma}(s),
\end{eqnarray*}
where we used Eq. (\ref{eq:isrFs}), the fact that $F^{(1)}$ only depends on $x_{0}$ such that
%\begin{\eqn*}
$F^{(1)}(x)\,=\,F^{(1)}(x_{0})$,
%\end{\eqn*}
 and 
\begin{\eqn}\label{eq:f1split}
-2\,F^{(1)}\,=\,-F^{(1)}(x_{1})\,-\,F^{(1)}(x_{2}).
\end{\eqn}
for the subtraction terms in $|\M_{eff}|^{2}$.\\
\\
We see that the $n\,=\,1$ term corresponds exactly to the $2\,\rightarrow\,2$ part of the leading log $\mO(\al)$ contribution of the fixed order calculation (\ref{eq:sigex}).\\ 
\\
For $\boldsymbol{n\,=\,2}$, we obtain the additional terms
\begin{eqnarray}\label{eq:2gcont}
&& \lb F^{(2)}(x_{2})\,\sigma(s) \,+\, \int^{x_{0}}_{0}\,dx_{2} F_{2}(x_{2})\,\sigma(x_{2}\,s)\rb \,+\,\lb x_{1}\,\leftrightarrow\,x_{2}\rb \nonumber\\
&&\,+\, \int^{x_{0}}_{0}\,dx_{1}\,\int^{x_{0}}_{0}\,dx_{2}\,F_{1}(x_{1})\,F_{1}(x_{2})\,\sigma(x_{1}\,x_{2}\,s)\,+\,\int^{x_{0}}_{0}\,dx_{1} F_{1}(x_{1})\,F^{(1)}(x_{2})\,\sigma(x_{1}\,s)\nonumber\\
&&
+ \int^{x_{0}}_{0}\,dx_{2} F_{1}(x_{2})\,F^{(1)}(x_{1})\,\sigma(x_{2}\,s)\,+\,F^{(1)}(x_{1})\,F^{(1)}(x_{2})\,\sigma(s) \nonumber \\
&&\lb \,+\,\int^{x_{0}}_{0}\,dx_{2} F_{1}(x_{2})\,\lb \hat{\sigma}(x_{2}\,s) - 2\,F^{(1)}\,\sigma(x_{2}\,s)\rb \right.\nonumber\\
&& \left. +\,F^{(1)}(x_{2})\,\lb \hat{\sigma}(s) - 2\,F^{(1)}\,\sigma(s)\rb \rb\,+\lb x_{1}\,\leftrightarrow\,x_{2}\rb,
\end{eqnarray}
where we again used Eq. (\ref{eq:isrFs}). To really understand the Feynman diagrams to which the different contributions in (\ref{eq:2gcont}) correspond and the cancellations, we have to take Eq. (\ref{eq:f1split}) into account.
We first consider the case of two photons radiating off the same incoming particle, e.g. the one depending on $x_{1}$. The relevant contributions are given by
\begin{eqnarray}\label{eq:2gsamep}
&& F^{(2)}(x_{1})\,\sigma(s) \,+\, \int^{x_{0}}_{0}\,dx_{1} F_{2}(x_{1})\,\sigma(x_{1}\,s)\nonumber\\
&&\,+\,\int^{x_{0}}_{0}\,dx_{1} F_{1}(x_{1})\,\lb \hat{\sigma}(x_{1}\,s) - \,F^{(1)}(x_{1})\,\sigma(x_{1}\,s)\rb 
\, +\,F^{(1)}(x_{1})\,\lb \hat{\sigma}(s) - F^{(1)}(x_{1})\,\sigma(s)\rb. \nonumber\\
&&
\end{eqnarray}
A comparison shows that this exactly corresponds to the $n\,=\,2$ case of only 1 particle radiating off photons (cf. Eq. (\ref{eq:1galln})). Therefore, the description of photon radiation given for this case also applies here.\\
\\
The mixed case is more complicated. The relevant terms are given by
\begin{eqnarray}\label{eq:2g2p}
&&\int^{x_{0}}_{0}\,dx_{1}\,\int^{x_{0}}_{0}\,dx_{2}\,F_{1}(x_{1})\,F_{1}(x_{2})\,\sigma(x_{1}\,x_{2}\,s)\,+\,\int^{x_{0}}_{0}\,dx_{1} F_{1}(x_{1})\,F^{(1)}(x_{2})\,\sigma(x_{1}\,s)\nonumber\\
&&
+ \int^{x_{0}}_{0}\,dx_{2} F_{1}(x_{2})\,F^{(1)}(x_{1})\,\sigma(x_{2}\,s)\,+\,F^{(1)}(x_{1})\,F^{(1)}(x_{2})\,\sigma(s)\nonumber  \\
&&\,+\,\int^{x_{0}}_{0}\,dx_{2} F_{1}(x_{2})\,\lb \hat{\sigma}(x_{2}\,s) - F^{(1)}(x_{1})\,\sigma(x_{2}\,s)\rb \,+\,F^{(1)}(x_{2})\,\lb \hat{\sigma}(s) - F^{(1)}(x_{1})\,\sigma(s)\rb 
\nonumber\\&&\,+\,\int^{x_{0}}_{0}\,dx_{1} F_{1}(x_{1})\,\lb \hat{\sigma}(x_{1}\,s) - F^{(1)}(x_{2})\,\sigma(x_{1}\,s)\rb \,+\,F^{(1)}(x_{1})\,\lb \hat{\sigma}(s) - F^{(1)}(x_{2})\,\sigma(s)\rb. \nonumber\\
\end{eqnarray}
If at least one of the radiated photons is hard, the relevant terms are given by (for e.g. $x_{1}$ hard and $x_{2}$ hard or soft):
\begin{eqnarray*}
&&\int^{x_{0}}_{0}\,dx_{1}\,\int^{x_{0}}_{0}\,dx_{2}\,F_{1}(x_{1})\,F_{1}(x_{2})\,\sigma(x_{1}\,x_{2}\,s)\,+\,\int^{x_{0}}_{0}\,dx_{1} F_{1}(x_{1})\,F^{(1)}(x_{2})\,\sigma(x_{1}\,s)\\
&&\,+\,\int^{x_{0}}_{0}\,dx_{1} F_{1}(x_{1})\,\lb \hat{\sigma}(x_{1}\,s) - F^{(1)}(x_{2})\,\sigma(x_{1}\,s)\rb \\
&&\,=\,\int^{x_{0}}_{0}\,dx_{1}\,\int^{x_{0}}_{0}\,dx_{2}\,F_{1}(x_{1})\,F_{1}(x_{2})\,\sigma(x_{1}\,x_{2}\,s)\,+\,\int^{x_{0}}_{0}\,dx_{1} F_{1}(x_{1})\, \hat{\sigma}(x_{1}\,s),
\end{eqnarray*}
i.e. we obtain the usual description (hard photon: leading log, soft photon: soft approximation, virtual photon: one-loop description from interference term). However, if both photons are soft, we have to closer investigate the phase space slicing. The soft-soft terms are given by
\begin{\eqn}\label{eq:softsoft}
F^{(1)}(x_{2})\,\hat{\sigma}(s)\,+\,F^{(1)}(x_{1})\,\hat{\sigma}(s)\,-\,F^{(1)}(x_{1})\,\,F^{(1)}(x_{2})\,\sigma(s).
\end{\eqn}
We now split $\hat{\sigma}$ into
\begin{\eqn}\label{eq:sigsplit}
\hat{\sigma}\,=\,F^{(1)}\,\sigma\,+\Delta\,F^{(1)}\sigma\,+\,\hat{\sigma}_{rest},
\end{\eqn}
where $\hat{\sigma}_{rest}$ contains all non-collinear virtual and non-photonic (ie, weak and SUSY) contributions, and $\Delta F^{(1)}$ is the difference between the soft approximation $f_{soft}$ (\ref{eq:f-soft}) and the first order contributions of the integrated leading log structure function $f_{soft,ISR}$ (\ref{eq:fsoftisr}):
\begin{\eqn}\label{eq:deltaf}
\Delta F^{(1)}\,=\,f_{soft}^{1\gamma}-f_{soft,ISR}.
\end{\eqn}
 In an ideal case, we would have
\begin{\eqn*}
(F^{(1)}+\Delta\,F^{(1)})(x_{1})\,(F^{(1)}+\Delta\,F^{(1)})(x_{2})\,\sigma.
\end{\eqn*}
Up to $\mO\lb(\Delta\,F^{(1)})^{2}\rb$ terms, this corresponds to eq. (\ref{eq:softsoft}). In this accuracy, both photons are described by $f_{soft}$. The same holds for virtual-soft and virtual-virtual emissions, where the virtual part is (up to similar corrections) described by the interference term.
\\
Finally, we have to consider whether these contributions correspond to the photon-photon diagrams as given in Figure \ref{fig:emitwophotons}. We obtain 6 diagrams (for each diagram, there is one with the outgoing photons crossed). However, as we have two indistinguishable photons in the final state, we obtain an additional factor $\frac{1}{2}$ so that the approximation above directly corresponds to the contributions we expect for the process $e^{+}\,e^{-}\,\longrightarrow\,\wt{\chi}\,\wt{\chi}\,\gamma\,\gamma$. The same holds of course for any convoluting of the Born cross section with $\int\,dx_{1}\,\int\,dx_{2}\,D^{NS}(x_{1})\,D^{NS}(x_{2})$.\\
\\
As in the case of only one particle emitting photons (Eq. (\ref{eq:1grest})), (\ref{eq:resummeds}) also includes collinear photonic
corrections to the Born/one-loop interference $\hat{\sigma}_{rest}$ in leading log accuracy. The corresponding convolution with $f_{ISR}$ is completely unaffected by the subtraction mechanism, which only accounts for the soft parts.\\
\\
 The complete result is supplemented by
the $2\to 3$ part,
\begin{align}\label{eq:sresummnog}
   \sigma_\text{tot,ISR}(s,m_e^2) &= \sigma_\text{v+s,ISR}
  +
  \int_{\Delta E_\gamma,\Delta\theta_\gamma} d\Gamma_3\,|\ME_{2\to 3}(s)|^2.
\end{align}
A final improvement is to also convolute the $2\to 3$ part with
the ISR structure function which defines
\begin{align}\label{eq:sresumm}
  \lefteqn{ \sigma_\text{tot,ISR+}(s,m_e^2)}
\nonumber\\
  &=
  \int dx_{1}\,f_\text{ISR}(x_{1};\Delta\theta_\gamma,\tfrac{m_e^2}{s})\,
  \int dx_{2}\,f_\text{ISR}(x_{2};\Delta\theta_\gamma,\tfrac{m_e^2}{s})\,
\nonumber\\
  &\quad\times
  \left(\int d\Gamma_2\,
    |\widetilde\ME_\text{eff}(\hat s;\Delta E_\gamma,\Delta\theta_\gamma,m_e^2)|^2
  + 
  \int_{\Delta E_\gamma,\Delta\theta_\gamma} d\Gamma_3\,|\ME_{2\to 3}(\hat s)|^2
  \right).
\end{align}
This introduces another set of higher-order corrections, namely those
where after an arbitrary number of collinear photons, one hard
non-collinear photon is emitted.  This additional resummation does not
double-count.  It catches logarithmic higher-order contributions where
ordering in transverse momentum can be applied. Other,
logarithmically subleading contributions are missed. In $\mO(\al)$, $\sigma_{tot}$ (\ref{eq:sigex}) is exactly reproduced by $\sigma_{tot,ISR}$; only the helicity flip-part of the hard-collinear radiation is not taken into account (which can become important for completely polarized initial particle states). For $n\,=\,2$, $\sigma_{tot,ISR}$ now also contains diagrams where the last radiated photon is hard, non-collinear, and obeying or not obeying the strong $k_{\perp}$ ordering.\\
\subsubsection*{Leading and higher-orders: summary}
We can therefore summarize that $\sigma_{tot,ISR+}$ (\ref{eq:sresumm}) reproduces $\sigma_{tot}$ (\ref{eq:sigex}) up to $\mO(\al)$. For terms $\mO(\al^{2})$ and higher, $\sigma_{tot,ISR+}$ contains all contributions of $\hat{\sigma}_{rest}$ convoluted with $f_{ISR}$ for both incoming particles, i.e. the radiation of infinitely summed up soft or virtual collinear photons and up to 3 hard-collinear photons off all non-photonic and non-collinear virtual contributions of the interference term. The same holds if the last radiated photon is non-collinear and hard, as the contribution is then given by the $2\,\rightarrow\,3$ part of (\ref{eq:sresumm}). For the emission of a last soft or virtual collinear photon, we carefully have to check the contributions resulting from the subtraction. For $\mO(\al^{2})$, at least one of the
photons is always described by the ISR structure function. But when
the Born term is convoluted with the ISR function, there are also two-photon
contributions described solely by the ISR. We have to distinguish
between the cases where (i) the two photons are attached to the same
or (ii) to different incoming particles.\\
 In case (i), we consider the
three terms (cf. Eq. (\ref{eq:2gsamep}))
\begin{equation}\label{caseone}
  \mO(\al^{2})_\text{ISR} \, - \, \mO(\al)_\text{ISR}\,\mO(\al)^\text{soft}_\text{ISR} \,
  + \, \mO(\al)_\text{ISR}\,\mO(\al)_\text{soft}.   
\end{equation}
The first term contains all pairs of collinear photons from the ISR,
$k_\perp$-ordered; the last term contains a first photon from ISR and a
second one from the soft-photon factor or the interference term. The
term in the middle is the subtraction to avoid double-counting of soft
photons. Here both photons
are from the ISR, the first one with arbitrary energy, the second one
real soft or virtual.  

If the second of the considered photons is real soft or virtual, and both are
$k_{\perp}$-ordered, then there is an exact cancellation between the
first two terms. For non $k_\perp$-ordered photons, the first term
gives no contribution, and there is a cancellation between the second
and third term, which results in a difference between the soft approximation/ interference term contribution expression and the ISR leading logarithmic approximation term given by $\Delta F^{(1)}$ (\ref{eq:deltaf}).\\ 
In the case (ii), we write the terms schematically as (cf. Eq. (\ref{eq:2g2p}))
\begin{equation*}
%  \label{casetwo}
  \mO(\al)_{1,\text{ISR}} \mO(\al)_{2,\text{ISR}} +
  \mO(\al)_{1,\text{ISR}}\,\lb\mO(\al)_{2,\text{soft}} - 
  \mO(\al)_{2,\text{ISR}}^\text{soft}\rb + \lb\mO(\al)_{1,\text{soft}} -
  \mO(\al)^\text{soft}_{1,\text{ISR}}\rb\mO(\al)_{2,\text{ISR}}.
\end{equation*}
Since there are always two different structure functions
involved, $k_\perp$-ordering is absent, and after a cancellation of
soft terms one is left with
\begin{equation*}
  \Delta F^{(1)}(x_{1})  \mO(\al)_{2,\text{ISR}} + \mO(\al)_{1,\text{ISR}} \Delta F^{(1)}(x_{2}) +
  \mO(\al)_{1,\text{ISR}} \mO(\al)_{2,\text{ISR}},
\end{equation*}
which is up to the missing terms $\Delta F^{(1)}(x_{1})\,  \Delta F^{(1)}(x_{2})$ equivalent to a soft approximation/ interference term description for both legs.
\subsubsection*{Remark: Soft approximation for matrix elements}
In the soft approximation as well as the exponentiation in structure functions, it is assumed that
\begin{\eqn}\label{eq:sigsofta}
\M(x\,\geq\,x_{0})\,\approx\,\M(x\,=\,1)\;,\;\sigma(x\,\geq\,x_{0})\,\approx\,\sigma(x\,=\,1)
\end{\eqn}
and therefore (cf. Eq. (\ref{eq:fsisr}))
\begin{\eqn}\label{eq:msoftapprox}
\int^{1}_{x_{0}}dx\,f_{ISR}(x)\,\sigma(x,s)\,\approx\,\lb \int^{1}_{x_{0}}dx\,f_{ISR}(x)\rb\,\sigma(s)\,=\,(1+F^{(1)})\,\sigma(s)\,+\,\mO(\eta^{2})
\end{\eqn}  
for the exponentiation; $f_{soft}$ is derived similarly.
This holds true up to errors proportional to $\int^{1}_{x_{0}}\,dx\,(\partial \sigma/\partial x)_{x\,=\,1}\,f_{ISR}(x)\,(1-x)$. Some of these contributions, however, are accounted for in the actual \whizard~ implementation of the resummation method; cf. Section \ref{sec:resumimp}.\\
  To take them into account, we have to substitute $f_{soft}\sigma(s')$ by
\begin{\eqn}\label{eq:msofteff}
f_{soft}\sigma(s')+\lb\int^{1}_{x_{0}}\,f_{ISR}(x)\,\sigma(x,s')\rb_{\mO(\al)}\,-\,f_{soft,ISR}\,\sigma(s')
\end{\eqn}
for any $s'$ in all expressions. The last two terms only cancel up to $\mO \lb \frac{\partial \sigma}{\partial x} |_{x=1}\,\Delta x \rb$. 
Equally, $F^{(l)}$ should be read as
\begin{\eqn}\label{eq:flreal}
F^{(l)}\,\sigma(s)\,\equiv\,\lb\int^{1}_{x_{0}}\,f_{ISR}(x)\,\sigma(x,s)\rb_{\mO(\al^{l})}
\end{\eqn}
for all higher-order terms.
\section{Implementation in \whizard~}\label{sec:resumimp}
The implementation of $\sigma_{tot,ISR+}$ in the Monte Carlo event generator \whizard~ is similar to the implementation of the fixed $\mO(\al)$ correction described in Section \ref{sec:fixinclusion}. We use $f_{ISR}$ (\ref{eq:f-ISR}) as a user-defined structure function for each incoming beam with the scale $Q\,=\,p_{0}\,\cos\theta$ such that only collinear photons are described (cf. Section \ref{sec:conlog}). We then integrate over the $x_{1},\,x_{2}$ dependent effective matrix element $\wt{\M}_{eff}$ (\ref{eq:meffisr}) according to Eq. (\ref{eq:resummeds}). Note that, in contrast to the fixed order method, there is no $(x_{1}, x_{2})$ phase space slicing involved. Therefore, in the integration over the $f_{ISR}$ soft region, the $x$-dependence of the matrix element is actually taken into account, so that in the code implementation additional contributions as given in (\ref{eq:msofteff}) and (\ref{eq:flreal}) appear. Note, however, that $f_{soft}$ (\ref{eq:f-soft}) in $\wt{\M}_{eff}$ still sets $x\,=\,1$ throughout the soft region. The effects of this can be seen in the $\Delta E_{\gamma}$ dependence of $\sigma_{tot,ISR(+)}$ as discussed in Chapter \ref{chap:resumres}.\\
\\
The $2\,\rightarrow\,3$ part of $\sigma_{tot,ISR+}$,
\begin{\eqn*}
\int dx_{1}\,f_\text{ISR}(x_{1};\Delta\theta_\gamma,\tfrac{m_e^2}{s})\,
  \int dx_{2}\,f_\text{ISR}(x_{2};\Delta\theta_\gamma,\tfrac{m_e^2}{s})\,
  \times
    \int_{\Delta E_\gamma,\Delta\theta_\gamma} d\Gamma_3\,|\ME_{2\to 3}(\hat s)|^2
\end{\eqn*}
 is again generated in a separate run of \whizard~, applying the soft and collinear cuts $\Delta E_{\gamma},\,\Delta \theta_{\gamma}$ and convoluted with the structure function $f_{ISR}$ for each incoming beam. The scale is set to $Q\,=\,p_{0}\,\cos\theta$ as given above. As in the fixed order method, both runs can be combined for integration and event simulation.
\section{Difference to fixed order method}
\subsection{Negative weights}\label{sec:noneg}
\begin{figure}
\begin{center}
  \includegraphics[width=.95\textwidth]{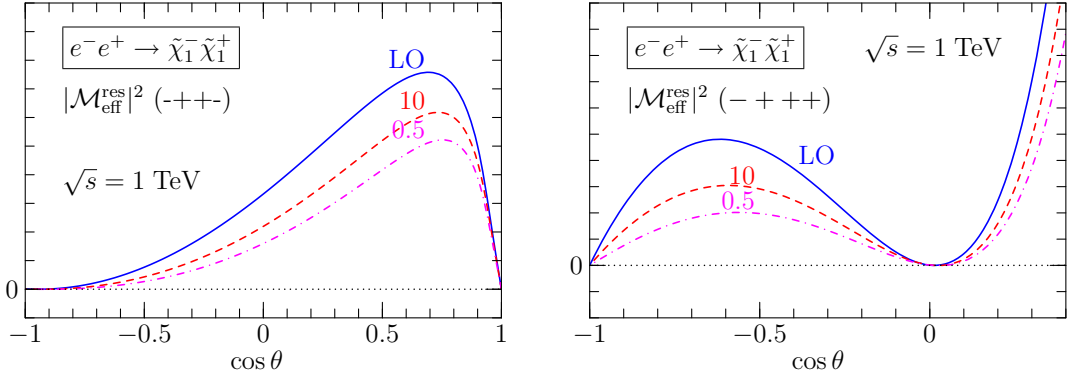}
\end{center}
\caption{Effective squared matrix element (arbitrary units) with
  the one-photon ISR part subtracted, for $e^-e^+\to\cham\chap$ as a
  function of the polar scattering angle~$\theta$ at
  $\sqrt{s}=1\;\TeV$.  Left figure: Helicity combination $-++-$; right
  figure: $-+++$.  Solid line: Born term; dashed: including virtual
  and soft contributions for $\Delta E_\gamma=10\;\GeV$; dotted: same with
  $\Delta E_\gamma=0.5\;\GeV$.  The collinear cutoff is fixed at
  $\Delta\theta_\gamma=1^\circ$.}
\label{fig:Meff-ISR}
\end{figure}

The resummation
approach does eliminate the problem of negative weights: shifting the
energy cutoff below the experimental resolution, such that photons are explicitly generated
whenever they can be resolved, the subtracted effective squared $2\to
2$ matrix element is still positive semidefinite in the whole phase
space.  
After resummation, the only potentially remaining source of negative
event weights is the soft-noncollinear region. Negative weights are
absent as long as 
\begin{equation*}
   \mathcal{O}(1) \times \frac{\alpha}{\pi} \ln \frac{(\Delta
   E_\gamma)^2}{s} \ln (\Delta \theta_\gamma)\,<\,1,
\end{equation*}
where the $\mathcal{O}(1)$ prefactor depends
on the specific process. For the cutoff and parameter ranges we
are considering here, this condition is fulfilled.\\
Figure \ref{fig:Meff-ISR} shows $|\wt{\M}_{eff}|^{2}$ for two different $\Delta E_{\gamma}$ cuts. We see that the positivity condition is fulfilled. Since neither the inclusion of the ISR structure function nor
the addition of the $2\to 3$ part introduces further sources of
negative weights, unweighting of generated events is now possible, so
this method allows for realistic simulation at NLO.
\subsection{Higher-order photon contributions}\label{sec:higho}
As the fixed order method discussed in Chapter \ref{chap:fixed} does not include higher-order contributions, the resummation method differs in all higher-order terms discussed in Section \ref{sec:resum}. Below, we give the analytic expression for these differences. In addition, we discuss the difference between the resummation method and the inclusion of ISR as discussed in Section \ref{sec:strfunisr}. Throughout the section, we neglect the contributions coming from the respective $2\,\rightarrow\,3$ integrations, i.e. processes where the last photon is hard and non-collinear. Here, differences between the methods are obvious from Section \ref{sec:resum}.\\
\subsubsection*{Radiation off one particle}
In this case, we again consider $\wt{\M}_{eff}^{(1)}$ (\ref{eq:mtild1}) and $\sigma_{v+s, ISR}^{(1)}$ (\ref{eq:sigeff1}), i.e. only subtract one factor of $f_{soft,ISR}$. We consider the soft and the hard region separately and again assume that, in the matrix element, $x$ can be set to 1 in the soft region (\ref{eq:sigsofta}).\\
\\
In the soft sector, we obtain
\begin{eqnarray}\label{eq:ds}
\lefteqn{\Delta_\text{soft}\,=\,\int^{1}_{x_{0}}\,dx\,f_\text{ISR}(x)\,|\wt{\M}_\text{eff}^{(1)}|^{2}(x,s)\,-(1+f_\text{soft})\,|\M_\text{Born}|^{2}(s)-2\,Re\,(\M_\text{Born}\,\M_\text{1-loop}^{*})(s)}\nonumber\\
&=&\lb f_\text{soft,ISR}\,(f_\text{soft}-f_\text{soft,ISR})+F^{(2)}\rb\,|\M_\text{Born}|^{2}(s)\,+\,2\,Re\,(\M_\text{Born}\,\M_\text{1-loop}^{*})(s)\,f_\text{soft,ISR}\nonumber\\
&+&\mO(\al^{3}).
\end{eqnarray}
\\
In the hard sector, we obtain
\begin{eqnarray}\label{eq:dh}
\Delta_\text{hard}&=&\int^{1}_{x_{0}}\,dx\,f_\text{ISR}(x)\,|\wt{\M}_\text{eff}^{(1)}|^{2}(x,s)\,-\,\sum_{\sigma\,=\,\pm}\,\int^{x_{0}}_{0}\,dx\,f^{\sigma}(x)\,|\M_\text{Born}^{\sigma}|^{2}(x,s)\nonumber\\
&=&\int^{x_{0}}_{0}\,dx\,\left\{\lb ( f^{+}(x)\,(f_{soft}-f_{soft,ISR})+F_{2}(x)\rb|\M_{Born}|^{2}(x,s)\right.\nonumber\\
&&\left. \,+\,f^{+}(x)\,2\,Re\,(\M_\text{Born}\,\M_\text{1-loop}^{*})(x,s) \right\}\,-,\int^{x_{0}}_{0}\,dx\,f^{-}(x)\,|\M^{-}_{Born}|^{2}(x,s)\nonumber\\
&&\,+\,\mO(\al^{3}).
\end{eqnarray}
These results correspond exactly to the $\mO(\al^{2})$ contributions discussed in Section \ref{sec:resum}.\\
\subsubsection*{Radiation off two particles}
We integrate now over $|\wt{\M}_{eff}|^{2}$ (\ref{eq:meffisr}) according to (\ref{eq:resummeds}).\\
\\
In the soft-soft sector, we obtain
\begin{eqnarray}\label{eq:dss}
\Delta_{ss}&=&\,\int^{1}_{x_{0}}\,dx_{1}\,\int^{1}_{x_{0}}\,dx_{2}\,f_{ISR}(x_{1})\,f_{ISR}(x_{2})\,|\wt{\M}_{eff}|^{2}(x_{1},x_{2},s)\nonumber \\
&&\,-\,(1+f_{soft}|\M_{Born}|^{2}(s)\,-\,2\,Re\,(\M_\text{Born}\,\M_\text{1-loop}^{*})(s)\nonumber\\
&=&\left\{ 2\,\lb f_{soft,ISR}\,(f_{soft}-f_{soft,ISR})+F^{(2)}\rb-f_{soft,ISR}\right\}|\M_{Born}|^{2}(s)\nonumber\\
&&+4\,f_{soft,ISR}\,Re\,(\M_\text{Born}\,\M_\text{1-loop}^{*})(s)\,+\,\mO(\al^{3})\nonumber\\
&=&2\,\Delta_{soft}-f_{soft,ISR}^{2}|\M_{Born}|^{2}(s)\,+\,\mO(\al^{3}).
\end{eqnarray}
\\
In the hard-soft section, we have
\begin{eqnarray}\label{eq:dsh}
\Delta_{sh}&=&\,\int^{1}_{x_{0}}\,dx_{1}\,\int^{x_{0}}_{0}\,dx_{2}\,f_{ISR}(x_{1})\,f_{ISR}(x_{2})\,|\wt{\M}_{eff}|^{2}(x_{1},x_{2},s)\nonumber\\
&&-\int^{x_{0}}_{0}dx_{2}\,\sum_{\sigma\,=\,\pm}\,f^{\sigma}(x_{2})|\M^{\sigma}_{Born}|^{2}(x_{2},s)\nonumber\\
&=&\int^{x_{0}}_{0}\,dx_{2}\,\left\{\left[ f^{+}(x_{2})\, (f_{soft}-2\,f_{soft,ISR})\,+\,F_{2}(x_{2})\right]|\M_{Born}|^{2}(x_{2},s)\right.\nonumber\\
&& \left.\,+\,f^{+}(x_{2})\,2\,Re\,(\M_\text{Born}\,\M_\text{1-loop}^{*})(x_{2},s)\right\}\,-\,\int^{x_{0}}_{0}\,dx_{2}\,f^{-}(x_{2})\,|\M_{Born}|^{2}(x_{2},s)\,+\,\mO(\al^{3})\nonumber\\
&=&\Delta_{hard}\,+\,\mO(\al^{3}).
\end{eqnarray}
\\
In the hard-hard sector,
\begin{eqnarray}\label{eq:dhh}
\Delta_{hh}&=&\int^{x_{0}}_{0}\,dx_{1}\,\int^{x_{0}}_{0}\,dx_{2}\,f_{ISR}(x_{1})\,f_{ISR}(x_{2})\,|\wt{\M}_{eff}|^{2}(x_{1},x_{2},s)\nonumber\\
&=&\int^{x_{0}}_{0}\,dx_{1}\,\int^{x_{0}}_{0}\,dx_{2}\,F_{1}(x_{1})\,F_{1}(x_{2})\,|\wt{\M}_{Born}|^{2}(x_{1},x_{2},s)\,+\,\mO(\al^{3}).
\end{eqnarray}
\\
We see that, in general, nearly all second order contributions for the simultaneous photon radiation of $e^{+}$ and $e^{-}$ can already be obtained from the second order terms for only one radiating particle. This only works in the approximation (\ref{eq:sigsofta}), as this approximation does not distinguish the emitting particles for soft photons. We will make use of this for a rough higher-order estimation in Section \ref{sec:highoest}. \\
\subsubsection*{Resumming only the Born matrix element}
We now consider the difference between $\sigma_{tot,ISR+}$ (\ref{eq:sresumm}) and $\sigma_{tot,ISR(b)}$ (\ref{eq:sigfin}); in the latter, only the Born cross section is convoluted with $f_{ISR}$. The leading and first order terms, which are already contained in the NLO corrected matrix element, are then subtracted to avoid double counting (cf. Eq. (\ref{eq:sigfin})). Up to $\mO(\al)$, $\sigma_{tot,ISR(b)}$ corresponds to the fixed order result $\sigma_{tot}$ (\ref{eq:sigex}). As in the last section, we only consider $\mO(\al^{2})$ contributions.\\
\\
In the soft-soft region, the second order effects from convoluting the Born cross section are given by
\begin{\eqn*}
(f^{2}_{soft,ISR}\,+\,2\,F^{(2)})\,|\M_{Born}|^{2}(s)
\end{\eqn*} 
and $\Delta_{ss}$ (\ref{eq:dss}) becomes
\begin{eqnarray*}
\Delta'_{ss}&=&
 2\,\lb f_{soft,ISR}\,(f_{soft}-2\,f_{soft,ISR})\rb\,|\M_{Born}|^{2}(s)\,+\,4\,f_{soft,ISR}\,Re\,(M_\text{Born}\,M_\text{1-loop}^{*})(s)\nonumber\\
&+&\mO(\al^{3}).
\end{eqnarray*}
In order to understand this expression, we again have to decompose $f_{soft,ISR}$ and $f_{soft}$ into parts radiating form the same or two different particles and include a decomposition as (\ref{eq:f1split}), obtaining
\begin{eqnarray*}
\lefteqn{2\,\lb f_{soft,ISR}\,(f_{soft}-2\,f_{soft,ISR})\rb\,\hat{=}}\\&&\lb F^{(1)}\,\Delta\,F^{(1)}\rb (x_{1})\,+\,\lb F^{(1)}\,\Delta\,F^{(1)}\rb (x_{2})\,+\,F^{(1)}(x_{1})\,\Delta\,F^{(1)}(x_{2})\,+\,F^{(1)}(x_{2})\,\Delta\,F^{(1)}(x_{1})
\end{eqnarray*}
This corresponds to the substitution of the $f_{ISR,soft}$ by $f_{soft}$ for the last soft-collinear photons in $\sigma_{tot,ISR+}$ and the substitution of the interference term description for virtual collinear last photons. In addition, $4\,f_{soft,ISR}\,Re\,(\M_\text{Born}\,\M_\text{1-loop}^{*})(s)$ gives all contributions of $\hat{\sigma}_{rest}$, i.e. virtual contributions untouched by the subtraction, combined with the radiation of a soft-collinear photon from either incoming particle. \\
\\
In the hard-soft region, the $\mO(\al^{2})$ contributions to $\sigma_{tot,ISR(b)}$ are given by
\begin{\eqn*}
\int^{x_{0}}_{0}\,dx_{2}\,\left[ f^{+}(x_{2})\,f_{soft,ISR}\,+\,F_{2}(x_{2})\right]|\M_{Born}|^{2}(x_{2},s)
\end{\eqn*}
and $\Delta_{sh}$ (\ref{eq:dsh}) becomes
\begin{eqnarray*}
\Delta'_{sh}&=&\int^{x_{0}}_{0}\,dx_{2}\, f^{+}(x_{2})\, (f_{soft}-2\,f_{soft,ISR})|\M_{Born}|^{2}(x_{2},s)\\
&& \left.\,+\,f^{+}(x_{2})\,2\,Re\,(M_\text{Born}\,M_\text{1-loop}^{*})(x_{2},s)\right\}\,-\,\int^{x_{0}}_{0}\,dx_{2}\,f^{-}(x_{2})\,|\M_{Born}|^{2}(x_{2},s)\,+\,\mO(\al^{3}).
\end{eqnarray*}
The hard-soft part of this is equivalent to
\begin{\eqn*}
F_{2}(x_{2})\,\lb (\Delta\,F^{(1)}(x_{1})\,+\,\Delta\,F^{(1)}(x_{2})\rb.
\end{\eqn*}
This corresponds again to the substitution of the $f_{ISR,soft}$ by $f_{soft}$ for the last soft-collinear photon (virtual last photon and interference term).\\
\\
In the hard-hard sector, the contribution to $\sigma_{tot,ISR(b)}$ up to $\mO(\al^{2})$ is given by $\Delta_{hh}$ (\ref{eq:dhh}).\\
\\
In summary, we can say that the difference between the $2\,\rightarrow\,2$ integration regions of $\sigma_{tot,ISR(+)}$ and $\sigma_{tot,ISR(b)}$ are the substitution of the $f_{ISR,soft}$ by $f_{soft}$ for soft-collinear and the interference term description for virtual collinear photons in $\sigma_{tot,ISR(+)}$. In addition, terms of the form
\begin{eqnarray}\label{eq:domseco}
&&\int^{1}_{0}dx_{1}\,\int^{1}_{0}dx_{2}\,f_{ISR}(x_{1})\,f_{ISR}(x_{2})\,\hat{\sigma}_{rest}(x_{1},\,x_{2},\,s)
\,-\,\hat{\sigma}_{rest}(s)
\end{eqnarray} 
are included, where $\hat{\sigma}_{rest}$ is defined as in Eq. (\ref{eq:sigsplit}). These are the dominant contributions, as all other differences between $\sigma_{tot,ISR}$ and $\sigma_{tot,ISR(b)}$ are proportional to $\Delta F^{(1)}$ and therefore subleading.
The $\mO(\al^{2})$ contribution to (\ref{eq:domseco}) is given by
\begin{\eqn}\label{eq:domsecoeff2}
2\,\lb \int^{x_{0}}_{0}\,dx\, f^{+}_{h}(x) \hat{\sigma}_{rest}(x,s)\, +\, f_{ISR, soft} \hat{\sigma}_{rest} (s)  \rb.
\end{\eqn} 
For $e^{+}e^{-}\,\rightarrow\,\wt{\chi}^{+}_{1}\,\wt{\chi}^{-}_{1}$, these contributions are in the $\%$ regime; cf. Section \ref{sec:resummtots}. The sign of the correction term depends on the specific process (note that $f_{soft,ISR}\,<\,0$, while $f^{+}(x)\,\ge\,0$).\\
\chapter{Results}\label{chap:resumres}
\section{Cut dependencies}\label{sec:cutresum}
\subsection{Energy cutoff dependence}\label{sec:ecutres}
\begin{figure}
\begin{center}
  \includegraphics[width=.95\textwidth]{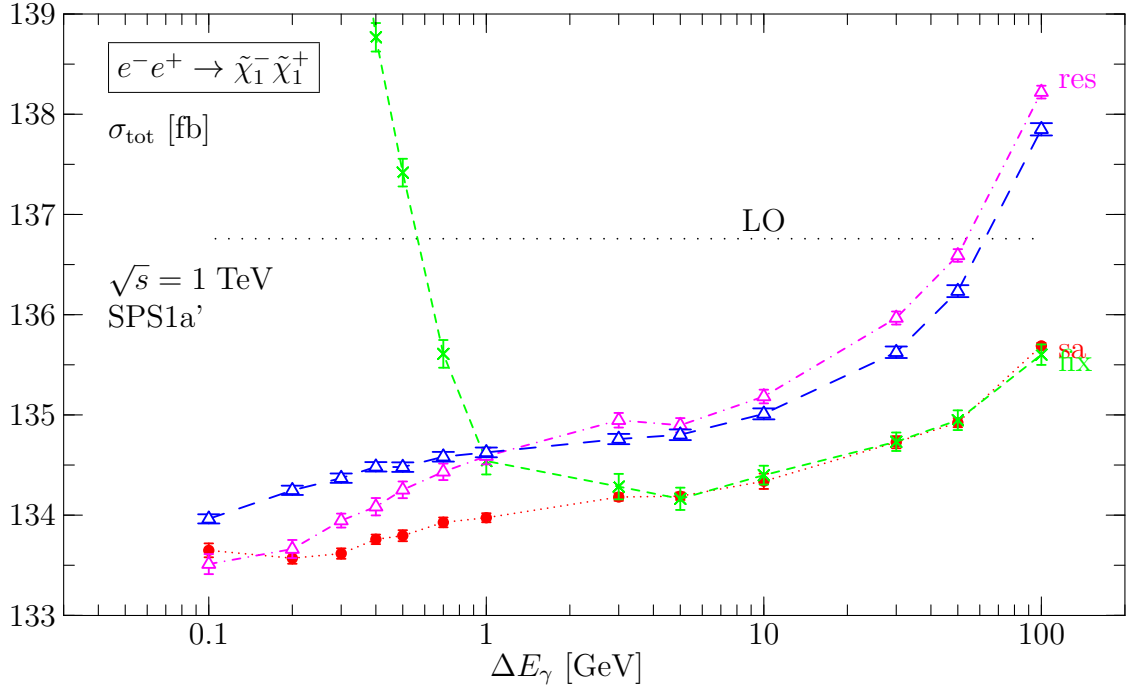}
  \vspace{\baselineskip}
  \caption{Total cross section dependence on the energy cutoff $\Delta
    E_\gamma$ using different calculational methods: {\rm `sa'} (red,
    dotted) = fixed-order semianalytic result  using
    \feynarts/\formcalc; {\rm `fix'} (green, dashed) = fixed-order
    Monte-Carlo result $\sigma_{tot}$ (\ref{eq:sigex}) using \whizard; {\rm `res'}
    (blue, long-dashed) = ISR-resummed Monte-Carlo
    result $\sigma_{tot,ISR+}$ (\ref{eq:sresumm}) using \whizard; (magenta, dash-dots) = same
    but resummation applied only to the $2\to 2$ part (\ref{eq:sresummnog}).
    Statistical Monte-Carlo integration errors are shown.  For the
    Monte-Carlo results, the collinear cutoff has been fixed to
    $\Delta\theta_\gamma=1^\circ$.  The Born cross section is
    indicated by the dotted horizontal line.}
\label{fig:edep}
\end{center}
\end{figure}

In Fig.~\ref{fig:edep} we compare the numerical results obtained using
the semianalytic fixed-order calculation with our Monte-Carlo
integration in the fixed-order and in the resummation schemes,
respectively.  Throughout this section, we set the process energy to
$\sqrt{s}=1\;\TeV$ and refer to the SUSY parameter point SPS1a'.  All
$2\to 2$ and $2\to 3$ contributions are included, so the results would
be cutoff-independent if there were no approximations involved. For the discussion of the $\Delta E_{\gamma}$ dependence of $\sigma_{tot}$, we refer to Section \ref{sec:excuts}.\\
The fully resummed result $\sigma_{tot,ISR+}$ shows an increase of about
$5\,\permil$ of the total cross section with respect to the
fixed-order result which stays roughly constant until $\Delta
E_\gamma>10\;\GeV$ where the soft approximation breaks down.  This
increase is a real effect; it is due to higher-order photon radiation
that is absent from the fixed-order calculation. \\
We now discuss the reshuffling of contributions
in the overlap region of the soft-collinear and hard-collinear (ISR)
descriptions for second order contributions. If we raise $\Delta E_\gamma$,  photons that have been hard now become soft.  In
the case of photons radiated from two different external
particles, a soft-collinear photon which has been described by $f_{ISR}$ is now described by $f_{soft}$ according to the soft approximation (\ref{eq:f-soft}). For the case of two photons radiated off the same particle, we have to
distinguish whether the two photons are $k_\perp$-ordered or not. If
they are, the description again changes from $f_{ISR}$ to $f_{soft}$. If
there is no $k_\perp$-ordering, either the first or the second radiated photon can change from hard to soft (or both). For the first photon,  it is a smooth transition where only the
last two terms of Eq.~(\ref{caseone}) are involved. If the second
photon changes to soft, new contributions of the form $ \Delta F^{(1)}$ (\ref{eq:deltaf}) appear.\\
The effects of reshuffling the photon descriptions are, up to $\Delta E_{\gamma}\,=\,10\, \GeV$, smaller than $2\,\permil$. If we only consider the dominant $k_{\perp}$ contributions, it would be preferable to raise $\Delta E_{\gamma}$ as long as the $\mO(\al)$ effects from the soft approximation are negligible. In our case, these effects reach the $2\permil$ level for $\Delta\,E_{\gamma}\,=\,1\,\GeV$ (cf. Section \ref{sec:excuts}). \\
Finally, we compare the difference between $\sigma_{tot,ISR}$ and $\sigma_{tot,ISR+}$. For the latter, the emission of a last hard non-collinear photon is also combined with multiple photon radiation described by $f_{ISR}$. We observe that for
$\Delta E_\gamma>1\;\GeV$ these higher-order contributions are caught
by ISR resummation of the $2\to 2$ part. For a lower cutoff, they have to be explicitly included in the $2\,\rightarrow\,3$ part. 
\subsection{Collinear cutoff dependence}\label{sec:collcutres}
\begin{figure}
\begin{center}
\includegraphics[width=0.9\textwidth]{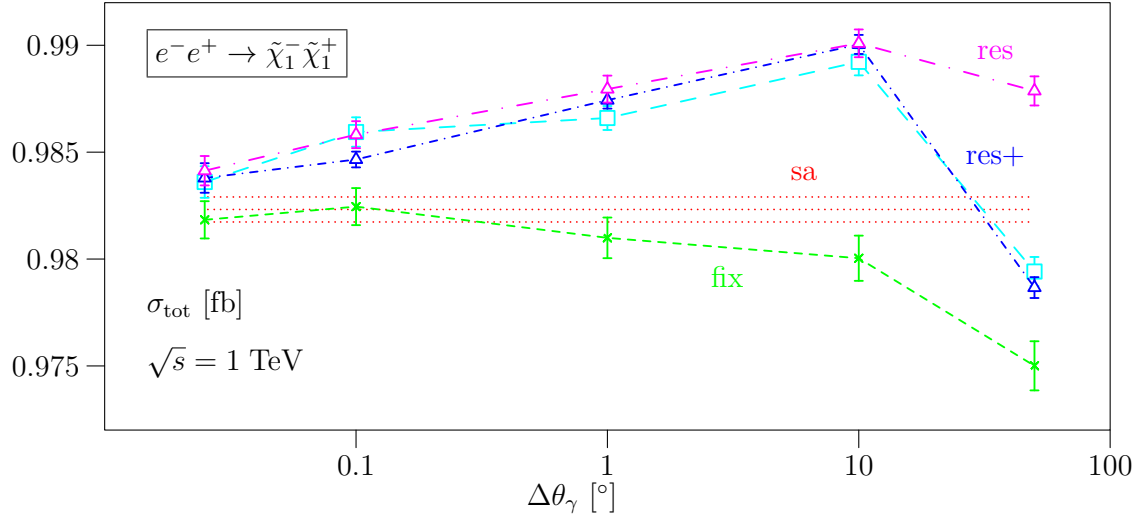}
  \vspace{10mm}
  \caption{Relative dependence of the cross section on the collinear cutoff
    $\Delta \theta_\gamma$ using different calculational
    methods: {\rm sa} (red, dotted) = fixed-order semianalytic result
    using \feynarts/\formcalc with respective integration errors; {\rm fix} (green, dashed) = fixed-order
    Monte-Carlo result $\sigma_{tot}$ (\ref{eq:sigex}) using \whizard; {\rm res} (blue, dash-dots) =
    third-order completely resummed Monte-Carlo result $\sigma_{tot,ISR+}$ (\ref{eq:sresumm}) using \whizard; (magenta, long dash-dotted) = third-order resummed Monte-Carlo result $\sigma_{tot,ISR}$ (\ref{eq:sresummnog}) using \whizard;  Statistical Monte-Carlo integration
    errors are shown.  The soft cutoff has been fixed to $\Delta
    E_\gamma=10\;\GeV$. All results are scaled to $\sigma_{Born}\,=\,1$.}
\label{fig:thdep}
\end{center}
\end{figure}

In $\mO(\al)$, the collinear angle dependence tests the validity regime of the collinear approximation used in Eq. (\ref{eq:sighardcoll}). This was already discussed in Section \ref{sec:excuts}, where we found the approximation is valid for $\Delta\theta_{\gamma}\,\leq\,1^{\circ}$. The difference between the fixed order result $\sigma_{tot}$ and the fully resummed result $\sigma_{ISR,+}$ are again the higher-order contributions discussed in Section \ref{sec:resum}.\\
Figure \ref{fig:thdep} shows the higher-order effects associated with raising the collinear cutoff $\Delta \theta_{\gamma}$.
For $\Delta\,\theta_{\gamma}\,=\,0.1^{\circ}\,(1^{\circ})$, the difference between $\sigma_{tot}$ and $\sigma_{tot,ISR+}$ is $4\,\permil\,(7\,\permil)$.
Raising $\Delta \theta_\gamma$ 
shuffles photons from a non-collinear to a collinear
description. First, this opens up phase space for all photons described by $f_{ISR}$. For the radiation off the same particle, raising the collinear cutoff also improves the description of $k_{\perp}$ ordered photons if the sum of their transverse momenta is now lower than the new transverse cutoff associated with $\Delta\theta_{\gamma}$. They are contained in the second order ISR contributions $F^{(2)},\,F_{2}$ and switch from a description $\propto\,\Delta \,F^{(1)}$ to terms of the form  (\ref{caseone}), where the second soft (virtual) collinear photon is given by $f_{soft}$ (the interference term).\\
Unfortunately, raising the collinear cutoff also enhances the phase space of contributions $\propto \Delta F^{(1)}$ which can appear when multiple (collinear) photons which are not $k_{\perp}$ ordered are emitted from the same particle; cf. Section \ref{sec:resum}. Similarly, raising the cutoff can worsen the description for  photons that lie in the
soft regime near the limit of the soft-collinear regime and change
into the latter. For non-$k_{\perp}$-ordered photons, the description switches from $f_{soft}$ (interference term) to contributions $\propto\,\Delta\,F^{(1)}$. However, these effects are subleading.\\
For the leading-log contributions, raising the cutoffs gives a better description as long as the collinear approximation is valid. From Figure \ref{fig:thdep}, we see that this is a $3-4\,\permil$ effect. The difference between $\sigma_{tot,ISR}$ and $\sigma_{tot,ISR+}$ is similarly small (about $1\,\permil$) in the validity regime of the soft approximation. The size of these contributions therefore mainly depends on the choice of the soft cut $\Delta E_{\gamma}$ (cf. Section \ref{sec:ecutres}). For the $\Delta E_{\gamma}\,=\,10\,\GeV$ as above, all leading contributions are already summed up in the ISR part of the $2\,\rightarrow\,2$ integration.
\section{Total cross section}\label{sec:resummtots}
\subsection{Leading and first order results}
\begin{figure}
\begin{center}
\includegraphics[width = 10cm]{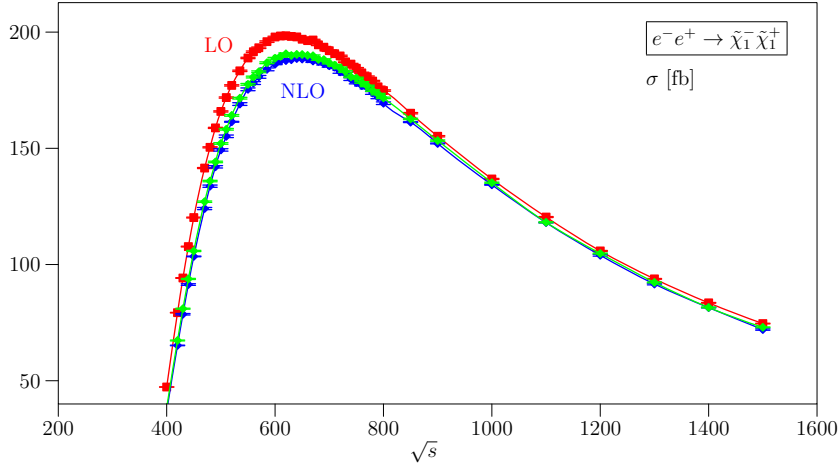}
\vspace{5mm}
\caption{\label{fig:sigresum} $\sigma_{Born},\,\sigma_{tot},\,\sigma_{ISR,+}$ as a function of $\sqrt{s}$. All results in fb }
\end{center}
\end{figure}

\begin{figure}
\begin{center}
\includegraphics[width = 10cm]{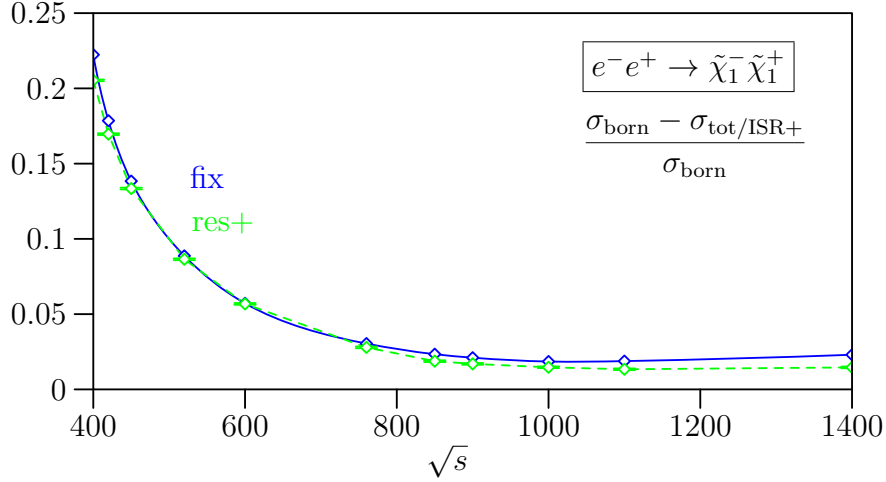}
\vspace{10mm}
\caption{\label{fig:resumreldiff} relative difference of $\sigma_{tot}$ (and $\sigma_{ISR+}$) to $\sigma_{Born}$, resulting from NLO (NLO and higher-order) corrections }
\end{center}
\end{figure}

\begin{figure}
\begin{center}
\includegraphics[width = 10cm]{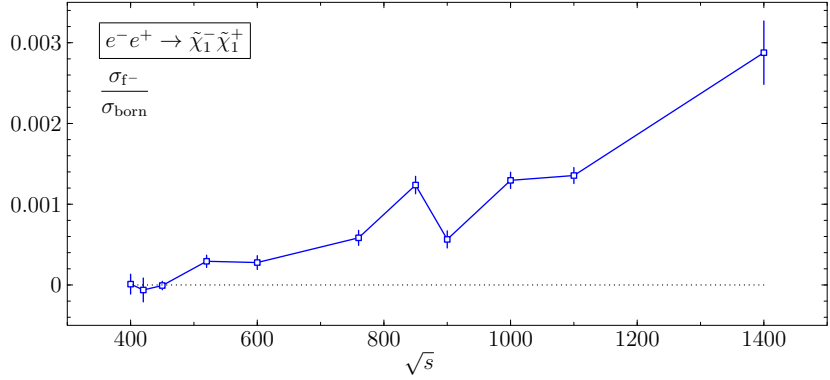}
\vspace{5mm}
\caption{\label{fig:fmineff} Contribution of the part of $\sigma_{tot}$ resulting from the helicity-flipping part $f^{-}$ (\ref{eq:hardcollfs}) of the hard-collinear approximation with respect to $\sigma_{Born}$. 1 $\sigma$ errors from \whizard~ integration are shown}
\end{center}
\end{figure}

In leading and next-to-leading order, the fixed order result $\sigma_{tot}$ (\ref{eq:sigex}) and the resummation result $\sigma_{tot, ISR(+)}$ (\ref{eq:sresummnog}), (\ref{eq:sresumm}) agree. All differences between the results are therefore due to higher-order effects. For the discussion of first order behavior of the total cross section, we therefore refer to the Sections \ref{sec:nloex} and \ref{sec:fixedres}.\\
In  Figures \ref{fig:sigresum} and \ref{fig:resumreldiff} we display the LO result together with the NLO results for the
fixed-order and resummed approach.  Near
the cross-section maximum, the relative correction in the 
resummed approach is about $-5.5\,\%$, approaching $-2\,\%$ at $\sqrt{s}=1\;\TeV$.  Near threshold and
at asymptotic energies, the relative NLO correction gets up to $-20\,\%$.\\
A short remark concerns the neglected helicity-flipping terms $f^{-}(x)$ (\ref{eq:hardcollfs}) in the hard part of $f_{ISR}$ which are in principle $\mO(\al)$, but do not contain collinear logarithms; cf. Section \ref{app:isr}. Figure \ref{fig:fmineff} shows the relative magnitude of these terms for unpolarized initial particles with respect to the  Born cross section. The contribution is in the per mille regime.
\subsection{Higher-order effects}\label{sec:highoest} 
\begin{figure}
\begin{center}
\includegraphics[width=10cm]{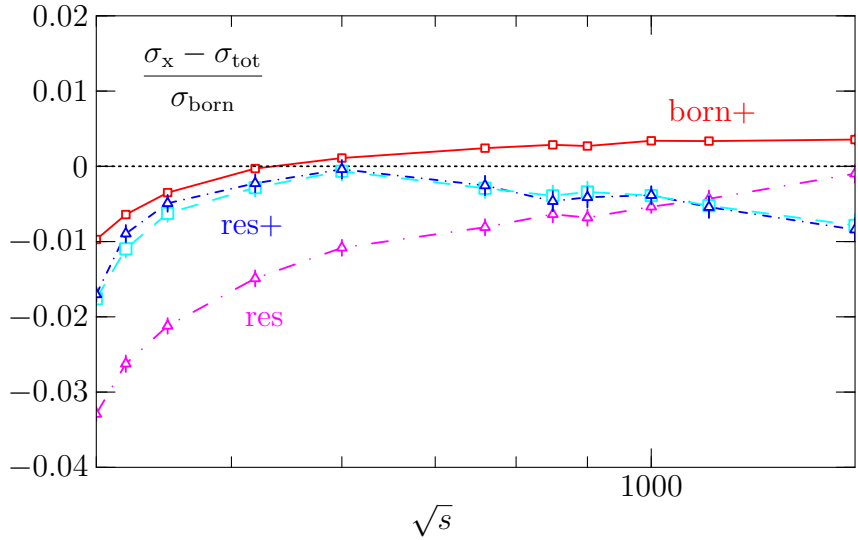}
\vspace{7mm}
\caption{Relative higher-order effects for different methods: (magenta, long dash dotted) = $\sigma_{tot,ISR}$ (\ref{eq:sresummnog}), (blue/ cyan and dash-dotted/ dashed) =  $\sigma_{tot,ISR+}$ (\ref{eq:sresumm}), and (red, solid) = $\sigma_{tot,ISR(b)}$ (\ref{eq:sigfin}) vs $\sigma_{tot}$ (\ref{eq:sigex}). $\Delta\,\theta_{\gamma}\,=\,1^{\circ},\,\Delta\,E_{\gamma}/\sqrt{s}\,=\,5\,\cdot\,10^{-3}$.}   
\label{fig:secoeff}
\end{center}
\end{figure} 
Figure \ref{fig:secoeff} shows the differences between $\sigma_{tot,ISR}$, $\sigma_{tot,ISR+}$, $\sigma_{tot,ISR(b)}$ and the fixed order result $\sigma_{tot}$ (\ref{eq:sigex}).  $\sigma_{tot,ISR}$, $\sigma_{tot,ISR+}$, and $\sigma_{tot,ISR(b)}$ all contain different kinds of higher-order contributions (cf. Section \ref{sec:higho}), which we will now consider in more detail.\\
The largest effects clearly result from the higher-order contributions in the transition from the fixed order result $\sigma_{tot}$ (\ref{eq:sigex}) to the resummation result $\sigma_{tot, ISR}$ (\ref{eq:sresummnog}) discussed in Section \ref{sec:higho}, i.e. the inclusion of multiple photon radiation off the Born as well as the interference term in the $2\,\rightarrow\,2$ integration region. They are up to a few percent for small center of mass energies. Including the convoluting of the $2\,\rightarrow\,3$ integration region of $\sigma_{tot, ISR}$ with $f_{ISR}$, i.e. including also diagrams where the last radiated photon is hard and non-collinear as in $\sigma_{tot,ISR+}$ (\ref{eq:sresumm}), significantly reduces these effects. These extra terms cannot be neglected. However, it was shown in Section \ref{sec:cutresum} that the magnitude of these corrections strongly depends on the soft cut $\Delta E_{\gamma}$. If the energy cutoff is high enough, the dominant contributions are included in the convolution of the effective matrix element in the $2\,\rightarrow\,2$ region with $f_{ISR}$; cf. Figure \ref{fig:edep}. Apparently, the cut value $\Delta E_{\gamma}/\sqrt{s}\,=\,5\,\cdot\,10^{-3}$ chosen for Figure \ref{fig:secoeff} is not sufficiently high for low $\sqrt{s}$ values to include all dominant contributions.\\
  Finally, we see that the difference between $\sigma_{tot,ISR(b)}$ and $\sigma_{tot, ISR+}$, which is mainly given by the terms $\propto\,\int\,\hat{\sigma}_{rest}$ (\ref{eq:domseco}), is similarly in the $\%$ range. Shuffling photon descriptions, in contrast, is a $\permil$ effect; cf. Section \ref{sec:cutresum}. The change of ISR order in the resummation method only concerns the hard regime and clearly shows no significant effect.
\subsubsection*{Third (and higher) order estimation}
\begin{figure}
\begin{center}
\includegraphics[width=10cm]{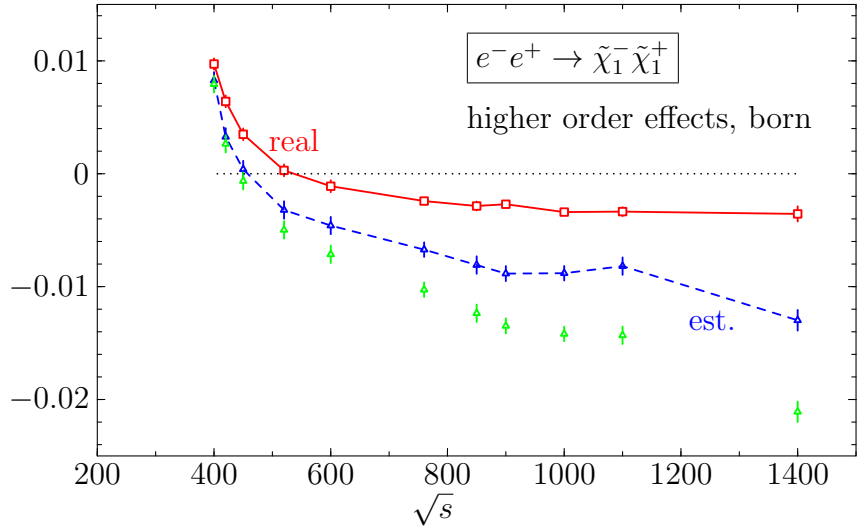}
\vspace{7mm}
\caption{Second and higher-order effects; estimate from 1 photon (blue, dashed) compared to actual result (red, solid) . Born only}
\label{fig:estimborn}
\end{center}
\end{figure}

\begin{figure}
\begin{center}
\includegraphics[width=10cm]{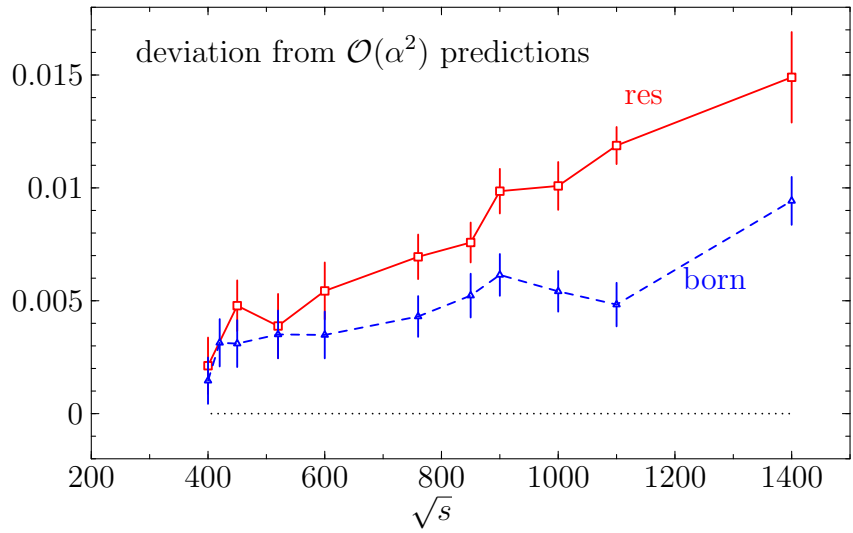}
\vspace{10mm}
\caption{third order effects from (\ref{eq:estimho}), Born (blue, dashed) only vs completely convoluted (red, solid) version}
\label{fig:estimvirt}
\end{center}
\end{figure}

Eqs. (\ref{eq:dss}), (\ref{eq:dsh}), and (\ref{eq:dhh}), show that we can in principle do a $\mO(\al^{3})$ and higher-order estimation when using the results from radiation off one particle $\Delta_{s}$ (\ref{eq:ds}), $\Delta_{h}$ (\ref{eq:dh}) to predict the $\mO(\al^{2})$ effects for the radiation off two particles. Analytically, we have
\begin{\eqn}\label{eq:estimho}
\Delta_{ss}\,+\,2\,\Delta_{sh}\,=\,2\,(\Delta_{s}\,+\,\Delta_{h})\,-\,f^{2}_{soft,ISR}\,|\M_{Born}|^{2}\,+\,\mO(\al^{3}).
\end{\eqn}
Deviations from the ``estimate'' on the right hand side can give us an idea of the order of magnitude of the higher-order corrections. Note, however, that numerically the calculation of these terms are very involved and limited by computer precision; similarly, errors resulting from setting $x\,=\,1$ in the matrix element the soft regime are neglected. This estimate also does not predict the actual sign of the higher-order corrections. It tests the numerical difference between expressions which are analytically equivalent up to $\mO(\al^{2})$.\\
Figure \ref{fig:estimborn} shows the estimated and actual $\mO(\al^{2})$ contributions, here for the convolution of $|\M_{Born}|^{2}$ as in $\sigma_{tot,ISR(b)}$ (\ref{eq:sigfin}) only where similar considerations can be done. We see that the deviation from the prediction is already in the $0.5 - 1 \%$ regime. Note that for a comparison we also added points where 
\begin{\eqn*}
\Delta^{Born}_{hh}\,=\,\int^{x_{0}}_{0}dx_{1}\,\int^{x_{0}}_{0}\,dx_{2}\,f_{ISR}(x_{1})\,f_{ISR}(x_{2})|\M_{Born}|^{2}
\end{\eqn*}
 was replaced by $\Delta_{hh}$ (\ref{eq:dhh}). These two expressions differ in third and higher-order terms involving the radiation of two and more hard-collinear photons off the virtual (+ soft-collinear) NLO term. They are of similar magnitude.\\
Figure \ref{fig:estimvirt} shows the magnitude of the $\mO(\al^{3})$ errors in Eq. (\ref{eq:estimho}) relative to the Born cross section and the equivalent contribution from $\sigma_{tot,ISR(b)}$ in Figure \ref{fig:estimborn}. We see that, for $\sqrt{s}\,\leq\,600\,\GeV$, the third order effects are of similar order for both methods. For larger cm energies, however, the estimate from $\sigma_{tot,ISR}$ deviates from the Born-related estimate and goes up to nearly $2\%$. This is related to second and higher-order effects resulting from the convoluting of the virtual part with $f_{ISR}$ for both beams. The importance of these contributions was already shown in $\mO(\al^{2})$ in Figure \ref{fig:secoeff} as well as the difference between $\Delta_{hh}^{Born}$ and $\Delta_{hh}$ in Figure \ref{fig:estimborn}.
Summarizing, we can say that according to the estimation done in Eq. (\ref{eq:estimho}), even third and higher-orders are important if the ILC precision of a few $\permil$ is reached.
\section{Event Generation}
\begin{figure}
\begin{center}
    \includegraphics[width=10cm]{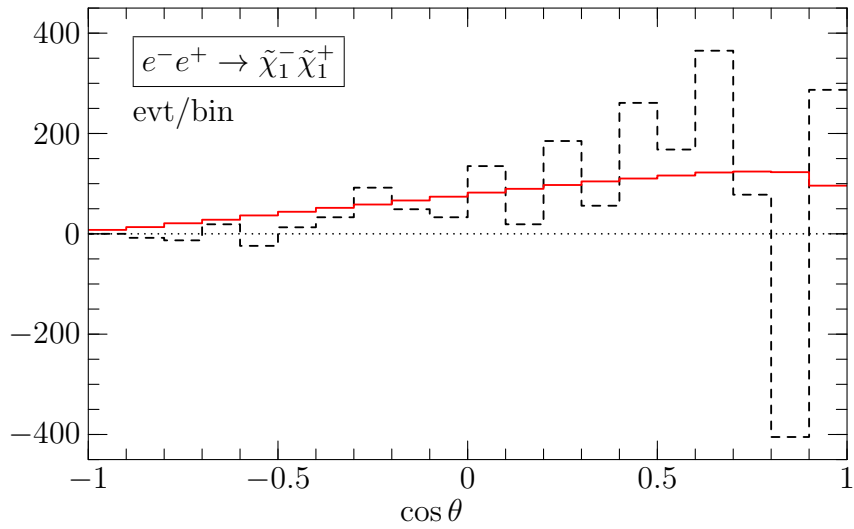}
\vspace{10mm}
  \caption{Polar scattering angle dependence of difference between events resulting from completely resummed and fixed order method: $N_{res}\,-\,N_{ex}$. Standard deviation from Born (solid, red) is shown}
  \label{fig:relcorrevnt}
\end{center}
\end{figure}

\begin{figure}
\begin{center}
    \includegraphics[height=6.5cm]{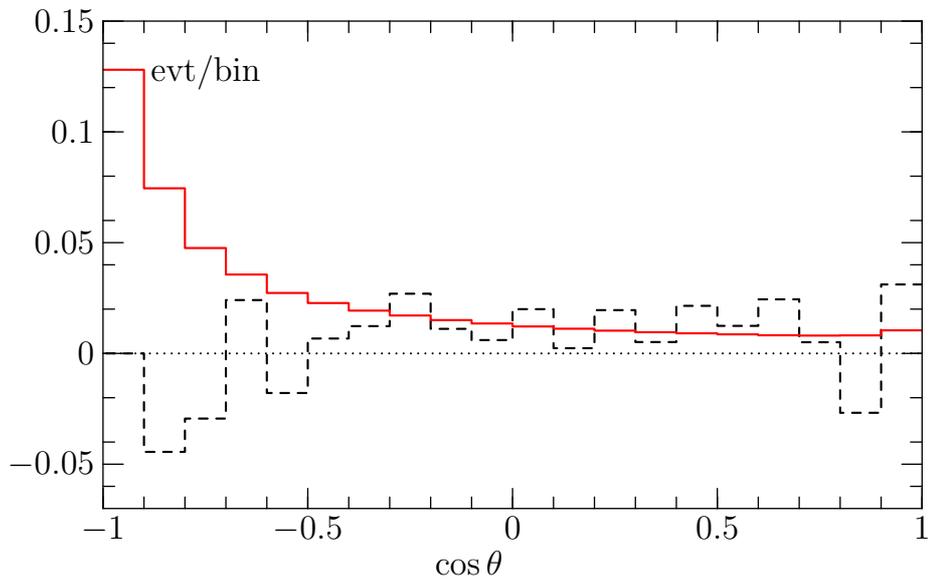}
\vspace{10mm}
  \caption{Same as Figure \ref{fig:relcorrevnt}, normalized to Born results}
  \label{fig:relcorrevntnorm}
\end{center}
\end{figure}

\begin{figure}
\begin{center}
    \includegraphics[height=6.5cm]{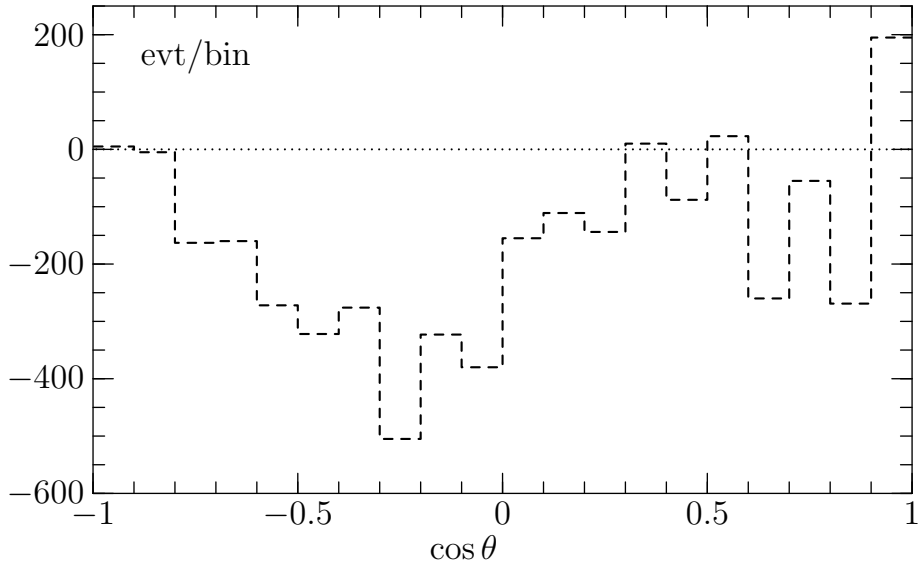}
\vspace{10mm}
  \caption{Difference between complete resummation and fixed order method for $\sqrt{s}\,=\,1\,\TeV$ with $\Delta\,E_{\gamma}\,=\,0.5\,\GeV,\,\Delta\theta_{\gamma}\,=\,0.5^{\circ}$. With the energy cut in the critical region (cf. Section \ref{sec:drawback}), effects of setting $|\M_{eff}|^{2}$ to zero become apparent for $\cos\theta\,<\,0$.}
  \label{fig:deltadethetlc}
\end{center}
\end{figure}

As for the total cross section, events generated using $\M_{eff}$ (\ref{eq:meff}) and $\wt{\M}_{eff}$ (\ref{eq:meffisr}) are equivalent up to and including $\mO(\al)$. For the discussion of the NLO effects, we therefore refer to Section \ref{sec:fixedres}. Differences between the two methods originate from higher-order contributions.\\
\begin{figure}
\begin{center}
  \includegraphics[width=.95\textwidth]{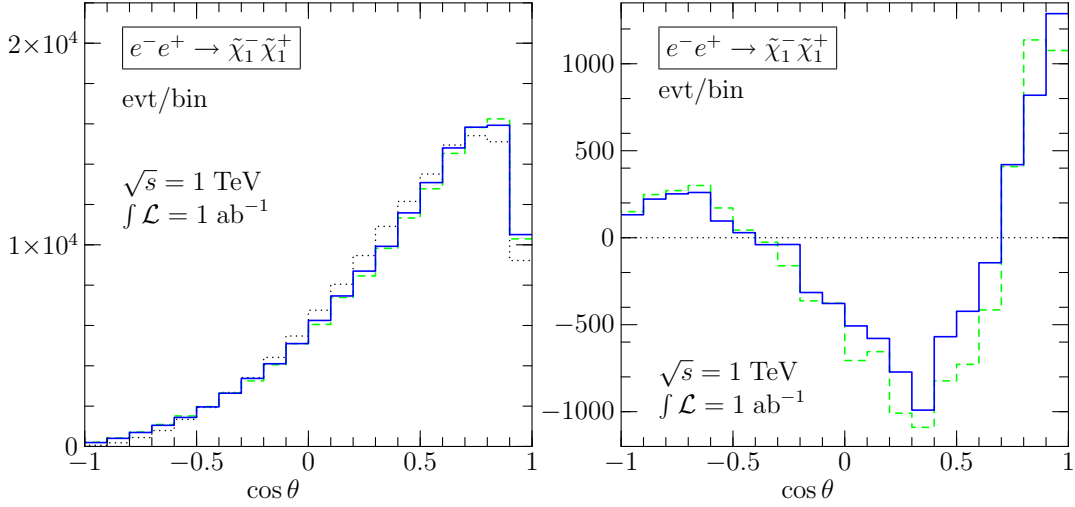}
  \vspace{\baselineskip}
  \caption{Polar scattering angle distribution for an integrated
    luminosity of $1\;\ab^{-1}$ at $\sqrt{s}=1\;\TeV$. Left: total
    number of events per bin; right: difference w.r.t.\ the Born
    distribution.  LO (black, dotted) = Born cross section without
    ISR; fix (green, dashed) = fixed-order approach; res (blue, full)
    = resummation approach.  Cutoffs: $\Delta E_\gamma=3\;\GeV$ and
    $\Delta\theta_\gamma=1^\circ$.}
\label{fig:histth}
\end{center}
\end{figure}
Figure \ref{fig:histth} shows the binned distribution of the chargino
production angle as in Figures \ref{fig:histthex1} and \ref{fig:histthex2}, together with the events generated using the resummation method. For a kinematically more accurate description of the NLO contributions, decreasing the cutoffs would be preferred, but choosing lower values would
invalidate the fixed-order approach for the comparison.\\
Figures \ref{fig:relcorrevnt} and \ref{fig:relcorrevntnorm} show the difference in the number of events between the angular distribution for fixed order and the resummation method as well as a $1\,\sigma$ effect from the Born cross section. The higher-order effects are not as striking. They are visible mostly in the
central-to-forward region. For $\cos\,\theta\,\geq\,0.5$, they are statistically significant. Going to higher luminosities would of course improve the significance for even lower values of $\cos\theta$. From Figure \ref{fig:relcorrevntnorm}, we can see that the integration over  $\cos\theta$ reproduces the $\sim\,6\permil$ effect shown in Figure \ref{fig:edep} (compare also to Figure \ref{fig:secoeff} with $\Delta\,E_{\gamma}\,=\,5\,\GeV$; from Figure \ref{fig:edep}, we know that the difference between the resummation and the fixed order result stays roughly the same up to $\Delta\,E_{\gamma}\,=\,10\,\GeV$).\\ 
Finally, we can test the effect of lowering $\Delta\,E_{\gamma}$ such that we get into the critical soft cut region discussed in Section \ref{sec:drawback}. Figure \ref{fig:deltadethetlc} shows the difference between the two methods of adding photons for $\Delta E_{\gamma}\,=\,0.5\,\GeV,\;\Delta\,\theta\,=\,0.5^{\circ}$. We clearly see the effects of setting the negative effective
squared matrix elements to zero where it becomes negative. The behaviour of $|\M_{eff}|^{2}$ for the (sub)dominant helicity combination was shown in Figures \ref{fig:meffsq, de0.5} and \ref{fig:twohelis}.  For $\cos\theta\,>\,0$, the differences between the resummation method and the fixed-order method have a behaviour as in the higher energy cut case, cf. Figure \ref{fig:relcorrevnt}. They are smaller in magnitude. As in this region of phase space, $|\M_{eff}|^{2}\,\geq\,0$ for the dominant helicity combination, this difference can be attributed to the lower angular cuts which decrease the difference between $\sigma_{tot}$ and $\sigma_{tot,ISR+}$ (cf. Section \ref{sec:collcutres}). 
\chapter{Summary and Outlook}\label{chap:sumup}
As an $e^{+}e^{-}$ collider with a cm energy of $500\,\GeV\,(1\,\TeV)$, the ILC will provide a clean environment for precision tests of physics in and beyond the Standard Model. In the chargino and neutralino sector of the MSSM, a small number of measurements already determines the supersymmetric parameters at the electroweak scale. The parameters at the SUSY breaking scale can be derived from renormalization group equations and eventually give information about the SUSY breaking mechanism.
The clean experimental environment of the ILC leads to small experimental errors which calls for matching accuracy of theoretical NLO and higher-order predictions. These effects need to be included in Monte Carlo event generators, the simulation tools that
are actually used in the experimental analyses.\\
\subsubsection*{Inclusion of NLO corrections in \whizard}
We have presented results obtained from implementing the NLO corrections for chargino pair production at the
ILC
into the Monte-Carlo event generator \whizard~.
On top of the genuine SUSY/electroweak corrections, we have
considered several approaches of including photon radiation, where a
strict fixed-order approach allows for comparison and consistency
checks with published semianalytic results in the literature. However, this approach suffers from negative event weights in certain points of phase space. Extending  approaches used in Monte Carlo generators for LEP analyses, we developed a method for including virtual-, soft- and hard-collinear resummation of photon radiation. It 
not just improves the numerical result, but actually is more
straightforward to implement and does not suffer from negative event
weights in or near the experimentally accessible part of phase space.
  The
generator accounts for all yet known higher-order effects, allows for small
cutoffs, and explicitly generates photons where they can be
resolved experimentally. In addition, the generated event samples allow for quick analyses of NLO effects in angular distributions, correlations, and other quantities.  We also interfaced \whizard~ with a (modified) \formcalc~ code which in principle allows for the inclusion of any other (NLO) process.\\
\subsubsection*{Magnitude of next-to-leading and higher-order corrections}
For the mSugra point SPS1a', the NLO corrections for chargino production and decay cross sections at the ILC are in the percent range.
Convolution of the virtual non-collinear part with the ISR structure function leads to additional percent effects for large $\sqrt{s}$. These contributions are not included in the standard treatment of ISR resummation, where multiple photon emissions are combined with the Born process only. \\
\\
Minor next-to-leading and higher-order uncertainties for the cross sections stem from the use of approximations in the soft and collinear regime, reshuffling of photon descriptions, enhancement of phase space for higher-order photons, and neglecting of the non-logarithmic $\mO(\al)$ photon radiation from the helicity flipping term in the hard-collinear approximation. 
A rough estimate shows that even higher-order contributions described by the resummed structure functions can become $\mO(\%)$ and are therefore important in a thorough analysis of the process and dominant with respect to the non-leading helicity flip contribution. The latter is $\mO(\permil)$ for unpolarized cross sections; it can in principle be easily implemented in the resummation method.\\
\subsubsection*{Cutoff dependencies}
In general, a careful analysis of the dependence on the technical cutoffs on
photon energy and angle is necessary, as it reveals
uncertainties related to higher-order radiation and breakdown of the
soft or collinear approximation. The soft approximation leads to a deviation from the exact solution of $\approx\,5\,\permil$ for a ratio of the soft photon cut to the center of mass energy $\Delta E_{\gamma}/\sqrt{s}\,=\,10^{-2}$. The collinear approximation breaks down for the collinear cutoff $\Delta\,\theta\,=\,1^{\circ}$. We compared the corrections resulting from the use of the approximations to the magnitude of the NLO effects as well as higher-order corrections.\\
For the precision reached at the ILC, both the soft and collinear approximations imply the use of low cuts. On the other hand, raising the cutoffs in principle enhances phase space for multiple photon emissions and reshuffles photon description from the leading log initial state radiation description to the soft approximation for soft and the exact virtual description for virtual photons at leading order. These reshuffling effects are in the $\permil$ regime, but smaller than the errors arising from raising the cutoffs for the first order approximations. This implies the use of low cutoffs for a more exact NLO description.
The $2\,\rightarrow\,3$ process describing the $\mO(\al)$ emission of hard non-collinear photons also needs to be convoluted with the ISR structure function, as these contributions also are in the $\permil$ regime. 
\subsubsection*{Cross sections and event generation}
With both the fixed order and the resummation method implemented in \whizard~, we reproduce the semianalytic literature results for chargino production at NLO. The angular event distributions reflect the size of the corrections for total cross sections discussed above. For a center of mass energy of $1\,\TeV$ and an integrated luminosity of $1\,\text{ab}^{-1}$, simulation results derived from both the NLO fixed order and the resummation method differ significantly from the Born distribution and each other. The corrections cannot be described by a constant proportionality factor K between the Born and the NLO result. The fixed and resummed distributions differ in higher-orders. This is most obvious in the forward-scattering region, where the distributions differ by more than three (Born) standard deviations.\\
\subsubsection*{Outlook}
The constructed generator should be regarded as a step
towards a complete NLO simulation of SUSY processes at the ILC.  If
charginos happen to be metastable, it already provides all necessary
ingredients.  Beam effects (beamstrahlung, beam energy spread,
polarization) are available for simulation and can easily be included.
However, charginos are metastable only for peculiar SUSY parameter
points. If the theoretical prediction should match the experimental accuracy, a full description of the end products seen in experiment needs to include the chargino decay at NLO as well as non-factorizing contributions to the $2\,\rightarrow\,n$ process, e.g. in the double-pole approximation. These we have to match with
off-shell and background effects, already available for simulation in
\whizard.  Furthermore, in the threshold region the Coulomb
singularity calls for resummation, not yet accounted for in the
program.  These lines of improvement will be pursued in future
work. 
\begin{appendix}
\chapter{Conventions, SUSY overview, MSSM Lagrangian}\label{app:susy}
There are various introductions to the formal construction of a $N\,=\,1$ SUSY Lagrangian, cf. \cite{Haber:1985rc,Sohnius:1985qm, Wess:1992cp, Bailin:1994qt, Drees:1996ca, Martin:1997ns}. We just cite the main results and refer to the literature for further details.
\section{Conventions}
In the following, we define $g_{\mu\nu}$ by
\begin{\eqn*}
g_{\mu\nu}\,=\,g^{\mu\nu}\,=\,\text{diag}(1,-1,-1,-1).
\end{\eqn*}
The Dirac matrices $\gamma^{\mu}$ obey
\begin{\eqn*}
\left\{\gamma_{\mu},\gamma_{\nu} \right\}\,=\,2\,g_{\mu\nu}\,\mathbbm{1}_{4\,\times\,4}.
\end{\eqn*}
In the Weyl representation, they read
\begin{\eqn}\label{eq:gmuweyl}
\gamma^{\mu}\,=\,\lb \begin{array}{cc} 0&\sigma^{\mu}\\ \bar{\sigma}^{\mu}&0\\ \end{array}\rb,
\end{\eqn}
with
\begin{\eqn}\label{eq:sigsim}
\sigma^{\mu}\,=\,\lb\,\mathbbm{1}_{2},\sigma^{i}\rb\;,\;\bar{\sigma}^{\mu}\,=\,\lb\,\mathbbm{1}_{2},-\sigma^{i}\rb,
\end{\eqn}
and the $\sigma^{i}$ being the Pauli matrices. In addition, we define
\begin{\eqn}\label{eq:sigdouble}
\sigma^{\mu\nu}\,=\,\frac{1}{4}\,\lb \sigma^{\mu}\,\bar{\sigma}^{\nu}-\sigma^{\nu}\,\bar{\sigma}^{\mu}\rb\;,\;\ol{\sigma}^{\mu\nu}\,=\,\frac{1}{4}\,\lb \bar{\sigma}^{\mu}\,\sigma^{\nu}-\bar{\sigma}^{\nu}\,\sigma^{\mu}\rb,
\end{\eqn}
and
\begin{\eqn*}
\Sigma_{\mu\nu}\,=\,\frac{i}{2}\,\left[\gamma_{\mu},\gamma_{\nu}\right].
\end{\eqn*}
The two-component Weyl spinors transforming under the $(\frac{1}{2},0)$ and $(0,\frac{1}{2})$ representation of the Lorentz-group are given by
\begin{\eqn*}
\psi_{\al}\;\; (\frac{1}{2},0) \;,\;\ol{\psi}^{\dot{\al}}\;\; (0,\frac{1}{2}).
\end{\eqn*}
The index-raising and lowering matrices are given by
\begin{eqnarray}\label{eq:epsilons}
&&\epsilon^{\al\,\be}\,=\,\epsilon_{\dot{\al}\,\dot{\be}}\,=\,\lb\begin{array}{cc}0&-1\\1&0\\ \end{array}\rb,\nonumber\\
&&\epsilon_{\al\,\be}\,=\,\epsilon^{\dot{\al}\,\dot{\be}}\,=\,\lb\begin{array}{cc}0&1\\-1&0\\ \end{array}\rb.
\end{eqnarray}
A four-component spinor is then given by
\begin{\eqn}\label{eq:spindef}
\Psi\,=\,{\psi_{\al} \choose \ol{\chi}^{\dot{\al}}}.
\end{\eqn}
The index structure of the $\sigma$ matrices (\ref{eq:sigsim}), (\ref{eq:sigdouble}) is
\begin{\eqn*}
(\sigma^{\mu})_{\al\dot{\al}}\,,\,(\bar{\sigma}^{\mu})^{\dot{\al}\al}\,,\,(\sigma^{\mu\nu})^{\;\;\;\be}_{\al}\;,\;(\ol{\sigma}^{\mu\nu})_{\;\;\;\dot{\be}}^{\dot{\al}}\,,
\end{\eqn*}
in accordance with Eqs. (\ref{eq:gmuweyl}) and (\ref{eq:spindef}).\\
Multiplication of Weyl spinors (as well as any Grassmann numbers) is defined by
\begin{\eqn*}
\chi\,\psi\,=\,\chi^{\al}\,\psi_{\al}\;,\;\ol{\chi}\,\ol{\psi}\,=\,\ol{\chi}_{\dot{\al}}\,\ol{\psi}^{\dot{\al}}.
\end{\eqn*}
Together with Eq. (\ref{eq:epsilons}), this leads to
\begin{eqnarray*}
&&\chi\,\psi\,=\,\psi\,\chi\,=\,-\chi_{\al}\,\psi^{\al},\\
&&\ol{\chi}\,\ol{\psi}\,=\,\ol{\psi}\,\ol{\chi}\,=\,-\ol{\chi}^{\dot{\al}}\,\ol{\psi}_{\dot{\al}}.
\end{eqnarray*}
\section{Poincar\'e and SUSY algebra}
We define the generators $P_{\mu}$ of the translation group and $\M_{\mu\nu}$ of the Lorentz group such that the infinitesimal transformations are 
\begin{eqnarray*}
x'_{\mu}\,=\,x_{\mu}\,+a_{\mu}\,=\,x_{\mu}-i\,a_{\lambda}(P^{\lambda})_{\mu}&&\text{(translation)},\\
x'^{\rho}\,=\,x^{\rho}+\omega^{\rho}_{\,\,\sigma}\,x^{\sigma}\,=\,x^{\rho}-\frac{i}{2}\,\omega_{\mu\nu}(M^{\mu\nu})^{\rho}_{\,\,\sigma}x^{\sigma}&&\text{(Lorentz transformation)} .
\end{eqnarray*}
The generators $P_{\mu},\,M_{\mu\nu}$ then obey the Poincar\'e algebra
\begin{eqnarray*}
\left[P^{\lambda},P^{\mu}\right]&=&0,\\
\left[M^{\mu\nu},P^{\lambda}\right]&=&i\,(g^{\nu\lambda} P^{\mu}\,-\,g^{\mu\lambda} P^{\nu}),\\
\left[M^{\mu\nu},M^{\rho\sigma}\right]&=&i\,(g^{\nu\rho}M^{\mu\sigma}\,+\,g^{\mu\sigma}M^{\nu\rho}\,-\,g^{\mu\rho}M^{\nu\sigma}\,-\,g^{\nu\sigma}M^{\mu\rho}).
\end{eqnarray*}
A supersymmetry transformation is an extension of the Poincar\'e Algebra relating bosonic and fermionic degrees of freedom. We define $Q_{\al},\,\ol{Q}^{\dot{\al}}$ to be the supersymmetry generators in the two-dimensional $(\frac{1}{2},0),\,(0,\frac{1}{2})$ representations of the Lorentz group. The properties under transformations and closure of the algebra require (anti-)commutation rules which in four-component notation read
\begin{\eqn*}
[M^{\mu\nu},Q]\,=\,0\;,\;\left\{Q,\ol{Q}\right\}\,=\,0\;,\;[M^{\mu\nu},Q]\,=\,-\frac{1}{2}\,\Sigma^{\mu\nu}\,Q,
\end{\eqn*}
where 
\begin{\eqn*}
Q\,=\,{Q_{\al} \choose \ol{Q}^{\dot{\al}}}.
\end{\eqn*}
In two-component notation, this reads
\begin{eqnarray}\label{eq:susyal}
&&[Q_{\al},P_{\mu}]\,=\,[\ol{Q}_{\dot{\al}},P_{\mu}]\,=\,0\;,\;[Q_{\al},M_{\mu\nu}]\,=\,-i\,(\sigma_{\mu\nu})_{\al}^{\;\;\;\be}\,Q_{\be},\nonumber\\
&&[\ol{Q}^{\dot{\al}},M_{\mu\nu}]\,=\,- i\,(\ol{\sigma}_{\mu\nu})_{\;\;\;\dot{\be}}^{\dot{\al}}\,\ol{Q}^{\dot{\be}}\;,\;\left\{ Q_{\al},\ol{Q}^{\dot{\be}}\right\}\,=\,2\,(\sigma^{\mu})_{\al}^{\;\;\;\dot{\be}}P_{\mu},\nonumber\\
&&\left\{Q_{\al},Q_{\be}\right\}\,=\,\left\{\ol{Q}^{\dot{\al}},\ol{Q}^{\dot{\be}}\right\}\,=\,0.
\end{eqnarray} 
\section{Expansion in component fields}
A SUSY transformation for a field operator $\phi(x)$ is given by
\begin{\eqn*}
\exp\,\lb i\,(\xi^{\al}Q_{\al}+\bar{\xi}_{\dot{\al}}\,\ol{Q}^{\dot{\al}})\rb\,\phi(x)\, \exp\,\lb -i\,(\xi^{\al}Q_{\al}+\bar{\xi}_{\dot{\al}}\,\ol{Q}^{\dot{\al}})\rb,
\end{\eqn*}
where $\xi$ are Grassmann-valued variables.
Infinitesimally, this becomes
\begin{\eqn*}
\phi(x)\,\rightarrow\,\phi(x)\,+\,\delta_{\xi}\,\phi(x)\;,\;\delta_{\xi}\,\phi(x)\,=\,\left[ i\,(\xi\,Q\,+\,\bar{\xi}\,\ol{Q}),\phi(x)\right].
\end{\eqn*}
Starting with a scalar field $\phi(x)$ and requiring the closure of the algebra, we obtain the transformation rules
\begin{eqnarray}\label{eq:scaltrans}
\delta_{\xi}\,\phi&=&\sqrt{2}\,\xi\,\psi,\nonumber\\
\delta_{\xi}\,\psi&=&i\,\sqrt{2}\,\sigma^{\mu}\,\bar{\xi}\,\partial_{\mu}\,\phi\,+\,\sqrt{2}\,\xi\,F,\nonumber\\
\delta_{\xi}\,F&=&i\,\sqrt{2}\,\bar{\xi}\,\ol{\sigma}^{\mu}\,\partial_{\mu}\,\psi,
\end{eqnarray}
where $\psi$ and $F$ are a spinor/ auxiliary field with the dimensions $\frac{3}{2}$/ $2$ respectively.\\
\\
From (\ref{eq:susyal}), we can determine the action of a SUSY transformation generator
\begin{\eqn*}
G(a^{\mu},\xi,\bar{\xi})\,=\,\exp\lb i\,(\xi\,Q+\bar{\xi}\bar{Q}-a^{\mu}P_{\mu})\rb\
\end{\eqn*}
on a superfield $S(x^{\mu},\theta,\bar{\theta})$:
\begin{\eqn*}
S(x^{\mu},\theta,\bar{\theta})\,\rightarrow\,\exp\lb i\,(\xi\,Q+\bar{\xi}\bar{Q}-a^{\mu}P_{\mu})\rb\,S(x^{\mu},\theta,\bar{\theta}).
\end{\eqn*}
$S(x^{\mu},\theta,\bar{\theta})$ can be expanded in components proportional to $\theta,\,\bar{\theta},\,\theta\theta, ...$; the generators are then given by
\begin{eqnarray*}
P_{\mu}&=&i\,\partial_{\mu},\\
i\,Q_{\al}&=&\frac{\partial}{\partial\theta^{\al}}-i\,\sigma^{\mu}\bar{\theta}\partial_{\mu},\\
i\,\ol{Q}_{\dot{\al}}&=&-\frac{\partial}{\partial\bar{\theta}^{\dot{\al}}}+i\,\theta\,\sigma^{\mu}\partial_{\mu}.
\end{eqnarray*}
Fermionic derivatives which anticommute with the $Q,\,\ol{Q}$ are then given by
\begin{\eqn*}
D_{\al}\,=\,\frac{\partial}{\partial\theta^{\al}}+i\sigma^{\mu}\bar{\theta}\partial_{\mu}\;,\;\ol{D}_{\dot{\al}}\,=\,-\frac{\partial}{\partial\ol{\theta}^{\dot{\al}}}-i\theta\sigma^{\mu}\partial_{\mu},
\end{\eqn*}
which obey
\begin{\eqn*}
\left\{D_{\al},\ol{D}_{\dot{\al}}\right\}\,=\,2\,i\,\sigma^{\mu}_{\al\dot{\al}}\partial_{\mu}.
\end{\eqn*}
All other anticommutators vanish.\\
\vspace{5mm}\\
Defining a {\bf chiral superfield} $\Phi$ by requiring
\begin{\eqn*}
\ol{D}_{\dot{\al}}\,\Phi\,=\,0,
\end{\eqn*}
we obtain
\begin{eqnarray*}
\Phi(x^{\mu},\theta,\bar{\theta})&=&\phi+\sqrt{2}\theta\psi+\theta\theta\,F\,+\,i\,\partial_{\mu}\,\phi\theta\sigma^{\mu}\bar{\theta}
-\frac{i}{\sqrt{2}}\,\theta\theta\partial_{\mu}\psi\,\sigma^{\mu}\bar{\theta}\,-\,\frac{1}{4}\partial_{\mu}\,\partial^{\mu}\,\phi\,\theta\theta\bar{\theta}\bar{\theta},
\end{eqnarray*}
with the component fields and their transformation rules given by Eq. (\ref{eq:scaltrans}).\\
The antichiral field $\Phi^{\dagger}$ is given by
\begin{eqnarray*}
\Phi^{\dagger}&=&\phi^{\dagger}+\sqrt{2}\bar{\theta}\bar{\psi}+\bar{\theta}\bar{\theta}\,F^{\dagger}\,-\,i\,\partial_{\mu}\,\phi^{\dagger}\theta\sigma^{\mu}\bar{\theta}
+\frac{i}{\sqrt{2}}\,\bar{\theta}\bar{\theta}\theta\sigma^{\mu}\partial_{\mu}\bar{\psi}\,-\,\frac{1}{4}\partial_{\mu}\,\partial^{\mu}\,\phi^{\dagger}\,\theta\theta\bar{\theta}\bar{\theta},
\end{eqnarray*}
which obeys
\begin{\eqn*}
D_{\al}\,\Phi^{\dagger}\,=\,0.
\end{\eqn*}
\vspace{5mm}\\
A {\bf vector superfield} V is defined by requiring
\begin{\eqn*}
V\,=\,V^{\dagger}.
\end{\eqn*}
This leads to
\begin{eqnarray*}
V(x,\theta,\bar{\theta})&=&C(x)+i\,\theta\,\chi-i\bar{\theta}\bar{\chi}+\frac{1}{2}\,i\,\theta\theta[M(x)+N(x)]-\frac{1}{2}\,i\,\bar{\theta}\bar{\theta}[M(x)-iN(x)]\\
&&+\,\theta\sigma^{\mu}\bar{\theta}V_{\mu}(x)+i\theta\theta\bar{\theta}\left[ \bar{\lambda}(x)+\frac{i}{2}\bar{\sigma}^{\mu}\partial_{\mu}\,\chi(x)\right]-i\bar{\theta}\bar{\theta}\theta\left[ \lambda(x)+\frac{i}{2}\sigma^{\mu}\partial_{\mu}\,\bar{\chi}(x)\right]\\
&&+\frac{1}{2}\,\theta\theta\bar{\theta}\bar{\theta}\left[D-\frac{1}{2}\partial_{\mu}\partial^{\mu}C\right],
\end{eqnarray*}
with $C,M,N,D$ being real scalar fields, $\chi,\,\lambda$ Weyl spinor fields, and $V_{\mu}$ a real vector field. Note that there is a gauge invariance with respect to
\begin{\eqn*}
V\,\rightarrow\,V+i\,[\Phi-\Phi^{\dagger}],
\end{\eqn*}
which helps to reduce $V$ to
\begin{\eqn*}
V_{WZ}(x,\theta,\bar{\theta})\,=\,
\,\theta\sigma^{\mu}\bar{\theta}V_{\mu}(x)+i\theta\theta\bar{\theta} \bar{\lambda}(x)-i\bar{\theta}\bar{\theta}\theta\lambda(x)\,+\,\frac{1}{2}\,\theta\theta\bar{\theta}\bar{\theta}\,D
\end{\eqn*}
in the Wess Zumino (WZ) gauge. The fields $V_{\mu},\lambda,\,D$ transform under SUSY transformation as
\begin{eqnarray*}
\delta\lambda_{\al}&=&-iD\xi_{\al}-\frac{1}{2}(\sigma^{\mu}\bar{\sigma}^{\nu})\,\xi\,V_{\mu\nu},\\
\delta V^{\mu}&=&i\,(\xi\sigma^{\mu}\bar{\lambda}-\lambda\sigma^{\mu}\bar{\xi})-\partial^{\mu}(\xi\chi+\bar{\chi}\bar{\xi}),\\
\delta D&=&\partial_{\mu}\,(-\xi\,\sigma^{\mu}\bar{\lambda}+\lambda\sigma^{\mu}\bar{\xi}),
\end{eqnarray*}
with the field strength $V_{\mu\nu}\,=\partial_{\mu}V_{\nu}-\partial_{\nu}V_{\mu}$. We see that supersymmetry is explicitly broken in the WZ gauge; however, it can be restored by a proper gauge transformation.\\
\vspace{5mm}\\
Combinations of (anti)chiral and vector superfields which are again (anti-)chiral and vector superfields are given in table \ref{tab:combis}. A more extensive list can be found in \cite{Sohnius:1985qm}.
\begin{table}
\begin{\eqn*}
\begin{array}{ll}
(\Phi_{1}\Phi_{2}),\,(\Phi_{1}\Phi_{2}\Phi_{3}),\,(\ol{D}^{2}D_{\al}V)&\text{chiral fields}\\
(\Phi_{1}^{\dagger}\Phi_{1}),\,i\,[\Phi_{1}-\Phi_{1}^{\dagger}]&\text{vector fields}
\end{array}
\end{\eqn*}
\caption{\label{tab:combis} superfield character of different superfield combinations }  
\end{table}
\section{General gauge-invariant Lagrangian and superpotential}
For the construction of a Lagrangian, we have to find components of superfields which are supersymmetric. It is easy to see that these are given by the $\theta\theta\,(\bar{\theta}\bar{\theta})$ components of the $\Phi\,(\Phi^{\dagger})$ (F-terms) and the $\theta\theta\bar{\theta}\bar{\theta}$ components of $V$ (D-term) respectively, as all of these transform up to a total derivative. Lagrangians are therefore constructed using the F-terms of chiral and the D-terms of vector superfields and the field combinations as given in table \ref{tab:combis}.\\
\subsubsection*{Gauge invariance}
For a general $U(1)$ invariant Lagrangian, we define the field-strength superfield
\begin{\eqn*}
W_{\al}(y,\,\theta)\,=\,4\,i\,\lambda_{\al}(y)-[4\,\delta^{\;\;\;\be}_{\al}D(y)+2\,i\,(\sigma^{\mu}\bar{\sigma}^{\nu})_{\al}^{\;\;\;\be}\,V_{\mu\nu}(y)]\theta_{\be}+4\,\theta\theta\sigma^{\mu}_{\al\dot{\al}}\partial_{\mu}\bar{\lambda}^{\dot{\al}},
\end{\eqn*}
where $y^{\mu}\,=\,x^{\mu}+i\theta\sigma^{\mu}\bar{\theta}$. This contains a spinor $\lambda_{\al}$  with the dimension $\frac{3}{2}$ and the field strength $V_{\mu\nu}$ with dimension $2$ as well as the auxiliary field $D$. The pure gauge term is then given by the F-term of $(W_{\al}W^{\al}\,+\,hc)$:
\begin{\eqn}\label{eq:Lgauge}
\mL_{V}\,=\frac{1}{64}(W^{\al}W_{\al}+\ol{W}_{\dot{\al}}\ol{W}^{\dot{\al}})_{F}\,=\,-\frac{1}{4}\,V^{\mu\nu}V_{\mu\nu}+\,i\,\lambda\,\sigma^{\mu}\,\partial_{\mu}\bar{\lambda}+\frac{1}{2}D^{2}.
\end{\eqn}
The interaction between matter and gauge bosons is then given by the D-term of the vector superfield $\Phi^{\dagger}\,e^{V}\,\Phi$:
\begin{eqnarray}\label{eq:Lint}
\mL_{\Phi}\,=\,[\Phi^{\dagger}\,e^{V}\,\Phi]_{D}&=&(D_{\mu}\phi)^{\dagger}(D^{\mu}\phi)+i\,\psi\,\sigma^{\mu}D_{\mu}^{\dagger}\ol{\psi}+F^{\dagger}F
+i\,\sqrt{2}\,g\,(\phi^{\dagger}\lambda\psi-\ol{\psi}\bar{\lambda}\phi)+g\phi^{\dagger}D\phi,\nonumber\\
&&
\end{eqnarray}
where $D_{\mu}\,=\,\partial_{\mu}+i\,g\,V_{\mu}$. Then, the total gauge-invariant Lagrangian is
\begin{\eqn*}
\mL_{inv}\,=\,\mL_{V}+\mL_{\Phi}.
\end{\eqn*}
For the construction of QED with massive fermions, we have to include left- and righthanded component fields; this can be done by defining the combinations
\begin{\eqn*}
\Phi_{+}\,=\,\frac{1}{\sqrt{2}}\,(\Phi_{1}\,+\,i\Phi_{2})\;,\;\Phi_{-}\,=\,\frac{1}{\sqrt{2}}\,(\Phi_{1}\,-\,i\Phi_{2}).
\end{\eqn*}
Their transformations are then given by
\begin{\eqn*}
\Phi_{+}\,\rightarrow\,\exp(-2\,i\,\Lambda)\Phi_{+}\;,\;\Phi_{-}\,\rightarrow\,\exp(2\,i\,\Lambda)\Phi_{-}\;,\;V\,\rightarrow\,V+\,i\,(\Lambda-\Lambda^{\dagger}).
\end{\eqn*}
In addition to $\mL_{\Phi_{+}},\,\mL_{\Phi_{-}}$, this allows the inclusion of mass generating terms of the form
\begin{\eqn*}
\mL_{\text{mass}}\,=\,m\,\lb (\Phi_{+}\Phi_{-})_{\theta\theta}\,+\,(\Phi^{\dagger}_{+}\Phi^{\dagger}_{-})_{\bar{\theta}\bar{\theta}}\rb,
\end{\eqn*}
so the total gauge-invariant Lagrangian including the coupling of massive particles to the gauge bosons is
\begin{\eqn*}
\mL_{inv}\,=\,\mL_{V}\,+\,\mL_{\Phi_{+}}\,+\,\mL_{\Phi_{-}}\,+\,\mL_{\text{mass}}.
\end{\eqn*}
This procedure can be extended to any (non)abelian gauge group. We have to add terms in the form of Eqs. (\ref{eq:Lgauge}), (\ref{eq:Lint}) for every new gauge group and modify $D_{\mu}$ accordingly. In the non-abelian case, the vector superfield is given by
\begin{\eqn*}
V^{a}_{\mu\nu}\,=\,\partial_{\mu}V^{a}_{\nu}-\partial_{\nu}V^{a}_{\mu}-g\,f^{abc}\,V_{\mu}^{b}V_{\nu}^{c},
\end{\eqn*}
where $f^{abc}$ are the structure constants of the non-abelian gauge group. Its transformation is given by
\begin{\eqn*}
V^{a}\,\rightarrow\,V^{a}+\,i\,(\Lambda^{a}-\Lambda^{a,\,\dagger})\,-\,g\,f^{abc}\,V^{b}\,(\Lambda^{c}+\Lambda^{\dagger,c})\,+\,\mO(g^{2}).
\end{\eqn*}
\subsubsection*{Superpotential}
The superpotential defines the possible self-interactions terms of chiral fields. Renormalizability requirements determine the most general SUSY-preserving potential 
\begin{\eqn*}
f(\Phi_{i})\,=\,\sum_{i}a_{i}\,(\Phi_{i})_{\theta\theta}\,+\,\frac{1}{2}\,\sum_{i,j}m_{i,j}\,(\Phi_{i}\Phi_{j})_{\theta\theta}\,+\,\frac{1}{3}\sum_{i,j,k}g_{ijk}(\Phi_{i}\Phi_{j}\Phi_{k})_{\theta\theta}.
\end{\eqn*}
\subsubsection*{Elimination of auxiliary fields}
The unphysical auxiliary fields $F_{i},D$ are generally eliminated by requiring
\begin{\eqn*}
\frac{\partial\,\mL}{\partial D}\,=\,\frac{\partial \mL}{\partial F_{i}}\,=\,0,
\end{\eqn*} 
which leads to
\begin{\eqn}\label{eq:FDsol}
F_{i}^{\dagger}\,=\,-\frac{\partial f}{\partial \phi_{i}}\;,\,D\,=\,-\sum_{i}\phi_{i}^{\dagger}g\,\phi_{i}.
\end{\eqn}
\section{MSSM field content and superpotential}
As the MSSM is the minimal supersymmetric extension of the Standard Model, it is designed to reproduce the SM particle content including as few extra fields as possible.\\
We therefore need
\begin{itemize}
\item{}Vector fields as in (\ref{eq:Lgauge}) for the $SU(2)\,\times\,U(1)$ electroweak as well as the $SU(3)$ strong gauge groups
\item{}gauge-particle interactions as in (\ref{eq:Lint}) for the Standard Model matter content as listed in  table \ref{tab:smmatter}.
\item{}a superpotential $f_{MSSM}$ giving masses to the (s)fermions. This requires a modified Higgs sector: instead of the SM Higgs, the two Higgs doublets given by Eq. (\ref{eq:higgses}) are introduced. They carry the hypercharges $Y=\frac{1}{2}\,(H_{u})$ and $Y=-\frac{1}{2}\,(H_{d})$.
\end{itemize}
\begin{table}
\begin{\eqn*}
{\renewcommand{\arraystretch}{1.5}
\begin{array}{l|l} \text{gauge group}& \text{non-singlet SM matter fields} \\ \hline
U(1)\,(Y)& u_{L},d_{L}\; (\frac{1}{6}),\;\nu_{l,L},l_{L}\;(-\frac{1}{2}),\;u_{R}\;(-\frac{2}{3}),\;d_{R}\;(\frac{1}{3}),\;l_{R}\;(1),\;H\;(\frac{1}{2})\\ 
SU(2)&{u_{L} \choose d_{L}},\;{\nu_{l,L}\choose l_{L}},H\\
SU(3)&u_{L,c},\;d_{L,c},\;u_{R,c},\;d_{R,c}
\end{array}}
\end{\eqn*}
\caption{\label{tab:smmatter} Standard Model matter content and gauge-group transformation properties. $u=(u,c,t),\,d=(d,s,b),\,l=(e,\mu,\tau)$ run over the respective families and $c$ is the color index. $Y$ is the hypercharge. Transformation properties of gauge fields are not listed.}
\end{table}
The minimal superpotential reproducing the Yukawa-terms of the Standard Model and preserving supersymmetry is 
\begin{\eqn}\label{eq:superpot}
f_{MSSM}(\phi)=\bar{u} y_{u} Q H_{u}-\bar{d} y_{d} Q H_{d}- \bar{e} y_{e}L H_{d}+\mu H_{u} H_{d},
\end{\eqn}
with
\begin{\eqn}\label{eq:higgses}
H_{u}={H^{+}_{u} \choose H^{0}_{u}}, H_{d}={H^{0}_{d} \choose H^{-}_{d}},
\end{\eqn}
and $\bar{u},\bar{d},\bar{e}$ denote the righthanded SU(2) quark and lepton singlets  and $Q,L$ the lefthanded  SU(2) quark doublets. $y_{i}$ are the respective Yukawa matrices.\\
\section{Symmetry breaking and Lagrangian of the MSSM}
\subsubsection*{SUSY and gauge breaking}
Supersymmetry breaking terms need to preserve the cancellation of the quadratic divergences of the corrections to the Higgs mass. The most general terms fulfilling this requirement are 
\begin{eqnarray}\label{eq:lbreak}
\lefteqn{\mL_{SUSY-breaking}\,=\,m_{\tilde{q}}^{2}|\tilde{q}_{L}|^{2}\,+\,m^{2}_{\tilde{u}}|\bar{\tilde{u}}_{R}|^{2}+m^{2}_{\tilde{d}}|\bar{\tilde{d}}_{R}|^{2}\,+\,m_{\tilde{l}}^{2}|\tilde{l}_{L}|^{2}+m^{2}_{\tilde{e}}|\bar{\tilde{e}}_{R}|^{2}}\nonumber\\
&+&\lb\,\lambda_{E}\,A_{E}\,H_{d}\tilde{l}_{L}\bar{\tilde{e}}_{R}\,+\,\lambda_{D}\,A_{D}\,H_{d}\tilde{q}_{L}\bar{\tilde{u}}_{R}\,+\,\lambda_{U}\,A_{U}\,H_{u}\tilde{q}_{L}\bar{\tilde{d}}_{R}\,+\,B\,\mu\,H_{u}\,H_{d}\,+\,h.c.\rb\nonumber\\
&+&m^{2}_{H_{u}}|H_{u}|^{2}\,+\,m^{2}_{H_{d}}\,|H_{d}|^{2}\,+\,\frac{1}{2}\,\lb M_{1}\,\wt{B}\,\wt{B}\,+\,M_{2}\,\wt{W}\,\wt{W}\,+\,M_{3}\,\tilde{g}\,\tilde{g}\,+\,h.c.\rb.\nonumber\\
&&
\end{eqnarray} 
Here, $m^{2}_{\tilde{q}},\,m^{2}_{\tilde{u}},\,m^{2}_{\tilde{d}},\,m^{2}_{\tilde{l}},\,m^{2}_{\tilde{e}}$ are general hermitian $3\, \times\, 3$ matrices and $\lambda_{E}\,A_{E},\,\lambda_{D}\,A_{D},\,\lambda_{U}\,A_{U}$ general $3\, \times\, 3$ matrices, $|\tilde{q}_{L}|^{2}\,=\,(u^{\dagger}_{L}\,u_{L}\,+\,d^{\dagger}_{L}\,d_{L})$ (same for $\tilde{l}_{L}$) and the multiplication of two $SU(2)$ doublets is given by
\begin{\eqn}\label{eq:doubmult}
D_{1}\,D_{2}\,=\,\epsilon^{ij}D_{1,i}\,D_{2,j}\,=\,D_{1,1}\,D_{2,2}-D_{2,1}\,D_{1,2}.
\end{\eqn}
As in the SM, the electroweak gauge symmetry is broken because the Higgs doublet aquire non-zero VEVs. However, the specific form of the MSSM Higgs potential requires both $\mu$ in $f_{MSSM}$ and $B\,\mu$ in $\mL_{SUSY-breaking}$ to be non-zero.\\
\subsubsection*{MSSM Lagrangian}
The complete MSSM Lagrangian is then given by
\begin{eqnarray*}
\lefteqn{\mL_{MSSM}\,=\,-\frac{1}{4}\,V_{g'}^{\mu\nu}V_{g',\mu\nu}+\,i\,\lambda_{g'}\,\sigma^{\mu}\,\partial_{\mu}\bar{\lambda}_{g'}+\frac{1}{2}D_{g'}^{2}\,-\frac{1}{4}\,V_{g}^{i,\mu\nu}V^{i}_{g,\mu\nu}+\,i\,\,T^{i}\,\lambda^{i}_{g}\,\sigma^{\mu}\,\partial_{\mu}\bar{\lambda}^{i}_{g}+\frac{1}{2}(D^{i}_{g})^{2}}\\
&-&\frac{1}{4}\,V_{g_{s}}^{a,\mu\nu}V^{a}_{g_{s},\mu\nu}+\,i\,\,T^{a}\,\lambda^{a}_{g}\,\sigma^{\mu}\,\partial_{\mu}\bar{\lambda}^{a}_{g_{s}}+\frac{1}{2}(D^{a}_{g})^{2}\\
&+&(D_{\mu}\phi_{I})^{\dagger}(D^{\mu}\phi_{I})+i\,\psi_{I}\,\sigma^{\mu}D_{\mu}^{\dagger}\ol{\psi}_{I}+F_{I}^{\dagger}F_{I}+i\,\sqrt{2}\,g_{k}\,(\phi_{I}^{\dagger}T^{k}\lambda^{k}\psi_{I}-\ol{\psi}_{I}\bar{\lambda}^{k}T^{k}\phi_{I})+g_{k}\phi_{I}^{\dagger}D_{I}\phi_{I}\\
&+&m_{\tilde{q}}^{2}|\tilde{q}_{L}|^{2}\,+\,m^{2}_{\tilde{u}}|\bar{\tilde{u}}_{R}|^{2}+m^{2}_{\tilde{d}}|\bar{\tilde{d}}_{R}|^{2}\,+\,m_{\tilde{l}}^{2}|\tilde{l}_{L}|^{2}+m^{2}_{\tilde{e}}|\bar{\tilde{e}}_{R}|^{2}\\
&+&\lb\,\lambda_{E}\,A_{E}\,H_{d}\tilde{l}_{L}\bar{\tilde{e}}_{R}\,+\,\lambda_{D}\,A_{D}\,H_{d}\tilde{q}_{L}\bar{\tilde{u}}_{R}\,+\,\lambda_{U}\,A_{U}\,H_{u}\tilde{q}_{L}\bar{\tilde{d}}_{R}\,+\,B\,\mu\,H_{u}\,H_{d}\,+\,h.c.\rb\\
&+&m^{2}_{H_{u}}|H_{u}|^{2}\,+\,m^{2}_{H_{d}}\,|H_{d}|^{2}\,+\,\frac{1}{2}\,\lb M_{1}\,\wt{B}\,\wt{B}\,+\,M_{2}\,\wt{W}\,\wt{W}\,+\,M_{3}\,\tilde{g}\,\tilde{g}\,+\,h.c.\rb\\
&&+\lb \bar{u} y_{u} Q H_{u}-\bar{d} y_{d} Q H_{d}- \bar{e} y_{e}L H_{d}+\mu H_{u} H_{d}\rb_{\theta\theta},
\end{eqnarray*}
where $i\,=1,2,3$ is the $SU(2)$ and $a=1,2,...,8$ the $SU(3)$ group index; the index $I$ in the gauge-matter interaction part (\ref{eq:Lint}) runs over all superfields corresponding to the SM particles listed in Table \ref{tab:smmatter} and the Higgs fields (\ref{eq:higgses}), where the index $k$ symbolizes that all respective gauge-groups as given in Table \ref{tab:smmatter} have to be taken into account. The field content from the superpotential is given by its $(\theta\theta)$ component and  can be derived using Eq. (\ref{eq:doubmult}) and
\begin{eqnarray*}
\Phi_{i} \Phi_{j}|_{(\theta \theta)}&=&-\psi_{i} \psi_{j},\\
\Phi_{i} \Phi_{j} \Phi_{k}|_{(\theta \theta)}&=& -(\psi_{i}\psi_{j}\phi_{k}+ \mbox{cyclic terms}). 
\end{eqnarray*}
Finally, $D$ and $F$ terms are eliminated using Eq. (\ref{eq:FDsol}).\\
\vspace{3mm}\\
The field content of the MSSM is then given in Table \ref{tab:MSSMmat}. Physically observed particles correspond to superpositions of these fields carrying the same quantum numbers which are rotated into mass-eigenstates. In addition, the two-component Weyl-spinors $\lambda,\,\psi$ have to be combined to obtain four-component Dirac- or Majorana spinors. For the gaugino/ Higgsino sector, this is done in Appendix \ref{app:chneutr}. For a more complete treatment, we refer to e.g. \cite{Haber:1985rc, Drees:1996ca, Martin:1997ns}.
\begin{table}[tb]
\vspace{5mm}
\begin{\eqn*}
{\renewcommand{\arraystretch}{1.5}
\begin{array}{ll|ll}\text{bosonic}&&\text{fermionic, spin = }\frac{1}{2}\\ \hline
\tilde{e}_{L,R},\tilde{\mu}_{L,R}\,\tilde{\tau}_{L,R}&\text{(sleptons, spin  = $0$)}&e_{L,R},\mu_{L,R}\,\tau_{L,R}&\text{(leptons)}\\
\tilde{u}_{L,R},\tilde{d}_{L,R}\,\tilde{c}_{L,R}\,\tilde{s}_{L,R},\tilde{t}_{L,R}\,\tilde{b}_{L,R}&\text{(squarks, spin = $0$)}&u_{L,R},d_{L,R}\,c_{L,R}\,s_{L,R}\,t_{L,R}\,\,b_{L,R}&\text{(quarks)}\\
B,\,W^{i},\,g^{a}&\text{(gauge bosons, spin = $1$)}&\wt{B},\,\wt{W}^{i},\,\tilde{g}^{a}&\text{(gauginos)}\\
H^{+}_{u},\,H^{0}_{u},\,H^{-}_{d},\,H^{0}_{d}&\text{(Higgs bosons, spin = $0$)}&\wt{H}^{+}_{u},\,\wt{H}^{0}_{u},\,\wt{H}^{-}_{d},\,\wt{H}^{0}_{d}&\text{(Higgsinos)}
\end{array}}
\end{\eqn*}
\caption{\label{tab:MSSMmat} field content in the MSSM}
\end{table}

\chapter{Chargino and Neutralino Sector of the MSSM}\label{app:chneutr}
Here, we sketch the derivation of the Feynman rules for any vertices including charginos and neutralinos. We hereby closely follow \cite{Haber:1985rc}. We point to differences in conventions if necessary.\\
\section{Chargino mass eigenstates}
In the MSSM, the EW symmetry is spontaneously broken by non-zero VEVs of the two neutral components of the Higgs doublets:
\begin{equation*}
<H^{0}_{d}>\,=\,v_{1}\;,\;<H^{0}_{u}>\,=\,v_{2}.
\end{equation*}
$\tan\beta$ is defined as the ratio of these VEVs\footnote{Note that the definition of $\tan\theta_{v}$ in \cite{Haber:1985rc} is given by  $\tan\theta_{v}\,=\,(\tan\beta)^{-1}$.}:
\begin{\eqn*}
\tan \beta\,=\,\frac{v_{2}}{v_{1}}
\end{\eqn*}
Taking this into account in the chargino-chargino-Higgs couplings and using
\begin{\eqn*}
\mu H^{+}_{u} H^{-}_{d}|_{\theta\theta,\psi}\,=\,-\mu \widetilde{H}^{+}_{u}\widetilde{H}^{-}_{d},
\end{\eqn*}
 we obtain the following mass terms in the Lagrangian: 
\begin{eqnarray*}
&&(M_{2}\widetilde{W}^{+}\widetilde{W}^{-}-\mu\wt{H}^{+}_{u}\wt{H}^{-}_{d})+h.c.-\imath g \overline{\widetilde{H}^{-}_{d}}\overline{\widetilde{W}^{-}}v_{1}+\imath g v_{1}^{*}\widetilde{H}^{-}_{d}\widetilde{W}^{-}
-\imath g \overline{\widetilde{H}^{+}_{u}}\overline{\widetilde{W}^{+}}v_{2}+\imath g v_{2}^{*}\widetilde{H}^{+}_{u}\widetilde{W}^{-}.
\end{eqnarray*}
We now define $\Psi^{-}$ and $\Psi^{+}$ by
\begin{\eqn}\label{eq:psis}
\Psi^{-}={-\imath\widetilde{W}^{-} \choose \widetilde{H}^{-}_{d}}\;,\;\Psi^{+}={-\imath\widetilde{W}^{+} \choose \widetilde{H}^{+}_{u}},
\end{\eqn}
and obtain
\begin{\eqn*}
-{\Psi^{-}}^{\top}\,X\,\Psi^{+}-(\overline{\Psi}^{+})^{\top}\,X^{\dagger}\,\overline{\Psi}^{-},
\end{\eqn*}
as mass terms of the Lagrangian with
\begin{\eqn*}
X=\left( \begin{array}{cc}
M_{2}&g\,v_{2}\\
g\,v_{1}&\mu\end{array}\right).
\end{\eqn*}
Using the relation 
\begin{\eqn*}
m_{w}^{2}\,=\,\frac{1}{2}\,g^{2}(v^{2}_{1}+v^{2}_{2}),
\end{\eqn*}
we can rewrite this as
\begin{\eqn}\label{eq:xdef}
X=\left( \begin{array}{cc}
M_{2}&\sqrt{2}m_{w}\sin\beta\\
\sqrt{2}m_{w}\cos\beta&\mu\end{array}\right).
\end{\eqn}
This matrix can now be diagonalized using 
\begin{\eqn*}
U_{-}\,X\,U_{+}^{-1}\,=\,M_{\mbox{diag}},
\end{\eqn*}
with
\begin{\eqn*}
U_{\pm}=\left( \begin{array}{cc}
\cos\Phi_{\pm}&\sin\Phi_{\pm}\\
-\sin\Phi_{\pm}&\cos\Phi_{\pm}\end{array}\right).
\end{\eqn*}
The angles $\cos\Phi_{\pm}$ and $\sin\Phi_{\pm}$ obey \cite{Choi:2000ta, Choi:1998ei}:
\begin{eqnarray*}
\cos\,2\Phi_{\pm}&=&-\left[M_{2}^{2}-|\mu|^{2}\,\pm\,2\,m_{W}^{2}\,\cos\,2\beta\right]/\Delta_{C},\\
\sin\,2\,\Phi_{\pm}&=&-2\,m_{W}\,\sqrt{M_{2}^{2}+|\mu|^{2}\,\mp(M^{2}_{2}-|\mu|^{2})\,\cos\,2\beta+2\,M_{2}|\mu|\sin\,2\beta}/\Delta_{C},\\
\Delta_{C}&=&\sqrt{(M^{2}_{2}-|\mu|^{2})^{2}+4\,m_{W}^{2}\,\cos^{2}\,2\beta+4\,m_{W}^{2}\,(M^{2}_{2}+|\mu|^{2})+8m_{W}^{2}\,M_{2}|\mu|\,\sin\,2\beta}.
\end{eqnarray*}
The two-component charged mass-eigenstates $\chi^{\pm}$ are then given by
\begin{\eqn*}
\chi^{+}\,=\,U_{+}\,\Psi^{+}\;,\;\chi^{-}\,=\,U^{*}_{-}\,\Psi^{-},
\end{\eqn*}
and the four-component Dirac spinors $\widetilde{\chi}_{i}^{+}$ by
\begin{\eqn*}
\widetilde{\chi}^{+}_{i}\,=\,{\chi^{+}_{i} \choose \overline{\chi^{-}_{i}}}
\;\;\text{or}\;\; 
\widetilde{\chi}^{-}_{i}\,=\,{\chi^{-}_{i} \choose \overline{\chi^{+}_{i}}},
\end{\eqn*}
depending on the particle/ antiparticle choice.
\\
\hspace{3mm}
Note that there are different notations for these matrices in the literature which have to be taken into account when comparing results. We here give a short overview:\\
We start with the form 
\begin{\eqn*}
(\Psi^{+}\;\Psi^{-})\lb\begin{array}{cc}0&X^{\top}\\X&0\end{array}\rb{\Psi^{+} \choose \Psi^{-}}
\end{\eqn*}
with $\Psi^{\pm}$ as by Eq. (\ref{eq:psis} and $X$ as by Eq. (\ref{eq:xdef}).
We now define the matrices $U,\,V$ such that
\begin{\eqn*}
\chi^{-}\,=\,U\Psi^{-}\;;\;\chi^{+}\,=\,V\Psi^{+}
\end{\eqn*}
and see that $X$ is diagonalized using
\begin{\eqn*}
U^{*}\,X\,V^{\dagger}.
\end{\eqn*}
Depending on the parameterization of $U$ and $V$, $m_{\wt{\chi}^{\pm}_{1}}$ can be larger or smaller than $m_{\wt{\chi}^{\pm}_{2}}$. Table \ref{tab:convics} gives an overview of the different conventions used in the literature.
\begin{table}\label{tab:convics}
\begin{\eqn*}
\begin{array}{l|c|c|c}\mbox{Source}&&&\mbox{diag.}\\ \hline \mbox{HK} \cite{Haber:1985rc} & V & U & U^{*}\,X\,V^{\dagger}\\ \mbox{here} & U_{+} & U^{*}_{-} & U_{-}\,X\,U^{\dagger}_{+}\\ \mbox{LH} \cite{Skands:2003cj}& V & U& U^{*}\,X\,V^{\dagger}\\ \mbox{Choi} \cite{Choi:1998ei}& U_{R} & U_{L}& U^{*}_{L}\,X\,U^{\dagger}_{R}\\
\mbox{FC} \cite{FormCalc} & V& U& U^{*}\,X\,V^{\dagger}\\ 
 \end{array}
\end{\eqn*}
\caption{Different conventions for chargino-diagonalization matrices in the literature}
\end{table}
\section{Neutralino Mass Eigenstates}
From $\mL_{MSSM}$, the neutralino mass terms are given by
\begin{eqnarray*}
 && \lb \frac{1}{2}M_{1}\wt{B}\wt{B}+\frac{1}{2}M_{2}\wt{W}^{0}\wt{W}^{0}+\mu\wt{H}^{0}_{d}\wt{H}^{0}_{u}\,-\,
\frac{\imath\,g}{\sqrt{2}}v_{2}\wt{H}^{0}_{u}\wt{W}^{0}+\frac{\imath\,g'}{\sqrt{2}}v_{2}\wt{H}^{0}_{u}\wt{B}\right.\\
&&\left.+\frac{\imath\,g}{\sqrt{2}}v_{1}\wt{H}^{0}_{d}\wt{W}^{0}-\frac{\imath\,g'}{\sqrt{2}}v_{2}\wt{H}^{0}_{d}\wt{B}\rb\;\;+  h.c.\\
\end{eqnarray*}
We define $\Psi_{L}^{0}$ by
\begin{\eqn*} 
\Psi_{L}^{0}\,=\,\left(\begin{array}{c}-\imath\wt{B}\\-\imath\wt{W^{0}}\\\wt{H^{0}_{d}}\\\wt{H^{0}_{u}}\end{array}\right)\,=\,\left(\begin{array}{c}\wt{B}_{L}\\\wt{W}^{0}_{L}\\(\wt{H^{d}})_{L}\\(\wt{H^{u}})_{L}\end{array}\right),
\end{\eqn*}
with the four-component Majorana-spinors 
\begin{\eqn*}
\wt{B}\,=\,{-\imath\wt{B} \choose \imath\,\ol{\wt{B}}}\;,\;\wt{W}^{0}\,=\,{-\imath\wt{B} \choose \imath\,\ol{\wt{W}^{0}}}\;,\;
\wt{H}^{u}\,=\,{H_{u}^{0} \choose \ol{H^{0}_{u}}}\;,\;\wt{H}^{d}\,=\,{H_{d}^{0} \choose \ol{H^{0}_{d}}}.
\end{\eqn*}
We can then rewrite the mass part of the Lagrangian as 
\begin{\eqn*}
-\frac{1}{2}{\Psi^{0}_{L}}^{\top}\,M\,\Psi^{0}_{L}\,+\,h.c.,
\end{\eqn*}
with
\begin{\eqn*}
M\,=\,\left(\begin{array}{cccc}M_{1}&0&-\frac{g'}{\sqrt{2}}\,v_{1}&\frac{g'}{\sqrt{2}}\,v_{2}\\
0&M_{2}&\frac{g}{\sqrt{2}}\,v_{1}&-\frac{g}{\sqrt{2}}\,v_{2}\\
-\frac{g'}{\sqrt{2}}\,v_{1}&\frac{g}{\sqrt{2}}\,v_{1}&0&-\mu\\
\frac{g'}{\sqrt{2}}\,v_{2}&-\frac{g}{\sqrt{2}}\,v_{2}&-\mu&0\\
\end{array}\right).
\end{\eqn*}
Using
\begin{\eqn*}
m^{2}_{z}=\frac{1}{2}(g^{2}+g'^{2})v^{2}\;;\;\frac{s_{w}}{c_{w}}=\frac{g'}{g}\;;\;\frac{s_{\beta}}{c_{\beta}}=\frac{v_{2}}{v_{1}}\;;\;v^{2}=v^{2}_{1}+v^{2}_{2},
\end{\eqn*}
we obtain
\begin{\eqn*}
M\,=\,\left(\begin{array}{cccc}M_{1}&0&-m_{Z}c_{\beta}s_{w}&m_{Z}s_{\beta}s_{w}\\
0&M_{2}&m_{Z}c_{\beta}c_{w}&-m_{Z}s_{\beta}c_{w}\\
-m_{Z}c_{\beta}s_{w}&m_{Z}c_{\beta}c_{w}&0&-\mu\\
m_{Z}s_{\beta}s_{w}&-m_{Z}s_{\beta}c_{w}&-\mu&0\\
\end{array}\right).
\end{\eqn*}
This can be diagonalized using a unitary matrix such that
\begin{\eqn*}
M_{diag}\,=N^{*}\,M\,N^{\dagger}.
\end{\eqn*}
We obtain $N$ from the diagonalization of $M\,M^{\dagger}$:
\begin{\eqn*}
M_{diag}M_{diag}^{\dagger}\,=N^{*}M\,M^{\dagger}N^{\top}.
\end{\eqn*}
The two-component mass eigenstates are then given by
\begin{\eqn*}
\chi^{0}_{i}=N_{ij}\Psi_{L,j}^{0},
\end{\eqn*}
and the four-component spinors by
\begin{\eqn*}
\wt{\chi}^{0}_{i}\,=\,{\chi^{0}_{i} \choose \ol{\chi^{0}_{i}}}.
\end{\eqn*}
When taking $N$ purely real, neutralino masses can come out negative. In this case, the physical field is given by $\gamma_{5}\wt{\chi}^{0}$. Note that this has to be taken into account when deriving the Feynman rules for the physical particles from $\mL$.\\
An analytic expression for the diagonalization matrix $N$ can be found in the literature, cf. e.g. \cite{Bartl:1989ms} and \cite{Choi:2001ww}. 
\section{Chargino-Gauge-boson couplings}
We will here and in the following switch between two-component Weyl spinors and four-component Dirac- and Majorana-spinors; a good introduction can e.g. be found in \cite{Bailin:1994qt}.\\
The part of $\mL_{MSSM}$ describing chargino-chargino-gauge boson coupling is given by
\begin{eqnarray*}
&&\frac{1}{2}g\overline{\widetilde{H}^{-}_{d}}\bar{\sigma}^{\mu}W_{\mu}^{0}\widetilde{H}^{-}_{d}\,+\, \frac{1}{2}g'\overline{\widetilde{H}^{-}_{d}}\bar{\sigma}^{\mu}B_{\mu}\widetilde{H}^{-}_{d}
\,-\,\frac{1}{2}g\overline{\widetilde{H}^{+}_{u}}\bar{\sigma}^{\mu}W_{\mu}^{0}\widetilde{H}^{+}_{u}\\
&&- \frac{1}{2}g'\overline{\widetilde{H}^{+}_{u}}\bar{\sigma}^{\mu}B_{\mu}\widetilde{H}^{+}_{u}
\,-\,g\overline{\widetilde{W}^{+}}\bar{\sigma}^{\mu}W_{\mu}^{0}\widetilde{W}^{+}_{u}\,+\, g\overline{\widetilde{W}^{-}}\bar{\sigma}^{\mu}W^{0}_{\mu}\widetilde{W}^{-}
\end{eqnarray*}
 Let us define the $\wt{\chi}^{+}$ as the particle. If we use
\begin{\eqn*}
\widetilde{W}\,=\,{-\imath\widetilde{W}^{+} \choose \overline{(-\imath\widetilde{W}^{-})}}\;,\;\widetilde{H}\,=\,{\widetilde{H}^{+}_{u} \choose \overline{\widetilde{H}^{-}_{d}}},
\end{\eqn*} 
the Lagrangian is given by
\begin{\eqn}\label{eq:charggauge}
-g\overline{\widetilde{W}}\gamma^{\mu}W^{0}_{\mu}\widetilde{W}-\frac{1}{2}g\ol{\wt{H}}\gamma^{\mu}W^{0}_{\mu}\wt{H}-\frac{1}{2}g'\ol{\wt{H}}\gamma^{\mu}B_{\mu}\wt{H}.
\end{\eqn}
We can now split both $\wt{W}$ and $\wt{H}$ into left- and right-handed parts as 
\begin{\eqn*}
\ol{\wt{W}}\gamma^{\mu}W_{\mu}^{0}\wt{W}\,=\,\ol{\wt{W}_{L}}\gamma^{\mu}W_{\mu}^{0}\wt{W}_{L}+\ol{\wt{W}_{R}}\gamma^{\mu}W_{\mu}^{0}\wt{W}_{R}
\end{\eqn*}
and same with $B_{\mu}$ and $\wt{H}$.
We then obtain the following respective relations between the left- and righthanded parts of $\wt{\chi}_{1,2}$ and $\wt{H},\wt{W^{0}}$:
\begin{eqnarray*}
{\wt{\chi}_{1\;L} \choose\wt{\chi}_{2\;L}}\,=\,U^{+} {\wt{W}_{L} \choose \wt{H}_{L}}&,&{\wt{\chi}_{1\;R} \choose\wt{\chi}_{2\;R}}\,=\,U^{-} {\wt{W}_{R} \choose \wt{H}_{R}}.
\end{eqnarray*}
The whole coupling term of the Lagrangian containing left-handed parts now reads
\begin{eqnarray*}
&&-g\ol{\wt{W}_{L}}\gamma^{\mu}W_{\mu}^{0}\wt{W}_{L}-\frac{1}{2}g\ol{\wt{H}_{L}}\gamma^{\mu}W_{\mu}^{0}\wt{H}_{L}-\frac{1}{2}g'\ol{\wt{H}_{L}}\gamma^{\mu}B_{\mu}\wt{H}_{L}\\
&=&-\ol{{\wt{W}_L} \choose \wt{H}_{L}}^{\top}\left(\begin{array}{cc}
g\gamma^{\mu}W_{\mu}^{0}&0\\0 & \frac{1}{2}(g\gamma^{\mu}W^{0}_{\mu}+g'\gamma^{\mu}B_{\mu}) \end{array}\right){\wt{W}_{L} \choose \wt{H}_{L}}.
\end{eqnarray*}
The same relation (with $L \leftrightarrow R$) holds for the right-handed part.\\
If we define the negatively charged charginos to be the particles, the derivation is similar; however, we then have to define
\begin{\eqn*}
\widetilde{W}'\,=\,{-\imath\widetilde{W}^{-} \choose \overline{(-\imath\widetilde{W}^{+})}}\;\;\widetilde{H}'\,=\,{\widetilde{H}^{-}_{d} \choose \overline{\widetilde{H}^{+}_{u}}}.
\end{\eqn*}
The terms in the Lagrangian describing the gauge-boson couplings are the same up to a sign change. The transformation rules for the right- and left-handed parts are given by 
\begin{eqnarray*}
{\wt{\chi}_{1\;L} \choose\wt{\chi}_{2\;L}}\,=\,(U^{-})^{*} {\wt{W}_{L} \choose \wt{H}_{L}}&;&{\wt{\chi}_{1\;R} \choose\wt{\chi}_{2\;R}}\,=\,(U^{+})^{*} {\wt{W}_{R} \choose \wt{H}_{R}}
\end{eqnarray*}
and $\wt{\chi}^{-}$ is defined as
\begin{\eqn*}
\widetilde{\chi}^{-}_{i}\,=\,{\chi^{-}_{i} \choose \overline{\chi^{+}_{i}}},
\end{\eqn*}
i.e. the charge-conjugate of $\wt{\chi}^{+}$ (as required).\\
\hspace{5mm}
Using the definition of the physical fields $Z_{\mu}$ and $A_{\mu}$, 
\begin{\eqn*}
{Z_{\mu} \choose A_{\mu}}\,=\,\left(\begin{array}{cc} \cos \theta_{w} & -\sin\theta_{w}\\ \sin\theta_{w}& \cos\theta_{w}\end{array}\right)\,{W_{\mu}^{0} \choose B_{\mu}},
\end{\eqn*}
with $\theta_{w}$ being the weak mixing angle, we obtain from Eq. (\ref{eq:charggauge}) for the left-handed parts of the $\wt{\chi}^{+}$:
\begin{eqnarray*}
-\ol{\wt{\chi}^{+}_{L}}^{\top}\wt{A_{L}}\gamma^{\mu}Z_{\mu}\wt{\chi}^{+}_{L}&=&-\ol{{\wt{W}_L} \choose \wt{H}_{L}}^{\top}\left(\begin{array}{cc} g\, \cos\theta_{w}&0\\0 & \frac{1}{2}(g\, \cos\theta_{w}-g'\, \sin\theta_{w})
 \end{array}\right)\gamma^{\mu}Z_{\mu} {\wt{W}_{L} \choose \wt{H}_{L}},\\
\end{eqnarray*}
with 
\begin{\eqn*}
\wt{A_{L}}\,=\,{U^{+}}^{-1}\left(\begin{array}{cc}g\, \cos\theta_{w}&0\\0 & \frac{1}{2}(g\, \cos\theta_{w}-g'\, \sin\theta_{w})\end{array}\right)\,U^{+}\,=\,\left(\begin{array}{cc}c_{11}&c_{12}\\c_{12}&c_{22}\end{array}\right).
\end{\eqn*}
The $c_{ij}$ are given by \cite{Choi:2000ta}:
\begin{eqnarray*}
&&c_{11}\,=\,-\frac{g_{W}}{c_{W}}\,\left[s^{2}_{W}-\frac{3}{4}-\frac{1}{4}\,\cos\,2\Phi_{+}\right],
c_{12}\,=\,\frac{g_{W}}{4\,c_{W}}\,\sin\,2\Phi_{+},\,\\
&&c_{22}\,=\,-\frac{g_{W}}{c_{W}}\,\left[s^{2}_{W}-\frac{3}{4}+\frac{1}{4}\,\cos\,2\Phi_{+}\right].
\end{eqnarray*}
For the coupling to the right-handed $\wt{\chi}^{+}$, we obtain
\begin{\eqn*}
-\ol{\wt{\chi}^{+}_{R}}^{\top}\,\left(\begin{array}{cc}c'_{11}&c'_{12}\\c'_{12}&c'_{22}\end{array}\right)\,\gamma^{\mu}Z_{\mu}\wt{\chi}^{+}_{R},
\end{\eqn*}
with
\begin{eqnarray*}
&&c'_{11}\,=\,-\frac{g_{W}}{c_{W}}\,\left[s^{2}_{W}-\frac{3}{4}-\frac{1}{4}\,\cos\,2\Phi_{-}\right],
c'_{12}\,=\,\frac{g_{W}}{4\,c_{W}}\,\sin\,2\Phi_{-},\\
&&c'_{22}\,=\,-\frac{g_{W}}{c_{W}}\,\left[s^{2}_{W}-\frac{3}{4}+\frac{1}{4}\,\cos\,2\Phi_{-}\right].
\end{eqnarray*}
Similarly, the coupling of both left- and righthanded parts of $\wt{\chi}^{+}$ are given by
\begin{\eqn*}
\left<\wt{\chi}^{+}_{i}|A|\wt{\chi}^{+}_{i}\right>\,=\,e
\end{\eqn*}
with the term in $\mL_{MSSM}$ given by
\begin{\eqn*}
-\delta_{ij}\,e\,\ol{\wt{\chi}}^{+}\gamma^{\mu}A_{\mu}\wt{\chi}^{+}.
\end{\eqn*}
If we choose the $\wt{\chi}^{-}$ to be the particle, we have to substitute
\begin{\eqn*}
\ol{\wt{\chi}}^{+}_{L/R}\gamma^{\mu}X_{\mu}\wt{\chi}^{+}_{L/R}\rightarrow -\ol{\wt{\chi}}^{-}_{R/L}\gamma^{\mu}X_{\mu}\wt{\chi}^{-}_{R/L}.
\end{\eqn*}
The Feynman rules for $\wt{\chi}\,\wt{\chi}$ gauge-boson couplings are given in Table \ref{tab:chichigauge}.
\begin{table}%\label{tab:chichigauge}
\begin{\eqn*}
\begin{array}{c|c}\text{coupling}& \text{Feynman rule} \\\hline
\wt{\chi}^{+}_{i,L}\,\wt{\chi}^{+}_{j,L}\,Z_{\mu}&-\imath\,c_{ij}\,\gamma_{\mu}\\
\wt{\chi}^{+}_{i,R}\,\wt{\chi}^{+}_{j,R}\,Z_{\mu}&-\imath\,c'_{ij}\,\gamma_{\mu}\\
\wt{\chi}^{+}_{i}\,\wt{\chi}^{+}_{j}\,A_{\mu}&-\imath\,e\,\delta_{ij}\,\gamma_{\mu}
\end{array}
\end{\eqn*}
\caption{\label{tab:chichigauge} Feynman rules for $\wt{\chi}^{+}\wt{\chi}^{+}$ gauge boson couplings for an incoming $\wt{\chi}_{j}$}
\end{table}
\section{Lepton-slepton and quark-squark chargino couplings} 
We will now derive the general Feynman rules for the coupling of a sample $SU(2)$ doublet to charginos. Feynman rules for MSSM particles can then be obtained by putting in the respective quantum numbers and the MSSM form of the superpotential.\\
For a sample doublet
\begin{\eqn*}
\phi={\wt{u} \choose \wt{d}}, \psi={u \choose d},
\end{\eqn*}
the following couplings to charged gauginos and Higgsinos appear:
\begin{itemize}
\item{gauginos}
\begin{\eqn*}
\imath g(\tilde{u}_{L}^{*}\widetilde{W}^{+} d_{L}+\tilde{d}_{L}^{*}\widetilde{W}^{-}u_{L})\,+\,h.c.
\end{\eqn*}
(all SU(2) doublets)
\item{Higgsinos}\\
terms of the form
\begin{\eqn*}
\tilde{u}_{R}y_{U}d_{L}\wt{H^{+}_{u}}+u_{R}y_{U}\tilde{d}_{L}\wt{H^{+}_{u}+}\tilde{d}_{R}y_{D}u_{L}\wt{H^{-}_{d}}+d_{R}y_{D}\tilde{u}_{L}\wt{H^{-}_{d}}+h.c.
\end{\eqn*}
\end{itemize}
Rewriting these terms using
\begin{\eqn*}
\bar{\Psi}_{1}P_{L}\Psi_{2}\,=\,\eta_{1}\chi_{2}\;,\;\bar{\Psi}_{1}P_{R}\Psi_{2}\,=\,\bar{\eta_{2}}\bar{\chi_{1}},
\end{\eqn*}
with $\Psi$ defined by \cite{Haber:1985rc}
\begin{\eqn*}
\Psi\,=\,{\chi_{\al} \choose \bar{\eta}^{\dot{\al}}},
\end{\eqn*}
we end up with
\begin{eqnarray*}
-g\left(\bar{\psi}_{u}P_{R}\wt{W}\tilde{d_{L}}+\bar{\psi}_{d}P_{R}\wt{W^{c}}\tilde{u_{L}}\right)&+&y_{U,ij}\left(\bar{\psi}_{d,j}P_{R}\wt{H^{c}}\tilde{u}_{R,i}^{*}+\tilde{d}_{L,j}\bar{\psi}_{u,i}P_{L}\wt{H}\right)\\+y_{D,ij}\left(\bar{\psi}_{u,j}P_{R}\wt{H}\tilde{d}_{R,i}^{*}+\tilde{u}_{L,j}\bar{\psi}_{d,i}P_{L}\wt{H^{c}}\right)&+&h.c.\\
\end{eqnarray*}
We use 
\begin{eqnarray*}
{P_{L}\wt{W} \choose P_{L}\wt{H}}\,=\,(U^{+})^{-1} {\wt{\chi_{1}}_{L} \choose \wt{\chi_{2}}_{L}}&,&{P_{R}\wt{W} \choose P_{R}\wt{H}}\,=\,(U^{-})^{-1} {\wt{\chi_{1}}_{R} \choose \wt{\chi_{2}}_{R}},\\
{P_{L}\wt{W}^{c} \choose P_{L}\wt{H}^{c}}\,=\,(U^{-})^{\top} {\wt{\chi_{1}}^{c}_{L} \choose \wt{\chi_{2}}^{c}_{L}}&,&{P_{R}\wt{W}^{c} \choose P_{R}\wt{H}^{c}}\,=\,(U^{+})^{\top} {\wt{\chi_{1}}^{c}_{R} \choose \wt{\chi_{2}}^{c}_{R}},
\end{eqnarray*}
with
\begin{\eqn*}
\widetilde{W}^{c}\,=\,{-\imath\widetilde{W}^{-} \choose \overline{(-\imath\widetilde{W}^{+})}}\;,\;\widetilde{H}^{c}\,=\,{\widetilde{H}^{-}_{d} \choose \overline{\widetilde{H}^{+}_{u}}}\;,\;\widetilde{\chi_{i}}^{c}\,=\,{\chi^{-}_{i} \choose \overline{\chi^{+}_{i}}}.
\end{\eqn*} 
This yields
\begin{itemize}
\item[(a)]{coupling to left-handed charginos}
\begin{\eqn*}\bar{\psi}_{u}y_{U}\tilde{d}_{L}(U^{+}_{i2})^{*}\wt{\chi_{i}}_{L}\end{\eqn*}
\item[(b)]{coupling to right-handed charginos}
\begin{\eqn*}
(-g\tilde{d}_{L}\bar{\psi}_{u}(U^{-}_{i1})^{*}+\tilde{d}_{R}^{*}y_{D}\bar{\psi}_{u}(U^{-}_{i2})^{*})\wt{\chi_{i}}_{R}
\end{\eqn*}
\item[(c)]{coupling to left-handed charge-conjugated charginos}
\begin{\eqn*}
\bar{\psi}_{d}y_{D}\tilde{u}_{L}U^{-}_{i2}\wt{\chi_{i}}^{c}_{L}
\end{\eqn*}
\item[(d)]{coupling to right-handed charge-conjugated charginos}
\begin{\eqn*}
(-g\tilde{u}_{L}\bar{\psi}_{d}U^{+}_{i1}+\tilde{u}_{R}^{*}\bar{\psi}_{d}y_{U}U^{+}_{i2})\wt{\chi_{i}}^{c}_{R}
\end{\eqn*}
\end{itemize}
(+ $h.c.$ for all terms).\\
They describe the following in- and outcoming charginos
\begin{itemize}
\item{incoming $+$, outgoing $-$}
\begin{displaymath}
(a),\,(b),\,(c)^{\dagger},\,(d)^{\dagger}
\end{displaymath}
\item{incoming $-$, outgoing $+$}
\begin{displaymath}
(a)^{\dagger},\,(b)^{\dagger},\,(c),\,(d)
\end{displaymath}
\end{itemize}
independent of the particle/ antiparticle choice. 
\\From the explicit for of the MSSM potential (\ref{eq:superpot})
\begin{\eqn*}
f_{MSSM}=\bar{u} y_{u} Q H_{u}-\bar{d} y_{d} Q H_{d}- \bar{e} y_{e}L H_{d}+\mu H_{u} H_{d},
\end{\eqn*}
we assign the following values to the sample matrices $y_{D}, y_{U}$ used above:
\begin{displaymath}
\begin{array}{l|c|c}
&y_{U}&y_{D}\\ \hline
\text{quarks}&y_{u}&y_{d}\\
\text{leptons}&0&y_{e}\\
\text{Higgs}&0&0\\
\end{array}
\end{displaymath}
In the calculation, we assumed $y_{U}$ and $y_{D}$ to be real (neglecting CP violating phases). In the high-energy limit, they are usually taken to be nonzero only for the third generations:
\begin{\eqn*}
y_{U}\propto\delta_{33},\;y_{D}\propto\delta_{33}.
\end{\eqn*}
The Feynman rules for an incoming $\wt{\chi}^{+}_{i}$ are given in Table \ref{tab:chileslep}.
\begin{table}
\begin{\eqn*}
\begin{array}{c|c|c|cc} \text{(s)quarks}&\text{(s)leptons}&\wt{\chi}_{i}^{+}\,\text{pol.}&\text{Feynman rule}&\\ \hline
(t\,\tilde{b}_{L})&&L&\imath\,y_{t}\,(U^{+}_{i2})^{*}&\text{(a)}\\
(u\,\tilde{d}_{L}),(c\,\tilde{s}_{L}),(t\,\tilde{b}_{L})&(\nu_{i}\,\tilde{l}_{i,L})&R&-\imath\,g\,(U^{-}_{i1})^{*}&\text{(b)}\\
(t\,\tilde{b}_{R})&(\nu_{\tau}\,\tilde{\tau}_{R})&R&\imath\,y_{b/\tau}\,(U^{-}_{i2})^{*}&\text{(b)}\\
(b\,\tilde{t}_{L})&(\tau\,\tilde{\nu}_{\tau,L})&R&\imath\,y_{b\,/\tau}(U^{-}_{i2})^{*}&\text{(c)}\\
(d\,\tilde{u}_{L}),(s\,\tilde{c}_{L}),(b\,\tilde{t}_{L})&(L_{i}\,\tilde{\nu}_{i,L})&L&-\imath\,g\,(U^{+}_{i1})^{*}&\text{(d)}\\
(b\,\tilde{t}_{R})&&L&\imath\,y_{t}\,(U^{+}_{i2})^{*}&\text{(d)}
\end{array}
\end{\eqn*}
\caption{\label{tab:chileslep} quark-squark and lepton-slepton couplings to $\wt{\chi}^{+}_{i}$. $y_{t/b/\tau}$ are the (3,3) indices of the respective CKM matrices, $l_{i}\,=\,e,\mu,\,\tau$.}
\end{table}
\section{Chargino-Neutralino-Gauge boson and Chargino-Neutralino-Higgs couplings}
The part of $\mL_{MSSM}$ describing the chargino-neutralino-gauge boson couplings is given by
\begin{eqnarray*}
&&\overline{\widetilde{W}}^{+}\bar{\sigma}^{\mu}W^{+}_{\mu}\widetilde{W}^{0}\,-\,\overline{\widetilde{W}}^{0}\bar{\sigma}^{\mu}W^{+}_{\mu}\widetilde{W}^{-}
\,-\,\overline{\widetilde{W}}^{-}\bar{\sigma}^{\mu}W^{-}_{\mu}\widetilde{W}^{0}\,+\,\overline{\widetilde{W}}^{0}\bar{\sigma}^{\mu}W^{-}_{\mu}\widetilde{W}^{+}\\
\end{eqnarray*}
from the gaugino sector and
\begin{eqnarray*}
&&-\frac{g}{\sqrt{2}}\lb\overline{\widetilde{H}^{0}_{d}}\bar{\sigma}^{\mu} W^{+}_{\mu} \widetilde{H}^{-}_{d}\,+\,\overline{\widetilde{H}^{+}_{u}}\bar{\sigma}^{\mu} W^{+}_{\mu} \widetilde{H}^{0}_{u}
\,+\,\overline{\widetilde{H}^{-}_{d}}\bar{\sigma}^{\mu} W^{-}_{\mu} \widetilde{H}^{0}_{d}\,+\,\overline{\widetilde{H}^{0}_{u}}\bar{\sigma}^{\mu} W^{-}_{\mu} \widetilde{H}^{+}_{u}\rb\\
\end{eqnarray*}
from the Higgsino sector.\\
Rewriting this in four-component notation using the notations of the previous sections, we obtain
\footnote{We use 
\begin{\eqn*}
(\bar{\psi}_{1}\gamma^{\mu}W^{-}_{\mu}P_{L/R}\psi_{2})^{\dagger}=\bar{\psi}_{2}\gamma^{\mu}W^{+}_{\mu}P_{R/L}\psi_{1}
\end{\eqn*}
and
\begin{\eqn*}
(\bar{\psi_{1}}\gamma^{\mu}W^{-}_{\mu}\psi_{2})^{\dagger}=\bar{\psi_{2}}\gamma^{\mu}W^{+}_{\mu}\psi_{1}.
\end{\eqn*}}
\begin{\eqn*}
g\lb\ol{\wt{W}^{0}}\gamma^{\mu}W^{-}_{\mu}\wt{W}+\frac{1}{\sqrt{2}}(\ol{\wt{H_{d}}}\gamma^{\mu}W^{-}_{\mu}P_{R}\wt{H}-\ol{\wt{H_{u}}}\gamma^{\mu}W^{-}_{\mu}P_{L}\wt{H}) \rb+h.c.
\end{\eqn*}
\hspace{5mm}\\
The chargino-neutralino-Higgs coupling terms arise from the gauge part of the Lagrangian
\begin{\eqn*}
-\sqrt{2}\imath g(\bar{\psi}T^{a}\bar{\lambda}^{a}\phi-\phi^{*}T^{a}\lambda^{a}\psi).
\end{\eqn*}
Taking the quantum numbers of the Higgs doublets into account, they are given explicitly by
\begin{itemize}
\item{}$H_{u}$ terms:
\begin{eqnarray*}
\imath \lb g\left[(H^{+}_{u})^{*}\wt{W}^{+}\wt{H}^{0}_{u}\right.\right.&+&\left.(H^{0}_{u})^{*}\wt{W}^{-}\wt{H}^{+}_{u}\right] \\
+\frac{g}{\sqrt{2}}\left[(H^{+}_{u})^{*}\wt{W}^{0}\wt{H}^{+}_{u}\right.&-&\left.(H^{0}_{u})^{*}\wt{W}^{0}\wt{H}^{0}_{u}\right]\\
+\frac{g'}{\sqrt{2}}\left[(H^{+}_{u})^{*}\wt{B}\wt{H}^{+}_{u}\right.&+&\left.\left.(H^{0}_{u})^{*}\wt{B}\wt{H}^{0}_{u}\right]\rb\;+\;h.c.
\end{eqnarray*}
\item{}$H_{d}$ terms:
\begin{eqnarray*}
\imath\lb g\left[(H^{0}_{d})^{*}\wt{W}^{+}\wt{H}^{-}_{d}\right.\right.&+&\left.(H^{-}_{d})^{*}\wt{W}^{-}\wt{H}^{0}_{d}\right] \\
+\frac{g}{\sqrt{2}}\left[(H^{0}_{d})^{*}\wt{W}^{0}\wt{H}^{0}_{d}\right.&-&\left.(H^{-}_{d})^{*}\wt{W}^{0}\wt{H}^{-}_{d}\right]\\
-\frac{g'}{\sqrt{2}}\left[(H^{0}_{d})^{*}\wt{B}\wt{H}^{0}_{d}\right.&+&\left.\left.(H^{-}_{d})^{*}\wt{B}\wt{H}^{-}_{d}\right]\rb\;+\;h.c.
\end{eqnarray*}
\end{itemize}
We can split this up into chargino-chargino-Higgs, neutralino-neutralino-Higgs, and chargino-neutralino-Higgs couplings. For actual calculations, we need to transform the Higgs bosons into Higgs mass eigenstates. We postpone this and here only sketch the coupling structure coming from the chargino/ neutralino sector. More details can be found in \cite{Gunion:1986yn}.\\
\hspace{7mm}\\
{\bf Chargino-Chargino-Higgs couplings}\\
\\
The respective terms in the Lagrangian are given by
\begin{\eqn*}
\imath g((H^{0}_{u})^{*}\wt{W}^{-}\wt{H}^{+}_{u}+(H^{0}_{d})^{*}\wt{W}^{+}\wt{H}^{-}_{d})+h.c.
\end{\eqn*}
In four-component notation, this reads
\begin{\eqn*}
-g(\ol{\wt{W}}_{R}(H^{0}_{u})^{*}\wt{H}_{L}+\ol{\wt{H}}_{R}(H^{0}_{d})^{*}\wt{W}_{L})+h.c.
\end{\eqn*}
Transforming this into the chargino eigenstates, we obtain
\begin{\eqn*}
-g\lb\ol{\wt{\chi}}^{+}_{i,R}\left[U^{-}_{i1}(H^{0}_{u})^{*}(U^{+}_{j2})^{*}+U^{-}_{i2}(H^{0}_{d})^{*}(U^{+}_{j1})^{*}\right]\wt{\chi}^{+}_{j,L}\rb+h.c.
\end{\eqn*}
for the couplings of left- and right-handed charginos to Higgs bosons. Apart from the rotation into the Higgs mass eigenstates, these couplings agree with the ones given in \cite{Gunion:1986yn}.\\
\hspace{3mm}\\
{\bf Chargino-neutralino-Higgs couplings}\\
\\
Chargino-neutralino-Higgs couplings are given by 
\begin{eqnarray*}
\imath\Big((H^{+}_{u})^{*} (g\wt{W}^{+}\wt{H}^{0}_{u}+\frac{g}{\sqrt{2}}\wt{W}^{0}\wt{H}^{+}_{u}+\frac{g'}{\sqrt{2}}\wt{B}\wt{H}^{+}_{u})&&\\
+(H^{-}_{d})^{*} (g\wt{W}^{-}\wt{H}^{0}_{d}-\frac{g}{\sqrt{2}}\wt{W}^{0}\wt{H}^{-}_{d}-\frac{g'}{\sqrt{2}}\wt{B}\wt{H}^{-}_{d})\Big) + h.c.&&
\end{eqnarray*}
In four-component notation, we obtain
\begin{eqnarray*}
\Big((H^{+}_{u})^{*} (-g\ol{\wt{H}}_{u,R}\wt{W}_{L}-\frac{g}{\sqrt{2}}\ol{\wt{W}}^{0}_{R}\wt{H}_{L}-\frac{g'}{\sqrt{2}}\ol{\wt{B}}_{R}\wt{H}_{L})&&\\
+(H^{-}_{d})(-g\ol{\wt{H}}_{d,L}\wt{W}_{R}+\frac{g}{\sqrt{2}}\ol{\wt{W}}^{0}_{L}\wt{H}_{R}+\frac{g'}{\sqrt{2}}\ol{\wt{B}}_{L}\wt{H}_{R})\Big) + h.c.&&
\end{eqnarray*}
Rotating this into the mass eigenstate basis, we obtain
\begin{itemize}
\item{}Coupling of right-handed neutralinos to left-handed charginos
\begin{\eqn*}
\left[-(H^{+}_{u})^{*}\,\ol{\wt{\chi}}^{0}_{R,i} \lb\frac{g'}{\sqrt{2}}N^{*}_{i1}(U^{+}_{j2})^{*}+\frac{g}{\sqrt{2}}N^{*}_{i2}(U^{+}_{j2})^{*}+g\,N^{*}_{i4}(U^{+}_{j1})^{*}\rb\wt{\chi}^{+}_{L,j}\,\right]+h.c.
\end{\eqn*}
\item{}Coupling of left-handed neutralinos to right-handed charginos
\begin{\eqn*}
\left[- H^{-}_{d}\,\ol{\wt{\chi}}^{0}_{L,i} \lb\,-\frac{g'}{\sqrt{2}}N_{i1}\,(U^{-}_{j2})^{*}-\frac{g}{\sqrt{2}}N_{i2}(U^{-}_{j2})^{*}+g\,N_{i3}(U^{-}_{j1})^{*}\rb\wt{\chi}^{+}_{R,j}\,\right]+h.c.
\end{\eqn*}
\end{itemize}
Apart from the rotation into Higgs mass eigenstates, this again agrees with the couplings as given in \cite{Gunion:1986yn}.\\
The Feynman rules for an incoming $\wt{\chi}^{+}$ are given in Table \ref{tab:higgsch}
\begin{table}
\begin{\eqn*}
\begin{array}{c|c}\text{coupling}&\text{Feynman rule}\\ \hline
\wt{\chi}^{+}_{i,R}\,\wt{\chi}^{-}_{j,L}\,H^{0}_{u}&\imath\,g\, (U^{-}_{i1})^{*}\,U^{+}_{j2}\\
\wt{\chi}^{+}_{i,R}\,\wt{\chi}^{-}_{j,L}\,H^{0}_{d}&\imath\,g\, (U^{-}_{i2})^{*}\,U^{+}_{j1}\\
\wt{\chi}^{+}_{i,L}\,\wt{\chi}^{0}_{j,R}\,H^{+}_{u}&- \imath\,\lb\frac{g'}{\sqrt{2}}N^{*}_{j1}(U^{+}_{i2})^{*}+\frac{g}{\sqrt{2}}N^{*}_{j2}(U^{+}_{i2})^{*}+g\,N^{*}_{j4}(U^{+}_{i1})^{*}\rb\\
\wt{\chi}^{+}_{i,R}\,\wt{\chi}^{0}_{j,L}\,H^{-}_{d}&\imath\,\lb\,\frac{g'}{\sqrt{2}}N_{j1}\,(U^{-}_{i2})^{*}+\frac{g}{\sqrt{2}}N_{j2}(U^{-}_{i2})^{*}-g\,N_{j3}(U^{-}_{i1})^{*}\rb
\end{array}
\end{\eqn*}
\caption{\label{tab:higgsch} Feynman rules for an incoming $\wt{\chi}^{+}$}
\end{table} 
\chapter{Point SPS1a'}\label{app:sps1asl}
\includegraphics[height=10cm]{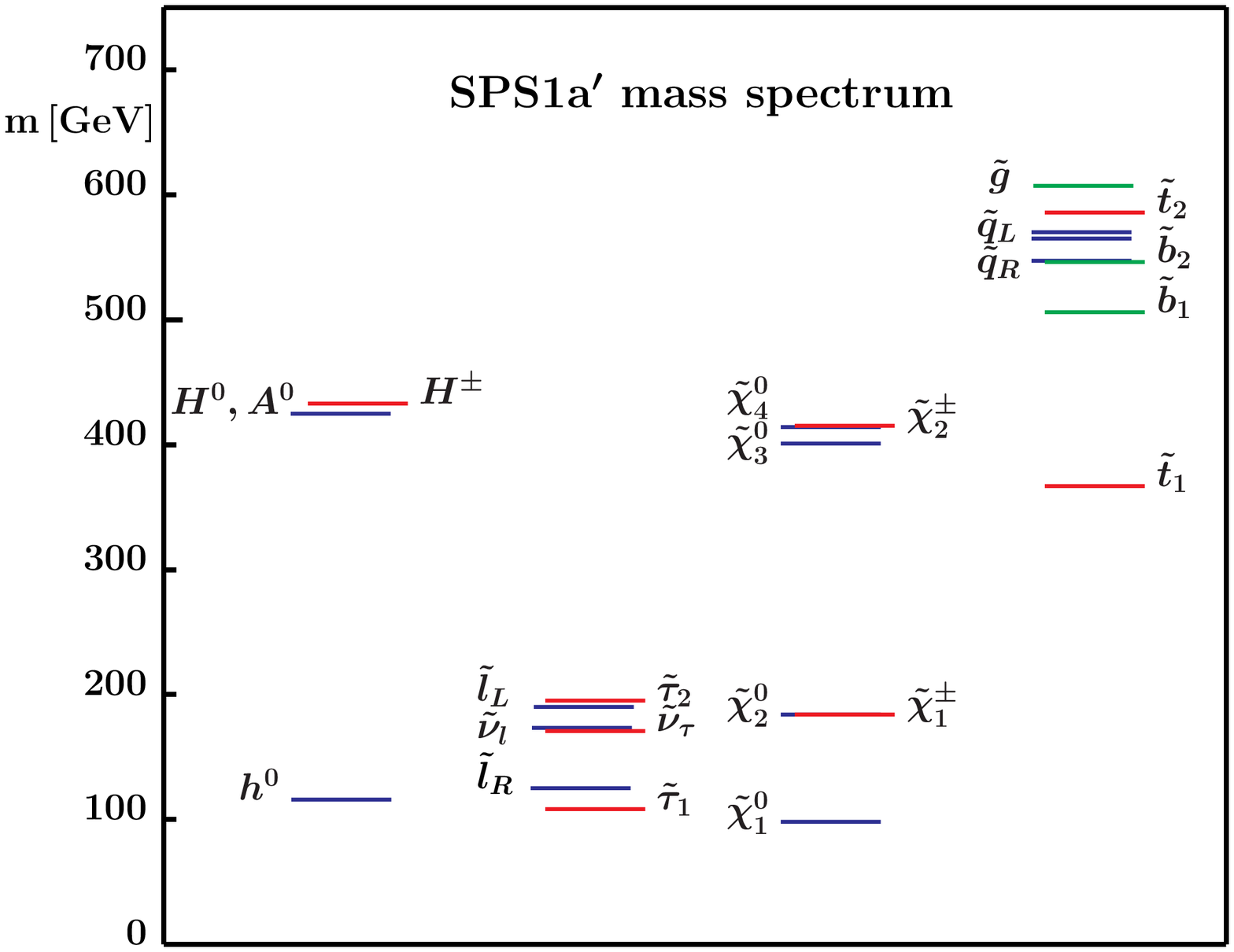}
\vspace{10mm}
\\
The mSugra point SPS1a' used in this analysis is defined in \cite{Allanach:2002nj, Aguilar-Saavedra:2005pw}. It corresponds to a ``typical'' mSugra scenario. The respective mSugra parameters are
\begin{\eqn*}
M_{1/2}\,=\,250\,\GeV\,,\,M_{0}\,=\,70\,\GeV,\,A_{0}\,=\,-300\,\GeV,\,\text{sign}(\mu)\,=\,+1,\,\tan\beta\,=\,10.
\end{\eqn*} 
All sparticles masses in the point SPS1a' are smaller than $1\,\TeV$ and therefore within the reach of future high energy colliders (LHC/ ILC). Branching ratios for decays can be found in \cite{Aguilar-Saavedra:2005pw}.
\chapter{Helicity eigenstates for massive fermions}\label{app:helmass}
In the Dirac formalism, the helicity operator is given by
\begin{\eqn*}
h=\Sigma^{k} \hat{e}_{k}\;,\;\Sigma^{k}=\lb\begin{array}{cc}\sigma^{k}&0\\0&\sigma^{k}\end{array}\rb
\end{\eqn*}
for the projection of the spin along the $\hat{e}_{k}$ direction (see e.g. \cite{Schwabl:1997gf}). Choosing the direction of flight along the z-axis, we can therefore define a spin up/ down projector by
\begin{\eqn*}
P_{u/d}=\frac{1\pm\sigma \hat{e}_{z}}{2}.
\end{\eqn*}
As the helicity operator and the Hamiltonian commute, the Dirac
spinors in the Dirac theory can be chosen to be eigenstates of the
helicity operator and the Hamilton operator simultaneously; however,
the helicity operator is not Lorentz-invariant\footnote{This only
  holds for massive particles; for massless particles, the helicity
  projector coincides with the chirality projector which is of course
  Lorentz-invariant.} and therefore frame-dependent. Samples of
eigenstates can be found in \cite{Schwabl:1997gf} or
\cite{Bjorken:1966dk}.\\
The projector can be brought into a covariant form
\begin{\eqn*}
\Sigma(s)\,=\,\frac{1+\gamma_{5}\slashed{s}}{2}
\end{\eqn*}
for a spin vector $s$ obeying
\begin{\eqn*}
p^{\mu}s_{\mu}=0
\end{\eqn*}
(see \cite{Bjorken:1966dk} or \cite{Itzykson:1980rh} for more details).\\It can easily be shown that
\begin{\eqn*}
\Sigma(\pm s)\Sigma(\pm s)\,=\,\Sigma(\pm s)\;\;\mbox{while}\;\;\Sigma(\pm s)\Sigma(\mp s)\,= 0.
\end{\eqn*}
The helicity projection operator as well as the energy projection operator then determine the eigenstates $u_{\pm}(p),\,v_{\mp}(p)$ which solve the Dirac equation. For massless states $(p^{2}=0)$, the helicity projector reduces to the chirality projector $P_{R/L}=\frac{1\pm\gamma^{5}}{2}$.\\
There are different approaches in the literature to deal with helicity states for massive fermions (see e.g. \cite{Kleiss:1985yh} for an original work or \cite{Haber:1994pe} and \cite{Dittmaier:1998nn} for reviews on massive fermion treatment/ spinor techniques in general). However, we stick to the formalism introduced in \cite{Hagiwara:1985yu}. Here, the authors construct $u_{\pm}(p),\,v_{\pm}(p)$ which satisfy the Dirac equation and are eigenstates of the helicity operator. They then reformulate the 4-component S-matrix elements into spinor-matrix elements using Fierz identities, which they evaluate explicitly in dependence of the respective momenta in a given lab frame. Here, we just repeat the formulas used in the calculation of the helicity amplitudes and refer to the original work for further details.\\
A four component matrix element can be reduced to a two component one using
\begin{\eqn*}
\bar{\psi}_{1}\gamma^{\mu}P_{R/L}\psi_{2}\,=\,\bar{\psi}_{1\,\pm}\sigma^{\mu}_{\pm}\psi_{2\,\pm}.
\end{\eqn*}
Here,
\begin{\eqn*}
\psi_{i}={\psi_{i\,-}\choose \psi_{i\,+}}.
\end{\eqn*}
Fierz identities give
\begin{eqnarray*}
(\psi^{\dagger}_{1})_{\alpha}\sigma^{\mu}_{\pm}(\psi_{2})_{\beta}(\psi^{\dagger}_{3})_{\gamma}\sigma_{\mu\,\pm}(\psi_{4})_{\delta}&=&2(\psi^{\dagger}_{1})_{\alpha}(\psi_{4})_{\delta}\;(\psi^{\dagger}_{3})_{\gamma}(\psi_{2})_{\beta},\\
(\psi^{\dagger}_{1})_{\alpha}\sigma^{\mu}_{\pm}(\psi_{2})_{\beta}(\psi^{\dagger}_{3})_{\gamma}\sigma_{\mu\,\mp}(\psi_{4})_{\delta}&=&2(\psi^{\dagger}_{1})_{\alpha}(\psi_{2})_{\beta}\;(\psi^{\dagger}_{3})_{\gamma}(\psi_{4})_{\delta}\\
&&-2(\psi^{\dagger}_{1})_{\alpha}(\psi_{4})_{\delta}\;(\psi^{\dagger}_{3})_{\gamma}(\psi_{2})_{\beta}
\end{eqnarray*}
where $\alpha,\,\beta,\,\gamma,\,\delta\,=+,-$.
The product of two two-component spinors is given by
\begin{\eqn*}
(\psi^{\dagger}_{i})_{\alpha}(\psi_{j})_{\beta}\,=\,C_{i}C_{j}w_{\alpha\lambda_{i}}w_{\beta\lambda_{j}}S(p_{i},p_{j})_{\lambda_{i}\lambda_{j}},
\end{\eqn*}
where
\begin{eqnarray*}
&&C_{k}=1\;\mbox{for}\;(\psi_{k})_{\tau}=(u_{k})_{\tau},\\
&&C_{k}=\tau\;\mbox{for}\;(\psi_{k})_{\tau}=(v_{k})_{\tau},\\
&&w_{\pm}(p)\,=\,\sqrt{E\pm|\overrightarrow{p}|}.\\
\end{eqnarray*}
If $S$ only depends on the momenta of the external particles, it reduces to the scalar quantity $T$:
\begin{\eqn*}
S(p_{i}p_{j})_{\lambda_{i}\lambda_{j}}\,=T(p_{i}p_{j})_{\lambda_{i}\lambda_{j}}=\chi^{\dagger}_{\alpha}(p_{i})\chi_{\beta}(p_{j})
\end{\eqn*}
where $\chi_{\pm}$ are helicity eigenstate spinors. $\lambda_{i,j}$ denote the helicities of the resulting massive particles in the given frame while $\alpha,\beta$ denote the chiralities\footnote{$w_{\alpha\lambda_{i}}$ should be read as $w_{\alpha \times \lambda_{i}}$, i.e. $w_{+}$ for $\alpha\,=\,\lambda$ and $w_{-}$ for $\alpha\,\neq\,\lambda$. $w_{-}\,=\,0$ for massless particles.}. $T(p_{i}p_{j})$ are given as
\begin{eqnarray*}
T(p_{1} p_{2})_{++}&=&\frac{(|\ora{p_{1}}|+p_{1,z})(|\ora{p_{2}}|+p_{2,z})+(p_{1,x}-\imath p_{1,y})(p_{2,x}-\imath p_{2,y})}{2\sqrt{|\ora{p_{1}}|(|\ora{p_{1}}|+p_{1,z})|\ora{p_{2}}|(|\ora{p_{2}}|+p_{2,z})}},\\
T(p_{1} p_{2})_{+-}&=&\frac{-(|\ora{p_{1}}|+p_{1,z})(p_{2,x}-\imath p_{2,y})+(|\ora{p_{2}}|+p_{2,z})(p_{1,x}-\imath p_{1,y})}{2\sqrt{|\ora{p_{1}}|(|\ora{p_{1}}|+p_{1,z})|\ora{p_{2}}|(|\ora{p_{2}}|+p_{2,z})}},\\
T(p_{1}p_{2})_{-+}&=&-T^{*}(p_{1}p_{2})_{+-},\\
T(p_{1}p_{2})_{--}&=&T^{*}(p_{1}p_{2})_{++}.
\end{eqnarray*} 
If for example $p_{1,z}=-|\ora{p_{1}}|$ 
\begin{eqnarray*}
T(p_{1}p_{2})_{++}&=&\frac{p_{2,x}+\imath p_{2,y}}{\sqrt{2|\ora{p_{2}}|(|\ora{p_{2}}|+p_{2,z})}},\\
T(p_{1}p_{2})_{+-}&=&\frac{|\ora{p_{2}}|+p_{2,z}}{\sqrt{2|\ora{p_{2}}|(|\ora{p_{2}}|+p_{2,z})}}.
\end{eqnarray*}
Furthermore, it holds that
\begin{\eqn*}
T(p_{1}p_{2})_{\alpha\beta}\,=\,T^{*}(p_{2}p_{1})_{\beta\alpha},
\end{\eqn*}
which gives a complete determination of all possible $T_{\alpha\beta}$.\\
\hspace{2mm}
The definition of the eigenstates given here always guarantees that the helicity eigenstates projected in an arbitrary frame correspond to eigenstates of the spin polarization in the rest frame of the respective particle; see \cite{Hagiwara:1985yu} and \cite{Itzykson:1980rh} for further details\footnote{You just have to identify the $\chi_{\pm}$ in \cite{Hagiwara:1985yu} (3.19) with the $\Lambda_{\pm}$ in \cite{Itzykson:1980rh} (Eqn (2-50)).}.
\chapter{Generic diagrams contributing to NLO Chargino production}\label{app:diagramms}
In the following, we give the generic Feynman diagrams contributing to the process $e^{+}e^{-}\,\rightarrow\,\wt{\chi}^{+}_{1}\,\wt{\chi}^{-}_{1}$ at NLO. $S$ denotes scalar, $V$ vector particles, and $F$ fermions. All diagrams including tadpoles are omitted. The output was generated using \feynarts.
\begin{center}
%\vspace{-4cm}
\begin{minipage}{\textwidth}
\begin{center}
\includegraphics[width=0.7\textwidth]{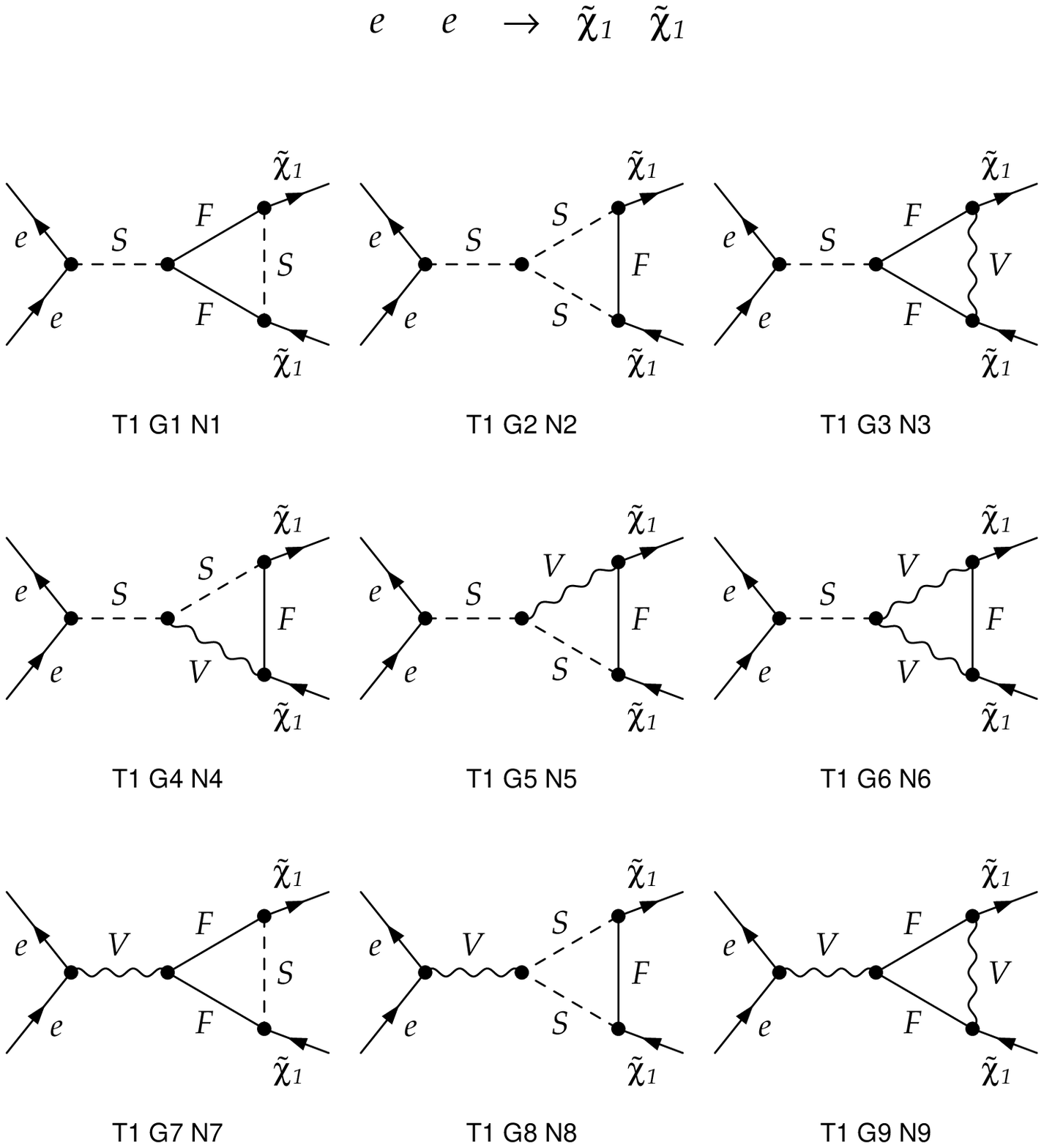}\
\end{center}
\end{minipage}
\newpage
\begin{minipage}{\textwidth}
\vspace{-4cm}
\centering
\includegraphics[width=0.75\textwidth]{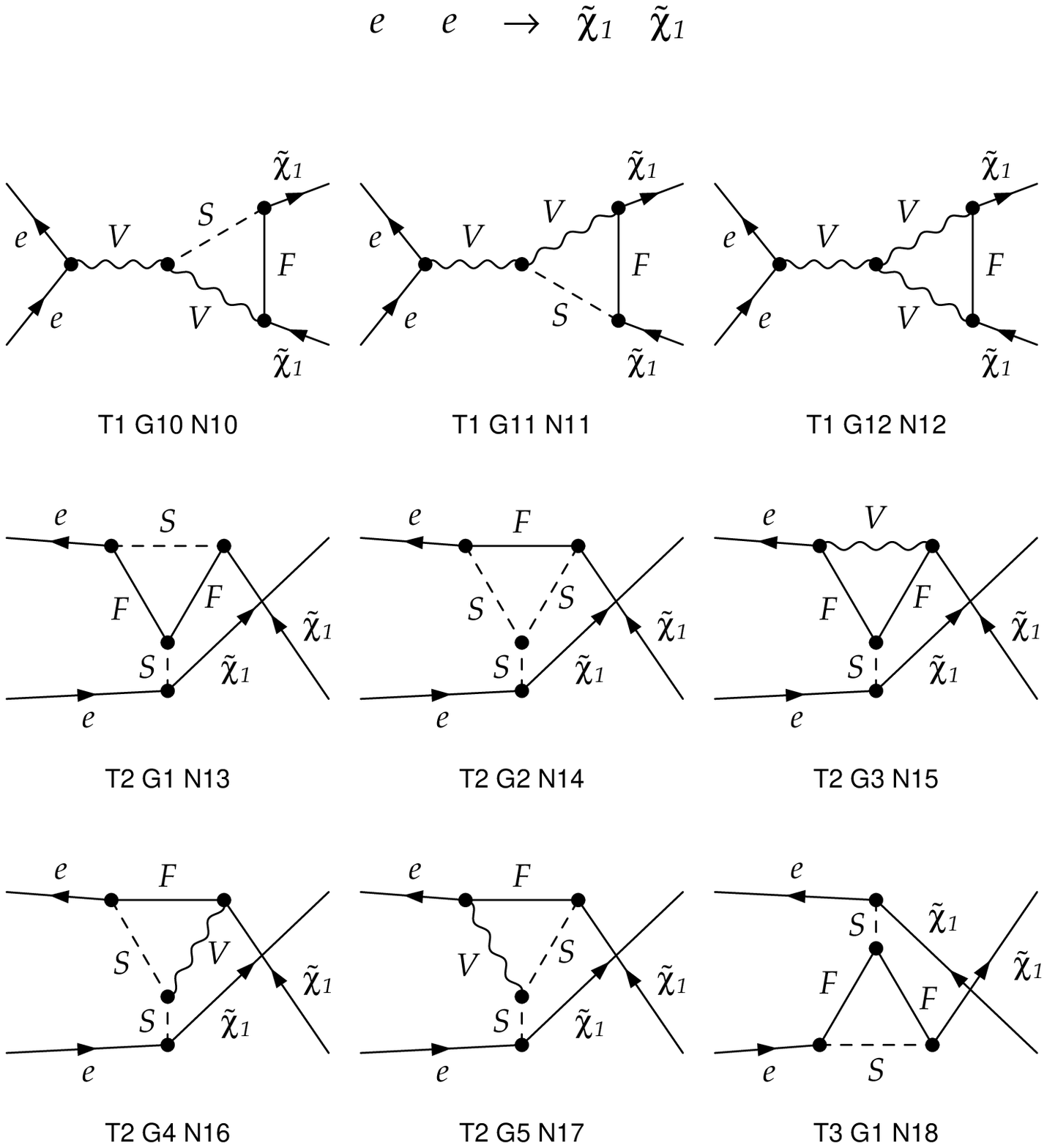}\vspace{-3cm}
\end{minipage}
\vspace{-6cm}
\begin{minipage}{\textwidth}
\centering
\includegraphics[width=0.75\textwidth]{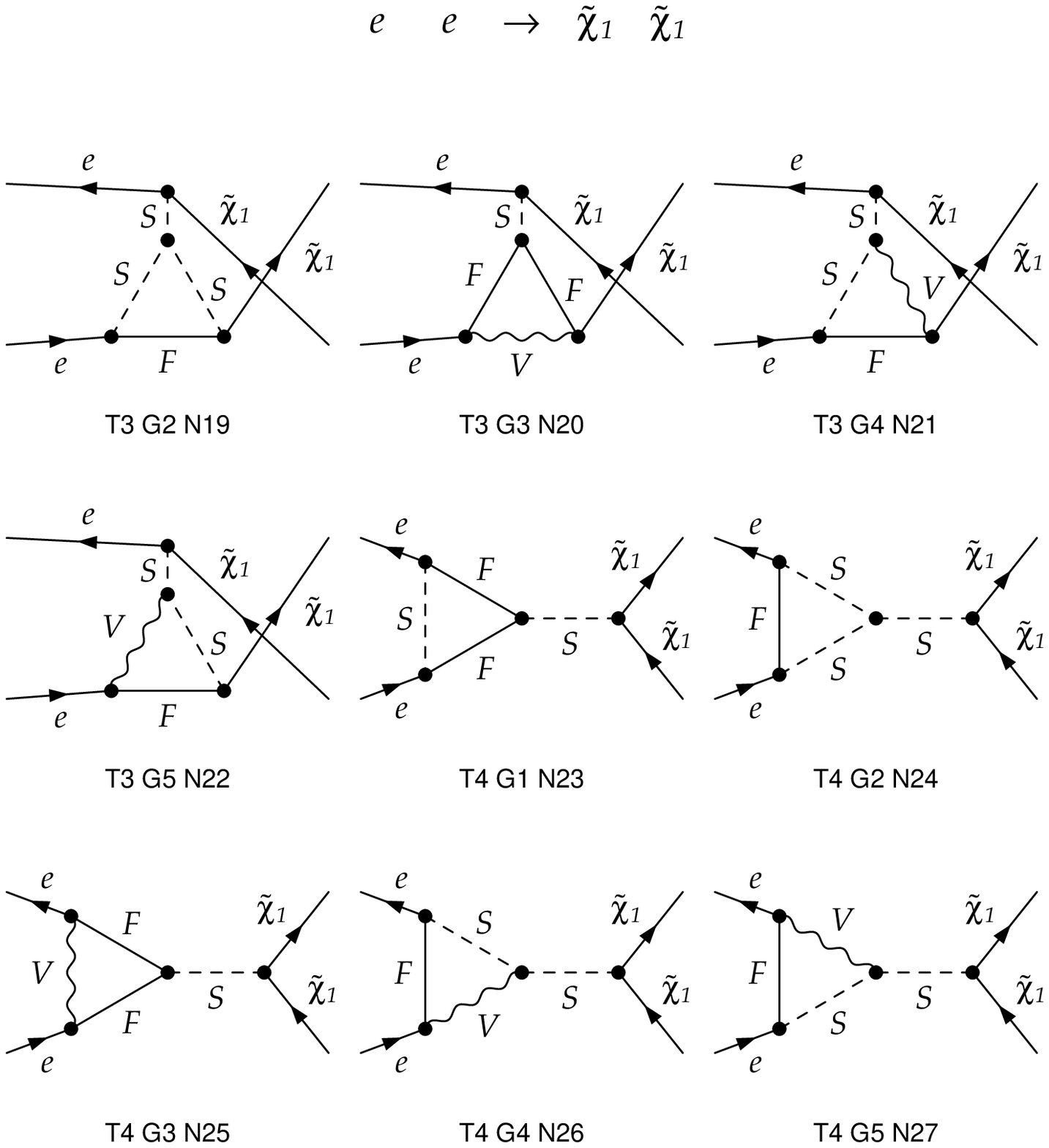}
\end{minipage}
\newpage
\begin{minipage}{\textwidth}
\vspace{-4cm}
\centering
\includegraphics[width=0.75\textwidth]{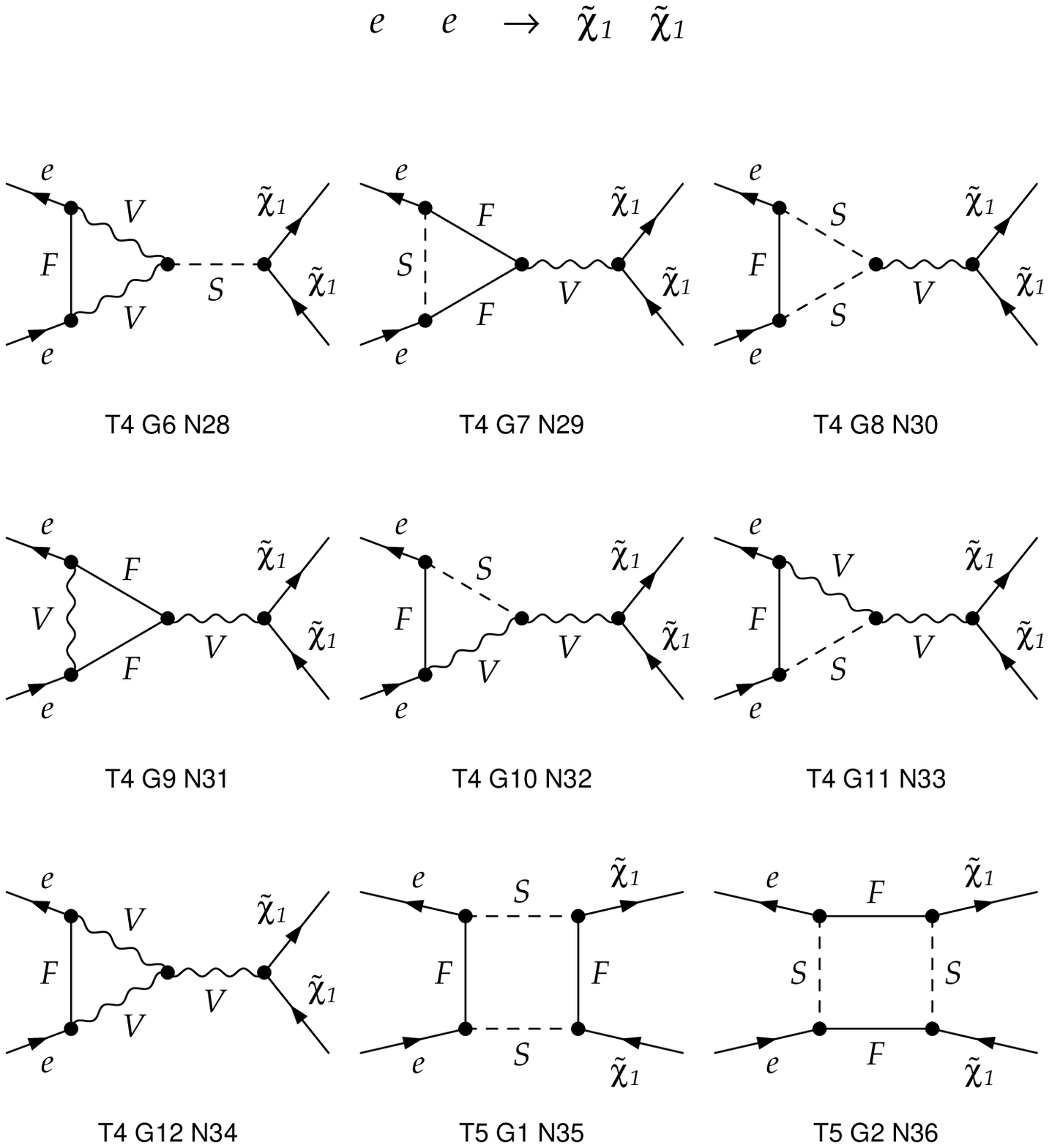}\vspace{-3cm}
\end{minipage}
\vspace{-6cm}
\begin{minipage}{\textwidth}
\centering
\includegraphics[width=0.75\textwidth]{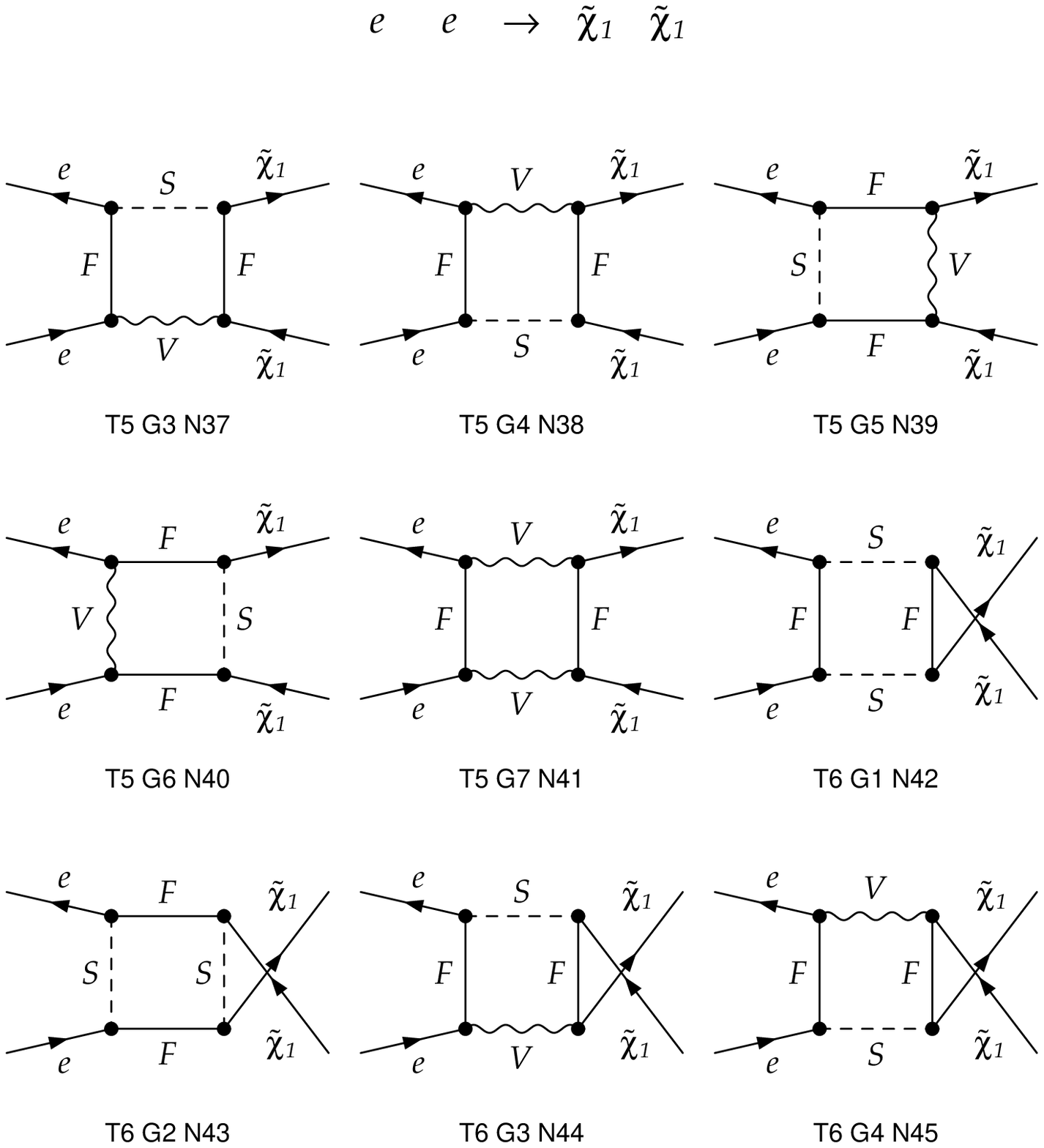}
\end{minipage}
\newpage
\begin{minipage}{\textwidth}
\vspace{-4cm}
\centering
\includegraphics[width=0.75\textwidth]{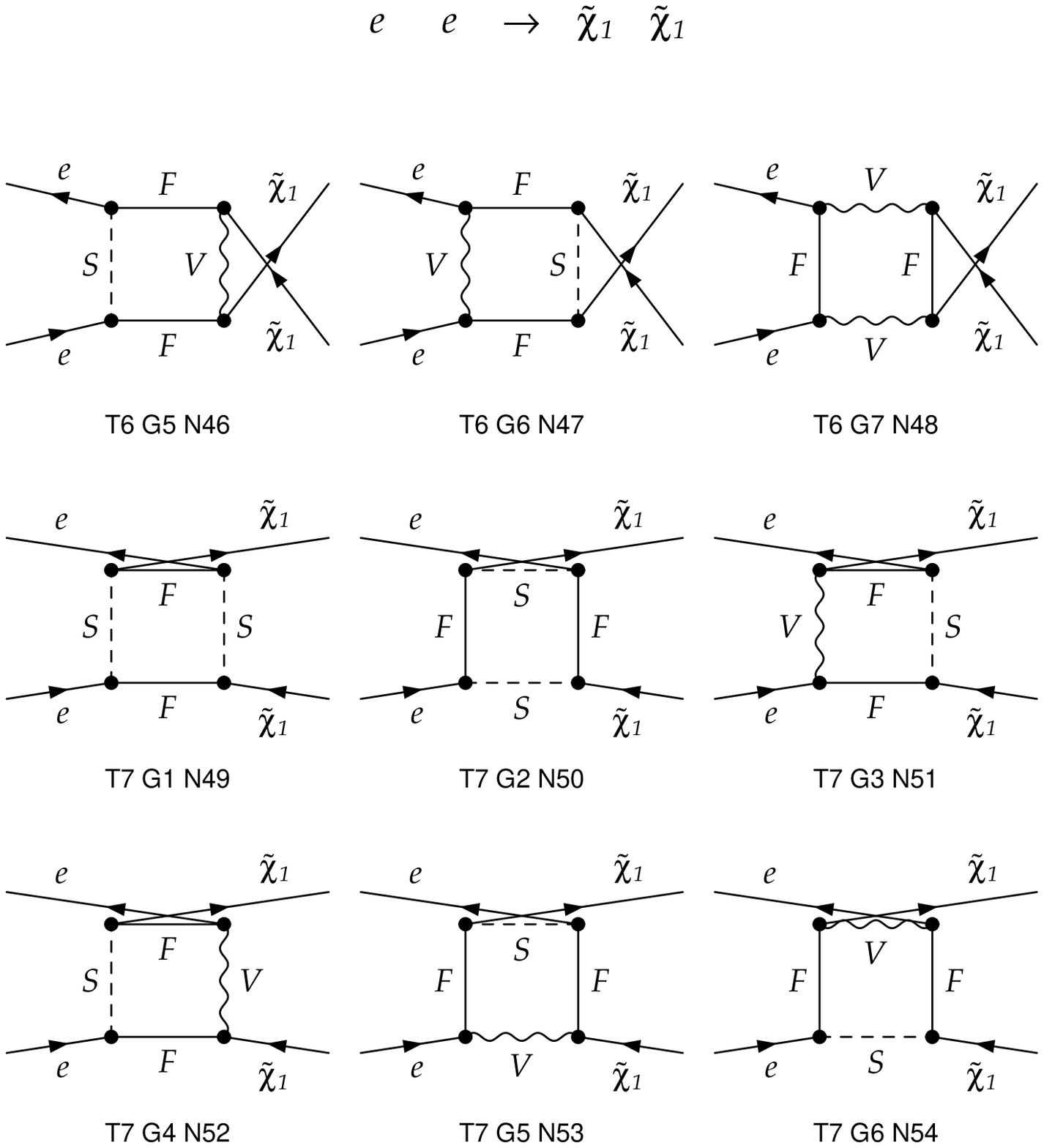}\vspace{-3cm}
\end{minipage}
\vspace{-6cm}
\begin{minipage}{\textwidth}
\centering
\includegraphics[width=0.75\textwidth]{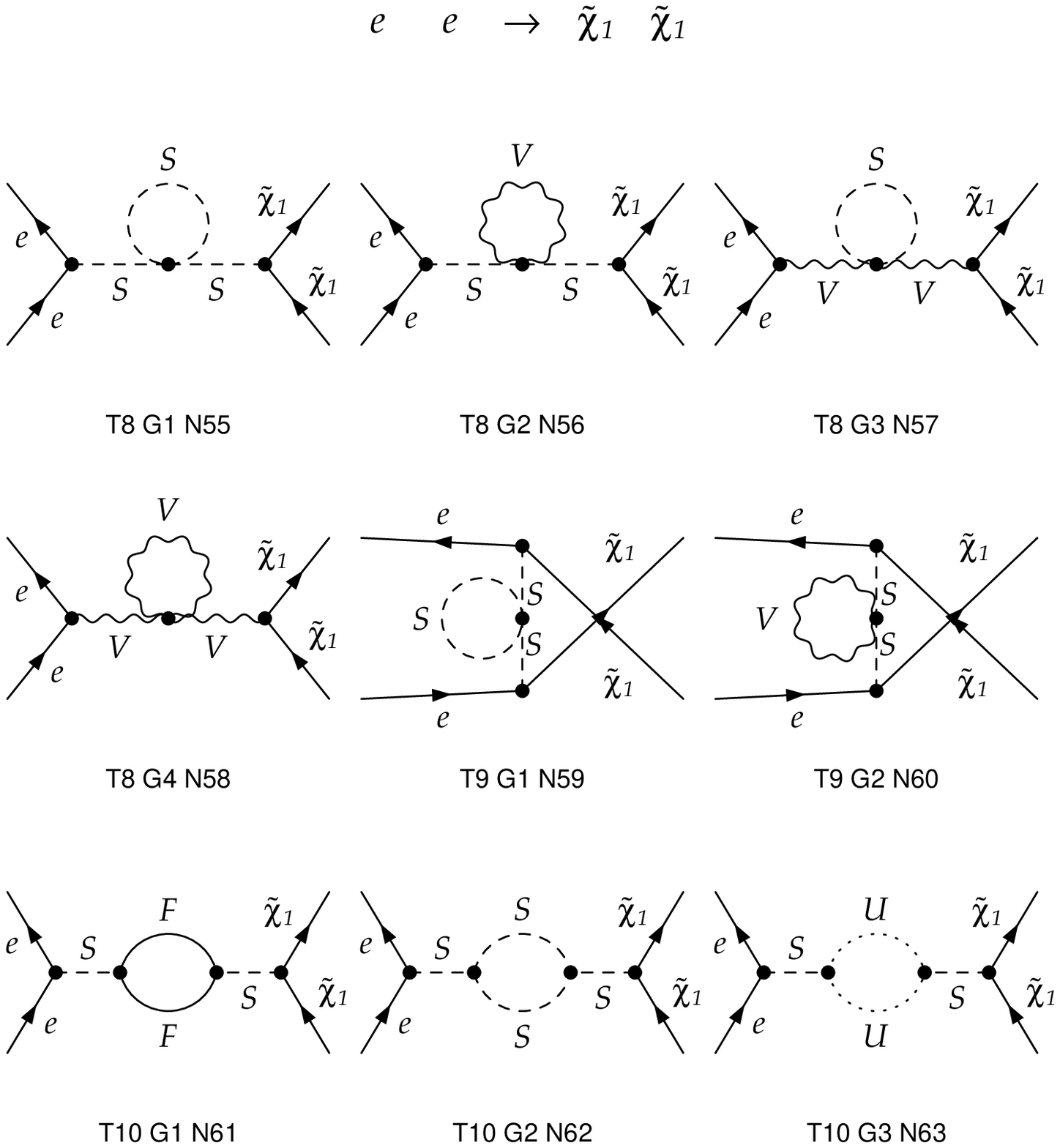}
\end{minipage}
\newpage
\begin{minipage}{\textwidth}
\vspace{-4cm}
\centering
\includegraphics[width=0.75\textwidth]{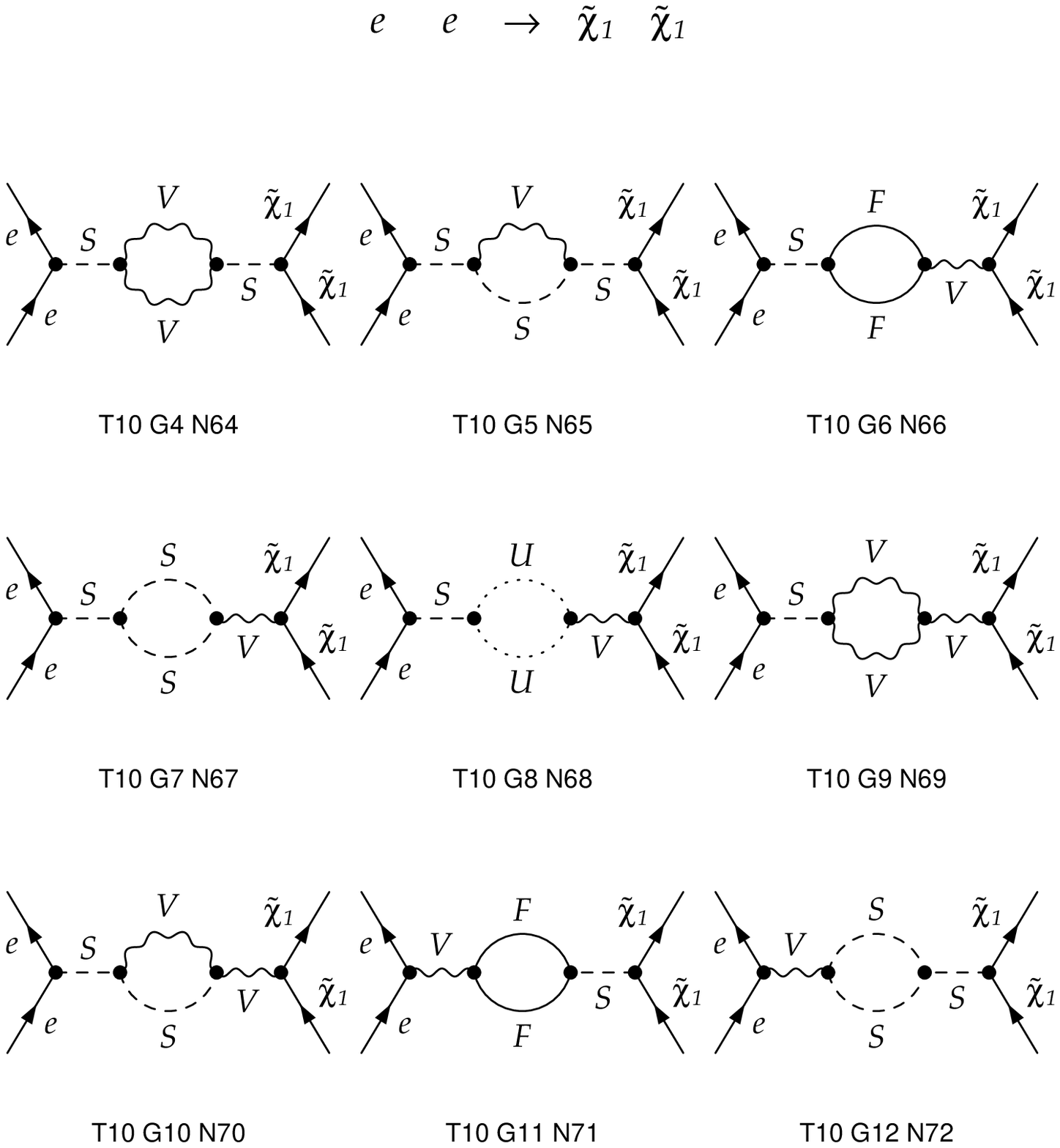}\vspace{-3cm}
\end{minipage}
\vspace{-6cm}
\begin{minipage}{\textwidth}
\centering
\includegraphics[width=0.75\textwidth]{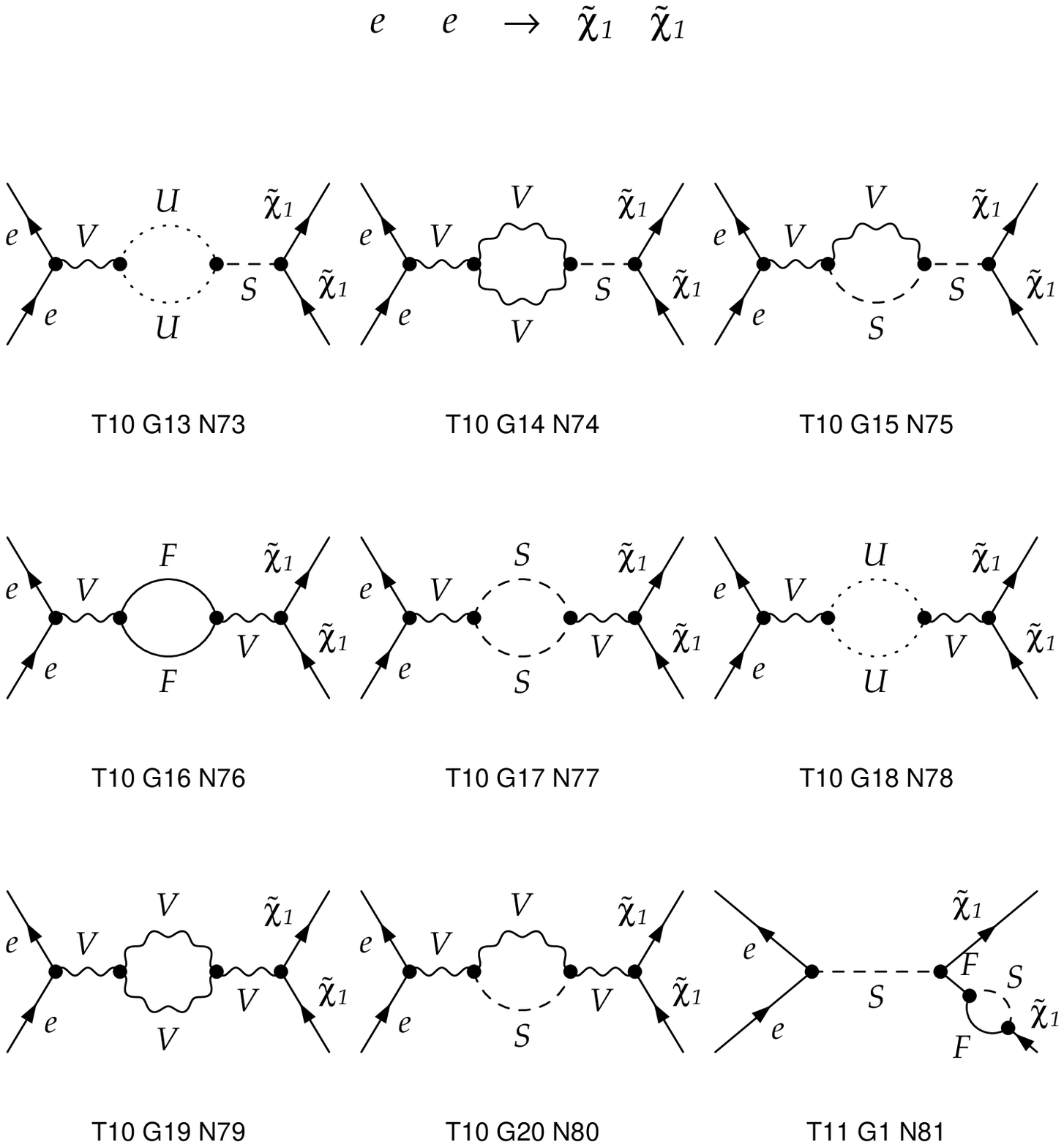}
\end{minipage}
\newpage
\begin{minipage}{\textwidth}
\vspace{-4cm}
\centering
\includegraphics[width=0.75\textwidth]{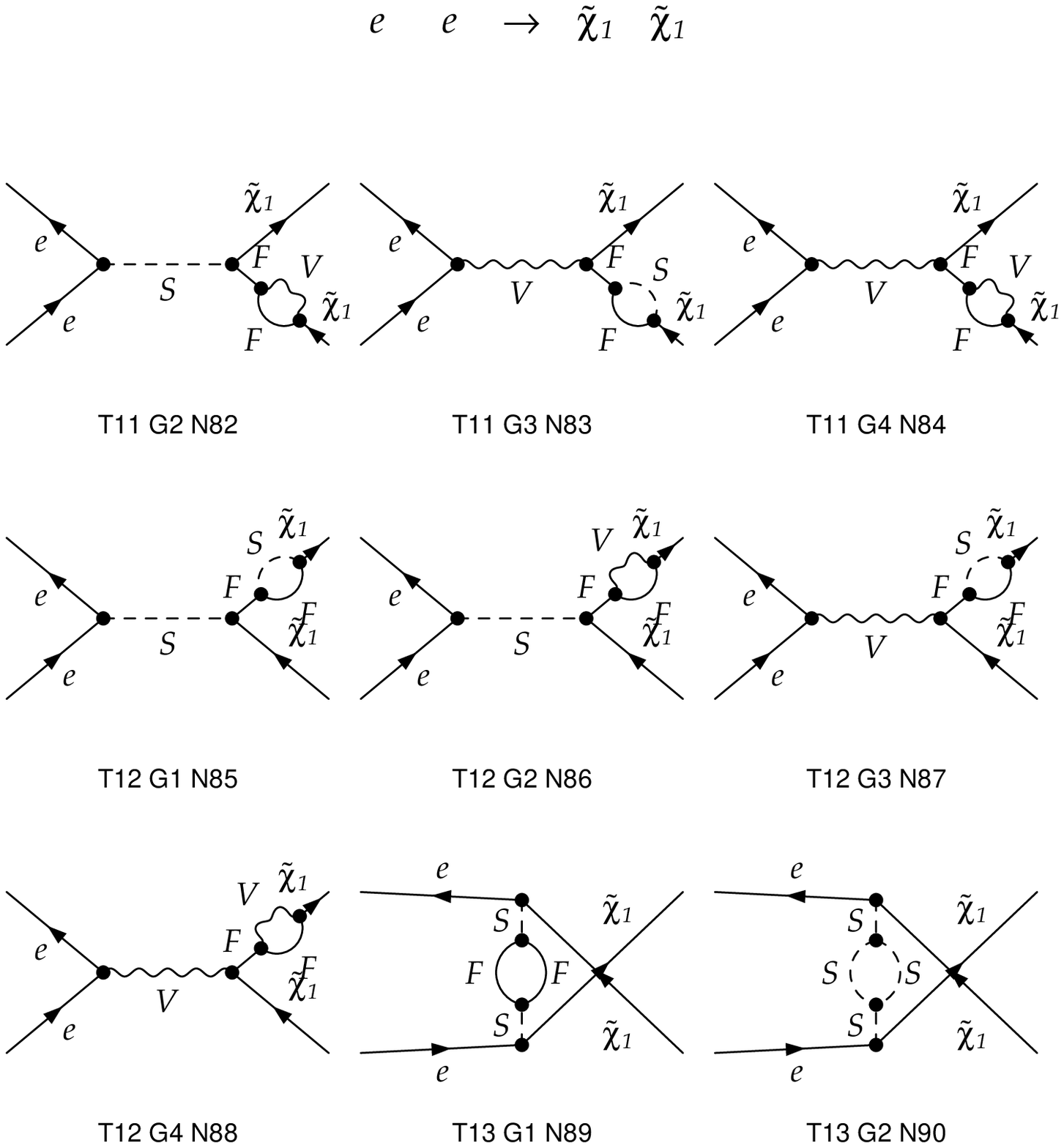}\vspace{-3cm}
\end{minipage}
\vspace{-6cm}
\begin{minipage}{\textwidth}
\centering
\includegraphics[width=0.75\textwidth]{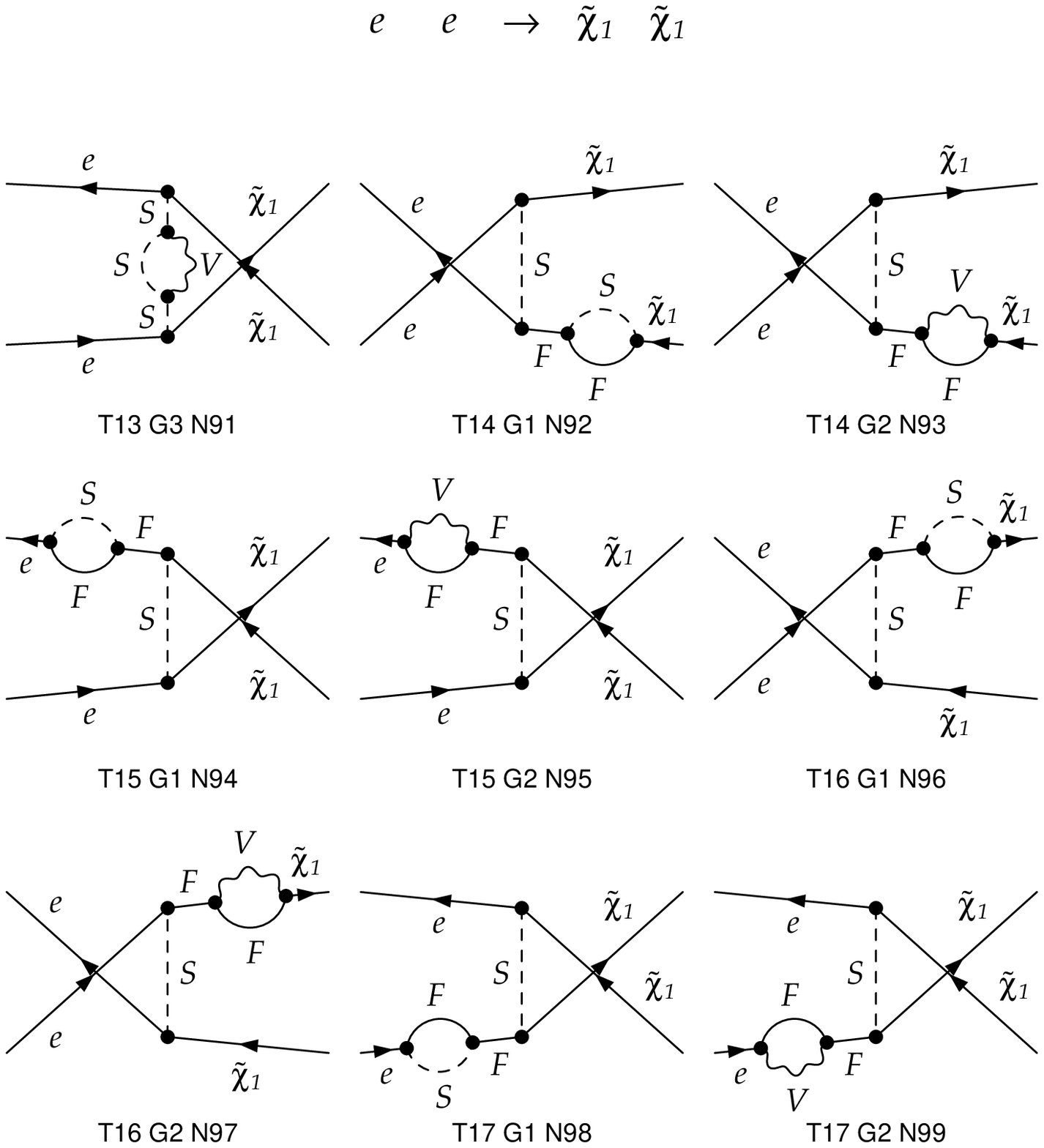}
\end{minipage}
\newpage
\begin{minipage}{\textwidth}
\vspace{-4cm}
\centering
\includegraphics[width=0.75\textwidth]{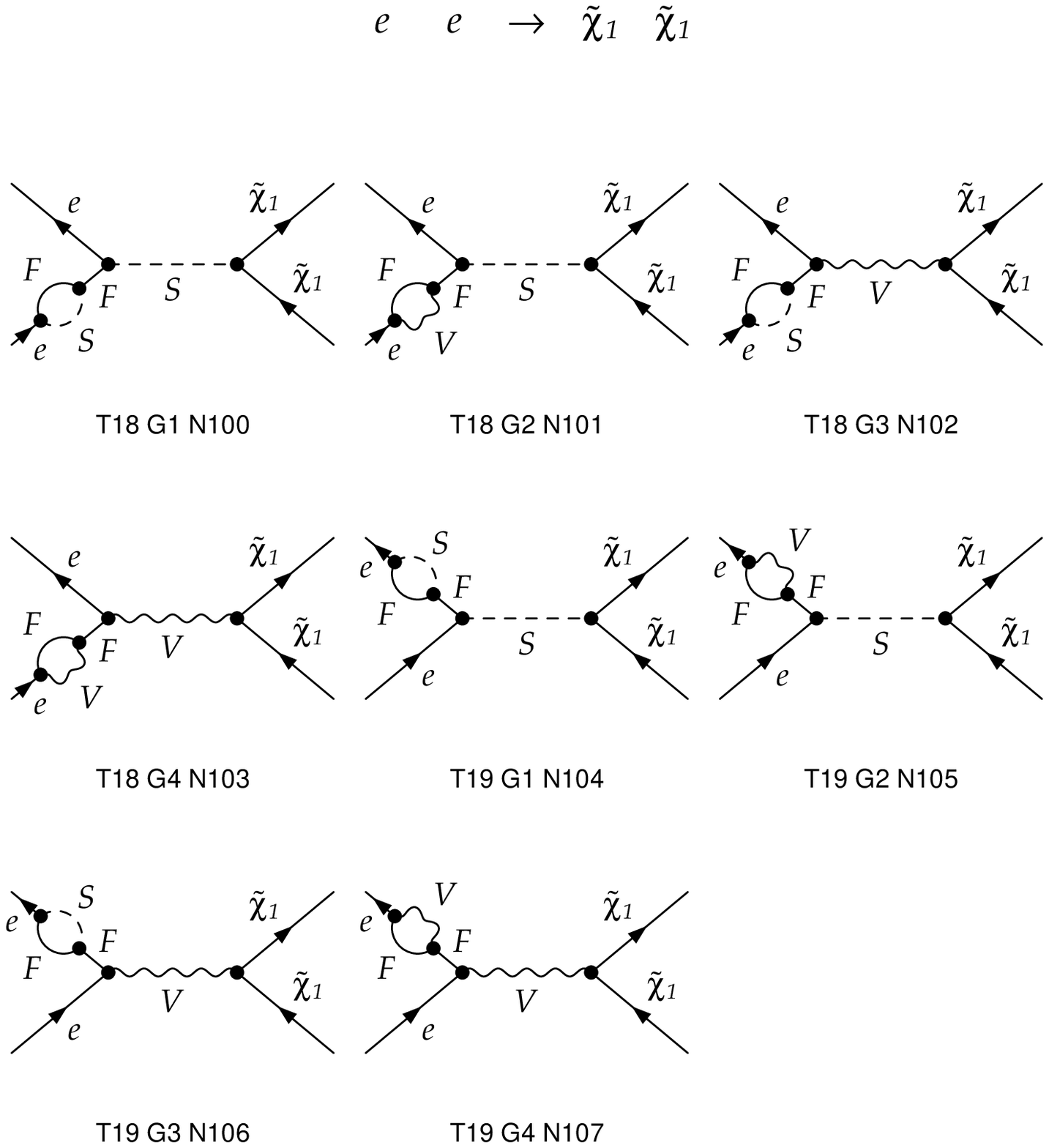}\vspace{-3cm}
\end{minipage}
\end{center}
\chapter{Soft and collinear approximation; ISR structure function}\label{app:photons}
\section{Soft Photon Factor}\label{app:softphot}
The soft photon factor is given by Eq.(\ref{eq:f-soft}), the integral appearing therein has been calculated in \cite{'tHooft:1978xw} and is given by
\begin{eqnarray*}
\lefteqn{
\int_{|\bf{k}|\leq\Delta E_\gamma}\frac{d^{3}k}{2\omega_{k}}\,\frac{2p_{i}p_{j}}{(p_{i}k)(p_{j}k)}\,=\, 4\,\pi\,\frac{\al\,p_{i}p_{j}}{(\al p_{i})^{2}-p_{j}^{2}}\,\left\{\frac{1}{2}\,\log\frac{(\al p_{i})^{2}}{p^{j}_{2}}\,\log\frac{4\Delta E_{\gamma}^{2}}{\lambda^{2}}\right.}\\
&+&\left. \left[ \frac{1}{4}\,\log^{2}\,\frac{u_{0}-|\bf{u}|}{u_{0}+|\bf{u}|}\,Li_{2}\lb 1-\frac{u_{0}+|\bf{u}|}{v}\rb\,+\,Li_{2}\lb 1-\frac{u_{0}-|\bf{u}|}{v}\rb\right]^{u=\al p_{i}}_{u=p_{j}}\right\},
\end{eqnarray*}
where
\begin{\eqn*}
v\,=\,\frac{(\al p_{i})^{2}-p^{2}_{j}}{2\,(\al\,p_{i0}-p_{j0})},
\end{\eqn*}
and $\al$ defined by
\begin{\eqn*}
\al^{2}p_{i}^{2}-2\al\,p_{i}p_{j}+p_{j}^{2}\,=\,0\;,\;\frac{\al p_{i0}-p_{j0}}{p_{j0}}\,>\,0.
\end{\eqn*}
$Li_{2}$ are dilogarithms. The integral is regulated by the photon mass $\lambda$.
\section{Hard-collinear approximation}\label{app:hardcol}
\subsection{Finite mass effects in collinear radiation}
If photons are radiated off a charged particle under a very small angle $\theta$, in the integration over this region of phase space logarithms $\propto \log m$ of the particle appear. The effects can be estimated by the collinear approximation, where in the squared matrix element only terms $\propto\,\frac{1}{m^{2}}$ are kept and higher-order terms in $\theta$ are neglected. Work along these lines has been done in \cite{Kleiss:1986ct}; we just sketch the derivation and point to this work for further reference. We will deal with the case of unpolarized initial state particles and initial particles in a well-defined helicity eigenstate.\\
\subsubsection*{Unpolarized initial particle}
We start with the general description of an amplitude where one of the
incoming particles radiates off a photon. Without the
radiation, the amplitude is given by
\begin{\eqn*}
\M\,=\,A(q_{i})\,u(p), 
\end{\eqn*}
where $q_{i}$ denote the momenta of the other particles involved in
the process. If we include radiation, this cross section is modified
to
\begin{\eqn*}
\M '\,=-\frac{e}{2\,pk}\,A(q_{i})\,(\slashed{p}-\slashed{k}+m)\slashed{\vareps}(k)\,u(p), 
\end{\eqn*}
where $k$ denotes the momentum of the photon and $\vareps$ its
polarization vector. Squaring the amplitude and summing over the
photon polarizations, we obtain
\begin{\eqn*}
|\M|^{2}\,=\sum_{\lambda} \frac{e^{2}}{4\,(pk)^{2}}A(q_{i})(\slashed{p}-\slashed{k}+m)\gamma^{\mu}(\slashed{p}+m)\gamma^{\nu}(\slashed{p}-\slashed{k}+m)\overline{A}(q_{i})\vareps^{\mu}_{\lambda}\vareps^{\nu}_{\lambda}
\end{\eqn*}
For the photon polarization sum, we take advantage of the gauge
freedom and choose
\begin{\eqn*}
\sum \vareps^{\mu}\vareps^{\nu}\,=\,-g^{\mu\nu}+\frac{k^{\mu}q^{\nu}+q^{\mu}k^{\nu}}{qk},
\end{\eqn*}
where $q$ is a general lightlike fourvector. 
We consider the $g^{\mu\nu}$ and the
$\frac{k^{\mu}q^{\nu}+q^{\mu}k^{\nu}}{qk}$ part separately.\\
From the $g^{\mu\nu}$ term, we obtain
\begin{eqnarray*}
|\M|^{2}&=&-\frac{e^{2}}{4\,(pk)^{2}}A(q_{i})(\slashed{p}-\slashed{k}+m)\gamma^{\mu}(\slashed{p}-m)\gamma_{\mu}(\slashed{p}-\slashed{k}+m)\overline{A}(q_{i})\\
&=&-\frac{e^{2}}{(pk)^{2}}\,A(q_{i})\,(m^{3}+(\slashed{p}-\slashed{k})m^{2}-kp\,(\slashed{k}+m))\overline{A}(q_{i}),
\end{eqnarray*}
and for the second part
\begin{eqnarray*}
|\M|^{2}&=&\frac{e^{2}}{4\,(pk)^{2}}A(q_{i})(\slashed{p}-\slashed{k}+m)\gamma_{\mu}(\slashed{p}-m)\gamma_{\nu}(\slashed{p}-\slashed{k}+m)\overline{A}(q_{i})\,\frac{k^{\mu}q^{\nu}+q^{\mu}k^{\nu}}{qk}\\
&=&\frac{e^{2}}{pk}\,A(q_{i})\,\lb\frac{\slashed{q}\,kp-\slashed{k}\,pq+2(m+\slashed{p})\,pq}{qk}-(m+\slashed{p})\rb\overline{A}(q_{i}).
\end{eqnarray*}
 In the
collinear approximation, we assume that
\begin{\eqn*}
\overrightarrow{k}\,\approx\,\frac{|\overrightarrow{k}|}{|\overrightarrow{p}|}\,\overrightarrow{p},
\end{\eqn*}
i.e. the angle $\theta$ between $\overrightarrow{k}$ and $\overrightarrow{p}$ is
small. We then obtain
\begin{\eqn*}
k^{\mu}\,=\,x\,\lb\begin{array}{c}p^{0}\\
|\overrightarrow{p}|\,(\theta+\mathcal{O}(\theta^{3}))\\ 0 \\ |\overrightarrow{p}|\,(1+\mathcal{O}(\theta^{2})) \end{array}\rb\,\approx\,x\,p^{\mu}(1+\mathcal{O}(\theta)),
\end{\eqn*} 
where
\begin{\eqn*}
x\,\equiv\,\frac{k^{0}}{p^{0}}.
\end{\eqn*}
Therefore,
\begin{\eqn*}
pk\,=\,p^{\mu}k_{\mu}\,=\,xm^{2}(1+\mathcal{O}(\theta)).
\end{\eqn*}
Now, we consider the terms $pq$ and $kq$. In \cite{Kleiss:1986ct}, it is assumed that the angles
$\alpha'$ between $\overrightarrow{k}$ and $\overrightarrow{q}$, and
$\beta'$ between $\overrightarrow{p}$ and $\overrightarrow{q}$ are very
large such that
\begin{eqnarray*}
\overrightarrow{k}\,\overrightarrow{q}&=&|\overrightarrow{k}|\,|\overrightarrow{q}|\,(\alpha+\mathcal{O}(\alpha^{3})),\\
\overrightarrow{p}\,\overrightarrow{q}&=&|\overrightarrow{p}|\,|\overrightarrow{q}|\,(\beta+\mathcal{O}(\beta^{3})),
\end{eqnarray*}
with 
\begin{\eqn*}
\alpha\,=\,\frac{\pi}{2}-\al'\;,\;\beta\,=\,\frac{\pi}{2}-\beta',
\end{\eqn*}
and therefore
\begin{\eqn*}
\frac{kq}{pq}\,=\,x\,\lb 1+\mathcal{O}(\al)+\mathcal{O}(\beta)+\mathcal{O}(\frac{m_{e}^{2}}{(p^{0})^{2}})\rb.
\end{\eqn*}
Note that this analysis also holds if $\overrightarrow{q}$ is nearly parallel to $\overrightarrow{k}$ and $\overrightarrow{p}$ as for
\begin{\eqn*}
q^{\mu}\,=\,\frac{1}{\sqrt{2}}\,{1\choose -\hat{k}}.
\end{\eqn*}
Then, one has
\begin{eqnarray*}
\overrightarrow{k}\,\overrightarrow{q}\,=\,-\frac{1}{\sqrt{2}}\,|\overrightarrow{k}|&,&
\overrightarrow{p}\,\overrightarrow{q}\,=\,-\frac{1}{\sqrt{2}}\,|\overrightarrow{p}|\,\cos\theta
\end{eqnarray*}
and therefore
\begin{\eqn*}
\frac{kq}{pq}\,=\,\frac{2\,k^{0}}{p^{0}+p^{0}\sqrt{(1-\frac{m^{2}}{(p^{0})^{2}})\,(1-\sin^{2}\theta)}}\,=\,x\,\lb 1+\mO(\theta)+\mO(\frac{m^{2}}{(p^{0})^{2}})\rb.
\end{\eqn*}
We can then continue as before and replace $\mO(\al)\,+\,\mO(\be)$ by $\mO(\theta)$ in the error considerations.\\
\\
Now we look at the single contributions, where we only keep terms of $\mO(m^{-2})$ and equally ignore higher-order terms in $\theta,\,\al,\,\be$. We obtain from the $-g^{\mu\nu}$ term:
\begin{eqnarray*}
\frac{m^{3}}{(pk)^2}&\longrightarrow&\frac{1}{x^{2}}\mO(m^{-1})\,(1+\mO(\theta));\\
\frac{m^{2}}{(pk)^{2}}\slashed{k}&\longrightarrow&x\,\frac{m^{2}}{(pk)^{2}}\,\slashed{p}+\frac{1}{x}\,\mO(m^{-2})\mO(\theta)\slashed{p};\\
\frac{1}{kp}\,\slashed{k}&\longrightarrow&\frac{1}{kp}\,x\,\slashed{p}+\mO(\theta)\mO(m^{-2})\slashed{p};\\
\frac{m}{pk}&\longrightarrow&\frac{1}{x} \mO(m^{-1})(1+\mO(\theta))
\end{eqnarray*}
and therefore
\begin{eqnarray*}
\frac{1}{(pk)^{2}}\,(m^{3}+(\slashed{p}-\slashed{k})m^{2}-kp\,(\slashed{k}+m))
&=&\lb \frac{m^{2}\,(1-x)}{(pk)^{2}}-\frac{x}{pk}\rb\slashed{p} \,+\,\mO\lb\frac{1}{m},\,\,\frac{\theta}{m^{2}}\rb.\\
\end{eqnarray*}
\\
For the $\frac{k^{\mu}q^{\nu}+q^{\mu}k^{\nu}}{qk}$ part, one has 
\begin{eqnarray*}
\frac{1}{qk}&\longrightarrow&\mO(1),\\
\slashed{k}\,\frac{pq}{kq\,pk}&\longrightarrow&\frac{1}{pk}\slashed{p}\,+\frac{1}{x}\,\mO(m^{-2})\lb\mO(\theta,\,\al\,\be) \rb\slashed{p},\\
\frac{m\,pq}{pk\,kq}&\longrightarrow&\frac{1}{x^{2}}\mO(m^{-1})(1+\mO(\theta,\,\al,\,\be)),\\
\slashed{p}\frac{pq}{kq\,pk}&\longrightarrow&\slashed{p}\,\frac{1}{x}\,\frac{1}{pk}+\frac{1}{x^{2}}\mO(m^{-2})\,\mO(\al,\, \be)\slashed{p},\\
\end{eqnarray*}
and therefore 
\begin{eqnarray*}
\frac{1}{pk}\lb\frac{\slashed{q}\,kp-\slashed{k}\,pq+2(m+\slashed{p})\,pq}{qk}-(m+\slashed{p})\rb
&=&\frac{1}{x\,pk}(-2x+2)\slashed{p}+\,\mO\lb\frac{1}{m},\,\frac{\theta}{m^{2}},\frac{\al}{m^{2}},\,\frac{\be}{m^{2}}\rb.\\
\end{eqnarray*}
Summing up everything, we arrive at
\begin{\eqn}\label{eq:hcone}
|\M|^{2}\,=\,\frac{e^{2}}{pk}\,A(q_{i})\lb\frac{2-2x+x^{2}}{x}-\frac{m^{2}(1-x)}{pk}\rb\,\slashed{p}\,\overline{A}(q_{i})
\end{\eqn}
as given in \cite{Kleiss:1986ct}.\\
\subsubsection*{Initial particle in definite helicity state}
If the radiating particle is in a definite helicity state, $\M'$ is given by
\begin{\eqn*}
\M '\,=-\frac{e}{2\,pk}\,A(q_{i})\,(\slashed{p}-\slashed{k}+m)\slashed{\vareps}(k)\,\frac{1}{2}(1\,\pm\,\gamma^{5}\slashed{s}) u(p), 
\end{\eqn*}
where $\overrightarrow{s}\,\parallel\overrightarrow{k}$ and $\frac{1}{2}(1\,\pm\,\gamma^{5}\slashed{s})$ projects out the positive/ negative helicity eigenstate. We parameterize $s$ according to
\begin{\eqn}\label{eq:spinvec}
s^{\mu}\,=\,P_{\parallel}(\frac{p^{\mu}}{m}-m\frac{q^{\mu}}{pq})+P_{\perp}s_{\perp}^{\mu}.
\end{\eqn}
For the helicity eigenstates, $P_{\perp}=0$, and we choose $P_{\parallel}\,\ge\,0$ and define positive/ negative helicity by the $\pm$ sign.\\
We now have to calculate
\begin{\eqn*}
|\M|^{2}\,=\sum_{\lambda} \frac{e^{2}}{4\,(pk)^{2}}A(q_{i})(\slashed{p}-\slashed{k}+m)\gamma_{\mu}(\slashed{p}+m)\frac{1}{2}(1\,\pm\,\gamma^{5}\slashed{s})\gamma_{\nu}(\slashed{p}-\slashed{k}+m)\overline{A}(q_{i})\vareps^{\mu}_{\lambda}\vareps^{\nu}_{\lambda}
\end{\eqn*}
with the same gauge choice as for the unpolarized case. We first consider the $g^{\mu\nu}$ contribution and the $p^{\mu}$ and $q^{\mu}$ part of $s^{\mu}$, separately. Furthermore, we only calculate the part depending on $P_{\parallel}$ (the part proportional to $1$ in the helicity projector corresponds to the unpolarized case multiplied by $\frac{1}{2}$). In the following, we use $Z$ defined by
\begin{\eqn*}
|\M'|^{2}\,\equiv\,e^{2}\,A(q_{i})\,Z\,\overline{A}(q_{i}).
\end{\eqn*}
We consider the contributions coming from the combination of two terms in the polarization sum and two terms in $s$ separately. In the following, the first term in the brackets denotes the term coming from the polarization sum $\sum\,\vareps^{\mu\nu}$, the second the term stemming from $s^{\rho}$.\\
\\
\centerline{\boldmath $\lb -g^{\mu\nu}, \frac{p^{\rho}}{m} \rb$}
\\
 \\
For the $\frac{p^{\mu}}{m}$ part, we obtain
\begin{\eqn*}
Z\,=-\frac{1}{4\,(pk)^2}\,\left( -2\,\Sk \cdot \gamma^{5}\,
     \left( m^2 - kp\right)  +
    m\,\left( \Sk \cdot
        \Sp \cdot
        \gamma^{5} -
       \Sp \cdot
        \Sk \cdot
        \gamma^{5} +
       4\,\gamma^{5}\,
        kp \right)  \right).
\end{\eqn*}
We first look at the $\mO(m^{2})$ terms in the numerator; they are given by
\begin{\eqn*}
-\frac{1}{2}\,\frac{m^{2}-kp}{(kp)^{2}}\,\gamma^{5}\,\Sk\;\longrightarrow\,-\frac{1}{2}\,\gamma^{5}\,\frac{m^{2}-kp}{(pk)^{2}}\,x\Sp+(1+\frac{1}{x})\,\mO(m^{-2})\,\mO(\theta).
\end{\eqn*}
The leftover terms are of order $\frac{1}{x}\mO(m^{-1})$.\\
\\
\centerline{\boldmath $\lb-g^{\mu\nu}, -m\frac{q^{\rho}}{pq} \rb$}
\\
\\
From the $-m\frac{q^{\mu}}{pq}$ part, we obtain
\begin{eqnarray*}
  Z&=&\frac{m}{4\,(pk)^{2}\,(pq)}\,\lb m^2\,\Sk\cdot \Sq\cdot       \gamma^{5} -     m^2\,\Sp \cdot 
  \Sq\cdot       \gamma^{5} -     m^2\,\Sq\cdot       \Sk\cdot 
  \gamma^{5} +    m^2\,\Sq\cdot       \Sp\cdot 
  \gamma^{5}  \right. \\  &-& 2\,\Sq\cdot       \gamma^{5}\,     \left( m^3 - m\,(kp)         \right) 
  \,+\, \left.    2\,m\,\Sk\cdot       \gamma^{5}\,     (kq) -   2\,m\,\Sp \cdot 
      \gamma^{5}\,      (kq) -     2\,m\,\Sk\cdot \gamma^{5}\,
      (pq) \right. \\ &+& \left.  2\,m\,\Sp\cdot \gamma^{5}\,(pq) + 
    4\,\gamma^{5}\,(kp)\,(pq) \rb .
\end{eqnarray*}
Here, the terms relevant for the approximation are given by
\begin{\eqn*}
-\frac{m^{2}}{2\,(pk)^{2}}\,\gamma^{5}\,(1-\frac{kq}{pq})(\Sp-\Sk)\;\longrightarrow\;-\frac{m^{2}}{2\,(pk)^{2}}\gamma^{5}\,(x^{2}-2x+1)\Sp+(1+\frac{1}{x})\,\mO(m^{-2})\,\mO(\theta,\,\al,\, \be) \,\Sp.
\end{\eqn*}
\\
\\
\centerline{\boldmath $\lb\frac{k^{\mu}q^{\nu}+q^{\mu}k^{\nu}}{kq}, \frac{p^{\rho}}{m}\rb $}
\\
\\
The $\frac{k^{\mu}q^{\nu}+q^{\mu}k^{\nu}}{kq}$ part of the polarization sum is given by
\begin{eqnarray*}
Z&=&\frac{1}{4\,(pk)^{2}\,qk}\,\left( m\,\Sp\cdot \Sq\cdot \gamma^{5}\,
      (kp) + 
    2\,m\,\Sq\cdot 
      \Sk\cdot 
      \gamma^{5}\,
      (kp) - 
    m\,\Sq\cdot 
      \Sp\cdot 
      \gamma^{5}\,
      (kp) \right. \\ & +& \left.
    2\,\Sq\cdot 
      \gamma^{5}\,
     \left( m^2 -  (kp) \right) \,
      (kp) - 
    2\,m^2\,\Sk\cdot 
      \gamma^{5}\,
      (kq) - 
    2\,m\,\Sp\cdot 
      \Sk\cdot 
      \gamma^{5}\,
      (kq) \right. \\ &+& \left.
    2\,m\,\gamma^{5}\,
      (kp)\,
      (kq) + 
    2\,\Sp\cdot 
      \gamma^{5}\,
      (kp)\,
      (kq) + 
    2\,m^2\,\Sk\cdot 
      \gamma^{5}\,
      (pq) - 
    m\,\Sk\cdot 
      \Sp\cdot 
      \gamma^{5}\,
      (pq)\right.\\& +& \left. 
    m\,\Sp\cdot 
      \Sk\cdot 
      \gamma^{5}\,
      (pq) + 
    2\,\Sk\cdot 
      \gamma^{5}\,
      (kp)\,
      (pq) - 
    4\,\Sp\cdot 
      \gamma^{5}\,
      (kp)\,
      (pq) \right). 
\end{eqnarray*}
The $\mO(m^{2})$ terms in the numerator are given by
\begin{eqnarray*}
&&-\frac{1}{2\,(pk)^{2}}\,\gamma^{5}\,\lb m^{2}\,(\frac{pq}{kq}-1)\Sk + (pk)\lb\Sp+\frac{pq}{kq}(\Sk-2\Sp)\rb\rb\\
&\longrightarrow&-\gamma^{5}\,\frac{1}{2}\,\lb \frac{m^{2}}{(pk)^{2}}(1-x)+\frac{2}{pk}\lb 1-\frac{1}{x}\rb\rb \Sp\,+\,\lb\frac{1}{x}+\frac{1}{x^{2}}\rb\,\,\mO(m^{-2})\lb\mO(\theta,\,\al,\, \be)\rb\Sp.
\end{eqnarray*}
\\
\\
\centerline{\boldmath $\lb\frac{k^{\mu}q^{\nu}+q^{\mu}k^{\nu}}{kq}, -\frac{m\,q^{\rho}}{pq}\rb $}
\\
\\
This part of $Z_{tot}$ is given by
\begin{eqnarray*}
Z&=&-\frac{m}{4\,(pk)^{2}\,kq}\,
  \lb m\,\Sp\cdot 
      \Sq\cdot 
      \Sk\cdot 
      \gamma^{5} - 
    m\,\Sq\cdot 
      \Sk\cdot 
      \Sp\cdot 
      \gamma^{5} + 
    2\,\Sq\cdot 
      \Sp\cdot 
      \gamma^{5}\,
      (kp) - 
    2\,m\,\Sk\cdot 
      \gamma^{5}\,
      (kq)\right.\\& +& \left. 
    2\,\Sk\cdot 
      \Sp\cdot 
      \gamma^{5}\,
      (kq) - 
    2\,\Sk\cdot 
      \Sp\cdot 
      \gamma^{5}\,
      (pq) \rb. 
\end{eqnarray*}
From this, we keep\footnote{Here we used that
\begin{\eqn*}
\Sp\Sq\Sk\,=\,2\,qk\,\Sp-\Sp\Sk\Sq\,=\,2\,qk\,\Sp-m^{2}(1+\mO(\theta))\,\Sq.
\end{\eqn*}}
\begin{eqnarray*}
-\frac{m^{2}}{2\,(pk)^{2}}\,\gamma^{5}\,(x-1)(\Sp-\Sk)&\longrightarrow&-\frac{m^{2}}{2\,(pk)^{2}}\gamma^{5}\,(x-1)\Sp+\frac{1}{x}\mO(m^{-2})\,\mO(\theta).
\end{eqnarray*}
\\
\\
Summing up all terms, we end up with
\begin{\eqn}\label{eq:hcpparr}
|\M'|^{2}\,=\,\frac{e^{2}}{2\,pk}\,A(q_{i})\lb\frac{2-2x+x^{2}}{x}-\frac{m^{2}(1-x+x^{2})}{pk}\rb\,\,\gamma^{5}\,\slashed{p}\,\overline{A}(q_{i})
\end{\eqn}
for the part proportional to $P_{\parallel}$ as given in \cite{Kleiss:1986ct}.
\subsection{Helicity dependent structure functions}\label{app:colstrfun}
In \cite{Dittmaier:1993jj, Bohm:1993qx, Dittmaier:1993da}, Eqs. (\ref{eq:hcone}) and (\ref{eq:hcpparr}) have been used to derive helicity-dependent structure functions which describe photon radiation integrated over the collinear region. Again, we only sketch the result and refer to these works for further reference.\\
Without photon radiation, the matrix element is given by
\begin{\eqn*}
\M\,=\,A(q_{i})\frac{1}{2}(1+\gamma^{5}\slashed{s})\,u(p),
\end{\eqn*}
and the square by
\begin{\eqn*}
|\M|^{2}\,=\,A(q_{i})\frac{1}{2}(1+\gamma^{5}\slashed{s})(\Sp+m)\,\overline{A}(q_{i}).
\end{\eqn*}
With $s^{\mu}$ given by Eq. (\ref{eq:spinvec}), we obtain
\begin{\eqn*}
\slashed{s}(\Sp+m)\,=P_{\parallel}\,\,\Sp+\mO(m^{2})\Sq
\end{\eqn*} 
and
\begin{\eqn*}
|\M|^{2}\,=\,A(q_{i})\lb\frac{1}{2}(1+\gamma^{5}\,P_{\parallel})\Sp\,+\mO(m)\rb\overline{A}(q_{i}).
\end{\eqn*}
We now have to equate this with
\begin{\eqn*}
|\M'|^{2}\,=\frac{e^{2}}{2\,pk}\,A(q_{i})(A_{1}+P_{\parallel}\,A_{2}\gamma^{5})\overline{A}(q_{i}),
\end{\eqn*}
where the $A_{i}$ were derived in the previous sections. A short calculation then gives
\begin{\eqn*}
|\M'|^{2}\,=\frac{e^{2}}{2\,pk}\,A(q_{i})\lb B\frac{1}{2}(1+\gamma^{5}P_{\parallel})+C\frac{1}{2}(1-\gamma^{5}P_{\parallel})\rb\Sp\overline{A}(q_{i}),
\end{\eqn*}
with
\begin{eqnarray}\label{eq:bc}
B&=&(\frac{2}{x}-\frac{m^{2}}{pk})\,(2-2x+x^{2}),\nonumber\\
C&=&\frac{m^{2}}{pk}\,x^{2}.
\end{eqnarray}
The next step is the approximation of $\frac{1}{pk}$ by
\begin{\eqn*}
\frac{1}{pk}\,\approx\,\frac{2}{x\,(m^{2}+4(p^{0})^{2}\,sin^{2}\frac{\theta}{2})}.
\end{\eqn*}
This approximation uses
\begin{\eqn*}
\overrightarrow{p}\overrightarrow{k}\,\approx\,p^{0}k^{0}\,\lb\cos\theta-\frac{1}{2}\,\frac{m^{2}}{(p^{0})^{2}}\rb+\mO(\theta^{2})\,\mO\lb\frac{m^{2}}{(p^{0})^{2}}\rb+\mO\lb\frac{m^{4}}{(p^{0})^{4}}\rb.
\end{\eqn*}
Next, we have to approximate the integrals
\begin{\eqn*}
\int_{0}^{\Delta\theta}\,\frac{\sin\theta}{pk}\,d\theta\;,\,\int_{0}^{\Delta\theta}\,\frac{\sin\theta}{(pk)^{2}}\,d\theta
\end{\eqn*}
in the collinear limit. Here, we use
\begin{eqnarray*}
\frac{1}{m^{2}+4\,(p^{0})^{2}\,\sin^{2}\frac{\theta}{2}}&\approx&\frac{1}{m^{2}+(p^{0})^{2}\,\theta^{2}}\,\lb 1+\mO(\theta^{2})\rb,\\
\sin\theta&\approx&\theta+\mO(\theta^{3}),\\
\ln\lb\frac{m^{2}+(p^{0})^{2}(\Delta\theta)^{2}}{m^{2}}\rb&\approx&\ln\lb\frac{(p^{0})^{2}(\Delta\theta)^{2}}{m^{2}}\rb+\mO\lb\frac{m^{2}}{(p^{0})^{2}\,(\Delta\theta)^{2}}\rb,\\
\lb\frac{1}{m^{2}}-\frac{1}{m^{2}+(p^{0})^{2}(\Delta\theta)^{2}}\rb&\approx&\frac{1}{m^{2}}\,\lb 1+\mO\lb\frac{m^{2}}{(p^{0})^{2}\,(\Delta\theta)^{2}}\rb\rb.
\end{eqnarray*}
As a next step, we approximate the integral over $|\overrightarrow{k_{1}}|$:
\begin{\eqn*}
\int\,dk^{0}_{1}|\overrightarrow{k_{1}}|\,=\,\int\,(p^{0})^{2}\,x (1+\mO(\theta))\,dx.
\end{\eqn*}
Similarly, we assume
\begin{eqnarray*}
&&\int\,d\Gamma_{1}\,d\Gamma_{2}\,d\Gamma_{3}\,\delta^{(4)}(p_{1}+p_{2}-k_{1}-k_{2}-k_{3})\,F(p_{1},p_{2},k_{1},k_{2},k_{3})\,=\,\\
&&\int\,\frac{|\overrightarrow{k_{1}}|\,dk^{0}_{1}}{(2\pi)^{3}\,2}\,d\Omega_{1}\,\int\,d\Gamma_{2}\,d\Gamma_{3}\,\delta^{(4)}(p'_{1}+p_{2}-k_{2}-k_{3})\,F(p'_{1},p_{2},k_{2},k_{3}),
\end{eqnarray*}
when
\begin{\eqn*}
p'_{1}\,=\,\lb p_{1}-k_{1})\,=\,( 1-x+\mO(\theta)\rb\,p_{1}.
\end{\eqn*}
The differential phase space is given by
\begin{\eqn*}
d\Gamma_{i}\,=\,\frac{d^{3}k_{i}}{(2\pi)^{3}\,2\,k_{i}^{0}}.
\end{\eqn*}
The phase space integration approximation is exact for $\theta\,=\,0$. \\
Taking all this into account, we then obtain for the integration over the factors $B$ and $C$ (Eq. (\ref{eq:bc})):
\begin{eqnarray*}
&&\int\,d\,\Gamma_{1}\frac{e^{2}}{2\,pk}\,(2-2x+x^{2})\,\lb \frac{2}{x}-\frac{m^{2}}{pk}\rb\,|\M^{+}|^{2}\;= \\
&&=\frac{\al}{2\,\pi}\,\lb\ln\lb\frac{s\,(\Delta\theta)^{2}}{4\,m^{2}}\rb-1\rb\,\int^{x_{max}}_{x_{min}}\,\frac{2-2x+x^{2}}{x}\,|\M^{+}|^{2}\,dx,\\
&&\int\,d\,\Gamma_{1}\frac{e^{2}\,m^{2}}{2\,(pk)^{2}}\,x^{2}|\M^{-}|^{2}\;=\;\frac{\al}{2\,\pi}\int^{x_{max}}_{x_{min}}\,x\,|\M^{-}|^{2}\,dx.
\end{eqnarray*}  
Here, $\M^{\pm}$ denote the same/flipped radiating particle helicity amplitudes for the $2\,\rightarrow\,2$ particle process and $x_{min}$ and $x_{max}$ are given by
\begin{eqnarray*}
x_{min}\,=\,\frac{2\,\Delta\,E_{\gamma}}{\sqrt{s}}\;\;\;\;\;\;\;\;\;\;\;\;\;\;\;\;&&\mbox{$\Delta E_{\gamma}$: soft photon cut}\\
x_{max}\,=\, 1-\frac{(m_{3}+m_{4})^{2}}{s} &&\mbox{kinematic limit for 3 body decay}.
\end{eqnarray*}
If we want to bring this now in the form of a structure function
\begin{\eqn*}
\sigma^{hard,coll}(s)\,=\,\int \lb f^{+}(x')\,\sigma^{+}(x's)+f^{-}(x')\,\sigma^{-}(x's)\rb\,dx',
\end{\eqn*}
we have to substitute
\begin{eqnarray*}
x\,\rightarrow\,x'\,=\,1-x&,&
\Sp\,=\,\frac{x'}{x'}\,\Sp
\end{eqnarray*}
and obtain
\begin{eqnarray}\label{eq:hardcollfs}
f^{+}(x)&=&\,\frac{\al}{2\,\pi}\,\frac{\Phi(sx)}{\Phi(s)}\,\frac{1+x^{2}}{x\,(1-x)}\,\lb\ln \lb \frac{s\,(\Delta\theta)^{2}}{4\,m^{2}}\rb-1 \rb,\nonumber \\
f^{-}(x)&=&\frac{\al}{2\,\pi}\,\frac{\Phi(sx)}{\Phi(s)} \,\frac{1-x}{x}.
\end{eqnarray}
The factor
\begin{\eqn*}
\frac{\Phi(sx)}{\Phi(s)}\,=\,x\,\lb1+\mO\lb\frac{m^{2}}{s}\rb\rb
\end{\eqn*}
takes the difference of fluxes in the calculation of the total cross sections into account.
\subsection{Connection to leading log expressions}\label{sec:conlog}
In structure functions, collinear logarithms of the form
\begin{\eqn*}
\ln\,\lb\frac{Q^{2}}{m^{2}}\rb
\end{\eqn*}
appear,
where $Q$ is the scale of the respective process. In these processes, $Q$ is often taken as an upper limit for $k_{\perp}$. We show that a change of variables from $\cos\theta$ to $p_{\perp}$ exactly reproduces the results derived in section \ref{app:colstrfun} and we can substitute
\begin{\eqn*}
\ln\lb\frac{Q^{2}}{m^{2}}\rb\,\longrightarrow\,\ln\lb\frac{(p^{0})^{2}\,(\Delta\theta)^{2}}{m^{2}}\rb
\end{\eqn*}
in the structure functions.
With $k$ being the photon four-momentum,
\begin{eqnarray*}
k_{\perp}&=&k^{0}\,\sin\theta,\\
\cos\theta&=&\sqrt{1-\sin^{2}\theta}\,=\,1-\frac{1}{2}\,\frac{k_{\perp}^{2}}{(k^{0})^{2}}+\mO\lb\frac{k_{\perp}^{4}}{(k^{0})^{4}}\rb,\\
\frac{d\cos\theta}{dk_{\perp}}&=&-\frac{k_{\perp}}{(k^{0})^{2}}.
\end{eqnarray*}
Furthermore, one has
\begin{\eqn*}
\frac{1}{pk}\,=\,\frac{1}{p^{0}\,k^{0}}\,\frac{1}{\frac{1}{2}\,\frac{k_{\perp}^{2}}{(k^{0})^{2}}+\frac{1}{2}\,\frac{m^{2}}{(p^{0})^{2}}}\lb 1 +\mO\lb\frac{k_{\perp}^{4}}{(k^{0})^{4}}\rb+\mO\lb\frac{m^{2}}{(p^{0})^{2}}\rb\rb.
\end{\eqn*}
Integrating
\begin{\eqn*}
I\,:=\,\int^{(x p^{0} \Delta\theta)}_{0}\,\frac{e^{2}}{2\,pk}\,\lb\frac{2}{x}-\frac{m^{2}}{pk}\rb\,(2-2x+x^{2})\,dk_{\perp},
\end{\eqn*}
we obtain
\begin{\eqn*}
I\,=\,\frac{e^{2}}{(p^{0})^{2}\,x^{2}}\,\lb\ln\lb\frac{(\Delta\theta)^{2}\,(p^{0})^{2}}{m^{2}}\rb-1\rb\,(2-2x+x^{2}),
\end{\eqn*}
i.e. exactly the result as given in the previous section. We therefore see that an upper limit
\begin{\eqn*}
k^{(max)}_{\perp}\,=\,k^{0}\Delta\theta
\end{\eqn*}
exactly gives the usual structure function result integrated up to a scale
\begin{\eqn*}
Q^{2}\,=\,(k^{0}\Delta\theta)^{2}.
\end{\eqn*}
\section{ISR structure function}\label{app:isr}
\subsection{Exact first order and soft solution}
In the following, we will consider the exact first order solution as well as the infinitely summed up soft solution for the electron structure function evolution equation (\ref{eq:llev})
\begin{\eqn*}
\frac{\partial}{\partial\eta}D^{NS}(x,\eta)\,=\,\frac{1}{4}\,\int^{1}_{x}\,\frac{dz}{z}\,P(z)D^{NS}\lb\frac{x}{z},\eta \rb
\end{\eqn*}
with 
\begin{\eqn*}
P_{ee}(x)\,=\,\lb\,\frac{1+x^{2}}{1-x}\rb_{+}\,,\,
\eta\,=\,\frac{2\,\al}{\pi}\,\log\lb\frac{Q^{2}}{m^{2}}\rb.
\end{\eqn*}
(cf. Eqs. (\ref{eq:elsplit}),(\ref{eq:eta})).
The $+$-distribution $(h(x))_{+}$ is defined by
\begin{\eqn}\label{eq:plusdef}
\lb h(x) \rb _{+}\,=\,\lim_{\vareps\,\rightarrow\,0}\,\left[\theta(1-x-\vareps)\,h(x)-\delta(1-x)\,\int^{1-\vareps}_{0}\,h(y)\,dy\right]
\end{\eqn}
leading to
\begin{\eqn*}
\int^{1}_{0}\,dx\,f(x)\,\lb h(x) \rb _{+}\,=\,\int^{1}_{0}\,dx\lb f(x)-f(1) \rb\,h(x).
\end{\eqn*}
$D^{NS}$ denotes the non-singlet part of the structure function \cite{Kuraev:1985hb} describing photon radiation only.
In the derivation of the soft regime solution, we will closely follow \cite{Chen:1990qz}.\\
The first order solution is given by
\begin{eqnarray}
D^{NS,\al}&=&\delta(1-x)+\lim_{\vareps\,\rightarrow\,0} \frac{\eta}{4}\,\lb \delta(1-x)\,\lb \frac{3}{2}\,+2\,\ln(\vareps)\rb+\theta(1-x-\vareps)\,\frac{1+x^{2}}{1-x}\rb,\nonumber\\
&&\nonumber\\
\int^{1}_{x_{0}}\,D^{NS,\al}(x)\,dx&=&1+\frac{\eta}{4}\,\lb 2\,\ln(1-x_{0})+x_{0}+\frac{1}{2}x_{0}^{2}\rb.\label{eq:intsoftisr}
\end{eqnarray}
The soft integrated $\mO(\eta)$ contribution (\ref{eq:intsoftisr}) is negative because, when integrating from $x_{0}$ to $1$, we do take all virtual, but not all real photon contributions into account. For $x_{0}\,=\,0$, the $\mO(\eta)$ contribution vanishes and
as expected, $\int^{1}_{0}\,D^{NS,\al}(x)\,dx\,=\,1$.\\
\\
For the derivation of the solution in the soft regime, we perform
a Mellin transform (cf. App \ref{app:trafos}) leading to
\begin{\eqn}\label{eq:evmellin}
D^{NS(\xi)}(\eta)\,=\,1+\frac{1}{4}\,\int^{\eta}_{0}\,d\be\,D^{NS(\xi)}(\be)P^{(\xi)},
\end{\eqn}
where
\begin{\eqn*}
P^{(\xi)}\,=\,\psi_{0}(1)+\psi_{0}(3)-\psi_{0}(\xi)-\psi_{0}(\xi+2).
\end{\eqn*}
and $\psi_{0}(x)$ being the Digamma function defined in Appendix \ref{app:digamma}.
The formal solution to Eq. (\ref{eq:evmellin}) is 
\begin{\eqn*}
D^{NS(\xi)}(\eta)\,=\,\exp\lb \frac{1}{4}\eta\,P^{(\xi)}\rb.
\end{\eqn*}
Performing the inverse Mellin transform then leads to
\begin{\eqn}\label{eq:dnsint}
D^{NS}(\eta,x)\,=\,\frac{1}{2\pi\,i}\,\exp\lb \frac{\eta}{2}(\frac{3}{4}-\gamma)\rb\,\int^{c+\,i\,\infty}_{c-\,i\,\infty}\,d\xi\,x^{-\xi}\,\exp\lb \frac{\eta}{4}\,(-\psi_{0}(\xi)\,-\psi_{0}(\xi+2))\rb.
\end{\eqn}
Note that the exponential in front of the integral corresponds to the virtual part solution (modulo some constant terms which disappear when the $\xi$ dependent digamma functions are subtracted).\\
We then use 
\begin{\eqn}\label{eq:psiapprox}
\psi_{0}(\xi)+\psi_{0}(\xi+2)\,=\,2\,\lb \ln(\xi)-\sum_{n=1}^{\infty}\frac{B_{2n}}{2\,n\,\xi^{2\,n}}\,+\frac{1}{2\,(\xi+1)}\rb,
\end{\eqn}
where $B_{2n}$ denote the Bernoulli numbers (cf. Appendix \ref{app:digamma}). Only the logarithmic term of Eq. (\ref{eq:psiapprox}) is kept. This results in
\begin{\eqn}\label{eq:dnsfin}
D^{NS}(\eta,x)\,\simeq\,\frac{1}{2\pi\,i}\,\exp\lb \frac{\eta}{2}(\frac{3}{4}-\gamma)\rb\,\int^{c+\,i\,\infty}_{c-\,i\,\infty}\,d\xi\,\frac{x^{-\xi}}{\xi^{\frac{\eta}{2}}}\,=\,\exp\lb \frac{\eta}{2}(\frac{3}{4}-\gamma)\rb\,\frac{\log\lb\frac{1}{x}\rb^{\frac{\eta}{2}-1}}{\Gamma(\frac{\eta}{2})}.
\end{\eqn}
Finally, we use 
\begin{\eqn}\label{eq:approxlog}
-\ln(y)\,=\,(1-y)+\frac{1}{2}(1-y)^{2}+\mO\lb(1-y)^{3}\rb
\end{\eqn}
and obtain
\begin{\eqn}\label{eq:dnsgrib}
D^{NS}(\eta,x)\,=\,\exp\lb \frac{\eta}{2}(\frac{3}{4}-\gamma)\rb\,\frac{\eta}{2}\,\frac{(1-x)^{\frac{\eta}{2}-1}}{\Gamma(\frac{\eta}{2}+1)}.
\end{\eqn}
This is the solution to Eq. (\ref{eq:llev}) in the soft regime; it contains all orders in $\eta$.\\
\\
Finally, we can ask which splitting function would exactly lead to Eq. (\ref{eq:dnsgrib}). We therefore look for a function $f(\xi)$ such that (cf. Eq. (\ref{eq:dnsint}))
\begin{\eqn*}
\frac{1}{2\pi\,i}\,\int^{c+i\,\infty}_{c-i\infty}\,d\xi\,x^{-\xi}\,\exp\,\lb \frac{\eta}{4}f(\xi)\rb\,=\,\frac{(1-x)^{\frac{\eta}{2}-1}}{\Gamma(\frac{\eta}{2})}.
\end{\eqn*}
This is given by
\begin{\eqn*}
f(\xi)\,=\,\frac{1}{\Gamma(\frac{\eta}{2})}\,\int^{\infty}_{0}\,\lb 1-e^{-x} \rb^{\frac{\eta}{2}-1}\,e^{-\xi\,x}\,dx\,=\,\frac{\Gamma(\xi)}{\Gamma(\xi+\frac{\eta}{2})}\,=\,1-\frac{\eta}{2}\,\psi_{0}(\xi)+\mO((\frac{\eta}{2})^{2}),
\end{\eqn*}
where for simplicity we used the Laplace instead of the Mellin transform.
Therefore, we see that the approximation for $y\,\approx\,1$ leads then to
\begin{\eqn*}
\psi_{0}(\xi)\,+\,\psi_{0}(\xi+2)\,\longrightarrow\,2\,\psi_{0}(\xi).
\end{\eqn*}
This corresponds to
\begin{\eqn*}
\frac{1+y^{2}}{1-y}\,\longrightarrow\frac{2}{1-y}
\end{\eqn*}
for the real photon part in the splitting function. In making this substitution, we omitted the term
\begin{\eqn*}
\int^{1}_{x_{0}}\,\lb\frac{1+y^{2}}{1-y}-\frac{2}{1-y}\rb\,dy\,=\,-\frac{3}{2}+x_{0}+\frac{1}{2}x_{0}^{2}\,=\,\mO(1-x_{0})
\end{\eqn*}
in the integration of the splitting function in the soft regime.
We see that the (dominant) logarithmic behavior is exactly reproduced when integrating the modified structure function in the soft region; the errors are $\mO(1-y)$.\\
\subsection{Exponentiation}\label{app:expo}
The basic idea of exponentiation is to combine the $\mO(\eta^{n}), n\,\rightarrow\,\infty$  emission of soft photons, described in a leading logarithmic approach, with explicit finite order contribution in the hard-collinear regime into structure functions \cite{Kuraev:1985hb, Kleiss:1989de,Skrzypek:1990qs, Skrzypek:1992vk,Jadach:1990vz, Jadach:1990fy} . Then,
\begin{itemize}
\item{} $D(x)$ in the hard regime and
\item{} $\int^{1}_{x_{0}}\,D(x)\,dx$ in the soft regime
\end{itemize}
correspond to the exact solutions up to a certain order of $\eta$. Comparisons of different approaches can be found in \cite{Skrzypek:1990qs}.
Following these lines, Skrzypek and Jadach obtained the following ISR structure function\\
\begin{align}\label{eq:f-ISR}
  f_\text{ISR}(x) &=
  \frac{\exp\left(-\tfrac12\eta\gamma_{E} + \tfrac{3}{8}\eta\right)}
       {\Gamma(1+\frac{\eta}{2})}\,
  \frac{\eta}{2}\,(1-x)^{(\frac{\eta}{2}-1)}
  - \frac{\eta}{4}\,(1+x)
\nonumber\\ 
  &\quad
  + \frac{\eta^{2}}{16}\left(
      -2(1+x)\ln(1-x) - \frac{2\ln x}{1-x} + \frac32(1+x)\ln x
      - \frac{x}{2} - \frac{5}{2}\right)
\nonumber\\
  &\quad
  + \frac{\eta^3}{8}\left[
       -\frac{1+x}{2} \left(
          \frac{9}{32} - \frac{\pi^{2}}{12} + \frac{3}{4}\ln(1-x)
          + \frac{1}{2}\ln^{2}(1-x) - \frac{1}{4}\ln x\ln(1-x)
        \right.\right.
\nonumber\\
  &\quad\left. \qquad\qquad\qquad\quad
          + \frac{1}{16}\ln^{2} x - \frac{1}{4}\mathrm{Li}_{2}(1-x)\right)
\nonumber\\
  &\quad\qquad\quad
       + \frac{1+x^{2}}{2(1-x)}\left(
          - \frac{3}{8}\ln x + \frac{1}{12}\ln^{2} x 
          - \frac{1}{2}\ln x \ln (1-x)\right)
\nonumber\\
  &\quad\qquad\left.\quad
       - \frac{1}{4}(1-x)\left(\ln(1-x)+\frac{1}{4}\right)
       + \frac{1}{32}(5-3x)\ln x\,+\,\mO(\eta^{4})\right.
\end{align}
Here, 
\begin{equation}\label{eq:etareal}
  \eta = \frac{2\alpha}{\pi}\left[\ln
    \left(\frac{Q^{2}}{m_e^2}\,\right)-1\right].
\end{equation}
This includes Eq. (\ref{eq:dnsfin}) as well as the exact solution for photon radiation up to $\mO(\eta^{3})$. 
We see that the definitions of $\eta$ in (\ref{eq:eta}) and (\ref{eq:etareal}) differ by a non-logarithmic factor; this corresponds to taking finite mass effects in the collinear approximation into account (cf. Appendix \ref{app:hardcol}).\\
Checking the requirements for exponentiation, we find that up to $\mO(\eta)$, 
\begin{\eqn}\label{eq:doal}
f_{ISR}(x)\,=\,\frac{\eta}{2}(1-x)^{\frac{\eta}{2}-1}\,\lb 1+\frac{3}{4}\,\eta\rb\,-\,\frac{\eta}{4}\,\lb 1+x\rb.
\end{\eqn} 
For the hard photon regime, we can expand this in $\eta$ and obtain
\begin{\eqn*}
f_{ISR}(x)\,=\,\frac{\eta}{2}\,\lb\frac{1}{1-x}\,-\,\frac{1}{2}\,(1+x)\rb \,=\,\frac{\eta}{4}\,\frac{1+x^{2}}{1-x}\,+\,\mO(\eta^{2}).
\end{\eqn*}
This reproduces $f^{+}$ (\ref{eq:hardcollfs}). The term missing in the structure function expansion corresponds to the helicity flip contribution $f^{-}$ which is subdominant.
Integrating Eq. (\ref{eq:doal}) over the soft region yields
\begin{eqnarray}\label{eq:intsoftisr}
\int^{1}_{x_{0}}\,f_{ISR}\,dx&=&\lb 1-x_{0} \rb^{\frac{\eta}{2}}\,\lb1+\frac{3}{8}\,\eta \rb\,-\frac{\eta}{4}\lb\frac{3}{2}-x_{0}-\frac{1}{2}x_{0}^{2}\rb\nonumber\\
&=&1+\,\frac{\eta}{4}\,\lb 2\,\ln(1-x_{0})+x_{0}+\frac{1}{2}\,x_{0}^{2}\rb+\mO(\eta^{2}).
\end{eqnarray}
We see that this exactly corresponds to the result (\ref{eq:intsoftisr}) obtained from integrating the the $\mO(\eta)$ splitting function $P_{ee}$. We define the $\mO(\al)$ part of Eq. (\ref{eq:intsoftisr}) $f_{soft,ISR}$:\\
\begin{equation}\label{eq:fsoftisr}
  f_\text{soft,ISR}(\Delta E_\gamma,\Delta\theta_\gamma,m_e^2) =
  \frac{\eta}{4}\left(2\ln(1-x_0) + x_0 + \frac12 x_0^2\right).
\end{equation}
\chapter{Transformations and (Di)Gamma functions}\label{app:trafos}
\section{The Fourier Transform}
The Fourier transform of a function $f(x)$ is given by
\begin{\eqn*}
\hat{f}(p)\,=\int^{\infty}_{-\infty}\,e^{-\imath\,p\,x}\,f(x)\,dx.
\end{\eqn*}
Its inverse is given by
\begin{\eqn*}
f(x)\,=\frac{1}{2\pi}\,\int^{\infty}_{-\infty}\,e^{\imath\,p\,x}\,\hat{f}(p)\,dp,
\end{\eqn*}
where the distribution of factors of $\frac{1}{\sqrt{2\,\pi}}$ is purely conventional.
Note that we used 
\begin{\eqn*}
\delta(x)\,=\,\frac{1}{2\pi}\,\int^{\infty}_{-\infty}\,e^{\imath\,p\,x}\,dx.
\end{\eqn*}
\section{The Laplace Transform}
The Laplace transform can be derived from the Fourier transform by taking the Fourier transform of
\begin{\eqn*}
g(x)\,=\,f(x)\,e^{-c\,x},
\end{\eqn*}
where we assume that
\begin{\eqn*}
f(x)\,=\,0\;\;\mbox{for $x\,<\,0$}
\end{\eqn*}
and additionally making the change of variables
\begin{\eqn*}
s\,=\,c+\imath\,p.
\end{\eqn*}
We end up with the Laplace transform
\begin{\eqn*}
\mathcal{L}(s)\,=\,\int^{\infty}_{0}\,e^{-s\,x}\,f(x)\,dx
\end{\eqn*}
and its inverse
\begin{\eqn*}
f(x)\,=\,\frac{1}{2\,\imath\,\pi}\,\int^{c+\imath\,\infty}_{c-\imath\,\infty}\,e^{s\,x}\,\mathcal{L}(s)\,ds.
\end{\eqn*}
Note that here the delta-function is given by
\begin{\eqn*}
\delta(x)\,=\,\frac{1}{2\,\imath\,\pi}\,\int^{c+\imath\infty}_{c-\imath\infty}\,e^{s\,x}\,ds.
\end{\eqn*}
An exhaustive list of Laplace transforms and their inverses can be found in \cite{Prudni:1992v4, Prudni:1992v5, Erdelyi:1954}.
\section{The Mellin Transform}
The Mellin transform can be derived from the Laplace transform by substituting
\begin{\eqn*}
y\,=\,e^{-x}\;;\;g(y)\,=\,f(-\ln(y)).
\end{\eqn*}
We then obtain
\begin{\eqn*}
\mathcal{M}(s)\,=\,\int^{UL}_{0}\,y^{s-1}\,g(y)\,dy
\end{\eqn*}
and
\begin{\eqn*}
g(y)\,=\,\frac{1}{2\,\imath\,\pi}\,\int^{c+\imath\,\infty}_{c-\imath\,\infty}\,y^{-s}\,\mathcal{M}(s)\,ds.
\end{\eqn*}
Note that the upper limit of the Mellin transform integral $UL$ depends on the value of $f(x)$ for $x<0$:
\begin{\eqn*}
UL=\left\{\begin{array}{ll}
1 &\mbox{for $f(x)\,=\,0$ for $x<0$}\\
\infty&\mbox{for $f(x)\,\neq\,0$ for $x<0$}.\end{array}\right. 
\end{\eqn*}
In the first case, we can easily get the Mellin transform from the Laplace transform by performing the change of variables $x\,=\,-\ln(y)$. The delta function is here given by
\begin{\eqn*}
\delta(x-x')\,=\,\frac{1}{2\,\imath\,\pi}\,\int^{c+\imath\,\infty}_{c-\imath\,\infty}\,x^{-s}\,x'^{s-1}\,ds.
\end{\eqn*}
An overview on Mellin transformations can be found in \cite{Szpankow:2001}.\section{The Euler Gamma function $\Gamma(x)$}
The Gamma function $\Gamma(x)$ is defined by
\begin{\eqn*}
\Gamma(x)\,=\,\int^{\infty}_{0}\,t^{x-1}\,e^{-t}\,dt
\end{\eqn*}
for $x>0$.
More useful representations, expansions, and relations to other functions can be found in \cite{Spanier:1987, Erdelyi:1953, Gradstein:1981}.
\section{The Digamma function $\psi_{0}(x)$}\label{app:digamma}
The digamma function is formally defined by
\begin{\eqn*}
\psi_{0}(x)\,=\,\frac{d}{dx}\,\ln\lb\Gamma(x)\rb.
\end{\eqn*}
A lot of useful representations, expansions, and relations can be equally found in the literature given in the last section. We here just quote a few properties useful in the ISR derivation:
\begin{eqnarray}
\psi_{0}(x)&=&-\gamma\,+\,\int^{1}_{0}\,\frac{1-t^{x-1}}{1-t}\,dt\;\;\;(x>1),\label{eq:intdig}\\
\psi_{0}(x+n)&=&\psi_{0}(x)+\sum^{n-1}_{k=0}\,\frac{1}{x+k},\nonumber\\
\psi_{0}(1)&=&-\gamma\nonumber,
\end{eqnarray}
where $\gamma$ is the Euler constant. Eq.(\ref{eq:intdig}) leads to
\begin{\eqn*}
\int^{1}_{0}\frac{1+y^{2}}{1-y}\,y^{\xi-1}\,dy\,=\,-\psi_{0}(\xi)-\psi_{0}(\xi+2)+2\,C,
\end{\eqn*} 
where
\begin{\eqn*}
C\,=\,-\gamma+\int^{1}_{0}\frac{1}{1-t}\,dt.
\end{\eqn*}
Another useful relation is given by
\begin{\eqn*}
\psi_{0}(\xi+1)\,=\,\ln(\xi)+\frac{1}{2\,\xi}-\sum^{\infty}_{n=1}\frac{B_{2n}}{2\,n\,\xi^{2\,n}},
\end{\eqn*}
where $B_{2n}$ are the Bernoulli numbers.
\chapter{References for Computer codes}\label{app:refs}
We list the main literature references for all computer codes mentioned in this work (in order of appearance)
\begin{itemize}
\item{\whizard~} \cite{Kilian:2001qz}
\item{\feynarts~/ \formcalc~} \cite{FormCalc,Hahn:1998yk,FeynArts,Hahn:2000kx, Hahn:2001rv}
\item{\looptools~} \cite{Hahn:1998yk}
\item{\comphep} \cite{Pukhov:1999gg}
\item{\madgraph~} \cite{Stelzer:1994ta}
\item{\oMega~} \cite{Moretti:2001zz}
\item{\circe~} \cite{Ohl:1996fi}
\item{\prog{pythia}~} \cite{Sjostrand:2006za}
%\item{\prog{sMadgraph}~} \cite{Cho:2006sx}
\item{\prog{Sherpa}~} \cite{Gleisberg:2003xi}
\item{\prog{Isajet}~} \cite{Paige:2003mg}
\item{\prog{Form}~} \cite{Vermaseren:1992vn,Vermaseren:2000nd}
\end{itemize}
\end{appendix}
\newpage
\bibliography{lit}

\begin{mcbibliography}{100}

\bibitem{Glashow:1961tr}
S.~L. Glashow, Nucl. Phys. {\bf 22},  579  (1961)\relax
\relax
\bibitem{Weinberg:1967tq}
S. Weinberg, Phys. Rev. Lett. {\bf 19},  1264  (1967)\relax
\relax
\bibitem{Salam:1968rm}
A. Salam, WEAK AND ELECTROMAGNETIC INTERACTIONS, originally printed in
  *Svartholm: Elementary Particle Theory, Proceedings Of The Nobel Symposium
  Held 1968 At Lerum, Sweden*, Stockholm 1968, 367-377\relax
\relax
\bibitem{Glashow:1970gm}
S.~L. Glashow, J. Iliopoulos, and L. Maiani, Phys. Rev. {\bf D2},  1285
  (1970)\relax
\relax
\bibitem{Fritzsch:1973pi}
H. Fritzsch, M. Gell-Mann, and H. Leutwyler, Phys. Lett. {\bf B47},  365
  (1973)\relax
\relax
\bibitem{Higgs:1964ia}
P.~W. Higgs, Phys. Lett. {\bf 12},  132  (1964)\relax
\relax
\bibitem{Englert:1964et}
F. Englert and R. Brout, Phys. Rev. Lett. {\bf 13},  321  (1964)\relax
\relax
\bibitem{Higgs:1964pj}
P.~W. Higgs, Phys. Rev. Lett. {\bf 13},  508  (1964)\relax
\relax
\bibitem{Higgs:1966ev}
P.~W. Higgs, Phys. Rev. {\bf 145},  1156  (1966)\relax
\relax
\bibitem{Kibble:1967sv}
T.~W.~B. Kibble, Phys. Rev. {\bf 155},  1554  (1967)\relax
\relax
\bibitem{Yao:2006px}
W.~M. Yao {\it et~al.}, J. Phys. {\bf G33},  1  (2006)\relax
\relax
\bibitem{Lee:1977eg}
B.~W. Lee, C. Quigg, and H.~B. Thacker, Phys. Rev. {\bf D16},  1519
  (1977)\relax
\relax
\bibitem{Lee:1977yc}
B.~W. Lee, C. Quigg, and H.~B. Thacker, Phys. Rev. Lett. {\bf 38},  883
  (1977)\relax
\relax
\bibitem{Cabibbo:1979ay}
N. Cabibbo, L. Maiani, G. Parisi, and R. Petronzio, Nucl. Phys. {\bf B158},
  295  (1979)\relax
\relax
\bibitem{Hambye:1996wb}
T. Hambye and K. Riesselmann, Phys. Rev. {\bf D55},  7255  (1997)\relax
\relax
\bibitem{Isidori:2001bm}
G. Isidori, G. Ridolfi, and A. Strumia, Nucl. Phys. {\bf B609},  387
  (2001)\relax
\relax
\bibitem{Witten:1981nf}
E. Witten, Nucl. Phys. {\bf B188},  513  (1981)\relax
\relax
\bibitem{Georgi:1974sy}
H. Georgi and S.~L. Glashow, Phys. Rev. Lett. {\bf 32},  438  (1974)\relax
\relax
\bibitem{Haag:1974qh}
R. Haag, J.~T. Lopuszanski, and M. Sohnius, Nucl. Phys. {\bf B88},  257
  (1975)\relax
\relax
\bibitem{Wess:1992cp}
J. Wess and J. Bagger, {\em Supersymmetry and supergravity} (Univ. Pr.,
  Princeton, USA, 1992)\relax
\relax
\bibitem{Chamseddine:1982jx}
A.~H. Chamseddine, R. Arnowitt, and P. Nath, Phys. Rev. Lett. {\bf 49},  970
  (1982)\relax
\relax
\bibitem{Nilles:1983ge}
H.~P. Nilles, Phys. Rept. {\bf 110},  1  (1984)\relax
\relax
\bibitem{Hall:1983iz}
L.~J. Hall, J.~D. Lykken, and S. Weinberg, Phys. Rev. {\bf D27},  2359
  (1983)\relax
\relax
\bibitem{Dine:1993yw}
M. Dine and A.~E. Nelson, Phys. Rev. {\bf D48},  1277  (1993)\relax
\relax
\bibitem{Dine:1994vc}
M. Dine, A.~E. Nelson, and Y. Shirman, Phys. Rev. {\bf D51},  1362
  (1995)\relax
\relax
\bibitem{Giudice:1998bp}
G.~F. Giudice and R. Rattazzi, Phys. Rept. {\bf 322},  419  (1999)\relax
\relax
\bibitem{Randall:1998uk}
L. Randall and R. Sundrum, Nucl. Phys. {\bf B557},  79  (1999)\relax
\relax
\bibitem{Giudice:1998xp}
G.~F. Giudice, M.~A. Luty, H. Murayama, and R. Rattazzi, JHEP {\bf 12},  027
  (1998)\relax
\relax
\bibitem{Dimopoulos:1981zb}
S. Dimopoulos and H. Georgi, Nucl. Phys. {\bf B193},  150  (1981)\relax
\relax
\bibitem{Haber:1985rc}
H.~E. Haber and G.~L. Kane, Phys. Rept. {\bf 117},  75  (1985)\relax
\relax
\bibitem{Barbieri:1987xf}
R. Barbieri, Riv. Nuovo Cim. {\bf 11N4},  1  (1988)\relax
\relax
\bibitem{Sohnius:1985qm}
M.~F. Sohnius, Phys. Rept. {\bf 128},  39  (1985)\relax
\relax
\bibitem{Drees:1996ca}
M. Drees, hep-ph/9611409\relax
\relax
\bibitem{Martin:1997ns}
S.~P. Martin, hep-ph/9709356\relax
\relax
\bibitem{Allanach:2004ud}
B.~C. Allanach {\it et~al.}, hep-ph/0403133\relax
\relax
\bibitem{Blair:2002pg}
G.~A. Blair, W. Porod, and P.~M. Zerwas, Eur. Phys. J. {\bf C27},  263
  (2003)\relax
\relax
\bibitem{Aguilar-Saavedra:2005pw}
J.~A. Aguilar-Saavedra {\it et~al.}, Eur. Phys. J. {\bf C46},  43  (2006)\relax
\relax
\bibitem{Schmitt:2004gz}
M. Schmitt, Supersymmetry, Experiment\relax
\relax
\bibitem{Weiglein:2004hn}
G. Weiglein {\it et~al.}, hep-ph/0410364\relax
\relax
\bibitem{Allanach:2006fy}
B.~C. Allanach {\it et~al.}, hep-ph/0602198\relax
\relax
\bibitem{Djouadi:2006be}
A. Djouadi, M. Drees, and J.-L. Kneur, JHEP {\bf 03},  033  (2006)\relax
\relax
\bibitem{Adloff:2001at}
C. Adloff {\it et~al.}, Eur. Phys. J. {\bf C20},  639  (2001)\relax
\relax
\bibitem{Kuze:2002vb}
M. Kuze and Y. Sirois, Prog. Part. Nucl. Phys. {\bf 50},  1  (2003)\relax
\relax
\bibitem{Adloff:2003jm}
C. Adloff {\it et~al.}, Phys. Lett. {\bf B568},  35  (2003)\relax
\relax
\bibitem{Ellis:2006ix}
J.~R. Ellis, S. Heinemeyer, K.~A. Olive, and G. Weiglein, JHEP {\bf 05},  005
  (2006)\relax
\relax
\bibitem{Allanach:2002nj}
B.~C. Allanach {\it et~al.}, Eur. Phys. J. {\bf C25},  113  (2002)\relax
\relax
\bibitem{Nojiri:2004hp}
M.~M. Nojiri, hep-ph/0411127\relax
\relax
\bibitem{Aguilar-Saavedra:2001rg}
J.~A. Aguilar-Saavedra {\it et~al.}, hep-ph/0106315\relax
\relax
\bibitem{Desch:2003vw}
K. Desch {\it et~al.}, JHEP {\bf 02},  035  (2004)\relax
\relax
\bibitem{Lafaye:2004cn}
R. Lafaye, T. Plehn, and D. Zerwas, hep-ph/0404282\relax
\relax
\bibitem{Bechtle:2005ns}
P. Bechtle, K. Desch, and P. Wienemann, hep-ph/0506244\relax
\relax
\bibitem{Kilian:2001qz}
W. Kilian, WHIZARD 1.0: A generic Monte-Carlo integration and event generation
  package for multi-particle processes. Manual, LC-TOOL-2001-039\relax
\relax
\bibitem{Gleisberg:2003xi}
T. Gleisberg {\it et~al.}, JHEP {\bf 02},  056  (2004)\relax
\relax
\bibitem{Paige:2003mg}
F.~E. Paige, S.~D. Protopopescu, H. Baer, and X. Tata,   (2003)\relax
\relax
\bibitem{Sjostrand:2006za}
T. Sjostrand, S. Mrenna, and P. Skands, JHEP {\bf 05},  026  (2006)\relax
\relax
\bibitem{Hagiwara:2005wg}
K. Hagiwara {\it et~al.}, Phys. Rev. {\bf D73},  055005  (2006)\relax
\relax
\bibitem{Kleiss:1989de}
R. Kleiss {\it et~al.},  in {\em Proceedings, Z physics at LEP 1, vol. 3}
  (Geneva, 1989)\relax
\relax
\bibitem{Was:1992mm}
Z. Was and S. Jadach, Physics Monte Carlo generators for LEP: A systematic
  approach, prepared for 2nd International Workshop on Software Engineering,
  Artificial Intelligence and Expert Systems for High-energy and Nuclear
  Physics, La Londe Les Maures, France, 13-18 Jan 1992\relax
\relax
\bibitem{lepewwg}
http://epewwg.web.cern.ch\relax
\relax
\bibitem{Denner:2000bj}
A. Denner, S. Dittmaier, M. Roth, and D. Wackeroth, Nucl. Phys. {\bf B587},  67
   (2000)\relax
\relax
\bibitem{Jadach:2001mp}
S. Jadach {\it et~al.}, Comput. Phys. Commun. {\bf 140},  475  (2001)\relax
\relax
\bibitem{Denner:1999dt}
A. Denner, S. Dittmaier, M. Roth, and D. Wackeroth, Eur. Phys. J. direct {\bf
  C2},  4  (2000)\relax
\relax
\bibitem{Denner:2000tw}
A. Denner, S. Dittmaier, M. Roth, and D. Wackeroth, hep-ph/0005074\relax
\relax
\bibitem{Fritzsche:2004nf}
T. Fritzsche and W. Hollik, Nucl. Phys. Proc. Suppl. {\bf 135},  102
  (2004)\relax
\relax
\bibitem{Oller:2004br}
W. Oller, H. Eberl, and W. Majerotto, Phys. Lett. {\bf B590},  273
  (2004)\relax
\relax
\bibitem{Oller:2005xg}
W. Oller, H. Eberl, and W. Majerotto, hep-ph/0504109\relax
\relax
\bibitem{Nicrosini:1986sm}
O. Nicrosini and L. Trentadue, Phys. Lett. {\bf B196},  551  (1987)\relax
\relax
\bibitem{Bonvicini:1988vv}
G. Bonvicini and L. Trentadue, Nucl. Phys. {\bf B323},  253  (1989)\relax
\relax
\bibitem{Nagy:2005aa}
Z. Nagy and D.~E. Soper, JHEP {\bf 10},  024  (2005)\relax
\relax
\bibitem{Kramer:2005hw}
M. Kramer, S. Mrenna, and D.~E. Soper, Phys. Rev. {\bf D73},  014022
  (2006)\relax
\relax
\bibitem{Catani:2001cc}
S. Catani, F. Krauss, R. Kuhn, and B.~R. Webber, JHEP {\bf 11},  063
  (2001)\relax
\relax
\bibitem{Frixione:2002ik}
S. Frixione and B.~R. Webber, JHEP {\bf 06},  029  (2002)\relax
\relax
\bibitem{Frixione:2006he}
S. Frixione and B.~R. Webber, hep-ph/0601192\relax
\relax
\bibitem{Kilian:2006cj}
W. Kilian, J. Reuter, and T. Robens, hep-ph/0607127, to appear in
Eur. Phys. J. C\relax
\relax
\bibitem{Skrzypek:1990qs}
M. Skrzypek and S. Jadach, Z. Phys. {\bf C49},  577  (1991)\relax
\relax
\bibitem{Choi:2000ta}
S.~Y. Choi {\it et~al.}, Eur. Phys. J. {\bf C14},  535  (2000)\relax
\relax
\bibitem{Hagiwara:1985yu}
K. Hagiwara and D. Zeppenfeld, Nucl. Phys. {\bf B274},  1  (1986)\relax
\relax
\bibitem{Choi:1998ei}
S.~Y. Choi, A. Djouadi, H.~S. Song, and P.~M. Zerwas, Eur. Phys. J. {\bf C8},
  669  (1999)\relax
\relax
\bibitem{Tilman:progr}
T. Plehn, private computer program for SUSY at linear colliders\relax
\relax
\bibitem{FormCalc}
T. Hahn, The FormCalc homepage, http://www.feynarts.de/formcalc/\relax
\relax
\bibitem{Hahn:1998yk}
T. Hahn and M. Perez-Victoria, Comput. Phys. Commun. {\bf 118},  153
  (1999)\relax
\relax
\bibitem{Ladinsky:1992bd}
G.~A. Ladinsky, Phys. Rev. {\bf D46},  2922  (1992)\relax
\relax
\bibitem{Bloch:1937pw}
F. Bloch and A. Nordsieck, Phys. Rev. {\bf 52},  54  (1937)\relax
\relax
\bibitem{thomdiss}
T. Fritzsche, Ph.D. Thesis, 2005\relax
\relax
\bibitem{FeynArts}
T. Hahn, The FeynArts homepage, http://www.feynarts.de/\relax
\relax
\bibitem{Hahn:2001rv}
T. Hahn and C. Schappacher, Comput. Phys. Commun. {\bf 143},  54  (2002)\relax
\relax
\bibitem{'tHooft:1978xw}
G. 't~Hooft and M.~J.~G. Veltman, Nucl. Phys. {\bf B153},  365  (1979)\relax
\relax
\bibitem{Bohm:1986fg}
M. Bohm, A. Denner, and W. Hollik, Nucl. Phys. {\bf B304},  687  (1988)\relax
\relax
\bibitem{Berends:1987jm}
F.~A. Berends, R. Kleiss, and W. Hollik, Nucl. Phys. {\bf B304},  712
  (1988)\relax
\relax
\bibitem{Denner:1991kt}
A. Denner, Fortschr. Phys. {\bf 41},  307  (1993)\relax
\relax
\bibitem{Dittmaier:1993jj}
S. Dittmaier, Ph.D. thesis, RX-1526 (WURZBURG)\relax
\relax
\bibitem{Bohm:1993qx}
M. Bohm and S. Dittmaier, Nucl. Phys. {\bf B409},  3  (1993)\relax
\relax
\bibitem{Dittmaier:1993da}
S. Dittmaier and M. Bohm, Nucl. Phys. {\bf B412},  39  (1994)\relax
\relax
\bibitem{Gribov:1972rt}
V.~N. Gribov and L.~N. Lipatov, Sov. J. Nucl. Phys. {\bf 15},  675
  (1972)\relax
\relax
\bibitem{Chen:1990qz}
F.-z. Chen, P. Wang, C.~M. Wu, and Y.-s. Zhu, Analytical approximation of
  radiatively corrected resonant cross-section, BIHEP-EP-90-01\relax
\relax
\bibitem{Skrzypek:1992vk}
M. Skrzypek, Acta Phys. Polon. {\bf B23},  135  (1992)\relax
\relax
\bibitem{Denner:2005es}
A. Denner, S. Dittmaier, M. Roth, and L.~H. Wieders, hep-ph/0502063\relax
\relax
\bibitem{Fadin:1993kt}
V.~S. Fadin, V.~A. Khoze, and A.~D. Martin, Phys. Lett. {\bf B320},  141
  (1994)\relax
\relax
\bibitem{Fadin:1993dz}
V.~S. Fadin, V.~A. Khoze, and A.~D. Martin, Phys. Rev. {\bf D49},  2247
  (1994)\relax
\relax
\bibitem{Aeppli:1993rs}
A. Aeppli, G.~J. van Oldenborgh, and D. Wyler, Nucl. Phys. {\bf B428},  126
  (1994)\relax
\relax
\bibitem{Roth:1999kk}
M. Roth, Ph.D. thesis, 1999, hep-ph/0008033\relax
\relax
\bibitem{Melnikov:1995fx}
K. Melnikov and O.~I. Yakovlev, Nucl. Phys. {\bf B471},  90  (1996)\relax
\relax
\bibitem{Beenakker:1997bp}
W. Beenakker, A.~P. Chapovsky, and F.~A. Berends, Phys. Lett. {\bf B411},  203
  (1997)\relax
\relax
\bibitem{Beenakker:1997ir}
W. Beenakker, A.~P. Chapovsky, and F.~A. Berends, Nucl. Phys. {\bf B508},  17
  (1997)\relax
\relax
\bibitem{Denner:1997ia}
A. Denner, S. Dittmaier, and M. Roth, Nucl. Phys. {\bf B519},  39  (1998)\relax
\relax
\bibitem{Jadach:1996hi}
S. Jadach {\it et~al.}, Phys. Lett. {\bf B417},  326  (1998)\relax
\relax
\bibitem{Jadach:1998tz}
S. Jadach {\it et~al.}, Phys. Rev. {\bf D61},  113010  (2000)\relax
\relax
\bibitem{Beenakker:1998gr}
W. Beenakker, F.~A. Berends, and A.~P. Chapovsky, Nucl. Phys. {\bf B548},  3
  (1999)\relax
\relax
\bibitem{Kurihara:1999ii}
Y. Kurihara, M. Kuroda, and D. Schildknecht, Nucl. Phys. {\bf B565},  49
  (2000)\relax
\relax
\bibitem{sommi}
A. Sommerfeld, {\em Atombau und Spektrallinien, Bd 2} (Vieweg, Braunschweig,
  1939)\relax
\relax
\bibitem{Sakharov:1948yq}
A.~D. Sakharov, Zh. Eksp. Teor. Fiz. {\bf 18},  631  (1948)\relax
\relax
\bibitem{Bardin:1993mc}
D.~Y. Bardin, W. Beenakker, and A. Denner, Phys. Lett. {\bf B317},  213
  (1993)\relax
\relax
\bibitem{Fadin:1993kg}
V.~S. Fadin, V.~A. Khoze, and A.~D. Martin, Phys. Lett. {\bf B311},  311
  (1993)\relax
\relax
\bibitem{Fadin:1994pm}
V.~S. Fadin, V.~A. Khoze, A.~D. Martin, and A. Chapovsky, Phys. Rev. {\bf D52},
   1377  (1995)\relax
\relax
\bibitem{Fadin:1995fp}
V.~S. Fadin, V.~A. Khoze, A.~D. Martin, and W.~J. Stirling, Phys. Lett. {\bf
  B363},  112  (1995)\relax
\relax
\bibitem{Freitas:2001zh}
A. Freitas, D.~J. Miller, and P.~M. Zerwas, Eur. Phys. J. {\bf C21},  361
  (2001)\relax
\relax
\bibitem{James:1980yn}
F. James, Rept. Prog. Phys. {\bf 43},  1145  (1980)\relax
\relax
\bibitem{Dobbs:2004qw}
M.~A. Dobbs {\it et~al.}, hep-ph/0403045\relax
\relax
\bibitem{Pukhov:1999gg}
A. Pukhov {\it et~al.}, hep-ph/9908288\relax
\relax
\bibitem{Stelzer:1994ta}
T. Stelzer and W.~F. Long, Comput. Phys. Commun. {\bf 81},  357  (1994)\relax
\relax
\bibitem{Moretti:2001zz}
M. Moretti, T. Ohl, and J. Reuter, hep-ph/0102195\relax
\relax
\bibitem{Ohl:1996fi}
T. Ohl, Comput. Phys. Commun. {\bf 101},  269  (1997)\relax
\relax
\bibitem{Reuter:2005us}
J. Reuter {\it et~al.}, ECONF {\bf C0508141},  ALCPG0323, hep-ph/0512012\relax
\relax
\bibitem{Hahn:2000kx}
T. Hahn, Comput. Phys. Commun. {\bf 140},  418  (2001)\relax
\relax
\bibitem{Vermaseren:1992vn}
J.~A.~M. Vermaseren, The Symbolic manipulation program FORM, KEK-TH-326\relax
\relax
\bibitem{Vermaseren:2000nd}
J.~A.~M. Vermaseren, math-ph/0010025\relax
\relax
\bibitem{Dittmaier:1998nn}
S. Dittmaier, Phys. Rev. {\bf D59},  016007  (1999)\relax
\relax
\bibitem{howto}
T. Robens,   (2006), internal DESY notes (in preparation)\relax
\relax
\bibitem{Skands:2003cj}
P. Skands {\it et~al.}, JHEP {\bf 07},  036  (2004)\relax
\relax
\bibitem{Alexander:1987be}
J.~P. Alexander, G. Bonvicini, P.~S. Drell, and R. Frey, Phys. Rev. {\bf D37},
  56  (1988)\relax
\relax
\bibitem{Catani:1996jh}
S. Catani and M.~H. Seymour, Phys. Lett. {\bf B378},  287  (1996)\relax
\relax
\bibitem{Catani:1996vz}
S. Catani and M.~H. Seymour, Nucl. Phys. {\bf B485},  291  (1997)\relax
\relax
\bibitem{Dittmaier:1999mb}
S. Dittmaier, Nucl. Phys. {\bf B565},  69  (2000)\relax
\relax
\bibitem{Denner:1999gp}
A. Denner, S. Dittmaier, M. Roth, and D. Wackeroth, Nucl. Phys. {\bf B560},  33
   (1999)\relax
\relax
\bibitem{Nagy:2003qn}
Z. Nagy and D.~E. Soper, JHEP {\bf 09},  055  (2003)\relax
\relax
\bibitem{Yennie:1961ad}
D.~R. Yennie, S.~C. Frautschi, and H. Suura, Ann. Phys. {\bf 13},  379
  (1961)\relax
\relax
\bibitem{Gribov:1972ri}
V.~N. Gribov and L.~N. Lipatov, Sov. J. Nucl. Phys. {\bf 15},  438
  (1972)\relax
\relax
\bibitem{Kuraev:1985hb}
E.~A. Kuraev and V.~S. Fadin, Sov. J. Nucl. Phys. {\bf 41},  466  (1985)\relax
\relax
\bibitem{Bailin:1994qt}
D. Bailin and A. Love, {\em Supersymmetric gauge field theory and string
  theory} (IOP, Bristol, UK, 1994)\relax
\relax
\bibitem{Bartl:1989ms}
A. Bartl, H. Fraas, W. Majerotto, and N. Oshimo, Phys. Rev. {\bf D40},  1594
  (1989)\relax
\relax
\bibitem{Choi:2001ww}
S.~Y. Choi, J. Kalinowski, G. Moortgat-Pick, and P.~M. Zerwas, Eur. Phys. J.
  {\bf C22},  563  (2001)\relax
\relax
\bibitem{Gunion:1986yn}
J.~F. Gunion and H.~E. Haber, Nucl. Phys. {\bf B272},  1  (1986)\relax
\relax
\bibitem{Schwabl:1997gf}
F. Schwabl, {\em Advanced quantum mechanics (QM II)} (Springer, Berlin,
  Germany, 1997)\relax
\relax
\bibitem{Bjorken:1966dk}
J.~D. Bjorken and S.~D. Drell, {\em RELATIVISTIC QUANTUM MECHANICS. (GERMAN
  TRANSLATION)} (Bibliograph.Inst. (B.I.- Hochschultaschenbuecher), Mannheim,
  1966)\relax
\relax
\bibitem{Itzykson:1980rh}
C. Itzykson and J.~B. Zuber, {\em QUANTUM FIELD THEORY} (McGraw-Hill, New York,
  USA, 1980)\relax
\relax
\bibitem{Kleiss:1985yh}
R. Kleiss and W.~J. Stirling, Nucl. Phys. {\bf B262},  235  (1985)\relax
\relax
\bibitem{Haber:1994pe}
H.~E. Haber, hep-ph/9405376\relax
\relax
\bibitem{Kleiss:1986ct}
R. Kleiss, Z. Phys. {\bf C33},  433  (1987)\relax
\relax
\bibitem{Jadach:1990vz}
S. Jadach, M. Skrzypek, and B.~F.~L. Ward, Phys. Lett. {\bf B257},  173
  (1991)\relax
\relax
\bibitem{Jadach:1990fy}
S. Jadach, M. Skrzypek, and B.~F.~L. Ward, EXPONENTIATION, HIGHER ORDERS AND
  LEADING LOGS, presented at 25th Rencontre de Moriond: Electroweak
  Interactions and Unified Theories, Les Arcs, France, Mar 4- 11, 1990\relax
\relax
\bibitem{Prudni:1992v4}
A. Prudnikov, Y. Brychkov, and O. Marichev, {\em Integrals and Series: Volume
  4} (Gordon and Breach Science Publishers, USA, 1992)\relax
\relax
\bibitem{Prudni:1992v5}
A. Prudnikov, Y. Brychkov, and O. Marichev, {\em Integrals and Series: Volume
  5} (Gordon and Breach Science Publishers, USA, 1992)\relax
\relax
\bibitem{Erdelyi:1954}
A. Erdelyi, W. Magnus, F. Oberhettinger, and F. Tricomi, {\em Tables of
  Integral Transforms} (McGraw-Hill Book Company, USA, 1954)\relax
\relax
\bibitem{Szpankow:2001}
W. Szpankowski, {\em Average Case Analysis of Algorithms on Sequences} (Wiley
  and Sons, USA, 2001)\relax
\relax
\bibitem{Spanier:1987}
J. Spanier and K. Oldham, {\em An atlas of functions} (Hemisphere Publishing
  Corporation, USA, 1987)\relax
\relax
\bibitem{Erdelyi:1953}
A. Erdelyi, W. Magnus, F. Oberhettinger, and F. Tricomi, {\em Higher
  transcendental Functions} (McGraw-Hill Book Company, USA, 1953)\relax
\relax
\bibitem{Gradstein:1981}
I. Gradstein and I. Ryshik, {\em Tafeln/ Tables (Volume 2)} (Verlag Harri
  Deutsch, Frankfurt, 1981)\relax
\relax
\end{mcbibliography}
\chapter*{Acknowledgements}
First of all, I want to thank my advisor Wolfgang Kilian for supervision and encouragement throughout my whole time at DESY. I furthermore thank J\"urgen Reuter, Peter Zerwas, Seong Youl Choi, Jan Kalinowski, Markos Maniatis, Markus Diehl, Alexander Westphal and Frank Steffen for help and discussions. I am grateful for the support I got from the theory group at MPI Munich and the provision of their code, especially Thomas Fritzsche, Wolfgang Hollik, Thomas Hahn, Markus Roth, and Stefan Dittmaier. Thanks for further help and advice goes to Ayres Freitas, Heidi Rzehak, and Michael Spira. Finally, I want to thank Birgit Eberle, Oleg Lebedev, Adam Falkowski, Yann Mambrini, Iman Benmachiche, Yvonne Wong, Koichi Hamaguchi, Paolo Merlatti, Riccardo Catena, Tobias Kleinschmidt, and the rest of the DESY theory group for an open, friendly, and encouraging period at DESY Hamburg. I thank J\"urgen Reuter, Frank Deppisch, and Tobias Kleinschmidt for carefully reading the manuscript.\\
I am greatly indebted to my parents for moral and financial support througout all of my studies. 
\end{document}